\documentclass[10pt,letterpaper]{article}
\usepackage[top=0.85in,left=1in,footskip=0.75in,marginparwidth=1in]{geometry}
\usepackage[utf8]{inputenc}
\usepackage{natbib}
\usepackage{cite}
\usepackage{nameref,hyperref}
\usepackage[right]{lineno}
\usepackage{microtype}
\DisableLigatures[f]{encoding = *, family = * }
\usepackage{changepage}
\usepackage[aboveskip=1pt,labelfont=bf,labelsep=period,singlelinecheck=off]{caption}
\makeatletter
\renewcommand{\@biblabel}[1]{\quad#1.}
\makeatother
\usepackage{lastpage,fancyhdr,graphicx}
\usepackage{epstopdf, epsfig}
\fancyheadoffset[L]{2.25in}
\fancyfootoffset[L]{2.25in}
\usepackage{color}
\definecolor{Gray}{gray}{.25}
\usepackage{sidecap}
\usepackage{wrapfig}
\usepackage[pscoord]{eso-pic}
\usepackage[fulladjust]{marginnote}
\reversemarginpar

\usepackage{mathabx}
\usepackage{morefloats}
\usepackage{upgreek}
\usepackage{ragged2e}

\input{definitions.sty}
\def\minf {\mathrm{M}_{\infty}}
\def\rec {\mathrm{Re}_c}

\begin{document}
\vspace*{0.35in}

\begin{flushleft}

{\Large
\textbf\newline{Bursting and reformation cycle of the laminar separation bubble over a NACA-0012 aerofoil: The dynamics of the f\/low-f\/ield}
}
\newline\\
Eltayeb M. ElJack\textsuperscript{1,*}, and
Julio Soria\textsuperscript{2,3}\\
\bigskip
\bf{1} Mechanical Engineering Department, University of Khartoum, Khartoum, Sudan\\
\bf{2} Laboratory for Turbulence Research in Aerospace and Combustion, Department of Mechanical and Aerospace Engineering, Monash University, Melbourne, Australia\\
\bf{3} Aeronautical Engineering Department, King Abdulaziz University, Jeddah, Saudi Arabia\\
\bigskip
* emeljack@uofk.edu

\end{flushleft}

\section*{Abstract}
Detailed f\/low dissection using Dynamic Mode Decomposition (DMD) and Proper Orthogonal Decomposition (POD) is carried out to investigate the dynamics of the f\/low-f\/ield around a NACA-0012 aerofoil at a Reynolds number of $5\times10^4$, Mach number of $0.4$, and at various angles of attack around the onset of stall. Three distinct dominant f\/low modes are identif\/ied by the DMD and the POD: 1) a globally oscillating f\/low mode at a low-frequency (LFO mode 1); 2) a locally oscillating f\/low mode on the suction surface of the aerofoil at a low-frequency (LFO mode 2); and 3) a locally oscillating f\/low mode along the wake of the aerofoil at a high-frequency (HFO mode). The LFO mode 1 features the globally oscillating-f\/low around the aerofoil, the oscillating-pressure along the aerofoil chord, and the process that creates and sustains the triad of vortices (two counter-rotating vortices (TCV) and a secondary vortex beneath them)~\citet{eljack2018bursting2}. The LFO mode 2 features the expansion and advection of the upstream vortex (UV) of the TCV. The life-cycle of the triad of vortices is perfectly synchronised with the LFO mode 1 and 2. Time histories of the lift and the drag coeff\/icients mimic the temporal evolution of the LFO mode 1 and 2, respectively. The POD temporal modes show that the LFO mode 1 leads the LFO mode 2 by a phase of $\pi/2$; in total agreement with the previously reported observations that the lift coeff\/icient leads the drag coeff\/icient by a phase of $\pi/2$. The reattachment location of the shear layer oscillates along the aerofoil chord in harmony with the lift coeff\/icient and the LFO mode 1 with a phase difference of $\pi$. The HFO mode originates at the aerofoil leading-edge and features the oscillating mode along the aerofoil wake. The wall-normal velocity component drives the HFO mode and plays a profound role in the dynamics of the f\/low. The HFO mode exists at all of the investigated angles of attack and causes a global oscillation in the f\/low-f\/ield. The global f\/low oscillation around the aerofoil interacts with the laminar portion of the separated shear layer in the vicinity of the leading-edge and triggers an inviscid absolute instability that creates and sustains the TCV. When the UV of the TCV expands, it advects downstream and energies the HFO mode. The LFO mode 1, the LFO mode 2, and the HFO mode mutually strengthen each other until the LFO mode 1 and 2 overtake the HFO mode at the onset of stall. At the angles of attack $9.7^{\circ} \leq \alpha \leq 10.0^{\circ}$, the streamwise velocity, the wall-normal velocity, and the pressure are dominated by the LFO mode 1; and the low-frequency f\/low oscillation (LFO) phenomenon is fully developed. At higher angles of attack, the HFO mode overtakes the LFO mode 2, and the aerofoil undergoes a full stall.

\section{Introduction}\label{sec:intorudction}
The stability of the Laminar Separation Bubble (LSB) and its associated Low-frequency F\/low Oscillation (LFO) phenomenon at the onset of an aerofoil stall have been investigated extensively numerically and experimentally. Understanding the dynamics of the f\/low-f\/ield is the major goal of most of the research work. The dynamics of any turbulent or transitional f\/low is dominated by an organized motion known as coherent structures. These are recognizable f\/low structures that survive dissipation by viscosity for a relatively long time and have a typical life-cycle. The origin of such coherent structures is an unstable f\/low mode that feeds on the mean f\/low. The temporal and spatial evolution of such organized f\/low motion is of major importance because the instantaneous solution of the f\/low is dependent on it. If the dynamics of the f\/low is well understood and described mathematically, researchers would develop better models that predict the phenomenon; thus, reduce the huge cost of modelling the phenomenon numerically using LES or experimentally in very expensive test rigs. Furthermore, such coherent motion contains most of the kinetic energy and acts as a primary mechanism that dissipates energy. If the dynamics is better understood, more eff\/icient and effective aerodynamic shapes can be engineered. Most importantly, such organized f\/low motion induces oscillations in the aerodynamic forces, vibrations, noise, and drag. Thus, suppressing or energizing such coherent structures greatly improves the aerodynamic performance of aerofoils. Hence, researchers and engineers would develop smart control means that remove the undesired effects of the phenomenon, utilize the phenomenon when it presents a control opportunity, and improve aerodynamic performance of aerofoils.\newline
The aforementioned organized motion is embedded into a stochastic spatiotemporal data and its extraction is not as easy and straightforward as it might seem. The low-order statistics characterize the f\/low-f\/ield in the mean sense. However, valuable information is lost in the averaging process. For instance, the spatial evolution of coherent structures in the f\/low-f\/ield cannot be captured using the low-order statistical moments. Furthermore, the dynamics of the f\/low and the evolution of the various f\/low modes in time cannot be described using low-order statistics. High order statistical methods like Dynamic Mode Decomposition (DMD) and Proper Orthogonal Decomposition (POD) are used to extract deterministic organized f\/low motion from a stochastic spatiotemporal data. The power of the DMD method is that it provides growth rates in addition to the shape of the dynamic modes of the f\/low at various frequencies. Whereas, the power of the POD method lies in the fact that the decomposition of the f\/low-f\/ield in the POD eigenfunctions converges optimally fast in the energy sense ($L^2$--norm). Thus, the POD recovers the most dominant f\/low modes based on their energy content. Combining both methods in an analysis would combine their strengths and provide most of the information needed to describe the dynamics of the f\/low.\newline
Most of the previous work on the aerofoil at near stall conditions were experimental. The measurements were mostly acquired at a single point or a line, and two-dimensional simultaneous measurements are rare. Since the DMD and the POD methods are applied primarily to plane data or three-dimensional data, neither the DMD nor the POD was used in investigating the LSB and its associated LFO phenomenon in the f\/low-f\/ield around an aerofoil at near stall conditions. However, recently~\citet{almutairi2015large} and~\citet{almutairi2017dynamics} applied the DMD method to the pressure f\/ield of the f\/low-f\/ield around a NACA-0012 aerofoil at $\rec = 1.3\times10^5$, $\minf = 0.4$, and at the angle of attack of $11.5^{\circ}$. The authors sampled the instantaneous pressure f\/ield on the $x$--$y$ plane at a non-dimensional frequency of $658$ and the data spans about one low-frequency cycle. The DMD identif\/ied two dominant modes; a low-frequency mode at Strouhal number of $0.008$ featuring the bursting and reformation cycle of the LSB, and a high-frequency mode featuring the trailing-edge shedding frequency. They concluded that the trailing-edge shedding induces acoustic waves that travel upstream and excite the separated shear layer via some receptivity mechanism, and thus forces it to undergo early transition and reattach the separated f\/low. However, the second ``peak'' in the DMD spectrum the authors referred to as a high-frequency mode is not of signif\/icant magnitude. If the data-set spanned more than one low-frequency cycles, the high-frequency peak would have been f\/iltered out in the averaging process, or at least would not have been recognized as a ``peak'' in the DMD spectrum. Nevertheless, there exists such a high-frequency dominant mode in the f\/low-f\/ield featuring the oscillating f\/low mode along the aerofoil wake. However, the existence of a such f\/low mode is a necessary but not a suff\/icient condition for the acoustic waves feedback mechanism. Therefore, their conclusion that there is a high-frequency dominating mode in the pressure f\/ield and this mode is responsible for the feedback mechanism via acoustic waves needs further investigation. As will be seen later, there are no signif\/icant high-frequency modes in the pressure f\/ield in all of the investigated angles of attack. However, there is a pronounced peak in the DMD spectrum of the wall-normal velocity component in all of the investigated angles of attack.\newline
\citet{eljack2018bursting1} used the data sets generated by~\citet{eljack2017high} to characterise the f\/low-f\/ield around the NACA-0012 aerofoil. The authors used a conditional time-averaging to describe the f\/low-f\/ield in detail and investigate the effects of the angle of attack on the characteristics of the f\/low-f\/ield. They reported the existence of three distinct angle-of-attack regimes. At relatively low angles of attack, f\/low and aerodynamic characteristics are not much affected by the LFO. At moderate angles of attack, the f\/low-f\/ield undergoes a transition regime in which the LFO develops until it reaches a quasi-periodic switching between separated and attached f\/low. At high angles of attack, the f\/low-f\/ield and the aerodynamic characteristics are overwhelmed by a quasi-periodic and self-sustained LFO until the aerofoil approaches the angle of full stall. The authors concluded that most of the observations reported in the literature about the LSB and its associated LFO are neither thresholds nor indicators for the inception of the instability, but rather are consequences of it.~\citet{eljack2018bursting2} utilized the conditional time-averaging used by~\citet{eljack2018bursting1} and a conditional phase-averaging to reveal the underlying mechanism that generates and sustains the LFO phenomenon. The authors shown that a triad of three vortices, two co-rotating vortices (TCV) and a secondary vortex counter-rotating with them, is behind the quasi-periodic self-sustained bursting and reformation of the LSB and its associated LFO phenomenon. They reported that a global oscillation in the f\/low-f\/ield around the aerofoil is observed in all of the investigated angles of attack, including at zero angle of attack. The authors found that when the direction of the oscillating-f\/low is clockwise, it adds momentum to the boundary layer and helps it to remain attached against the APG and vice versa. However, the dynamics of the f\/low and how various f\/low modes interact with each other are lost in the averaging processes. The objective of the present paper is to carry out a detailed dissection of the f\/low-f\/ield and shed some light on the dynamics of the f\/low. The DMD and the POD methods were applied to data sets sampled on the $x$--$y$ plane including the velocity components, the pressure, and the aerodynamic coeff\/icients. The data sets span four low-frequency cycles and were locally-time-averaged every $50$ time-steps and ensemble-averaged in the spanwise direction on the f\/ly before they were recorded. This has enhanced the statistics and provided a database far better than the pressure f\/ield sampled and used by~\citet{almutairi2015large} and~\citet{almutairi2017dynamics}.

\subsection{Dynamic Mode Decomposition (DMD)}\label{sec:DMD}
Since it was introduced in the f\/luid dynamics community, the DMD method has been used extensively to analyze transitional and turbulent f\/lows~\citep{Schmid2008dynamic, Rowley2009spectral, Schmid2010dynamic, Schmid2011application, Schmid2011applications, tu2014dynamic}. The power of the DMD method lies in the fact that it provides growth rates, frequencies, and their associated dynamic modes. Such information is hard to recover using any of the other higher statistical methods including the POD~\citep{lumley1967structure, holmes1996turbulence}.\newline
The DMD method does not require any ordering of the data in space or in a form of a matrix. All that matters is a sequence of snapshots in time $\mathbf{V}(:,t)$ regardless how they are ordered in space. Following~\citet{Schmid2010dynamic},~\citet{Schmid2011application}, and~\citet{Schmid2011applications}, consider a set of data consisting of $n$ snapshots sampled experimentally or numerically and ordered in time with a constant time step $\Delta t$:
\begin{equation}\label{dataset}
   \{ \mathbf{V}_1^n={v_1, v_2, v_3,\ldots, v_n}\}
\end{equation}
The snapshots are assumed to be linearly correlated, \textit{i.e.}, $v_j$ is linearly correlated with $v_{j+1}$ or $v_{j+1}=\mathbf{A} v_j$, and this linear mapping can be implemented to the whole data set $\mathbf{V}_1^n$ to obtain a set of the following form:
\begin{equation}
   \{ v, \mathbf{A}v, \mathbf{A}^2 v, \ldots \mathbf{A}^{n-1} v \}
\end{equation}
or
\begin{equation}
    \mathbf{V}_2^{n+1}= \mathbf{A} \mathbf{V}_1^n
\end{equation}
For a suff\/iciently large sequence, one can assume a linear relation between snapshots and construct the $(n+1)^{th}$ snapshot by a linear combination of the preceding $n$ snapshots, thus:
\begin{equation}
    \mathbf{V}_2^{n+1}= \mathbf{A} \mathbf{V}_1^n \approx \mathbf{V}_1^n \mathbf{S}
\end{equation}
$\mathbf{S}$ is a companion matrix that contains the coeff\/icients of the linear mapping. The problem now becomes a least-square problem to f\/ind $\mathbf{S}$ that approximate the linear mapping with a minimum error.
\begin{equation}
   [\mathbf{Q},\mathbf{R}]=qr(\mathbf{V}_1^{n-1})
\end{equation}
\begin{equation}
   \mathbf{S}= \mathbf{R}^{-1} \mathbf{Q}^H \mathbf{V}_2^n
\end{equation}
Once the companion matrix is calculated one can solve the eigenvalue problem and f\/ind the eigenvalues $\mathbf{\Gamma}$ and the eigenvectors $\mathbf{\Phi}$ of the matrix $\mathbf{S}$, \textit{i.e.},
\begin{equation}
   \mathbf{S}\mathbf{\Phi}= \mathbf{\Gamma} \mathbf{\Phi}
\end{equation}
The eigenvalues of $\mathbf{S}$ contain the growth rates and phase velocities of the f\/low modes, while the eigenvectors represent the shape of the dynamic modes.
\begin{equation}
   \lambda_j =  \frac{\log(\mathbf{\Gamma}_{jj})}{\Delta t}
\end{equation}
The real parts of $\lambda$ represent the growth rates, while the imaginary parts represent the corresponding phase velocities. The dynamic modes are then constructed by projecting the original snapshots onto the eigenvectors as follows:
\begin{equation}
   \label{DM}
    DM(j) = \mathbf{V}_1^{n-1}(:,:) \mathbf{\Phi}(:,j)
\end{equation}
Where $DM(j)$ represents the $j^{th}$ dynamic mode. The multiplication in the right-hand side of the equation is matrix multiplication which means that the f\/low-f\/ield $\mathbf{V}_1^{n-1}$ is projected onto the entire length of the $j^{th}$ eigenfunction $\mathbf{\Phi}$. However, the eigenfunctions are not normalized. Consequently, the sum of all of the projected f\/low modes does not reconstruct the original f\/low-f\/ield as it is the case in the POD method.

\subsection{Proper Orthogonal Decomposition (POD)}\label{sec:POD}
The POD method was proposed by Lumley to objectively recover the most energetic structures of a turbulent f\/low-f\/ield~\citep{lumley1967structure}. There are various methods to implement the POD. In the classical POD method, the ensemble-averaged two-point correlation matrix is estimated, and the eigenvalue problem is then solved for the POD eigenvalues and eigenvectors. The classical method is computationally demanding when applied to numerical simulation data.~\citet{sirovich1987turbulence} introduced the ``snapshot'' method to meet this problem. The method is closely related to the space-time symmetry. That is, the two-point correlation in space is equivalent to the two-point correlation in time. Thus, the two-point correlation matrix is formulated in time rather than in space, and the eigenvalue problem is formulated by taking the inner product of the f\/low variables in time. The reader is referred to~\citet{lumley1967structure, lumley1981coherent},~\citet{sirovich1987turbulence}, and~\citep{holmes1996turbulence} for more details on the theory of the POD method and its various implementations to numerical and experimental data.\newline
The snapshot POD method is formulated as follows:
\begin{equation}\label{POD}
  \mathbf{A}\mathbf{\varphi}=\mathbf{\Lambda}\mathbf{\varphi}
\end{equation}
$\mathbf{A}$ is the correlation matrix, $\mathbf{\varphi}^{(n)}$ are the POD eigenfunctions, and $\mathbf{\Lambda}^{(n)}$ are the POD eigenvalues. The oscillating streamwise velocity component, wall-normal velocity component, and the pressure ($\mathbf{u}\mydprime$, $\mathbf{v}\mydprime$, and $\mathbf{p}\mydprime$) are used to formulate the correlation matrix $\mathbf{A}$ as follows:
\begin{equation}\label{POD_A}
  A_{ij}=\frac{1}{N}\left( \mathbf{U}(\overrightarrow{x},t_i), \mathbf{U}(\overrightarrow{x},t_j) \right)
\end{equation}
\begin{eqnarray*}
  \mathbf{U} &=& \left[
  \begin{array}{c}
    \mathbf{u}\mydprime \\
    \mathbf{v}\mydprime \\
    \mathbf{p}\mydprime \\
  \end{array}
\right]
\end{eqnarray*}
Where $(\cdot, \cdot)$ represents the inner product process, and $N$ is the number of snapshots. The eigenvalues and eigenfunctions are used to construct the spatial POD modes, $\mathbf{\tilde{\Psi}}^{(n)}(\overrightarrow{x})$, as follows:
\begin{equation}\label{POD_Psi}
  \mathbf{\tilde{\Psi}}^{(n)}(\overrightarrow{x}) = \frac{1}{N \sqrt{\mathbf{\Lambda}^{(n)}}} \sum_{k=1}^{N} \mathbf{\varphi}^{(n)}(t_k) \mathbf{U}(\overrightarrow{x},t_k)
\end{equation}
The obtained spatial POD modes, $\mathbf{\tilde{\Psi}}^{(n)}(\overrightarrow{x})$, are orthogonal but not normalized. The spatial POD modes are duly normalized to obtain the orthonormal spatial POD modes $\mathbf{\Psi}$. The POD coeff\/icients, $a^{(n)}(t)$, are then determined using:
\begin{equation}
a^{(n)}(t)=\int_{Domain} \mathbf{U}(\overrightarrow{x},t)\mathbf{\Psi}^{(n)}(\overrightarrow{x})d(\overrightarrow{x})
\end{equation}
The POD coeff\/icients are uncorrelated and their mean values are the eigenvalues $\Lambda^{(n)}$
\begin{equation}
\Lambda^{(n)} = \overline{a^{(n)}(t) a^{(m)}(t)}\delta_{nm}
\end{equation}
The oscillating f\/low variables, $\mathbf{U}$, are then reconstructed from:
\begin{equation}
  \mathbf{U}(\overrightarrow{x},t)=\sum_{m=1}^{M} a^{(m)}(t) \mathbf{\Psi}^{(m)}(\overrightarrow{x})
\end{equation}
Where $M$ is the number of POD modes. The total kinetic energy of the oscillating-f\/low is the sum over all the POD modes. Thus, the percentage of the kinetic energy fraction of the oscillating-f\/low in each POD mode is given by
\begin{equation}\label{POD_xi}
  \xi^{(n)} = \frac{\Lambda^{(n)}}{\sum_{n=1}^{N} \Lambda^{(n)}}
\end{equation}
The cumulative kinetic energy of the oscillating-f\/low over $m$ POD modes is given by
\begin{equation}\label{POD_xi_t}
  \xi_t(m) = \sum_{n=1}^{m} \Lambda^{(n)}
\end{equation}
The percentage of the kinetic energy of the oscillating-f\/low in each POD mode at any instant in time is given by
\begin{equation}\label{eq:POD_zeta}
  \zeta^{(n)}(t) = \frac{\Lambda^{(n)} \mathbf{\varphi}^{(n)}(t) }{\sum_{n=1}^{N} \Lambda^{(n)} \mathbf{\varphi}^{(n)}(t)}
\end{equation}

\section{Results and discussion}
LESs were carried out for the f\/low around the NACA-0012 aerofoil at sixteen angles of attack ($\alpha = 9.0^{\circ}$--$10.1^{\circ}$ at increments of $0.1^{\circ}$ as well as $\alpha = 8.5^{\circ}$, $8.8^{\circ}$, $9.25^{\circ}$, and $10.5^{\circ}$). The simulation Reynolds number and Mach number were $\rec=5\times10^4$ and $\minf = 0.4$, respectively. The free-stream f\/low direction was set parallel to the horizontal axis for all simulations ($u=1$, $v=0$, and $w=0$). The entire domain was initialised using the free-stream conditions. The simulations were performed with a time step of $10^{-4}$ non-dimensional time units. The samples for statistics are collected once transient of simulation has decayed after $50$ f\/low-through times which is equivalent to $50$ non-dimensional time units. Aerodynamic coeff\/icients (lift coeff\/icient $(\mathrm{C}_{\!_\mathrm{L}})$, drag coeff\/icient $(\mathrm{C}_{\!_\mathrm{D}})$, skin-friction coeff\/icient $(\mathrm{C}_{\!_\mathrm{f}})$, and moment coeff\/icient $(\mathrm{C}_{\!_\mathrm{m}})$) were sampled for each angle of attack at a frequency of $10,000$ to generate two and a half million samples over a time period of $250$ non-dimensional time units. The locally-time-averaged and spanwise ensemble-averaged pressure, velocity components, and Reynolds stresses were sampled every $50$ time steps on the $x$--$y$ plane. A dataset of $20,000$ $x$--$y$ planes was recorded at a frequency of $204$ for each angle of attack. The reader is referred to~\citet{eljack2017high} and~\citet{eljack2018bursting1} for more details.

\subsection{Application of the DMD}
The locally-time-averaged and spanwise ensemble-averaged data was formulated into a two-dimensional matrix. The rows contain the streamwise velocity component $\mathbf{u}(x, y, t)$; the wall-normal velocity component $\mathbf{v}(x, y, t)$; the pressure $\mathbf{p}(x, y, t)$; and the aerodynamic coeff\/icients ($\mathrm{C}_{\!_\mathrm{L}}(t)$, $\mathrm{C}_{\!_\mathrm{D}}(t)$, $\mathrm{C}_{\!_\mathrm{m}}(t)$, and $\mathrm{C}_{\!_\mathrm{f}}(t)$). The columns contain the temporal variation of these variables in time which span $20,000$ data points, $100$ non-dimensional time units, or four low-frequency cycles. The companion matrix, $\mathbf{S}$, was then determined and the eigenvalue problem was solved for the eigenvalues, $\mathbf{\Gamma}$, and the eigenvectors, $\mathbf{\Phi}$. After that, the growth rates and phase velocities were calculated, and the dynamic modes were constructed. All of the utilized f\/low variables have the same eigenvectors (dynamic modes shapes), and eigenvalues (growth rates and phase velocities). However, the amplitude of each dynamic mode at different frequencies is dependent on the f\/low variable. Thus, despite the fact that all of the analyzed f\/low variables share the same eigenvalues and eigenvectors, each f\/low variable has its own spectrum.

\subsubsection{The DMD spectra}
\begin{figure}
\begin{center}
\begin{minipage}{420pt}
\centering
\includegraphics[width=420pt, trim={0mm 0mm 0mm 0mm}, clip]{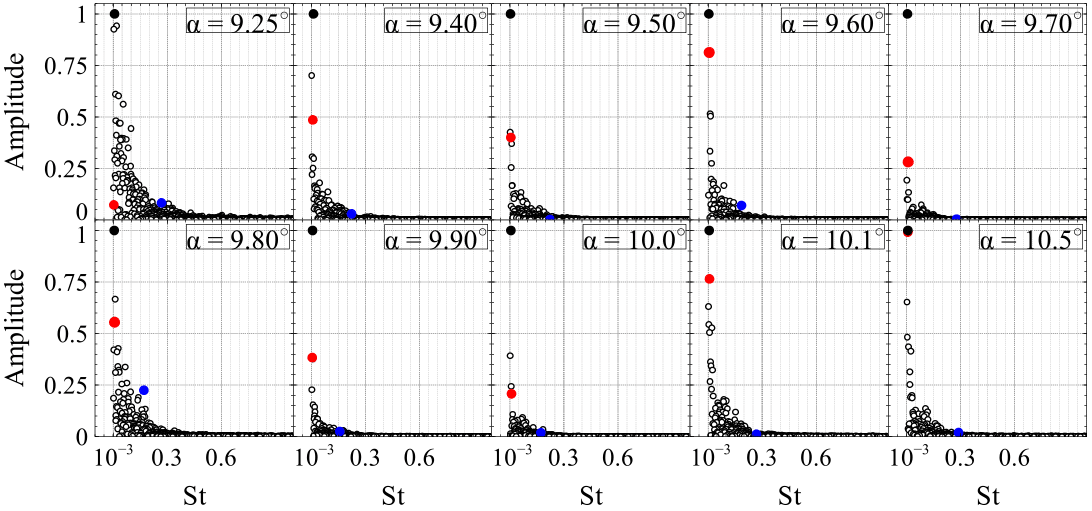}
\end{minipage}
\begin{minipage}{420pt}
\centering
\includegraphics[width=420pt, trim={0mm 0mm 0mm 0mm}, clip]{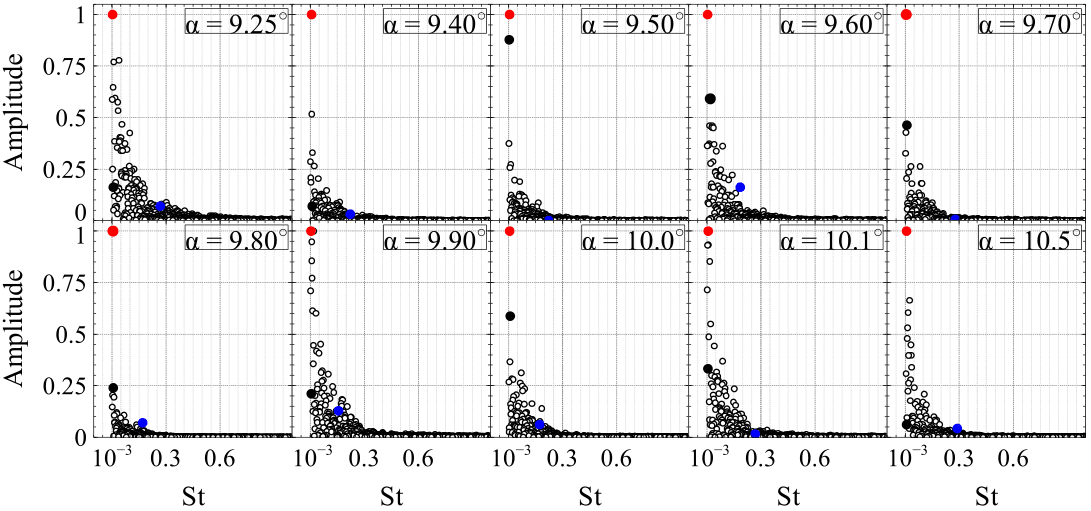}
\end{minipage}
\begin{minipage}{420pt}
\centering
\includegraphics[width=420pt, trim={0mm 0mm 0mm 0mm}, clip]{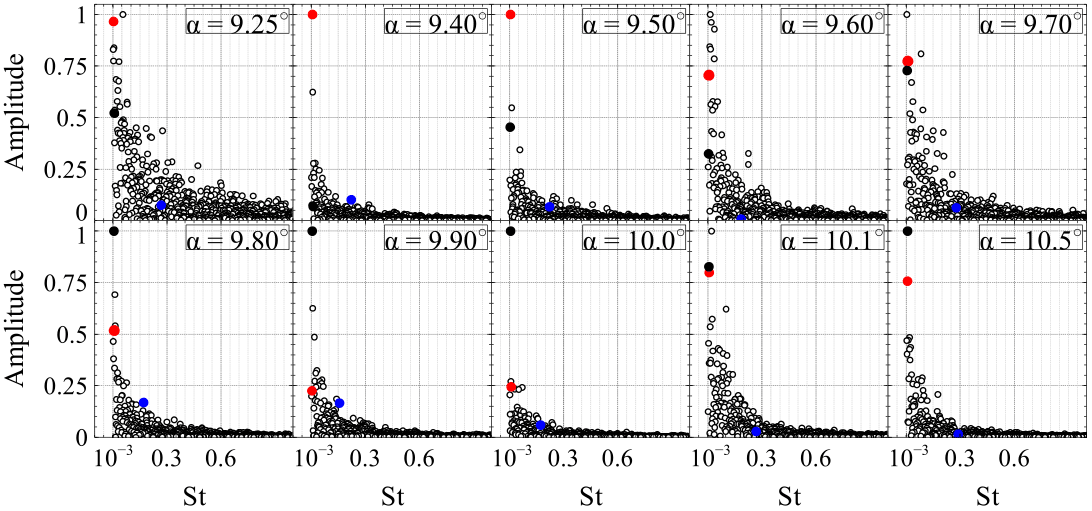}
\end{minipage}
\caption{DMD Spectra of the lift coeff\/icient (top), the drag coeff\/icient (medium), and the skin-friction coeff\/icient (bottom) for the angles of attack $\alpha = 9.25^{\circ}$--$10.5^{\circ}$. The f\/illed black circles denote the LFO mode 1, the f\/illed red circles display the LFO mode 2, and the f\/illed blue circles indicate the HFO mode.}
\label{DMD_spectra_cl_cd_cf}
\end{center}
\end{figure}

\begin{figure}
\begin{center}
\begin{minipage}{420pt}
\centering
\includegraphics[width=420pt, trim={0mm 0mm 0mm 0mm}, clip]{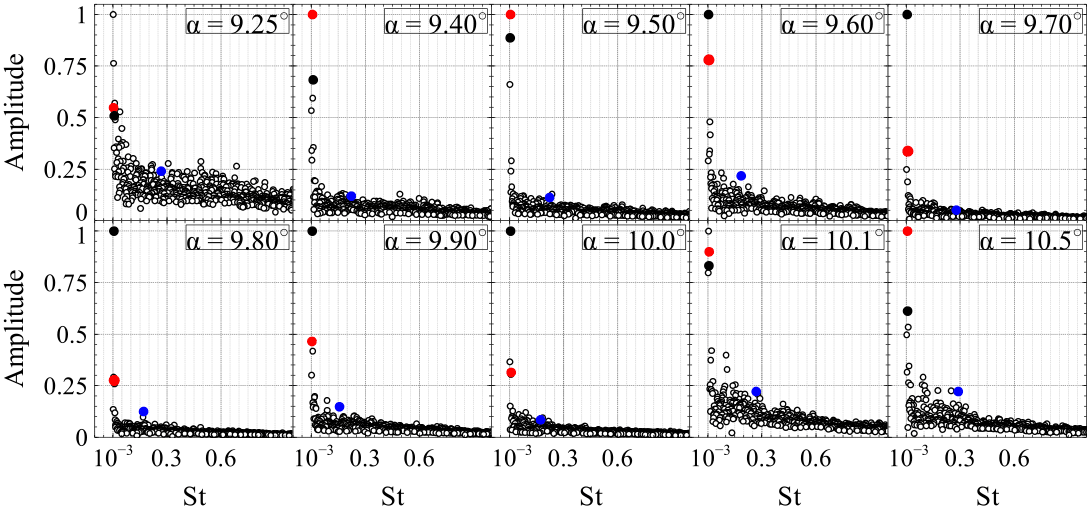}
\end{minipage}
\begin{minipage}{420pt}
\centering
\includegraphics[width=420pt, trim={0mm 0mm 0mm 0mm}, clip]{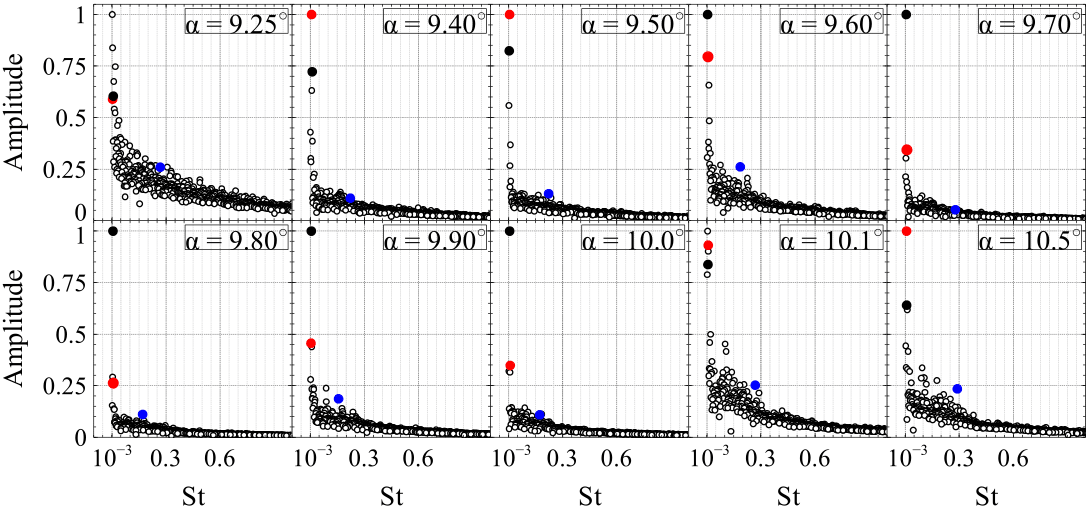}
\end{minipage}
\begin{minipage}{420pt}
\centering
\includegraphics[width=420pt, trim={0mm 0mm 0mm 0mm}, clip]{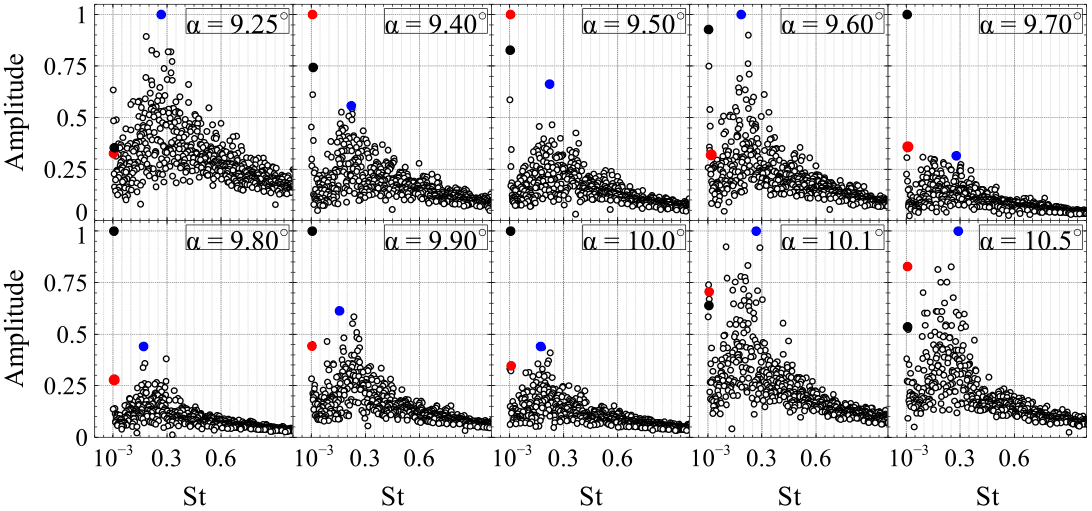}
\end{minipage}
\caption{DMD Spectra of the pressure (top), the streamwise velocity component (medium), and the wall-normal velocity component (bottom) for the angles of attack $\alpha = 9.25^{\circ}$--$10.5^{\circ}$. The f\/illed black circles denote the LFO mode 1, the f\/illed red circles display the LFO mode 2, and the f\/illed blue circles indicate the HFO mode.}
\label{DMD_spectra_p_u_v}
\end{center}
\end{figure}

\begin{figure}
\begin{center}
\begin{minipage}{220pt}
\centering
\includegraphics[width=220pt, trim={0mm 0mm 0mm 0mm}, clip]{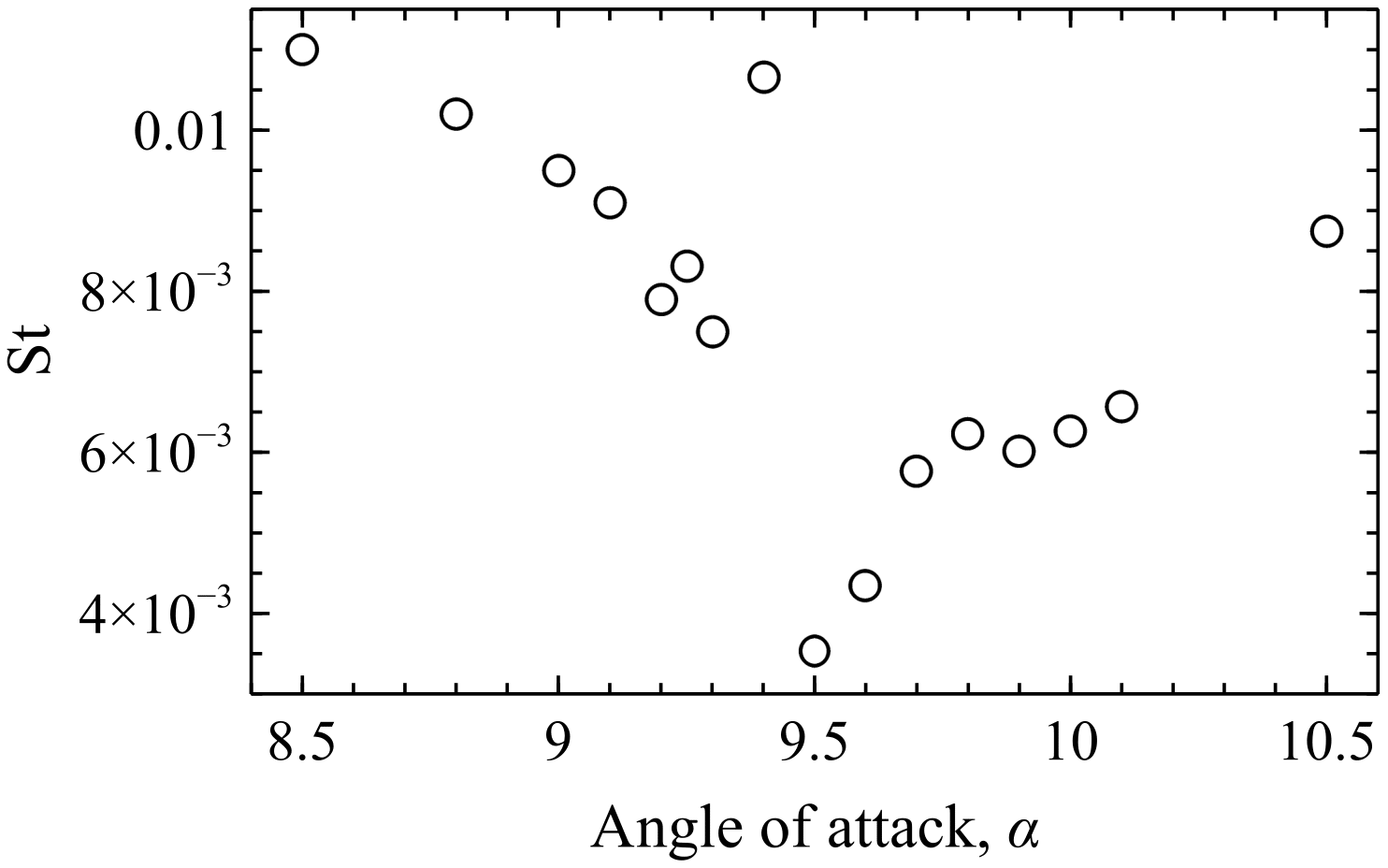}
\textit{(a) The lift coefficient (LFO mode 1)}
\end{minipage}
\medskip
\begin{minipage}{220pt}
\centering
\includegraphics[width=220pt, trim={0mm 0mm 0mm 0mm}, clip]{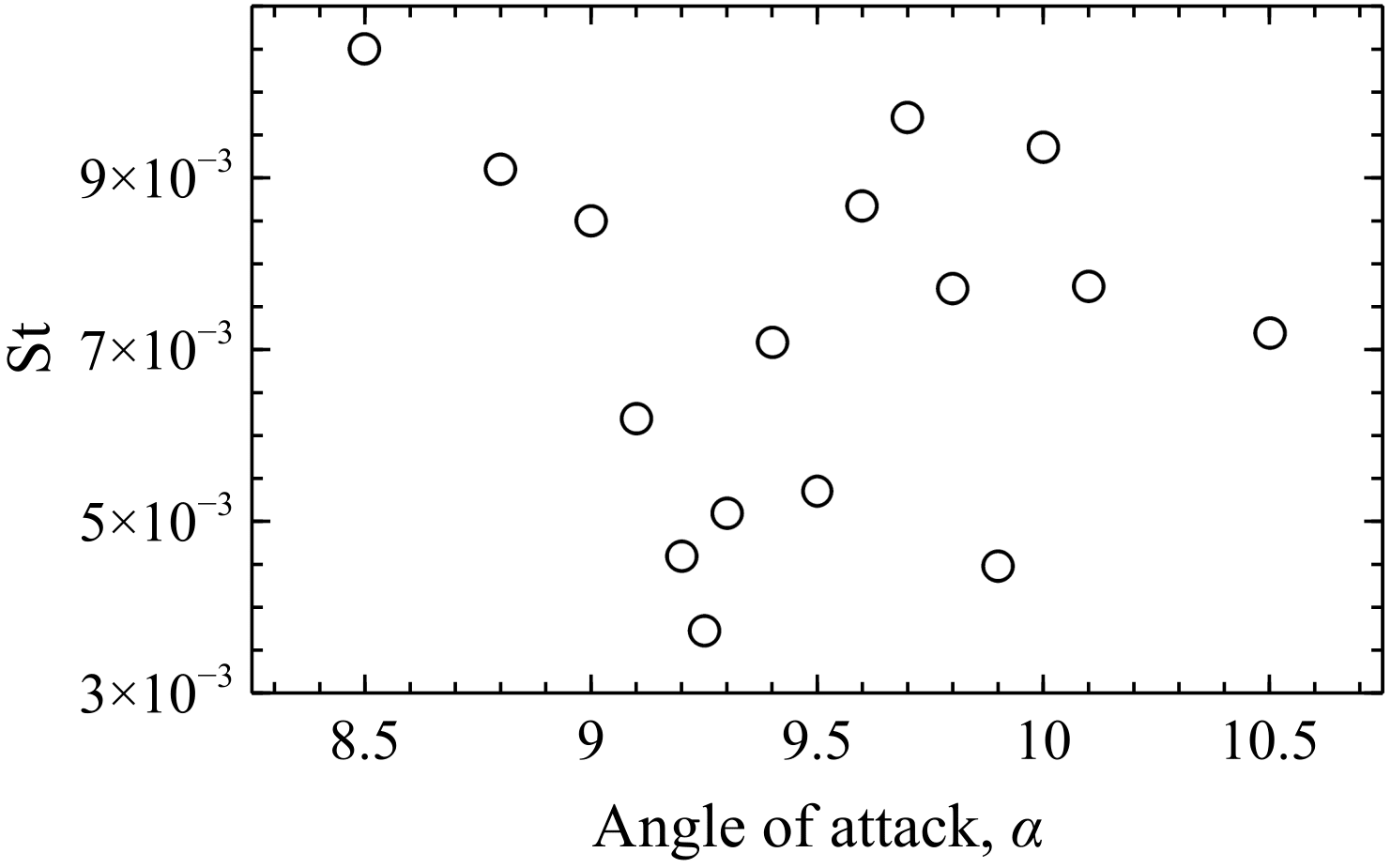}
\textit{(b) The drag coefficient (LFO mode 2)}
\end{minipage}
\medskip
\begin{minipage}{220pt}
\centering
\includegraphics[width=220pt, trim={0mm 0mm 0mm 0mm}, clip]{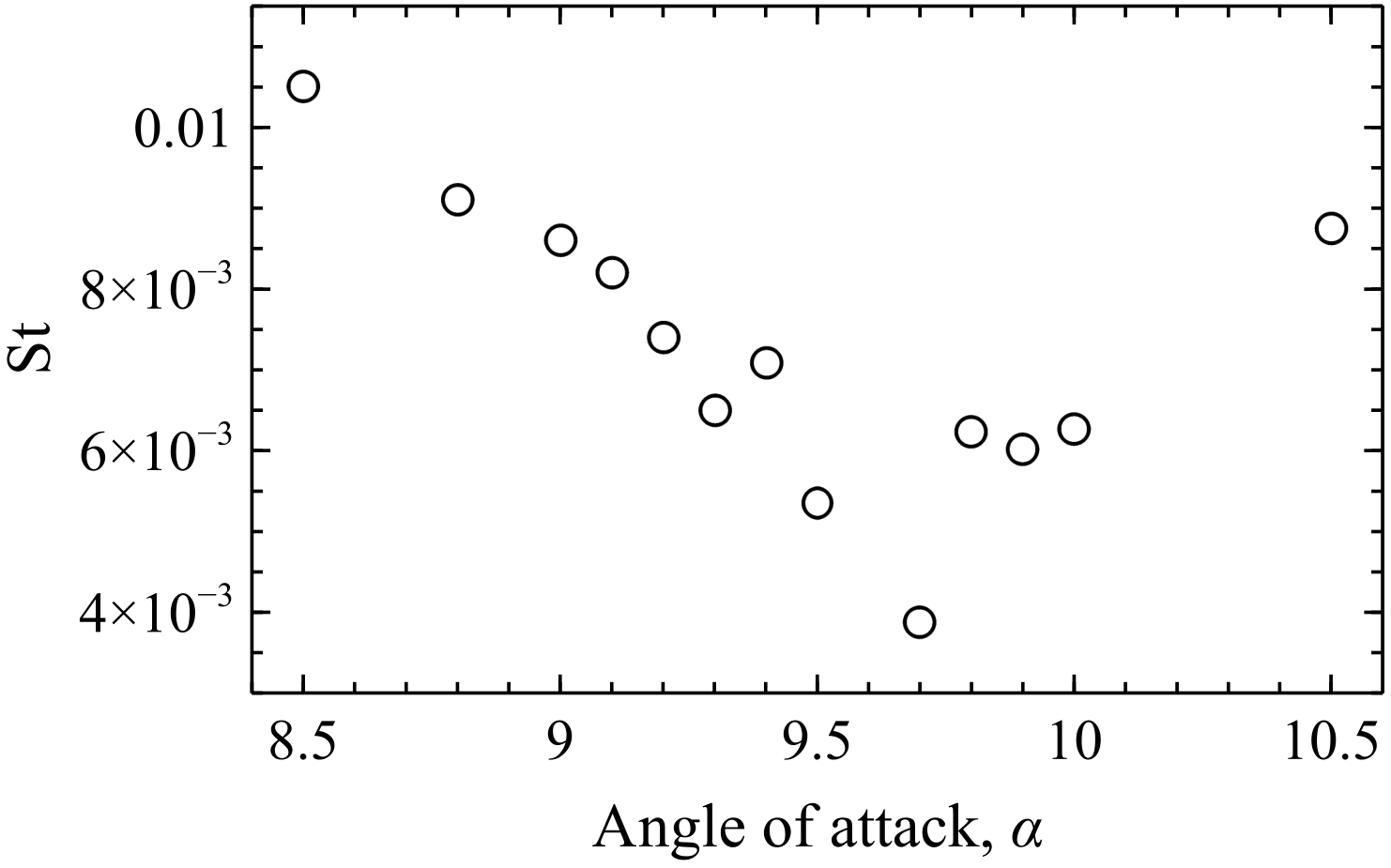}
\textit{(c) The skin-friction coefficient}
\end{minipage}
\medskip
\begin{minipage}{220pt}
\centering
\includegraphics[width=220pt, trim={0mm 0mm 0mm 0mm}, clip]{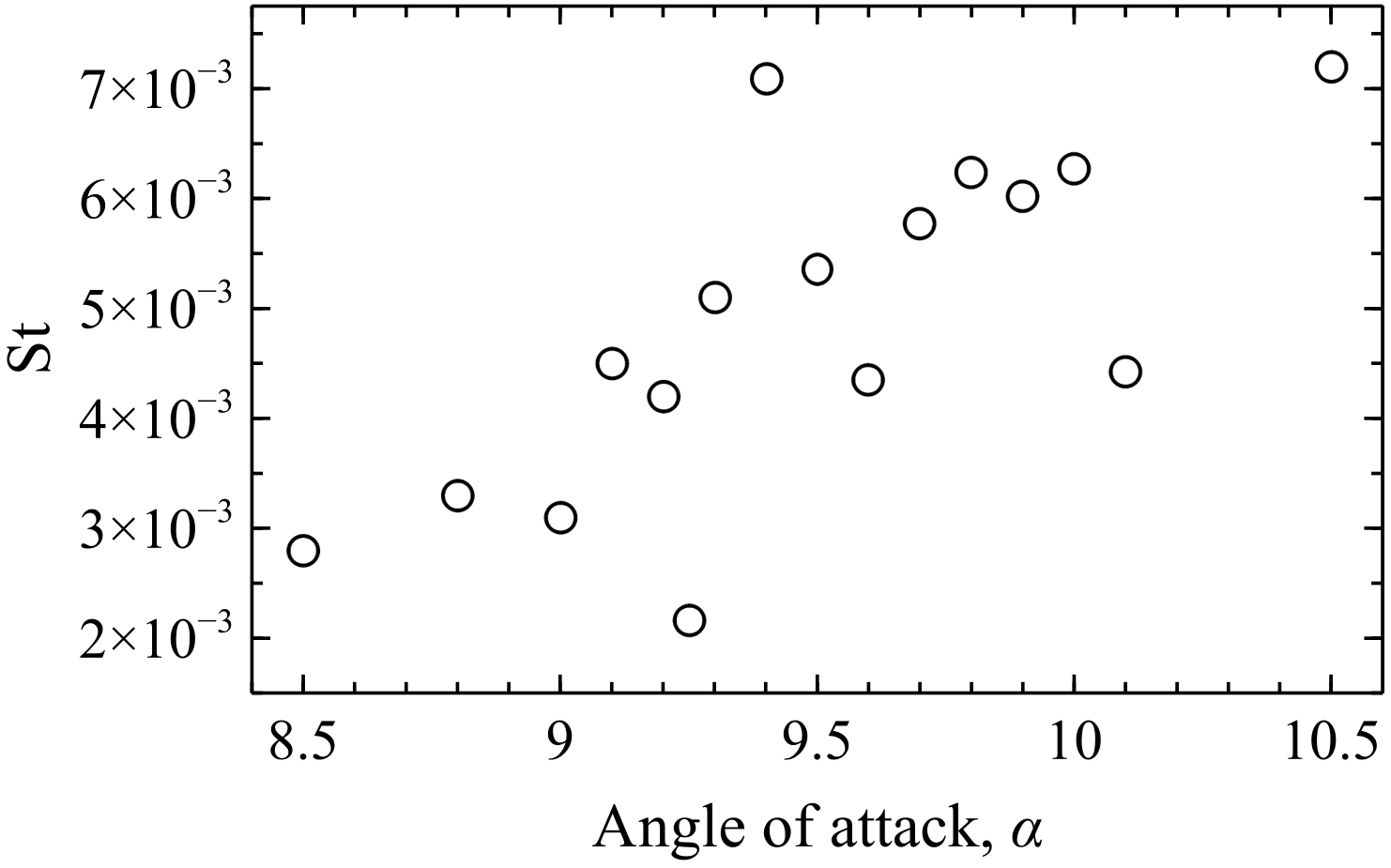}
\textit{(d) The pressure}
\end{minipage}
\medskip
\begin{minipage}{220pt}
\centering
\includegraphics[width=220pt, trim={0mm 0mm 0mm 0mm}, clip]{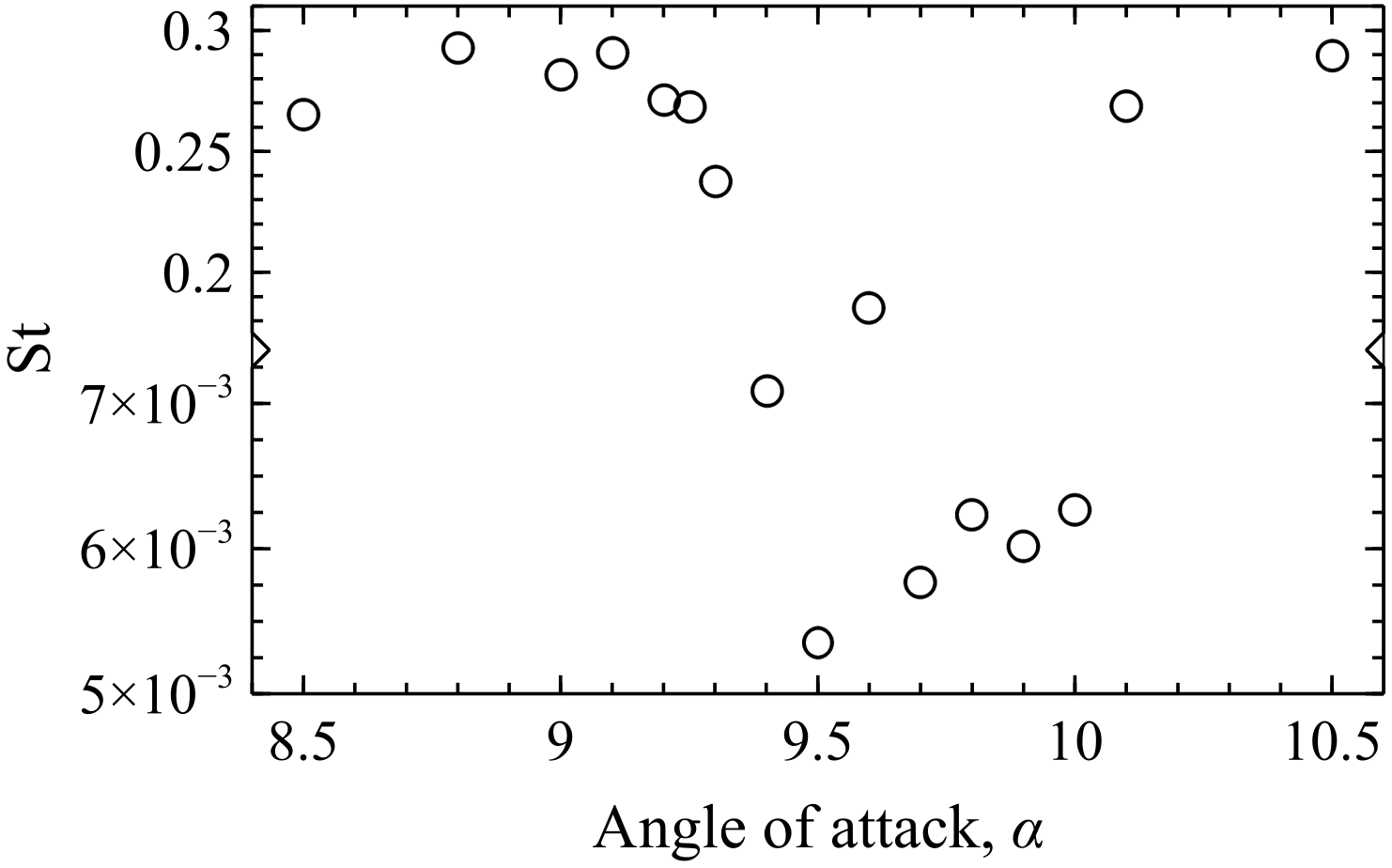}
\textit{(e) The wall-normal velocity component}
\end{minipage}
\begin{minipage}{220pt}
\centering
\includegraphics[width=220pt, trim={0mm 0mm 0mm 0mm}, clip]{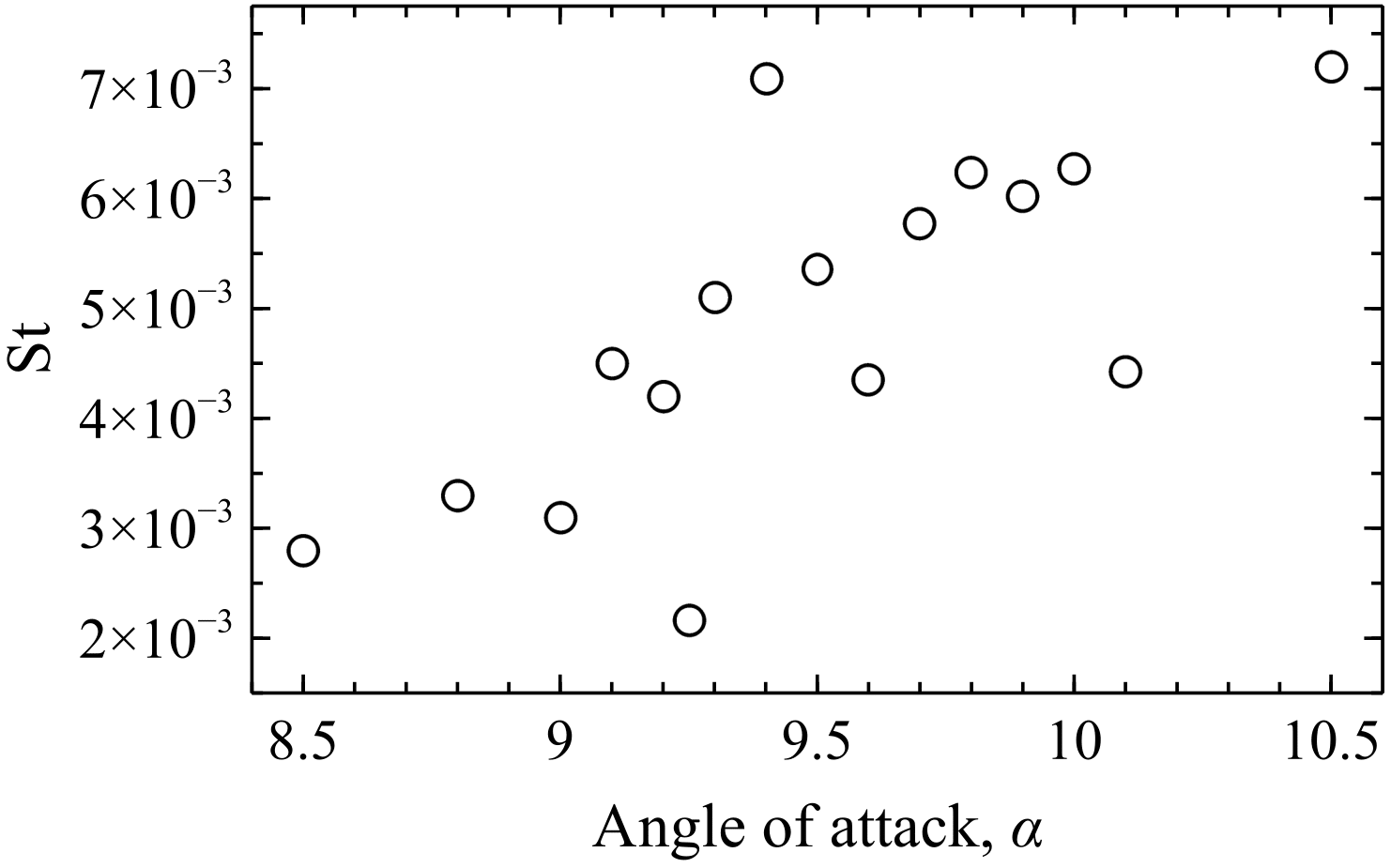}
\textit{(f) The streamwise velocity component}
\end{minipage}
\caption{Strouhal number $(St)$ of the most dominant dynamic mode plotted versus the angle of attack for the lift coeff\/icient, the drag coeff\/icient, the skin-friction coeff\/icient, the pressure, the wall-normal velocity, and the streamwise velocity. The vertical axis is broken into two ranges, $St = 0.005-0.0075$ and $St = 0.175-0.31$, in the plot for the wall-normal velocity.}
\label{DMD_Strouhal_alpha}
\end{center}
\end{figure}

The DMD spectra were estimated for the lift coeff\/icient, the drag coeff\/icient, the skin-friction coeff\/icient, the streamwise velocity, the wall-normal velocity, and the pressure. The DMD analysis decomposes the oscillating-f\/low into too many low-frequency modes. Whereas, the POD method recovers only two low-frequency modes that reconstruct the oscillating-f\/low favourably, as will be discussed later. The DMD spectra of all of the f\/low variables were used to identify the most dominant DMD f\/low mode. There is at least one low-frequency peak in the spectrum of each of the analyzed f\/low variables. It is found that there are two dominant low-frequency modes at all of the investigated angles of attack. It is noted that the f\/irst low-frequency mode (LFO mode 1) and the second low-frequency mode (LFO mode 2) dominate the lift coeff\/icient and the drag coeff\/icient, respectively, at all of the investigated angles of attack. Thus, the most dominant low-frequency mode in the DMD spectrum of the lift coeff\/icient and the drag coeff\/icient are considered to be the LFO mode 1 and the LFO mode 2 for all of the f\/low variables, respectively. The DMD spectrum of the wall-normal velocity is the only spectrum that peaks at a high frequency at all of the investigated angles of attack. Thus, the most dominant high-frequency mode in the DMD spectrum of the wall-normal velocity is considered to be the dominant high-frequency mode (HFO mode). Hence, the DMD spectra of the lift coeff\/icient, the drag coeff\/icient, and the wall-normal velocity are used to identify the LFO mode 1, the LFO mode 2, and the HFO mode, respectively. Once each of the three modes is identif\/ied, its frequency is used to locate its corresponding dynamic mode in the spectrum of the other f\/low variables. Thus, the DMD spectra of all of the f\/low variables show three dominant modes. However, each of the three dominant modes is only dominant in the DMD spectrum of the variable used to identify it.\newline
F\/igure~\ref{DMD_spectra_cl_cd_cf} (top) shows the DMD Spectra for the lift coeff\/icient for the angles of attack of $9.25^{\circ}$--$10.5^{\circ}$. The f\/illed black circles denote the LFO mode 1, the f\/illed red circles display the LFO mode 2, and the f\/illed blue circles indicate the HFO mode. As seen in the f\/igure, the LFO mode 1 dominates the spectra of the lift coeff\/icient in all of the investigated angles of attack. The LFO mode 2 is insignif\/icant at the angle of attack of $9.25^{\circ}$, and its relative amplitude increases as the angle of attack increases. At the angle of attack of $10.5^{\circ}$ both LFO mode 1 and 2 dominate the spectrum of the lift coeff\/icient with almost the same relative amplitude. The HFO mode has an insignif\/icant magnitude in the DMD spectra of the lift coeff\/icient at all of the investigated angles of attack. This is indicative that the HFO mode does not contribute, directly, to the oscillations in the lift coeff\/icient. The same can be said about the spectra of the drag coeff\/icient, f\/igure~\ref{DMD_spectra_cl_cd_cf} (medium); the spectra of the skin-friction coeff\/icient, f\/igure~\ref{DMD_spectra_cl_cd_cf} (bottom); the spectra of the pressure, f\/igure~\ref{DMD_spectra_p_u_v} (top); and the spectra of the streamwise velocity component, f\/igure~\ref{DMD_spectra_p_u_v} (medium). However, the LFO mode 2 dominates the spectra of the drag coeff\/icient, and the LFO mode 1 and 2 exchange the dominance of the spectra of the skin-friction coeff\/icient, the pressure, and the streamwise velocity. There is no signif\/icant high-frequency peak in any of the spectra of these variables.\newline
It is interesting to note that the DMD spectrum of each f\/low variable exhibits a single low-frequency peak. Whereas, the spectra of the lift
and the skin-friction coeff\/icients estimated using the fast Fourier transform algorithm exhibit two low-frequency peaks corresponding to the LFO mode 1 and the LFO mode 2. Furthermore, the two low-frequency peaks were more pronounced in the spectra of the lift and the skin-friction coeff\/icients as seen in f\/igures 33 and 34 in~\citet{eljack2018bursting1}. Two and half million samples spanning $250$ non-dimensional time units were used in estimating the lift and the skin-friction spectra; whereas, only $20,000$ data points spanning $100$ non-dimensional time units were used to estimate the DMD spectra. Thus, it is easy to identify multiple low-frequency peaks when the spectral resolution is high enough to resolve the gap between the two peaks. We argue that the DMD spectra exhibit two low-frequency peaks; however, the data points are not enough to resolve the two peaks. Thus, the two low-frequency peaks merged in one peak as shown in the f\/igures.\newline
The spectra of the wall-normal velocity are dominated by two low-frequency modes and a broad-band of high-frequency modes for each of the investigated angles of attack. At the angle of attack $9.25^{\circ}$ the spectra is dominated by a high-frequency mode as seen in f\/igure~\ref{DMD_spectra_p_u_v} (bottom). As the angle of attack increases above $9.25^{\circ}$, a sudden and a drastic change takes place and the dominant mode shifts to the LFO mode 1 and the LFO mode 2. The two low-frequency modes dominate the spectra of the wall-normal velocity component until the angle of attack is raised above $10.0^{\circ}$; then, the high-frequency mode dominates the spectra, again, as seen in the f\/igure.\newline
F\/igure~\ref{DMD_Strouhal_alpha} shows plots of Strouhal number of the most dominant f\/low modes versus the angle of attack. The most dominant low-frequency modes of the lift coeff\/icient (LFO mode 1) and the skin-friction coeff\/icient are similar to those obtained using the fast Fourier transform algorithm in~\citet{eljack2018bursting1}. The most dominant low-frequency mode of the drag coeff\/icient (LFO mode 2) is similar to that of the lift and the skin-friction coeff\/icients but does not exhibit a uniform pattern. The Strouhal number of the most dominant f\/low mode of the pressure and the streamwise velocity component increases almost linearly with the angle of attack as seen in the f\/igure. The wall-normal velocity component is dominated by a high-frequency f\/low mode at the angles of attack of $\alpha = 9.25^{\circ}$ and $\alpha \geq 10.1^{\circ}$, and dominated by a low-frequency mode at the angles of attack of $9.4^{\circ} \leq \alpha \leq 10.0^{\circ}$. The vertical axis is broken into two ranges, $St=0.005$--$0.0075$ and $St=0.175$--$0.31$, in the plot for the wall-normal velocity component to capture both the low-frequency and the high-frequency ranges that dominate it.

\subsubsection{The low-frequency modes}
\begin{figure}
\begin{center}
\begin{minipage}{220pt}
\centering
\includegraphics[width=220pt, trim={0mm 0mm 0mm 0mm}, clip]{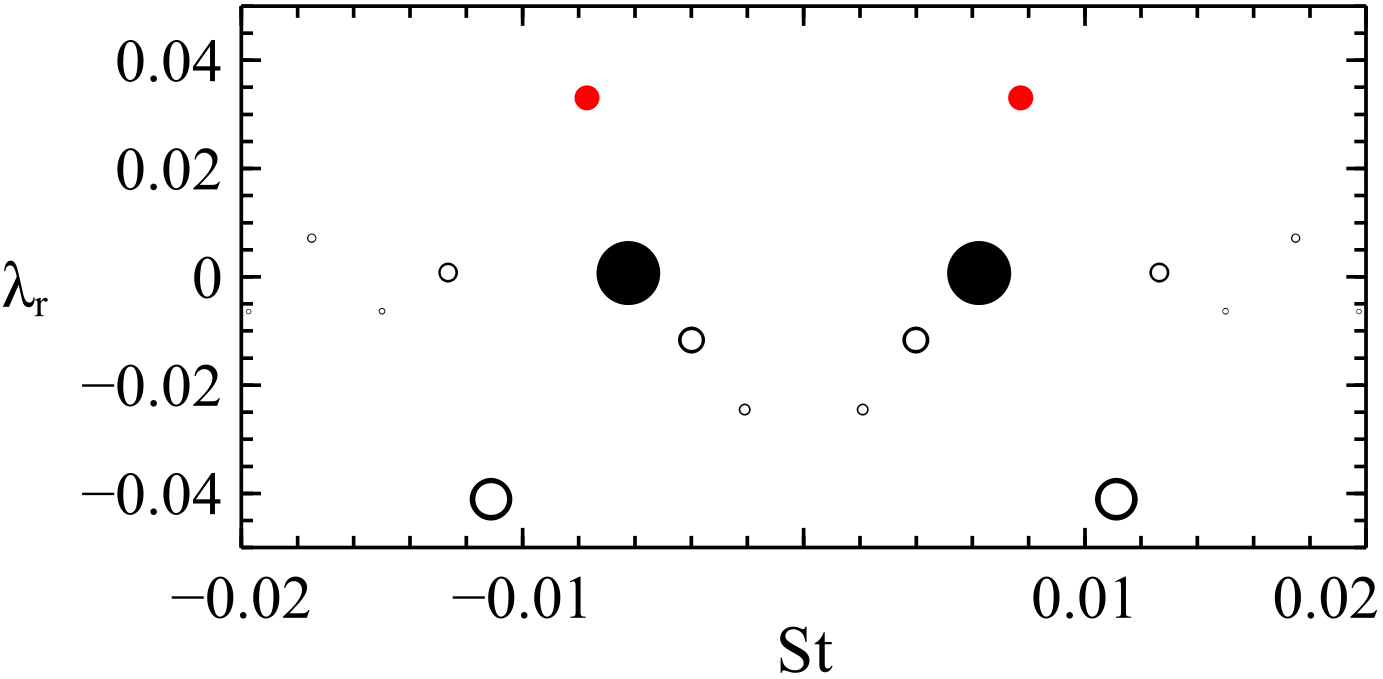}
\textit{$\alpha = 9.8^{\circ}$}
\end{minipage}
\begin{minipage}{220pt}
\centering
\includegraphics[width=215pt, trim={0mm 0mm 0mm 0mm}, clip]{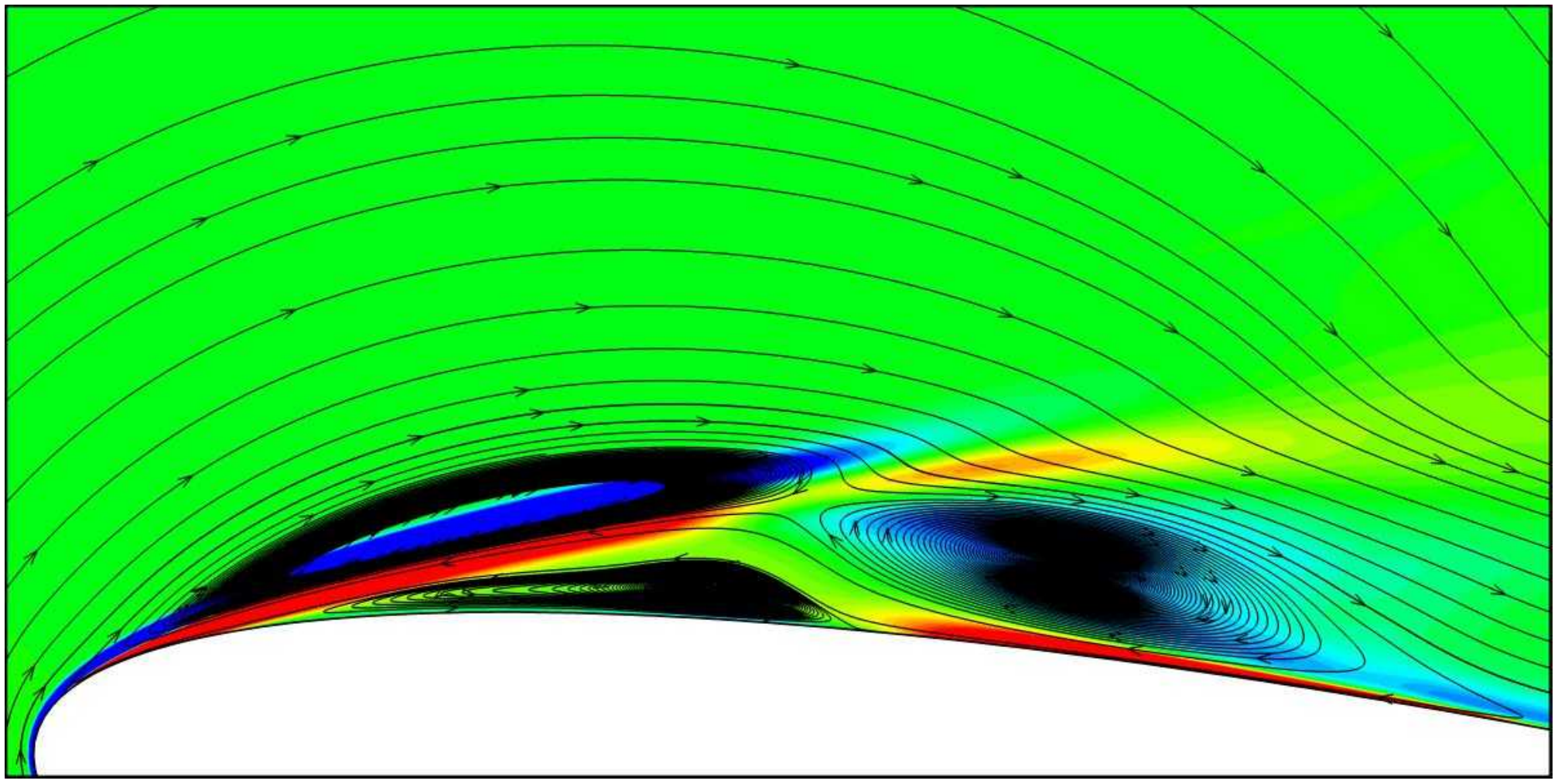}
\textit{$\alpha = 9.8^{\circ}$}
\end{minipage}
\caption{Low-frequency dynamic modes for the angle of attack $\alpha = 9.8^{\circ}$. Left: The spectra of the lift coeff\/icient, where $\lambda_r$ is the growth rate and $St$ is the Strouhal number. Right: Streamlines patterns superimposed on colour maps of the spanwise vorticity $\omega_z$ for the DMD construction of the oscillating-f\/low using the LFO mode 1.}
\label{DMD_streamlines_wz_LFO1}
\end{center}
\begin{center}
\begin{minipage}{220pt}
\centering
\includegraphics[width=220pt, trim={0mm 0mm 0mm 0mm}, clip]{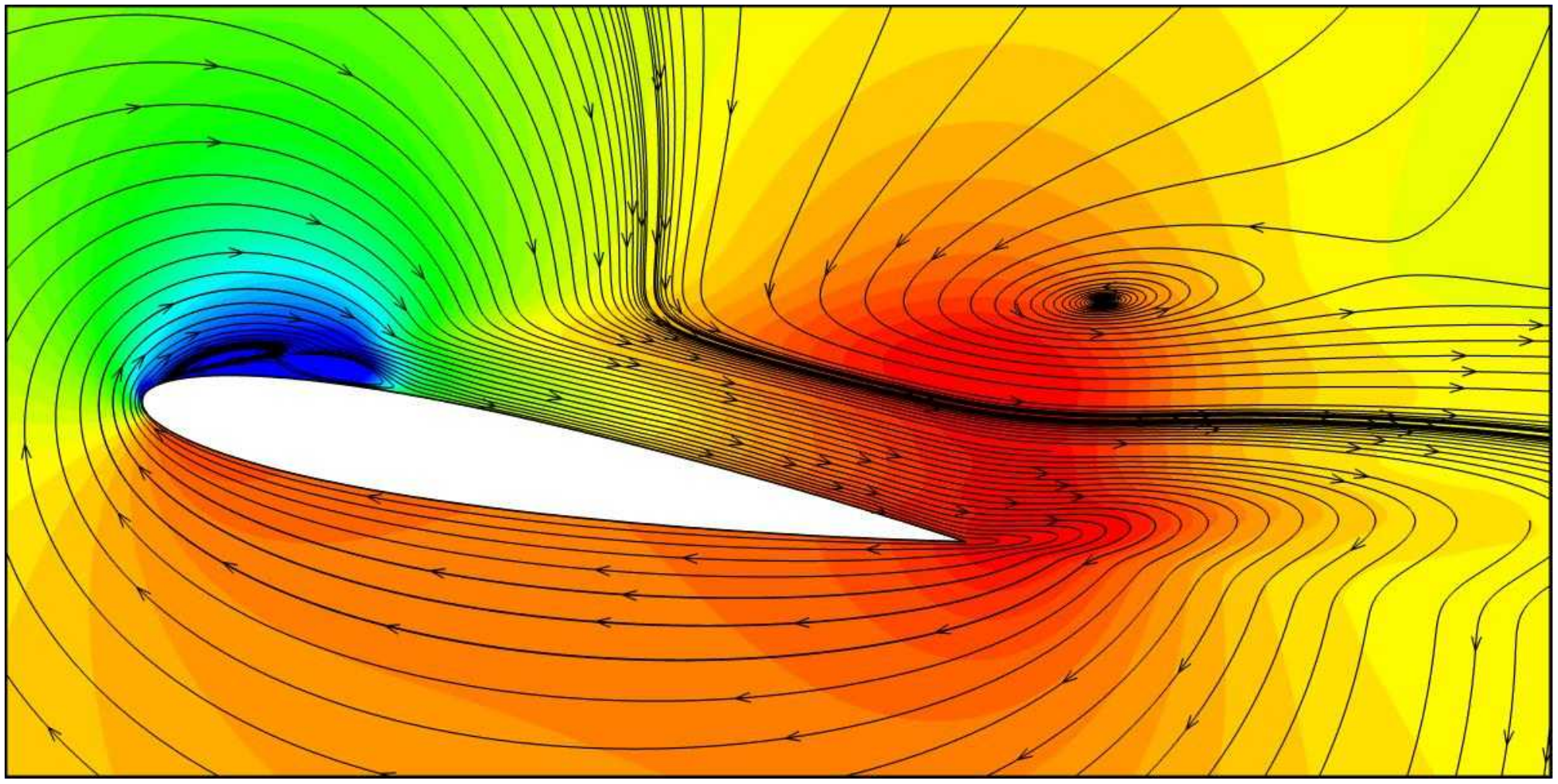}
\textit{LFO mode 1, $St = 0.0062$}
\end{minipage}
\begin{minipage}{220pt}
\centering
\includegraphics[width=220pt, trim={0mm 0mm 0mm 0mm}, clip]{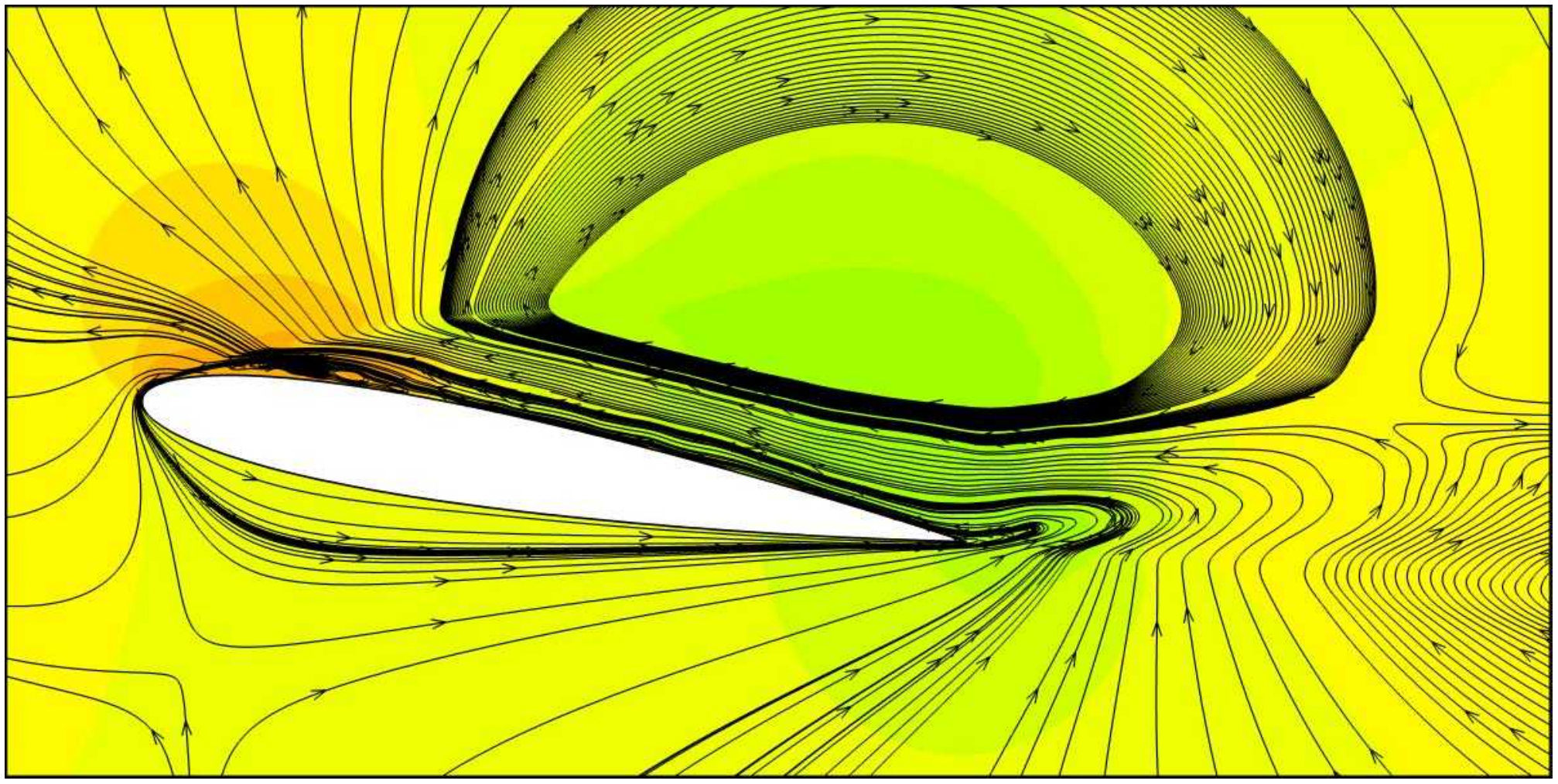}
\textit{LFO mode 2, $St = 0.0077$}
\end{minipage}
\caption{Streamlines patterns superimposed on colour maps of the pressure f\/ield for the DMD construction of the oscillating-f\/low using the LFO mode 1 and the LFO mode 2 for the angle of attack of $\alpha = 9.8^{\circ}$.}
\label{DMD_streamlines_P_LFO1_LFO2}
\end{center}
\end{figure}

F\/igure~\ref{DMD_streamlines_wz_LFO1} shows the DMD spectrum of the lift coeff\/icient (left) and streamlines patterns superimposed on colour maps of the spanwise vorticity $\omega_z$ for the LFO mode 1 (right) at the angle of attack of $9.8^{\circ}$. The DMD spectrum shows the growth rates of the LFO mode 1 and the LFO mode 2, indicated by black and red f\/illed circles, respectively. The LFO is quasi-periodic; however, the growth rates of the LFO mode 1 and the LFO mode 2 are of opposite signs in most of the investigated angles of attack. This is indicative that the two low-frequency modes drive the dynamics of the LFO, and play a contradicting role. That is, the LFO mode 1 decays to a minimum during which the LFO mode 2 grows to a maximum. Thus, when the LFO mode 1 vanishes and the f\/low reaches a temporary equilibrium, the LFO mode 2 amplif\/ies and starts a new disequilibrium. The process continues in a periodic manner. The size of the circles denotes the relative amplitude of each of the dynamic modes. The relative magnitude of the LFO mode 1 and the LFO mode 2 increases as the angle of attack is increased, and the relative amplitude of the higher modes diminishes. At the angle of attack of $9.25^{\circ}$, the relative amplitude of the oscillating-f\/low is distributed among various f\/low modes; thus, the size of the circles are almost of the same size and diminishes for higher modes. The relative magnitude of the LFO mode 1 and 2 becomes more pronounced as the angle of attack is increased. Consequently, only the two low-frequency modes and a few other low-frequency modes are shown in the f\/igure, and the higher modes become insignif\/icant. At the angle of attack of $9.8^{\circ}$, the LFO mode 1 and 2 contain more than $50\%$ of the total amplitude of oscillation and the remaining amplitude is almost equally distributed among higher modes. This is indicative that the LFO process is fully developed and the oscillating-f\/low is more pronounced in magnitude and more coherent in shape. This also implies that only the two low-frequency modes and a few other low-frequency modes play a major role in the dynamics of the LFO. As the angle of attack is further increased, the relative amplitude of the LFO mode 1 and 2 decreases while the relative amplitude of the higher modes increases. The f\/igure shows only the DMD spectra of the lift coeff\/icient; however, the DMD spectra of the pressure, the drag, the skin-friction, and the moment coeff\/icients are similar to that of the lift coeff\/icient.\newline
The LFO mode 1 features the triad of vortices at all of the investigated angles of attack as seen on the right-hand side of the f\/igure. The triad of vortices is driven by the oscillating-velocity across the laminar portion of the separated shear layer. The magnitude of the oscillating-velocity increases as the angle of attack increases. The oscillating-velocity drives the upstream vortex of the triad of vortices as discussed in~\citet{eljack2018bursting2}. However, in~\citet{eljack2018bursting2} the shape and evolution of the triad of vortices are obscured by higher f\/low modes. Whereas, the triad of vortices presented in f\/igure~\ref{DMD_streamlines_wz_LFO1} are evolving at a single frequency. The triad of vortices which governs and sustains the LFO phenomenon seem to exist at low angles of attack. Hence, the LFO phenomenon seems to exist at lower angles of attack but dominated by higher-frequency modes like vortex shedding and shear layer f\/lapping.\newline
F\/igure~\ref{DMD_streamlines_P_LFO1_LFO2} shows streamlines patterns superimposed on colour maps of the oscillating-pressure of the LFO mode 1 and 2 for the angle of attack of $9.8^{\circ}$. The left-hand side of the f\/igure shows the LFO mode 1, and the right-hand side shows the LFO mode 2. As seen in the f\/igure, the LFO mode 1 is a global f\/low mode that drives and sustains the triad of vortices. The LFO mode 1 seems to preserve shape over all of the investigated angles of attack. However, the magnitude of oscillation increases as the angle of attack increases. The LFO mode 2 originates and evolves on the suction surface of the aerofoil as seen in the f\/igure. The LFO mode 2 features the expansion and advection of the upstream vortex of the two co-rotating vortices. Therefore, these two low-frequency modes are interlinked and govern the dynamics of the LFO. The LFO mode 1 generates the triad of vortices and feeds energy into the upstream vortex of the TCV. When the upstream vortex of the triad of vortices is saturated with energy, it expands and the LFO mode 2 dominates the f\/low until the oscillating-f\/low switches its direction then the LFO mode 1 takes over again. This explains why the spectra of the lift and the skin-friction coeff\/icients, estimated using the Fast Fourier transform algorithm in~\citet{eljack2018bursting1}, have two low-frequency peaks. The two low-frequency peaks correspond to the LFO mode 1 and 2, and they strengthen each other. The Strouhal number of the LFO mode 1 and 2 is written at the bottom of each dynamic mode. Despite the fact that the LFO cycle is very disturbed at this relatively low Reynolds number, a pattern can be seen in the Strouhal numbers of the LFO mode 1 and 2. At the angle of attack of $9.25^{\circ}$, the LFO mode 1 oscillates at a low-frequency of $St = 0.0083$ while the LFO mode 2 oscillates at an exceptionally low frequency of $St = 0.0037$. This is indicative that while the f\/low oscillates globally, the UV of the TCV (the LFO mode 2) expands at a very low frequency. Thus, the f\/low remains attached with rare and occasional separations. The frequency of the LFO mode 2 increases as the angle of attack is increased. Hence, the f\/low-f\/ield separates more frequently. At the angle of attack of $9.8^{\circ}$ the LFO mode 1 and 2 oscillate at almost the same frequency. Thus, the UV of the TCV expands at the same rate at which the f\/low oscillates globally, and the LFO is fully developed.

\subsubsection{The high-frequency mode}
\begin{figure}
\begin{center}
\begin{minipage}{220pt}
\centering
\includegraphics[width=220pt, trim={0mm 0mm 0mm 0mm}, clip]{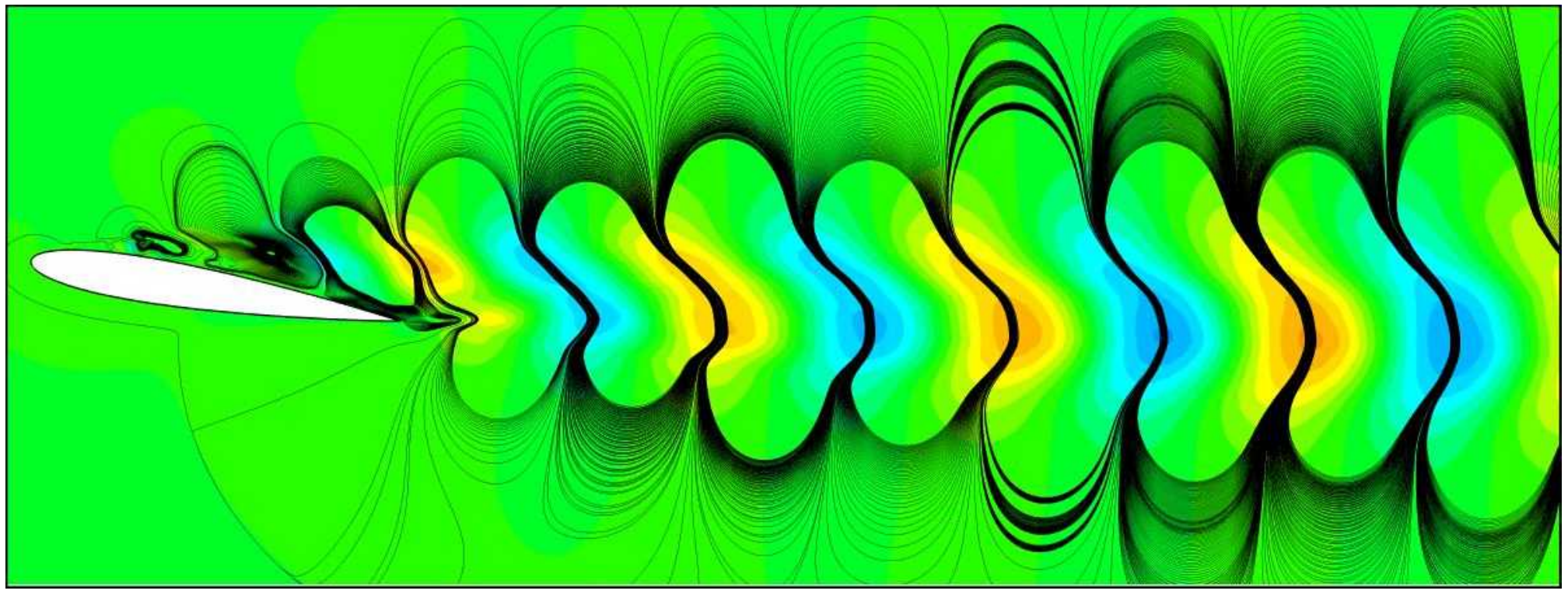}
\textit{$\alpha = 9.25^{\circ}$, $St = 0.2682$}
\end{minipage}
\medskip
\begin{minipage}{220pt}
\centering
\includegraphics[width=220pt, trim={0mm 0mm 0mm 0mm}, clip]{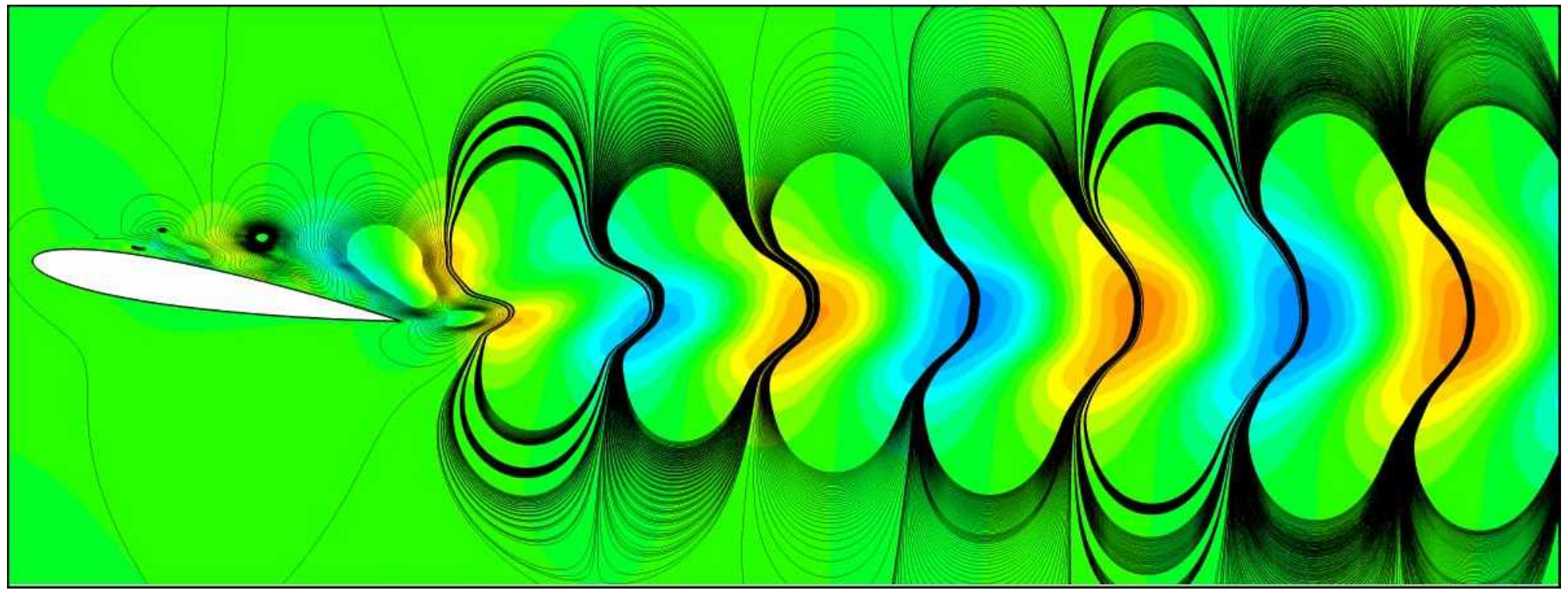}
\textit{$\alpha = 9.40^{\circ}$, $St = 0.2224$}
\end{minipage}
\medskip
\begin{minipage}{220pt}
\centering
\includegraphics[width=220pt, trim={0mm 0mm 0mm 0mm}, clip]{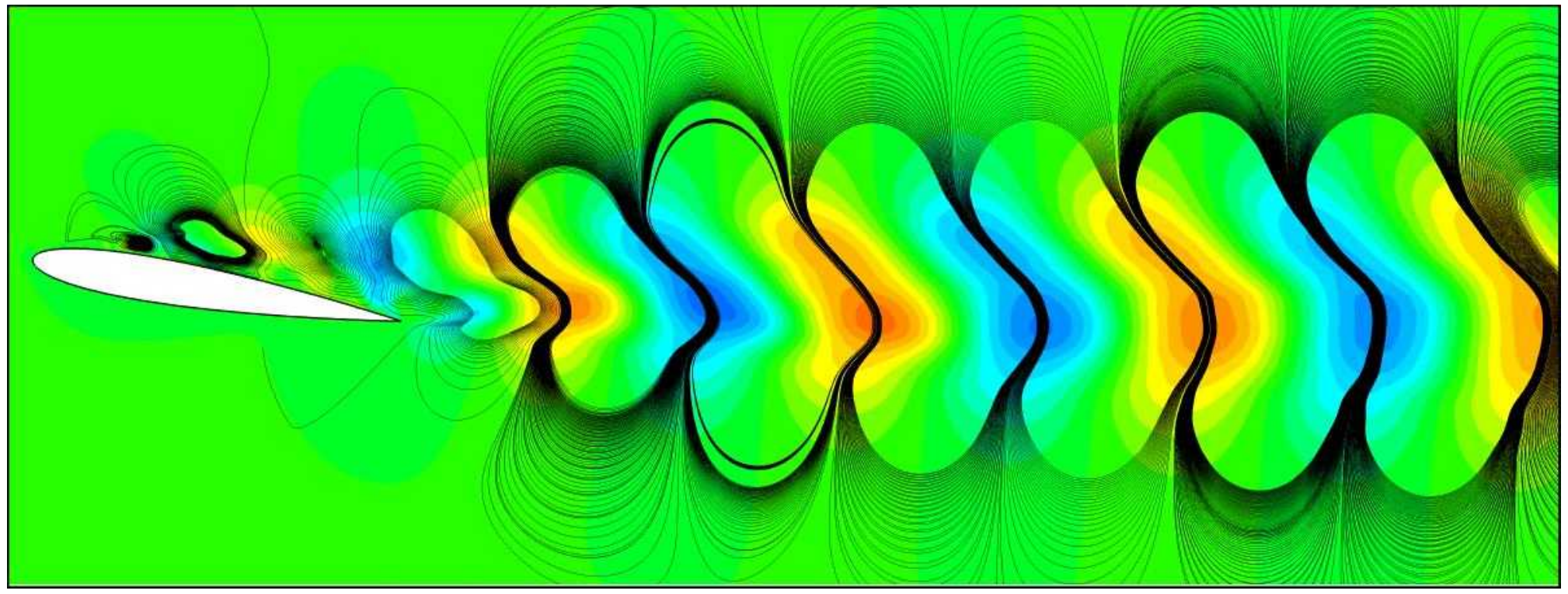}
\textit{$\alpha = 9.5^{\circ}$, $St = 0.2225$}
\end{minipage}
\medskip
\begin{minipage}{220pt}
\centering
\includegraphics[width=220pt, trim={0mm 0mm 0mm 0mm}, clip]{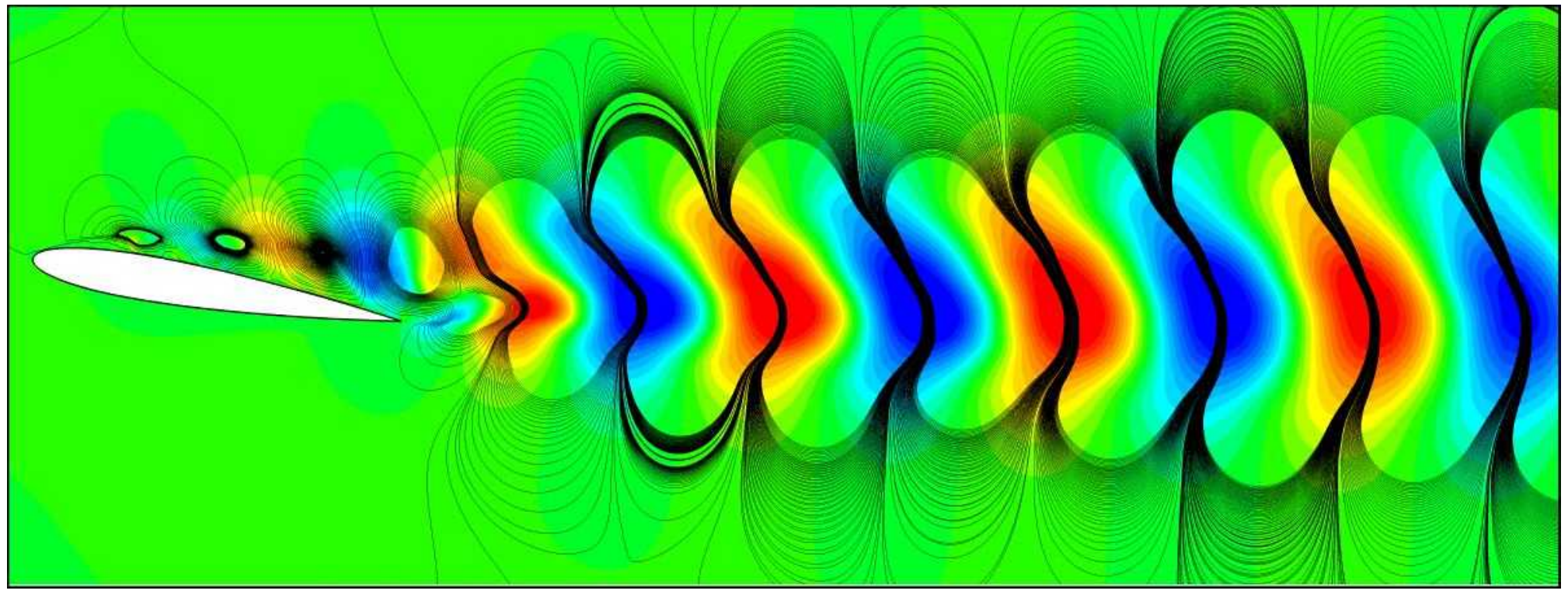}
\textit{$\alpha = 9.6^{\circ}$, $St = 0.1853$}
\end{minipage}
\medskip
\begin{minipage}{220pt}
\centering
\includegraphics[width=220pt, trim={0mm 0mm 0mm 0mm}, clip]{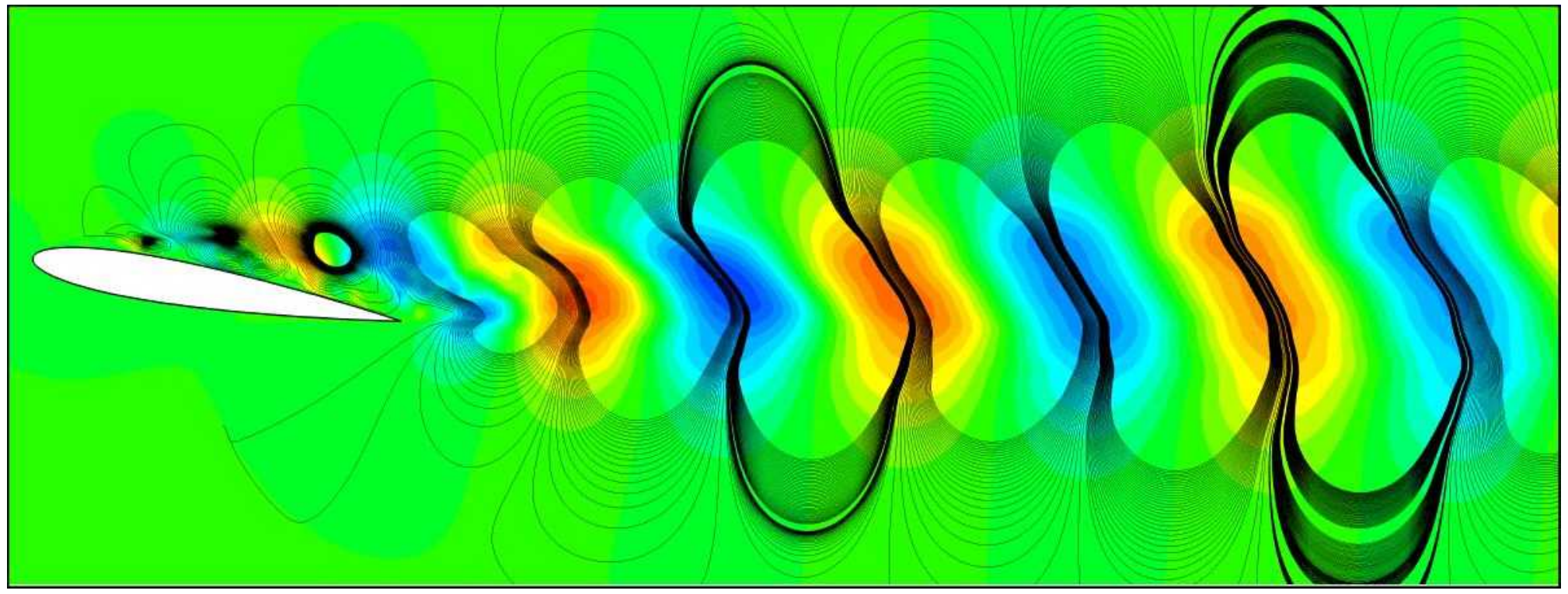}
\textit{$\alpha = 9.7^{\circ}$, $St = 0.2784$}
\end{minipage}
\medskip
\begin{minipage}{220pt}
\centering
\includegraphics[width=220pt, trim={0mm 0mm 0mm 0mm}, clip]{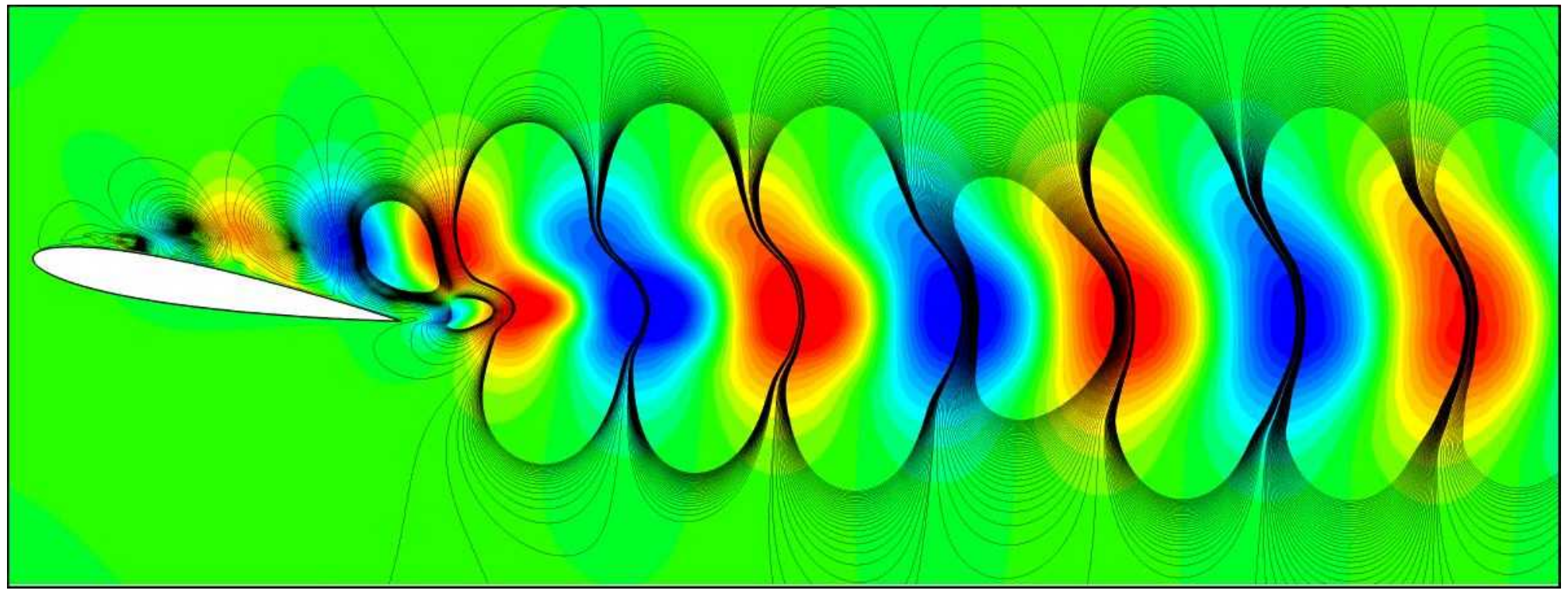}
\textit{$\alpha = 9.8^{\circ}$, $St = 0.1707$}
\end{minipage}
\medskip
\begin{minipage}{220pt}
\centering
\includegraphics[width=220pt, trim={0mm 0mm 0mm 0mm}, clip]{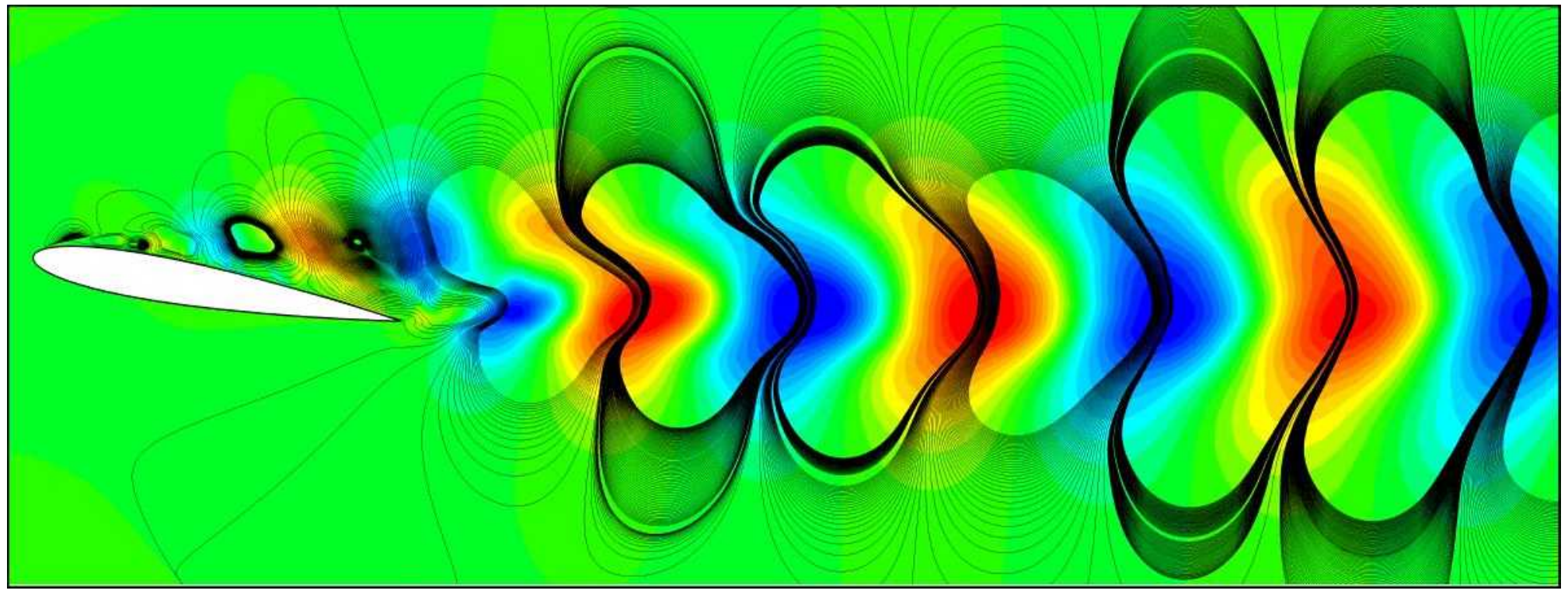}
\textit{$\alpha = 9.9^{\circ}$, $St = 0.1572$}
\end{minipage}
\medskip
\begin{minipage}{220pt}
\centering
\includegraphics[width=220pt, trim={0mm 0mm 0mm 0mm}, clip]{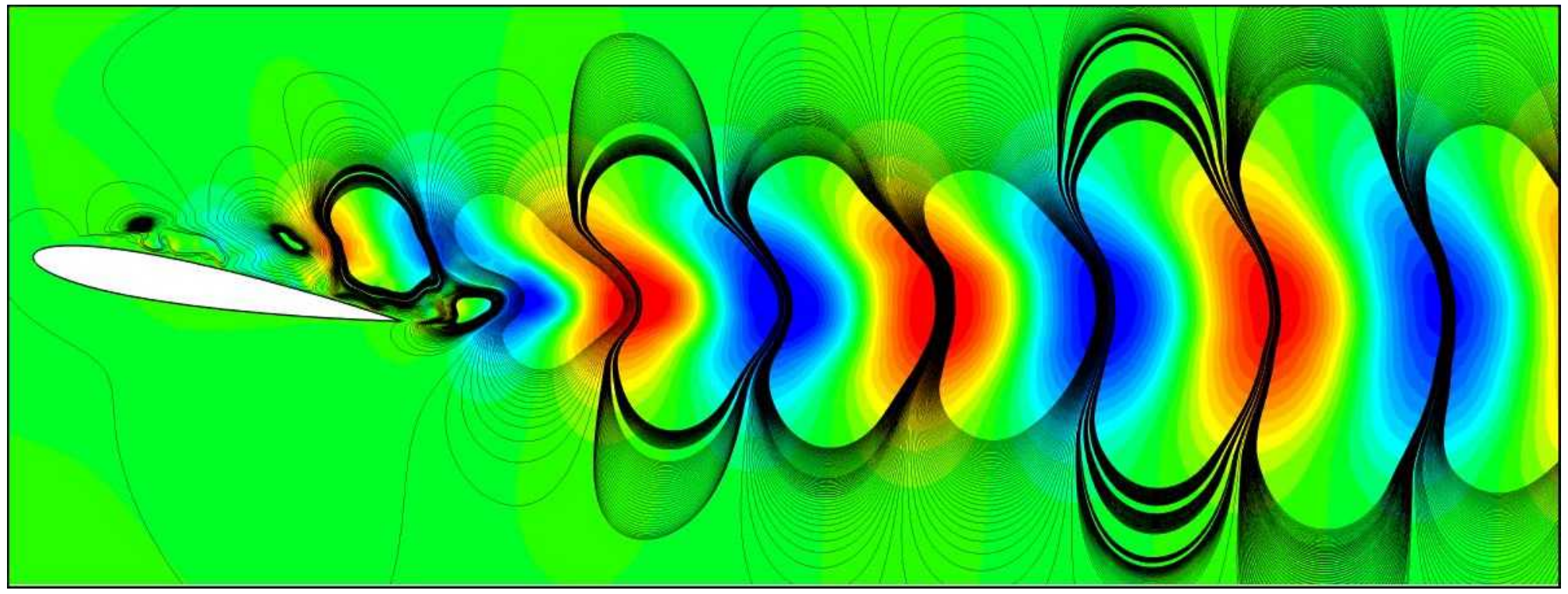}
\textit{$\alpha = 10.0^{\circ}$, $St = 0.1732$}
\end{minipage}
\medskip
\begin{minipage}{220pt}
\centering
\includegraphics[width=220pt, trim={0mm 0mm 0mm 0mm}, clip]{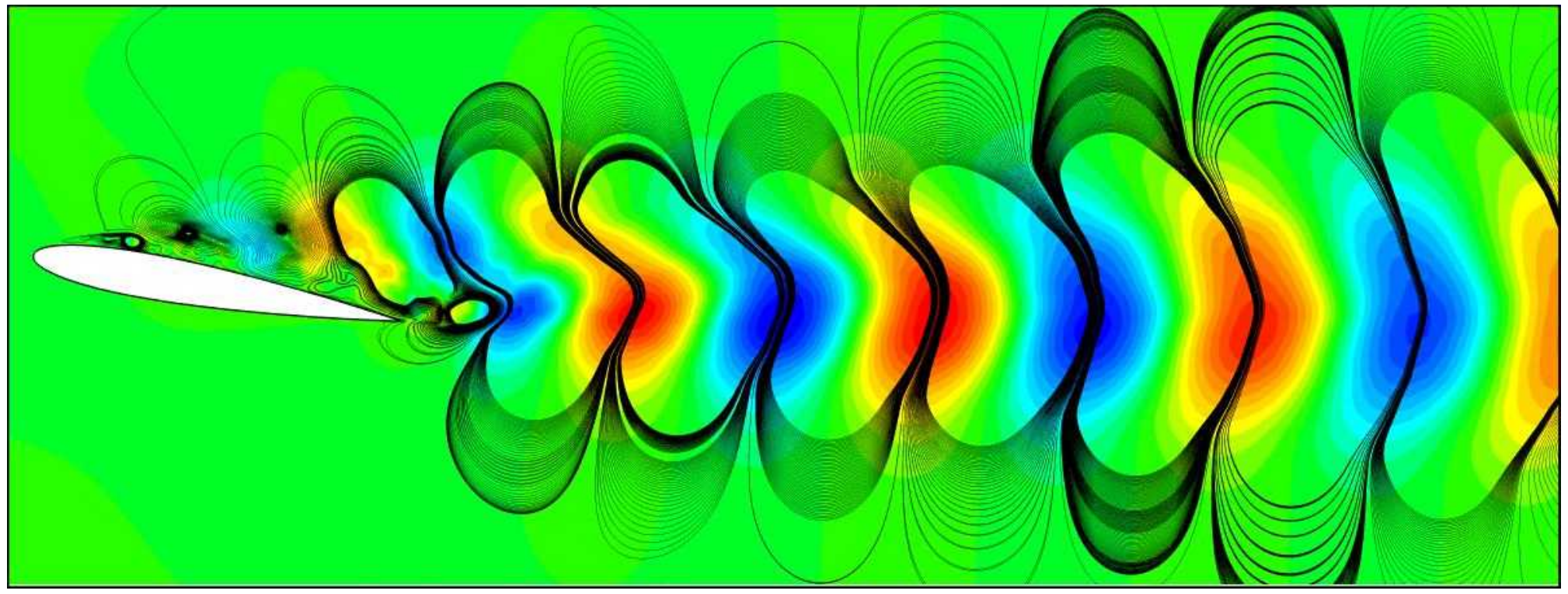}
\textit{$\alpha = 10.1^{\circ}$, $St = 0.2685$}
\end{minipage}
\begin{minipage}{220pt}
\centering
\includegraphics[width=220pt, trim={0mm 0mm 0mm 0mm}, clip]{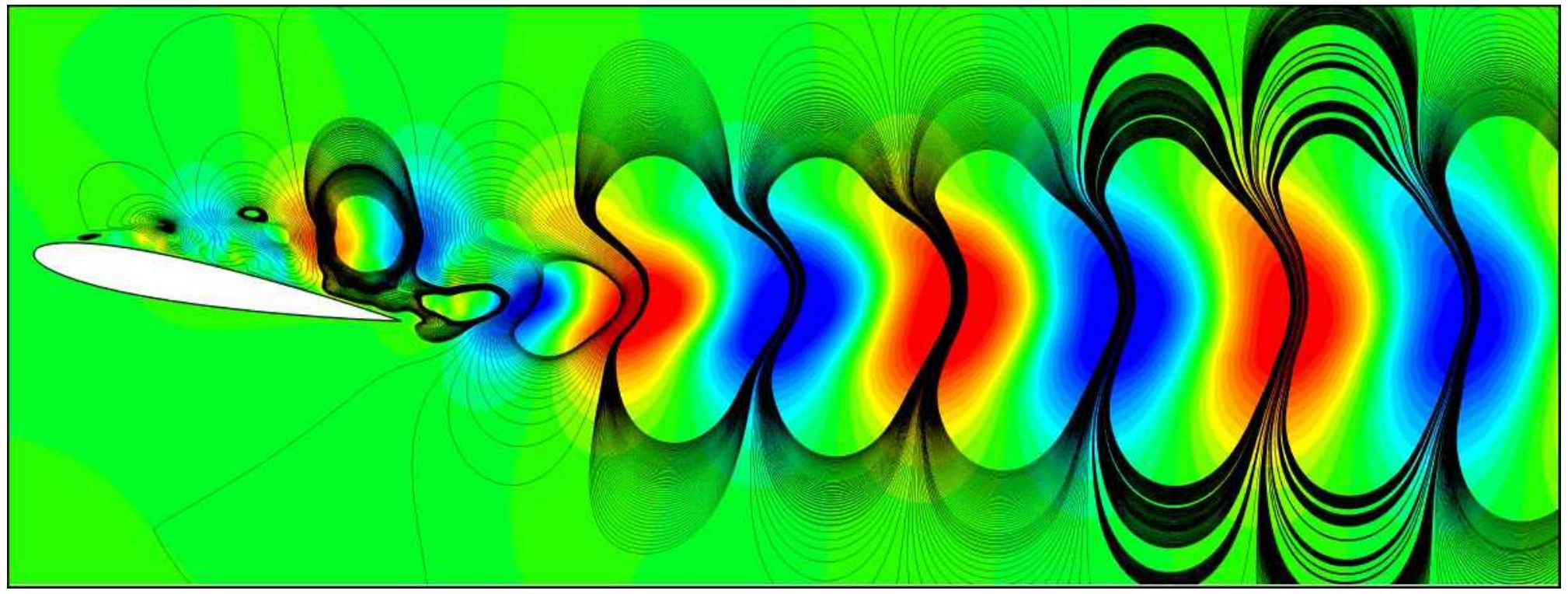}
\textit{$\alpha = 10.5^{\circ}$, $St = 0.2893$}
\end{minipage}
\caption{Streamlines patterns superimposed on colour maps of the wall-normal velocity component for the DMD construction of the oscillating-f\/low using the HFO mode for the angles of attack of $9.25^{\circ}$--$10.5^{\circ}$.}
\label{DMD_streamlines_V_HFO}
\end{center}
\end{figure}

F\/igure~\ref{DMD_streamlines_V_HFO} shows streamline patterns of the most dominant high-frequency mode superimposed on colour maps of its corresponding wall-normal velocity for the angles of attack of $9.25^{\circ}$--$10.5^{\circ}$. As seen in the f\/igure, the HFO mode features large oscillations that originate at the leading-edge and advects towards the trailing-edge. At the trailing-edge, the HFO mode interacts with the local f\/low instability, stretches by the strong shear, and energies the vortex shedding. The magnitude of the oscillation is proportional to the angle of attack. It is interesting to note that there is no high-frequency peak in the spectra of the lift, the drag, and the skin-friction coeff\/icients. This is due to the fact that the aerodynamic coeff\/icients are more sensitive to variations in the f\/low-f\/ield adjacent to the aerofoil surface. Thus, any important f\/low feature that dominates the f\/low-f\/ield away from the aerofoil surface, and does not affect the f\/low variables near the wall will not be ref\/lected in the aerodynamic forces. Consequently, such important f\/low feature will not show up in the spectra of the aerodynamic coeff\/icients.\newline
F\/igure~\ref{DMD_spectra_p_u_v} shows that the spectra of the wall-normal velocity peaks signif\/icantly at high-frequency. Whereas, the spectra of the streamwise velocity and the pressure show no signif\/icant peak at high-frequency. This is indicative that, the HFO is primarily driven by the wall-normal velocity component. However, the streamwise velocity and/or the pressure might indirectly affects the behaviour of the HFO mode. Furthermore, the spectra of the wall-normal velocity are interchangeably dominated by the HFO mode and the two low-frequency modes. Therefore, it seems that the HFO mode and the two low-frequency modes are interlinked, and play a profound role in the dynamics of the f\/low. The HFO mode originates in the vicinity of the leading-edge in a fashion similar to that of the LFO mode 2. Thus, the HFO mode is very likely to be a subharmonic of the LFO mode 2. Furthermore, there could be higher subharmonic of both the LFO mode 1 and 2 embedded in the high-frequency modes. However, at this relatively low Reynolds number the LFO cycle is very disturbed as mentioned before. Hence, further investigation at a higher Reynolds number is needed before such conclusion can be drawn.

\subsection{Application of the POD}\label{sec:POD_application}
Application of the DMD to the data revealed that there are distinct three f\/low modes, two low-frequency modes and a high-frequency mode, that dominate and control the dynamics of the f\/low. However, the evolution of each mode in time and details of the interaction among these three modes are not clear. Furthermore, the DMD method decomposed the oscillating-f\/low into too many low-frequency modes. Thus, it was diff\/icult to identify the dominant modes. Should the oscillating-f\/low being objectively recovered by a few low-frequency modes, identifying the most important low-frequency modes would have been straightforward. Therefore, the snapshot POD method was applied to the data to further investigate the dynamics of the f\/low. The locally-time-averaged and spanwise ensemble-averaged streamwise velocity, wall-normal velocity, and pressure are utilized. The two-point correlations of the three variable, in time, are duly estimated using $20,000$ data points which span $100$ non-dimensional time units, or four low-frequency cycles. The eigenvalue problem, $\mathbf{A}\varphi = \mathbf{\lambda} \mathbf{\varphi}$, was solved and the POD eigenvalues and eigenvectors are obtained, where $\mathbf{A}$ represents the correlation matrix.

\subsubsection{The POD modes}
\begin{figure}
\begin{center}
\begin{minipage}{220pt}
\centering
\includegraphics[width=220pt, trim={0mm 0mm 0mm 0mm}, clip]{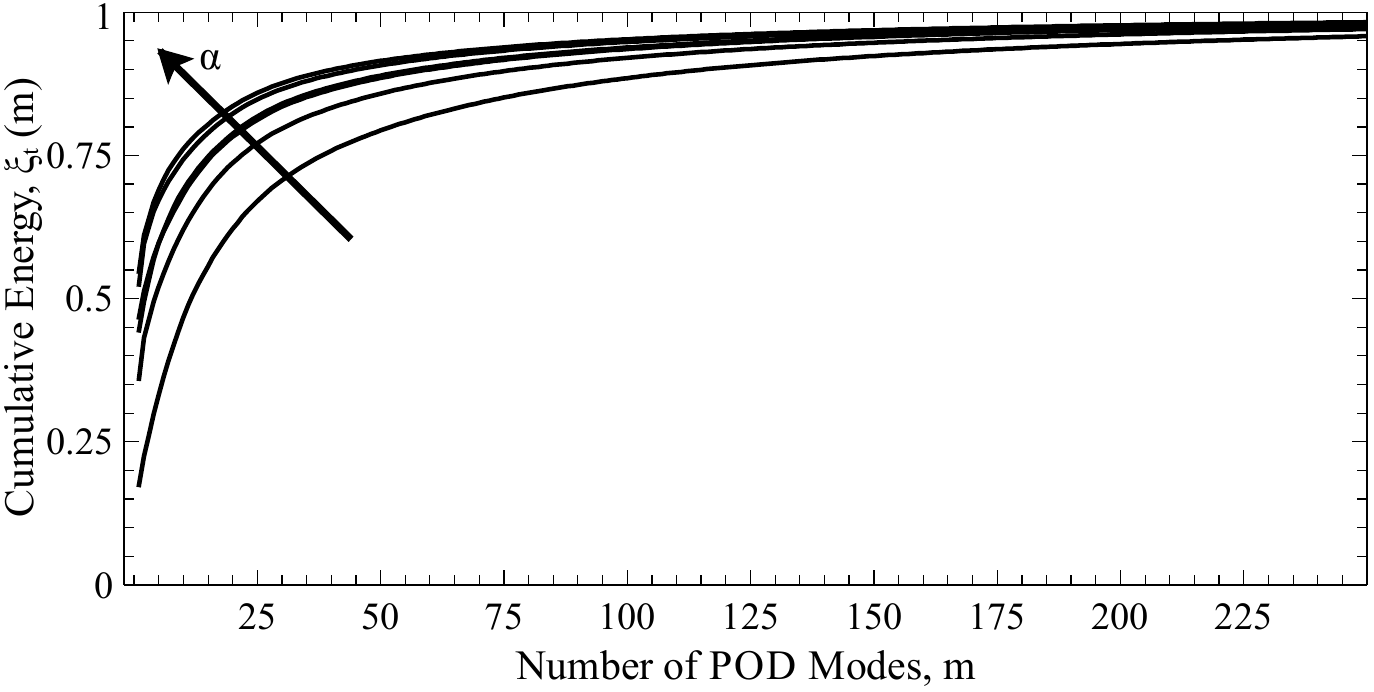}
\textit{$\alpha = 9.25^{\circ}$--$9.8^{\circ}$}
\end{minipage}
\begin{minipage}{220pt}
\centering
\includegraphics[width=220pt, trim={0mm 0mm 0mm 0mm}, clip]{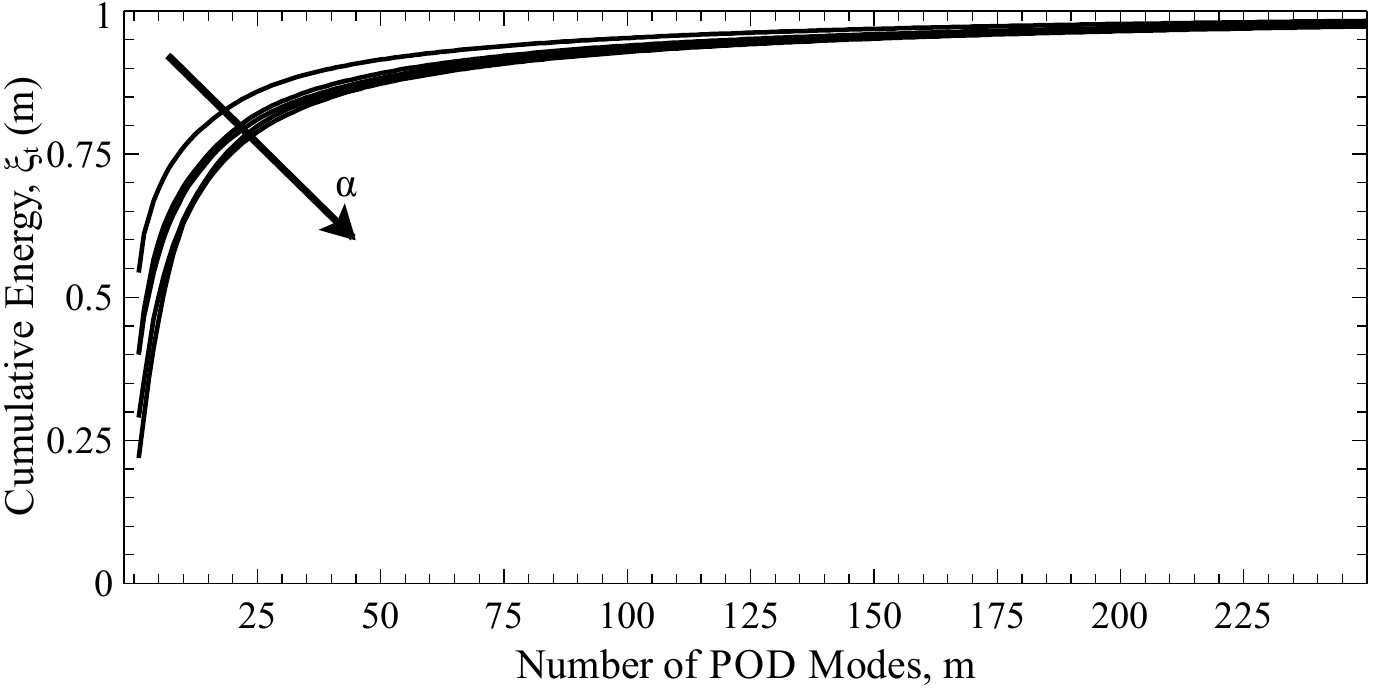}
\textit{$\alpha = 9.8^{\circ}$--$10.5^{\circ}$}
\end{minipage}
\caption{The cumulative energy of the POD modes, $\xi_t(m)$, estimated by summing the POD eigenvalues, $\lambda^{(n)}$, over $m$ POD modes. The arrow indicates the direction in which the angle of attack $\alpha$ increases.}
\label{POD_cumulative_energy}
\end{center}
\begin{center}
\includegraphics[width=420pt, trim={0mm 0mm 0mm 0mm}, clip]{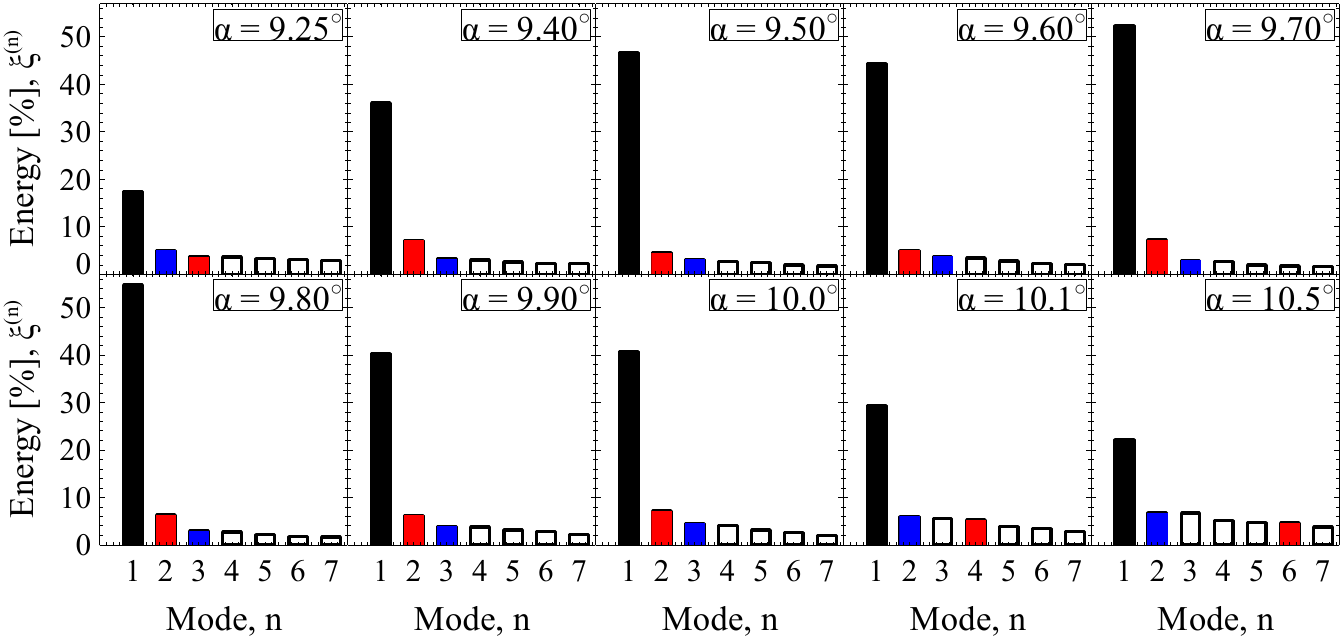}
\caption{The percentage of the energy content for POD mode $n$, $\xi^{(n)}$. The f\/illed black, red, and blue bars indicate the energy percentage of the LFO mode 1, the LFO mode 2, and the HFO mode, respectively.}
\label{POD_energy_percentage}
\end{center}
\end{figure}

\begin{figure}
\begin{center}
\begin{minipage}{220pt}
\centering
\includegraphics[width=220pt, trim={0mm 0mm 0mm 0mm}, clip]{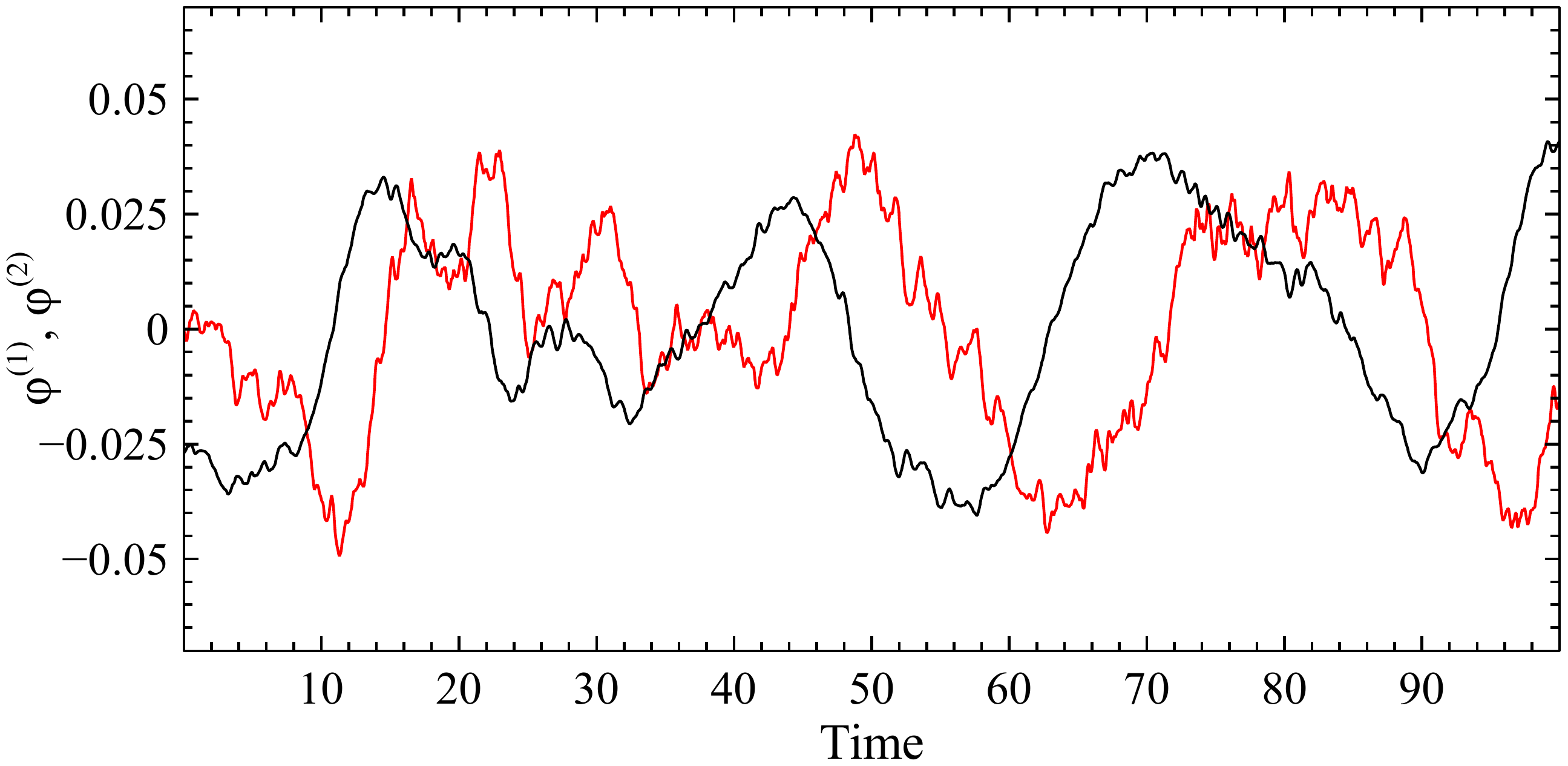}
\textit{$\alpha = 9.7^{\circ}$}
\end{minipage}
\medskip
\begin{minipage}{220pt}
\centering
\includegraphics[width=220pt, trim={0mm 0mm 0mm 0mm}, clip]{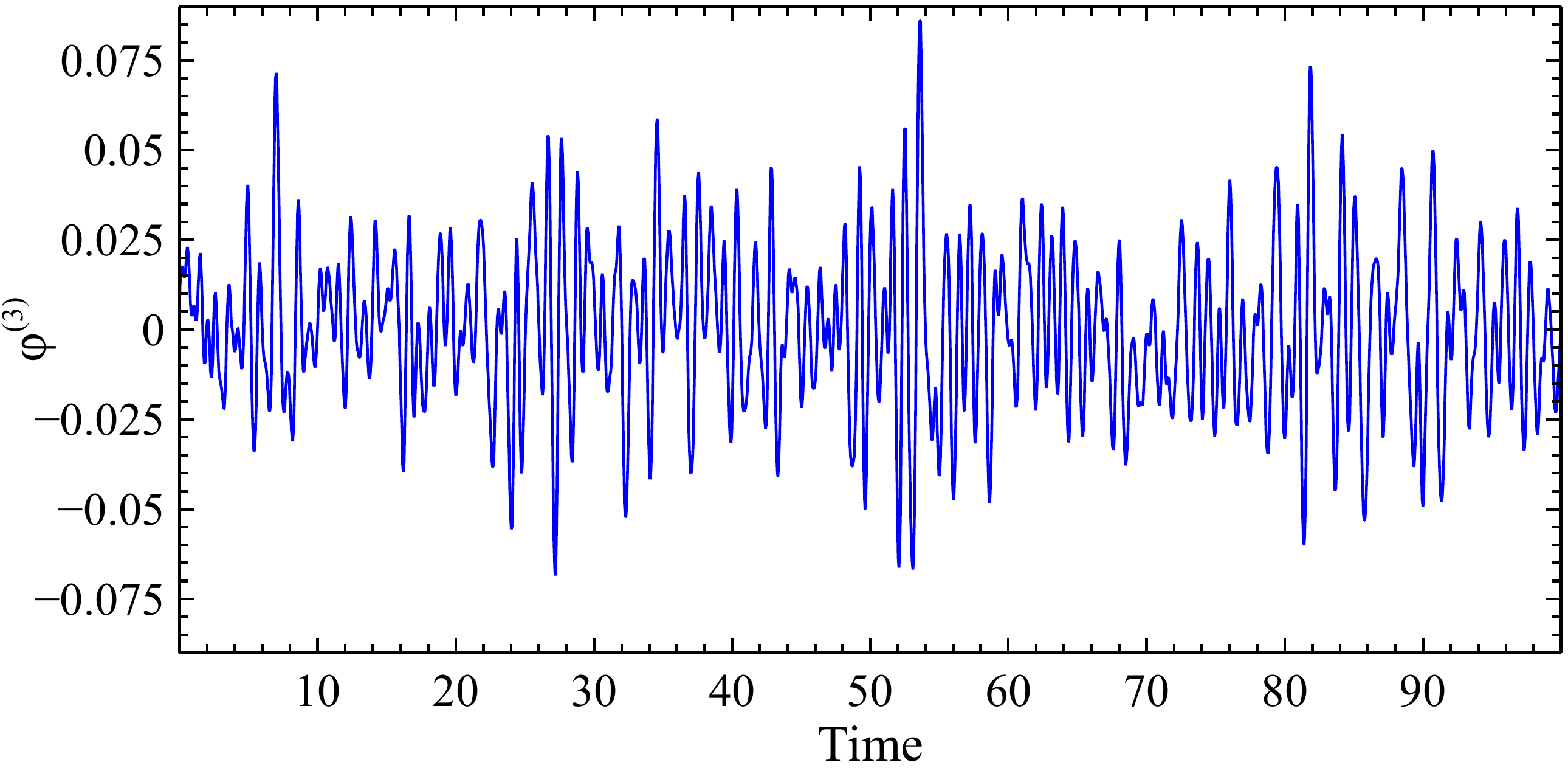}
\textit{$\alpha = 9.7^{\circ}$}
\end{minipage}
\medskip
\begin{minipage}{220pt}
\centering
\includegraphics[width=220pt, trim={0mm 0mm 0mm 0mm}, clip]{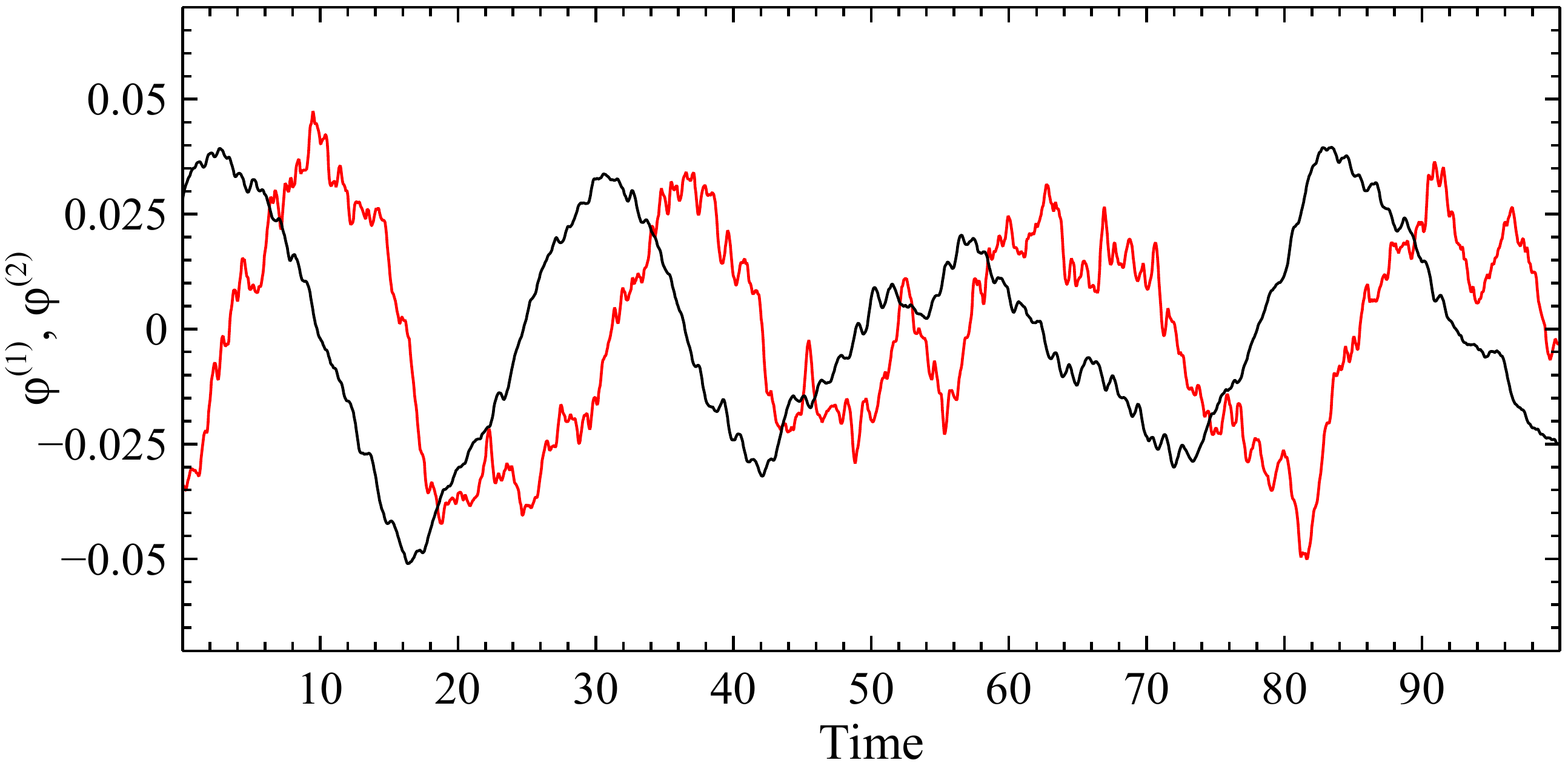}
\textit{$\alpha = 9.8^{\circ}$}
\end{minipage}
\medskip
\begin{minipage}{220pt}
\centering
\includegraphics[width=220pt, trim={0mm 0mm 0mm 0mm}, clip]{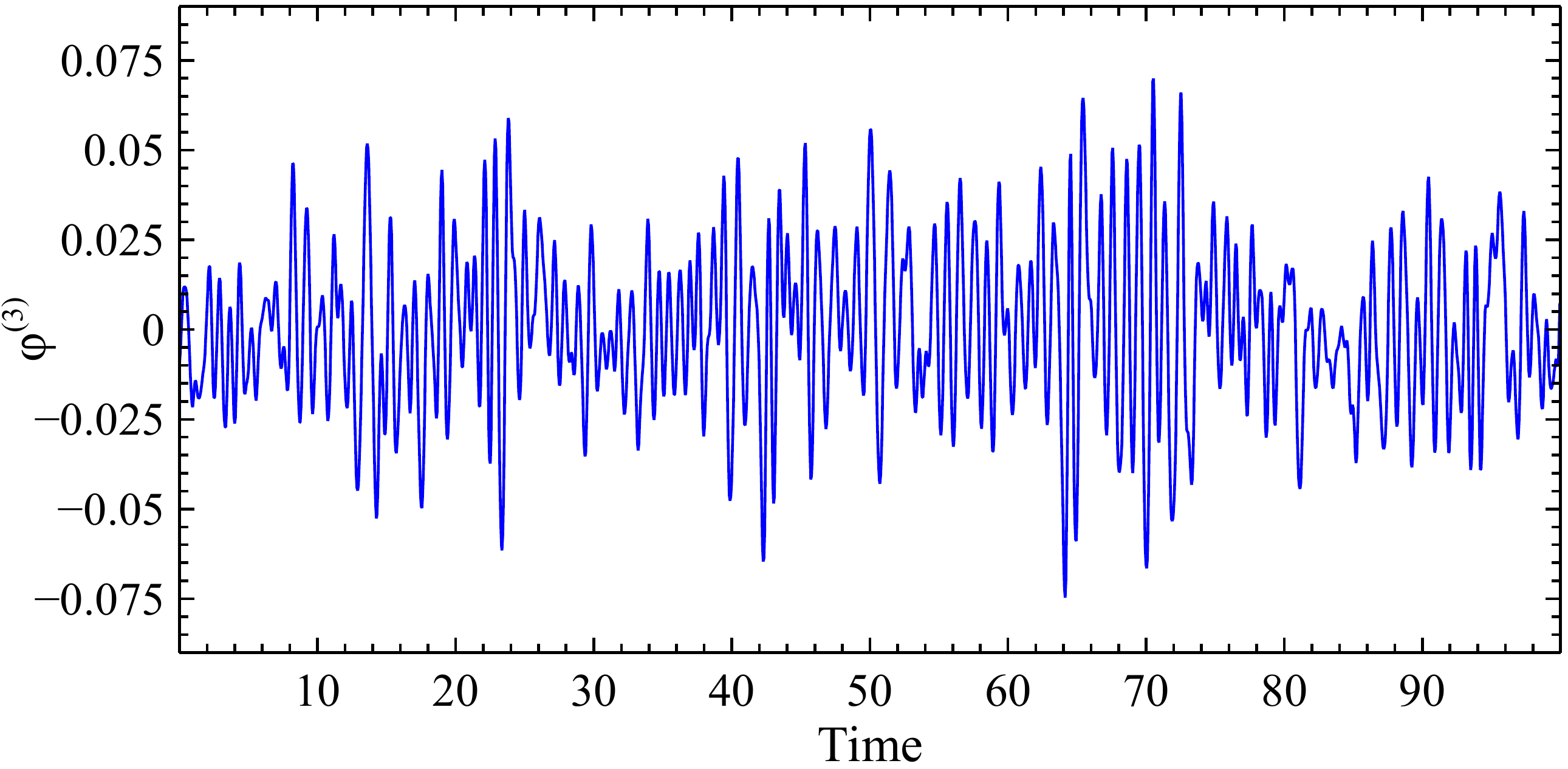}
\textit{$\alpha = 9.8^{\circ}$}
\end{minipage}
\medskip
\begin{minipage}{220pt}
\centering
\includegraphics[width=220pt, trim={0mm 0mm 0mm 0mm}, clip]{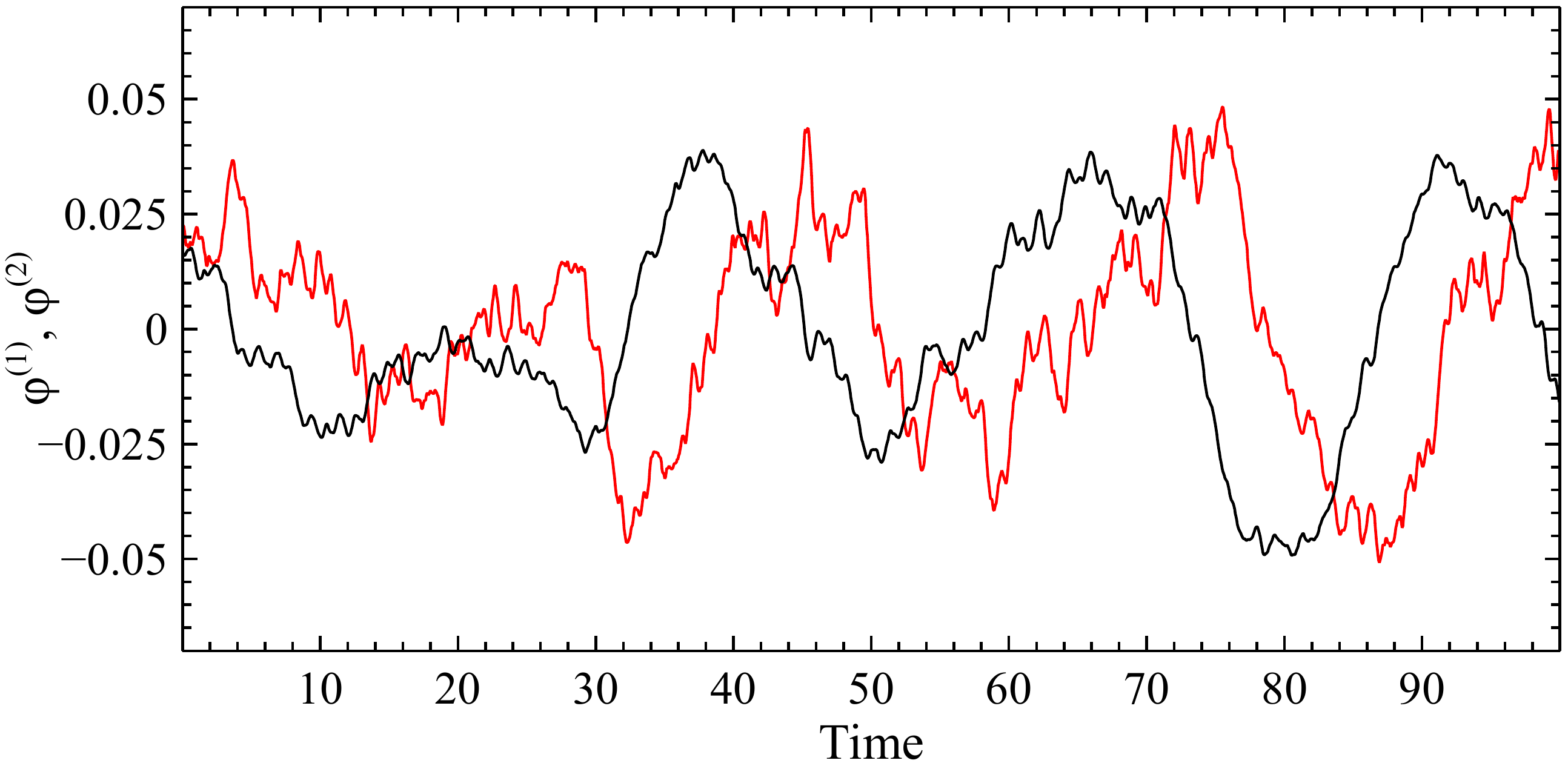}
\textit{$\alpha = 10.0^{\circ}$}
\end{minipage}
\begin{minipage}{220pt}
\centering
\includegraphics[width=220pt, trim={0mm 0mm 0mm 0mm}, clip]{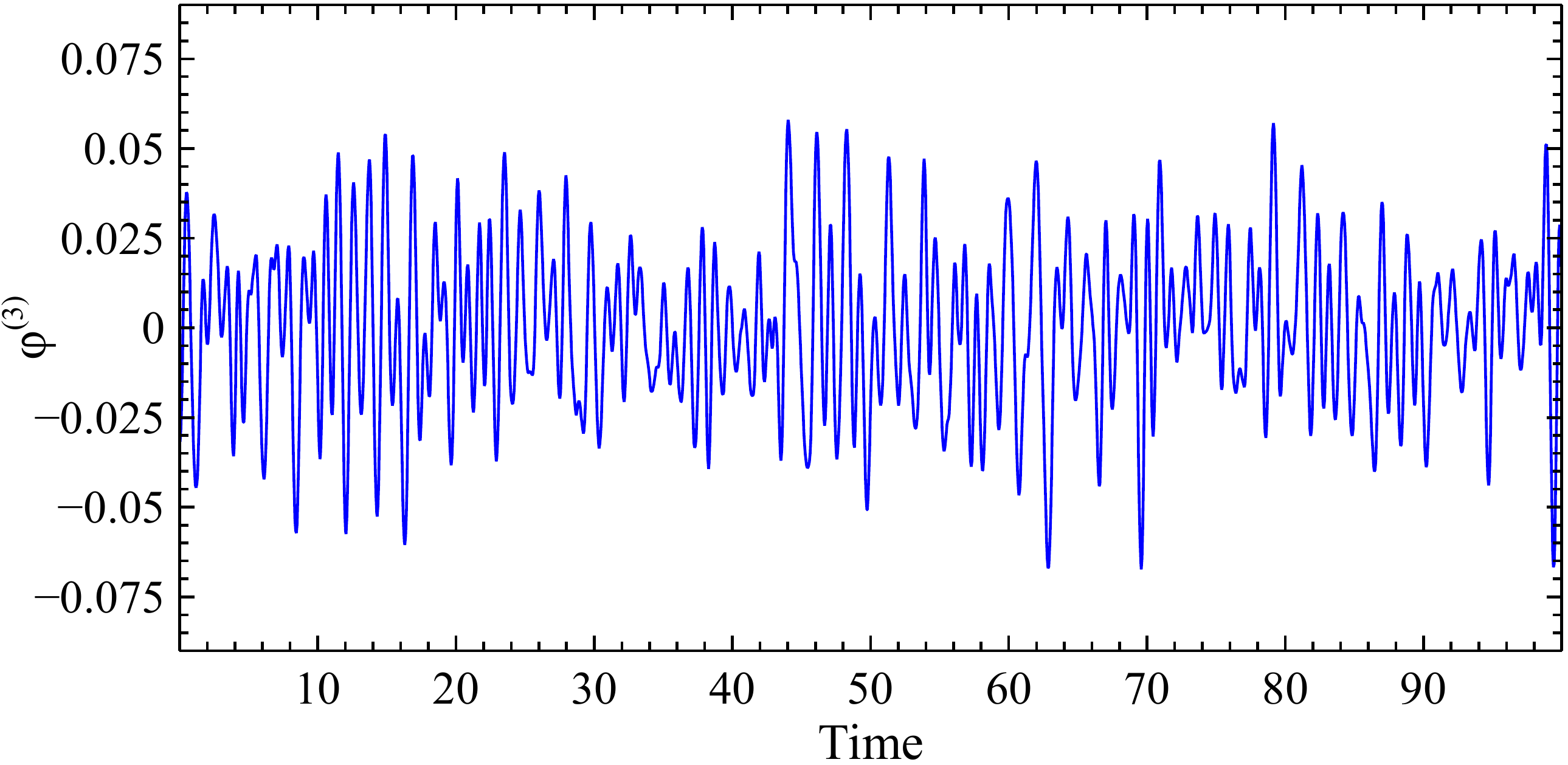}
\textit{$\alpha = 10.0^{\circ}$}
\end{minipage}
\caption{Temporal POD modes, $\varphi^{(n)}$, at the angles of attack of $9.7^{\circ}$--$10.0^{\circ}$. Left: the LFO mode 1 (black) and the LFO mode 2 (red), $\varphi^{(1)}$ and $\varphi^{(2)}$, respectively. Right: The HFO mode, $\varphi^{(3)}$.}
\label{POD_Vectors}
\end{center}
\end{figure}

\begin{figure}
\begin{center}
\begin{minipage}{220pt}
\centering
\includegraphics[width=220pt, trim={0mm 0mm 0mm 0mm}, clip]{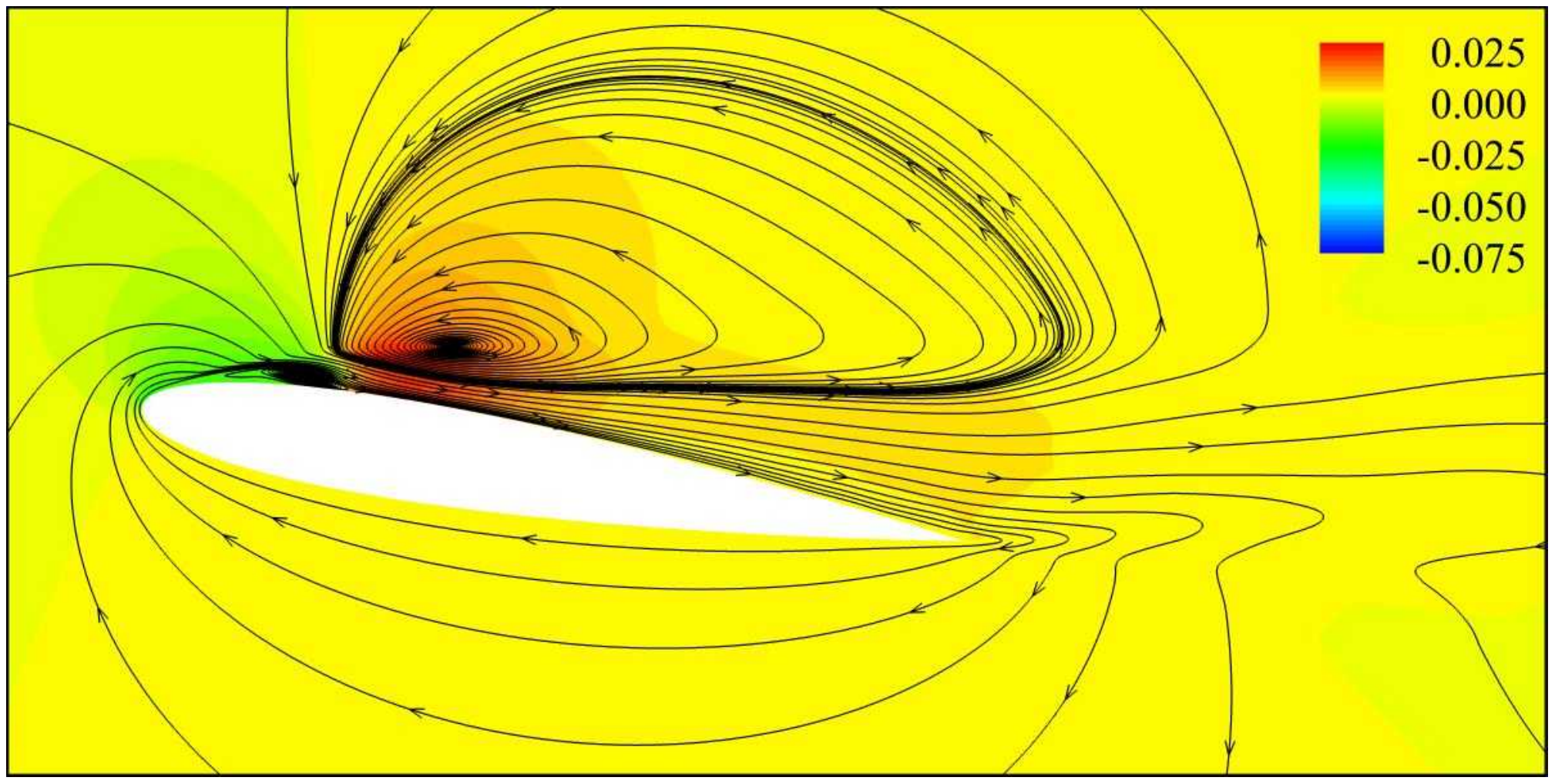}
\textit{$\alpha=9.25^{\circ}$, LFO 1}
\end{minipage}
\begin{minipage}{220pt}
\centering
\includegraphics[width=220pt, trim={0mm 0mm 0mm 0mm}, clip]{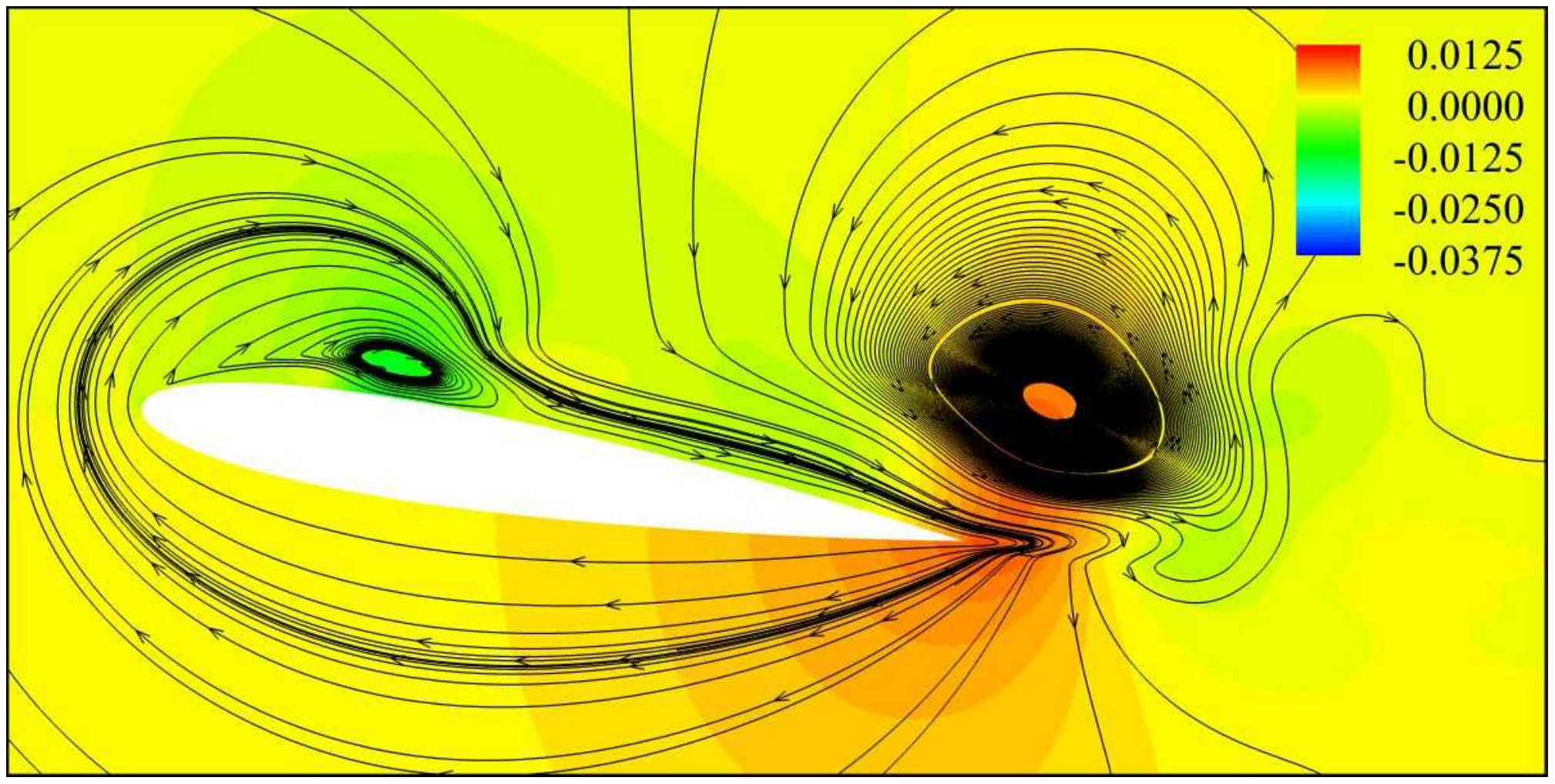}
\textit{$\alpha=9.25^{\circ}$, LFO 2}
\end{minipage}
\begin{minipage}{220pt}
\centering
\includegraphics[width=220pt, trim={0mm 0mm 0mm 0mm}, clip]{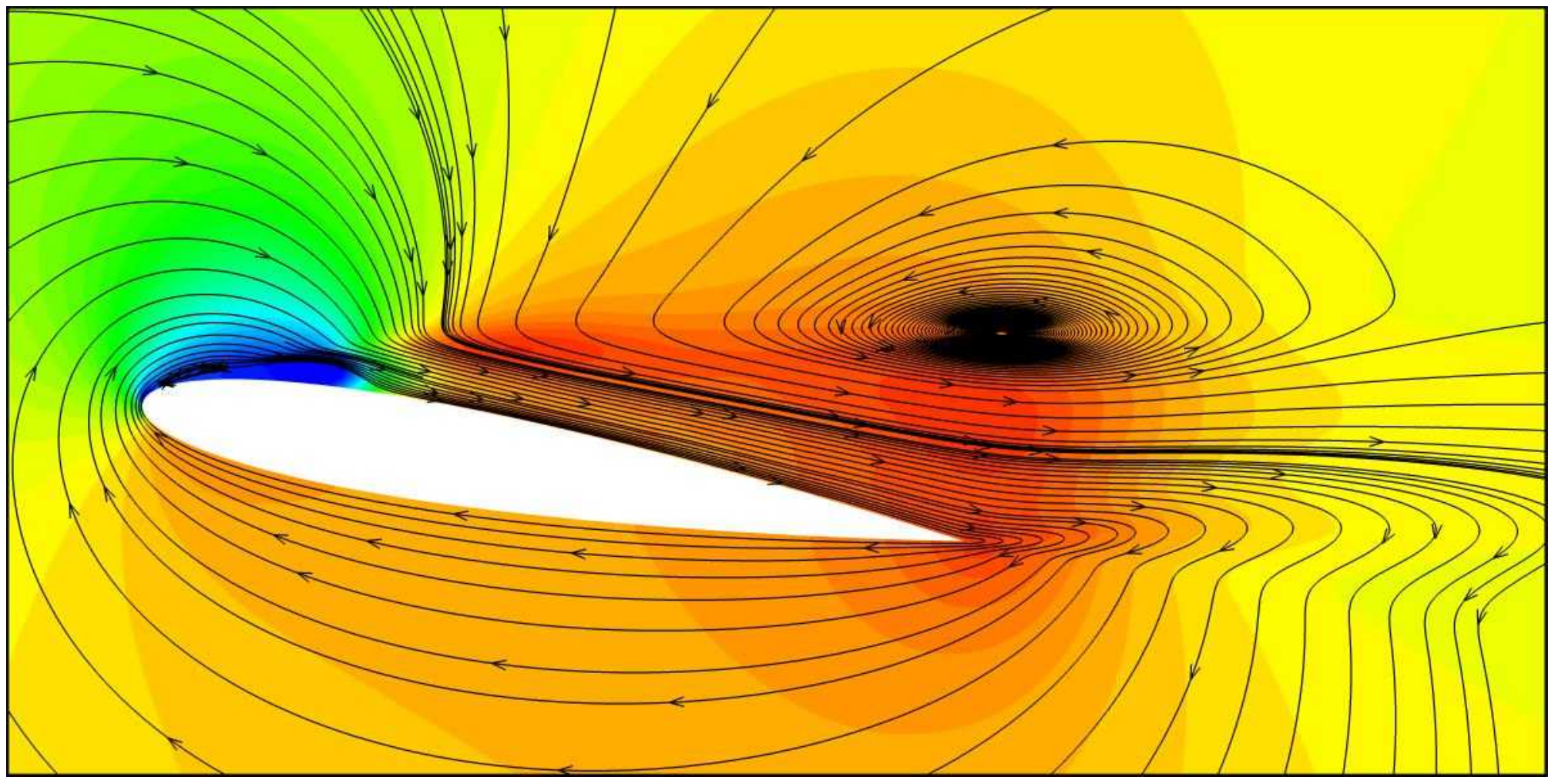}
\textit{$\alpha=9.5^{\circ}$, LFO 1}
\end{minipage}
\begin{minipage}{220pt}
\centering
\includegraphics[width=220pt, trim={0mm 0mm 0mm 0mm}, clip]{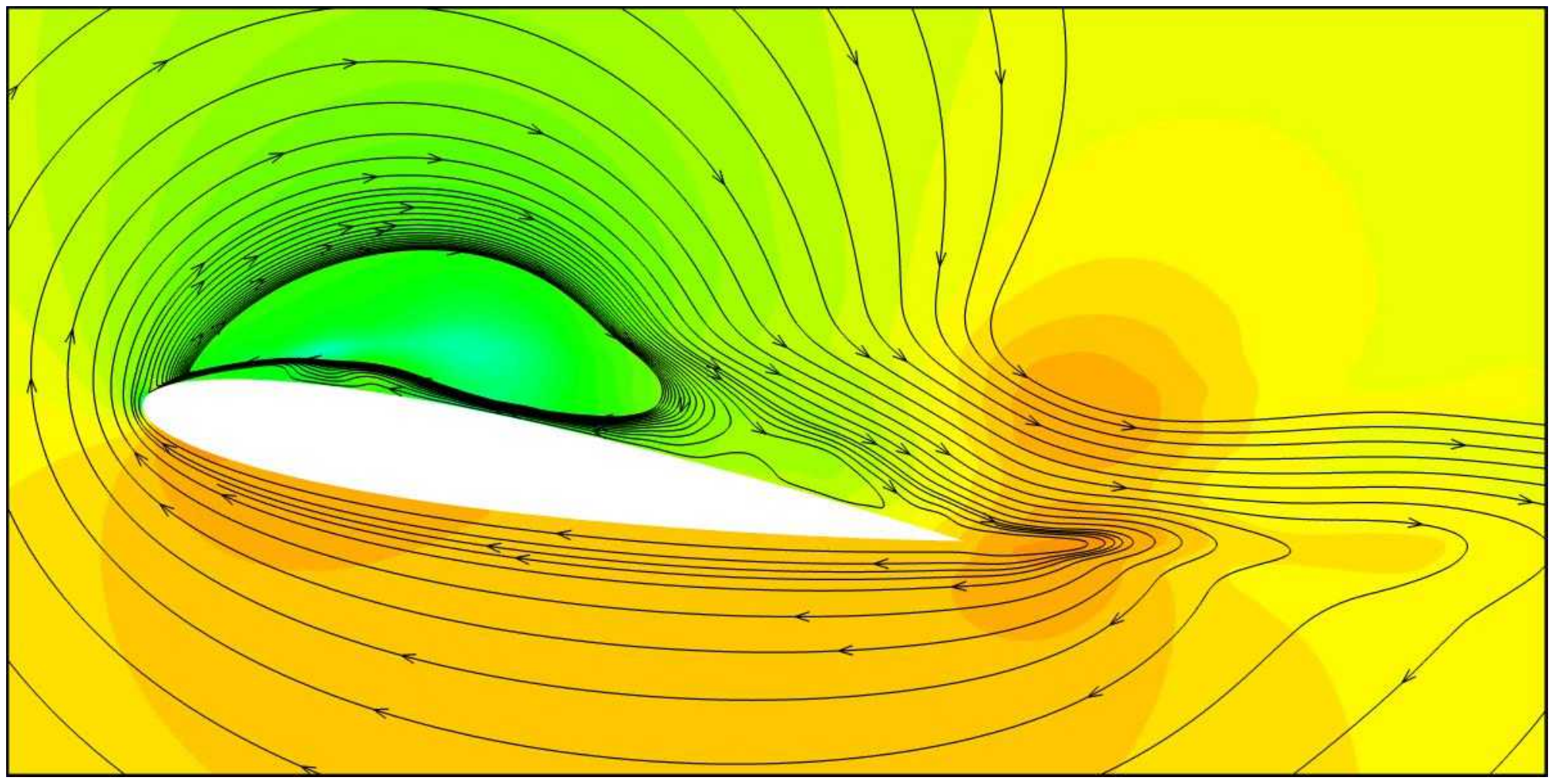}
\textit{$\alpha=9.5^{\circ}$, LFO 2}
\end{minipage}
\begin{minipage}{220pt}
\centering
\includegraphics[width=220pt, trim={0mm 0mm 0mm 0mm}, clip]{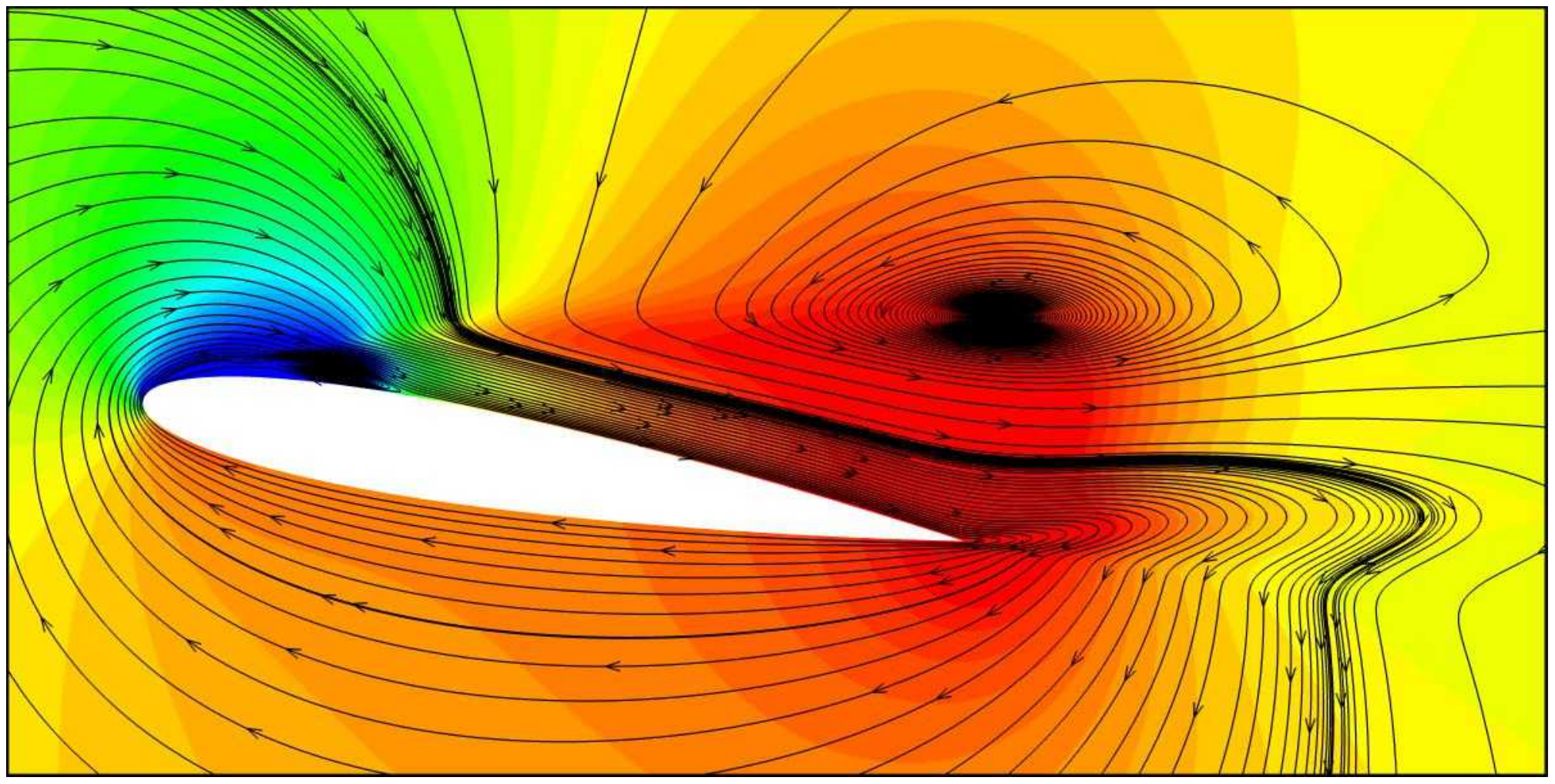}
\textit{$\alpha=9.7^{\circ}$, LFO 1}
\end{minipage}
\begin{minipage}{220pt}
\centering
\includegraphics[width=220pt, trim={0mm 0mm 0mm 0mm}, clip]{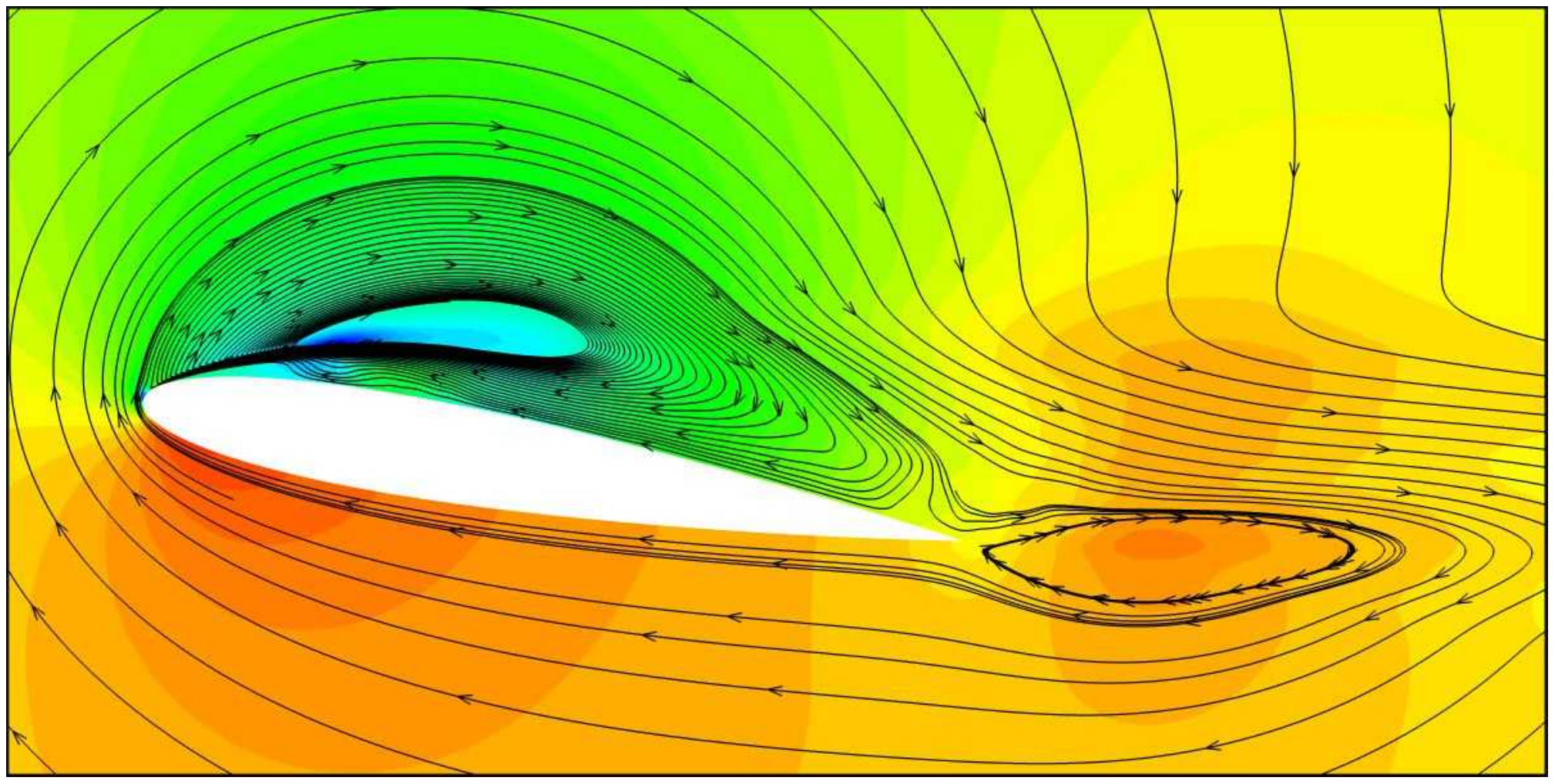}
\textit{$\alpha=9.7^{\circ}$, LFO 2}
\end{minipage}
\caption{Streamlines patterns superimposed on colour maps of the pressure f\/ield for the POD reconstruction of the oscillating-f\/low using the LFO mode 1 and the LFO mode 2 for the angles of attack of $9.25^{\circ}$--$9.7^{\circ}$.}
\label{POD_LFO1_LFO2_1}
\end{center}
\end{figure}

\begin{figure}
\begin{center}
\begin{minipage}{220pt}
\centering
\includegraphics[width=220pt, trim={0mm 0mm 0mm 0mm}, clip]{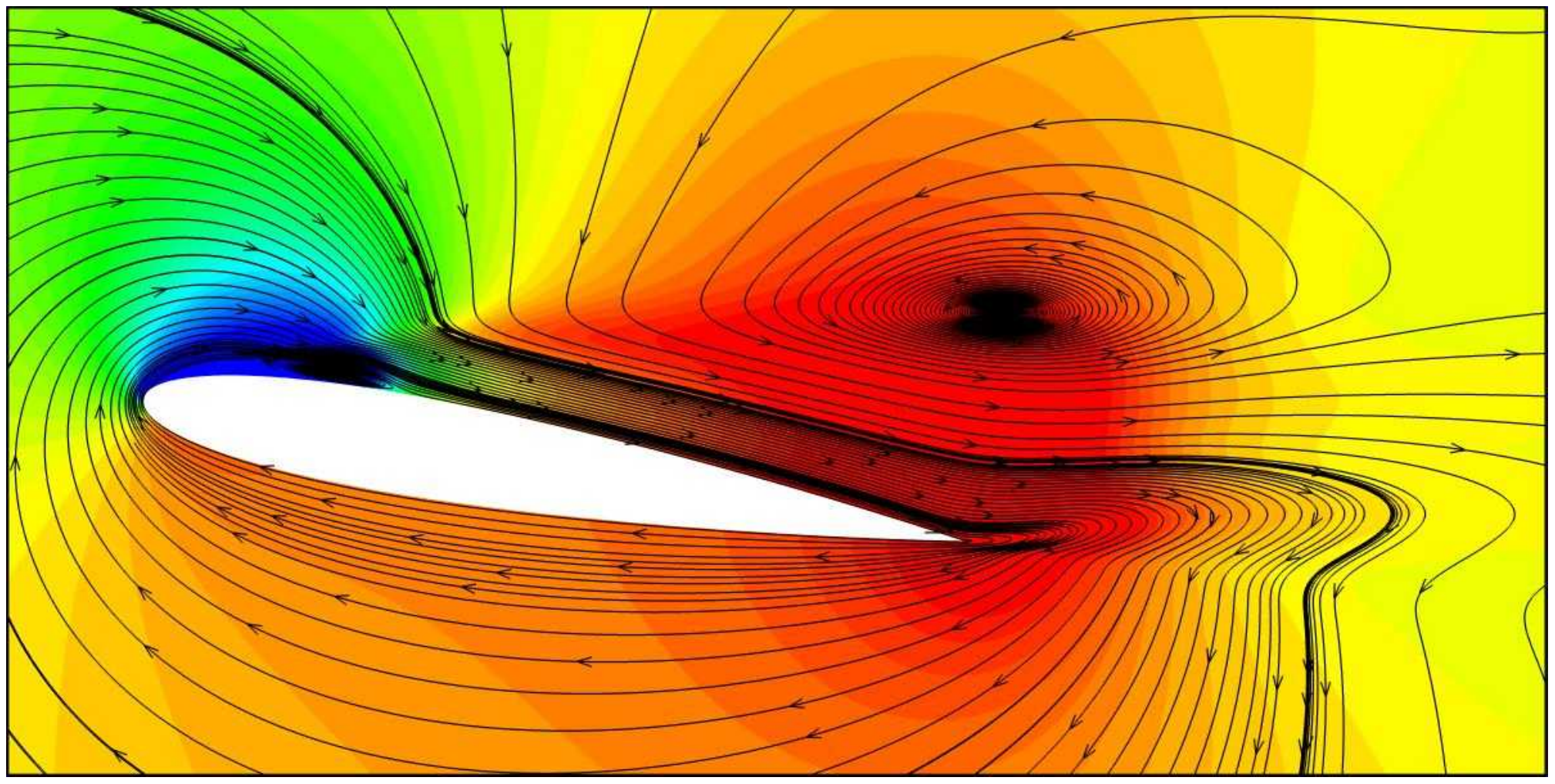}
\textit{$\alpha=9.8^{\circ}$, LFO 1}
\end{minipage}
\begin{minipage}{220pt}
\centering
\includegraphics[width=220pt, trim={0mm 0mm 0mm 0mm}, clip]{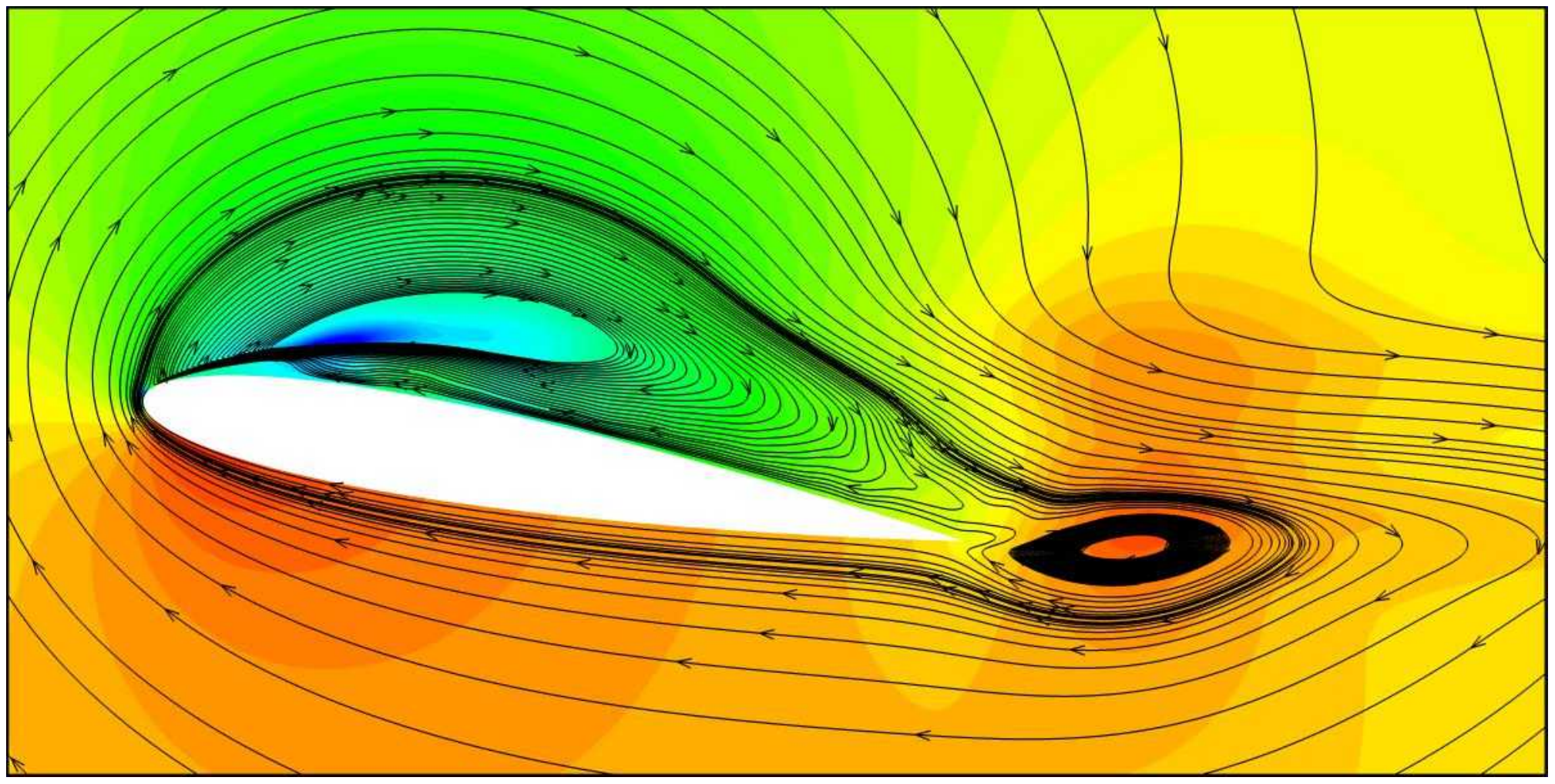}
\textit{$\alpha=9.8^{\circ}$, LFO 2}
\end{minipage}
\begin{minipage}{220pt}
\centering
\includegraphics[width=220pt, trim={0mm 0mm 0mm 0mm}, clip]{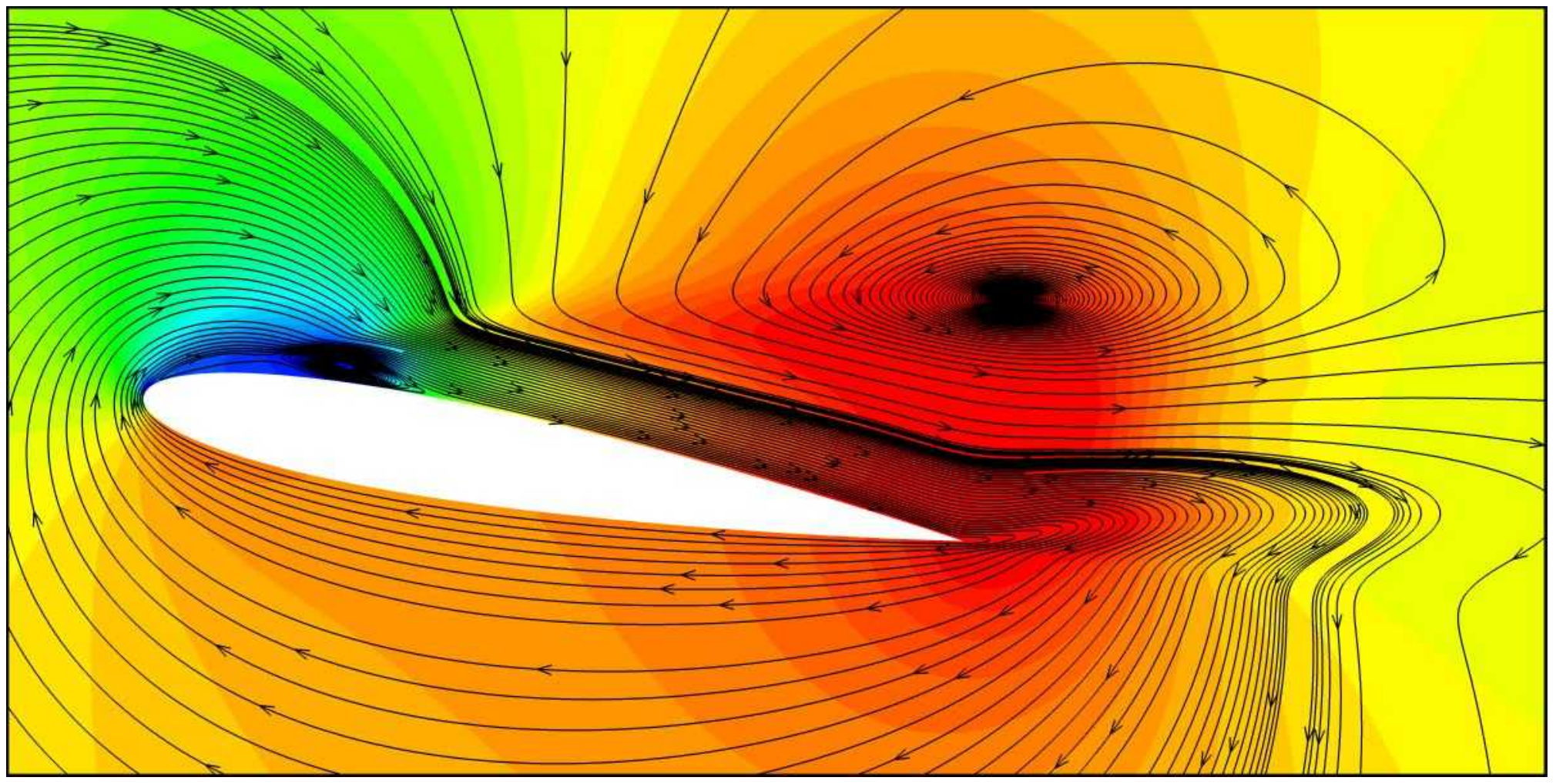}
\textit{$\alpha=10.0^{\circ}$, LFO 1}
\end{minipage}
\begin{minipage}{220pt}
\centering
\includegraphics[width=220pt, trim={0mm 0mm 0mm 0mm}, clip]{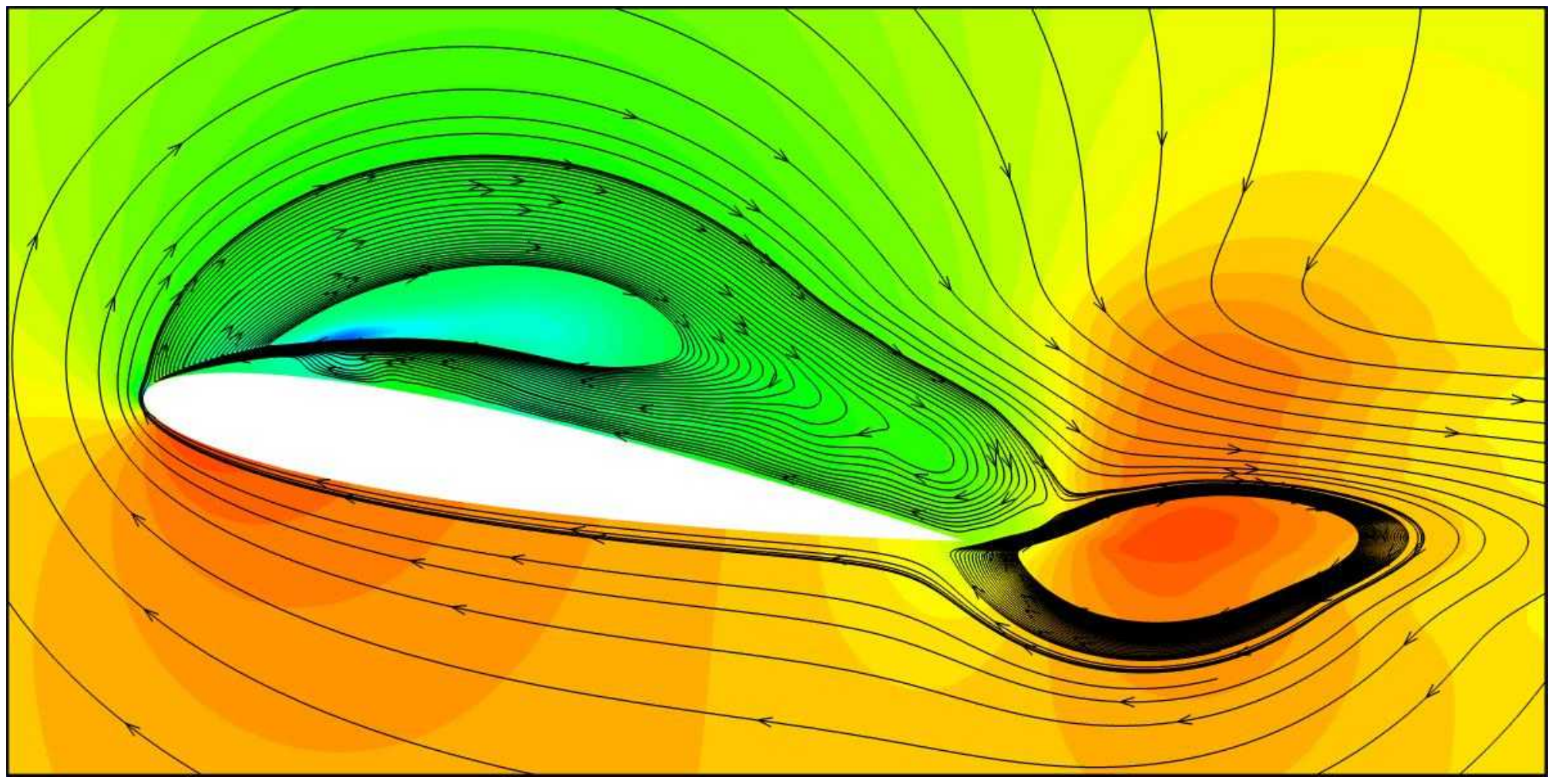}
\textit{$\alpha=10.0^{\circ}$, LFO 2}
\end{minipage}
\begin{minipage}{220pt}
\centering
\includegraphics[width=220pt, trim={0mm 0mm 0mm 0mm}, clip]{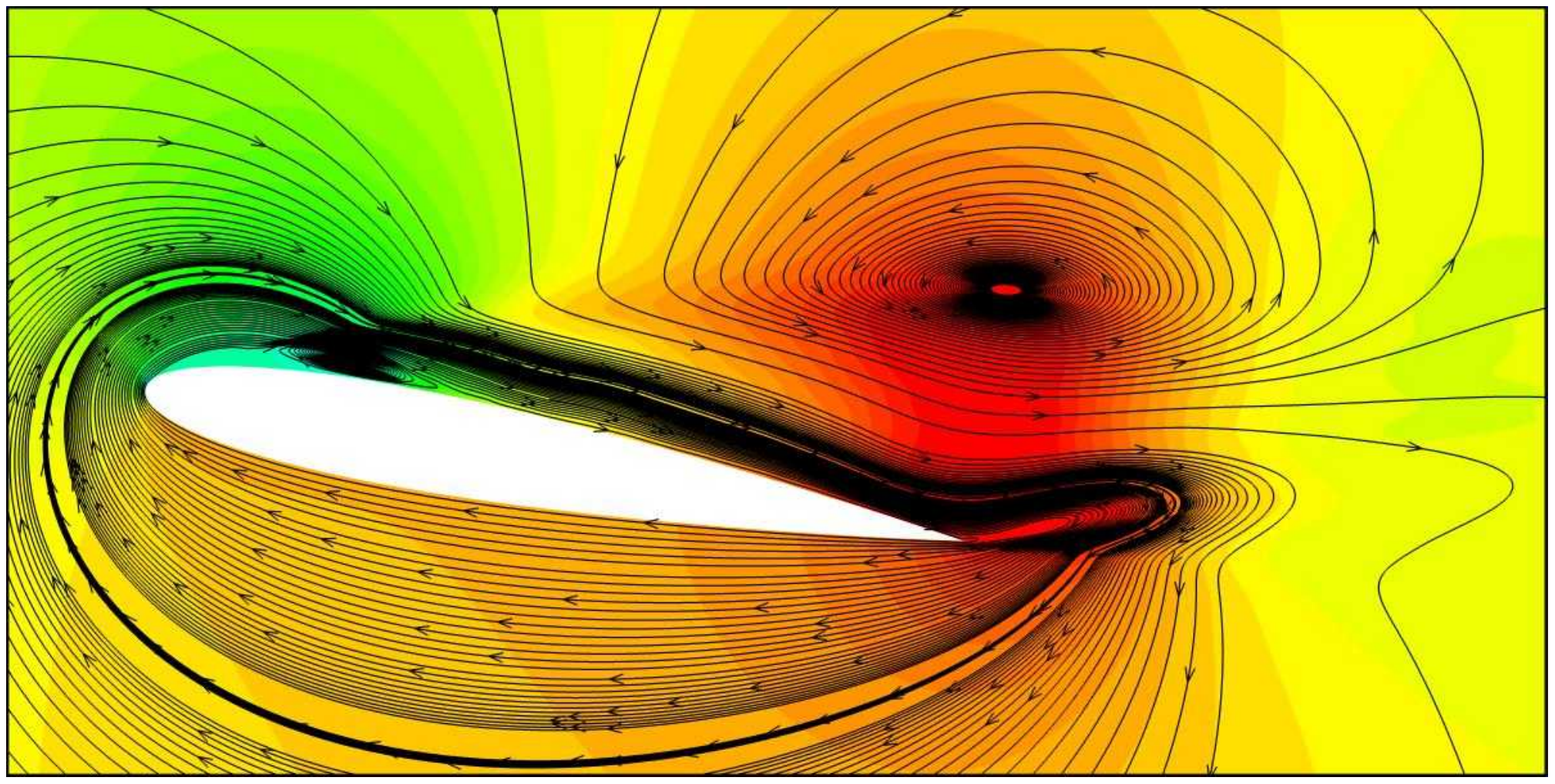}
\textit{$\alpha = 10.5^{\circ}$, LFO 1}
\end{minipage}
\begin{minipage}{220pt}
\centering
\includegraphics[width=220pt, trim={0mm 0mm 0mm 0mm}, clip]{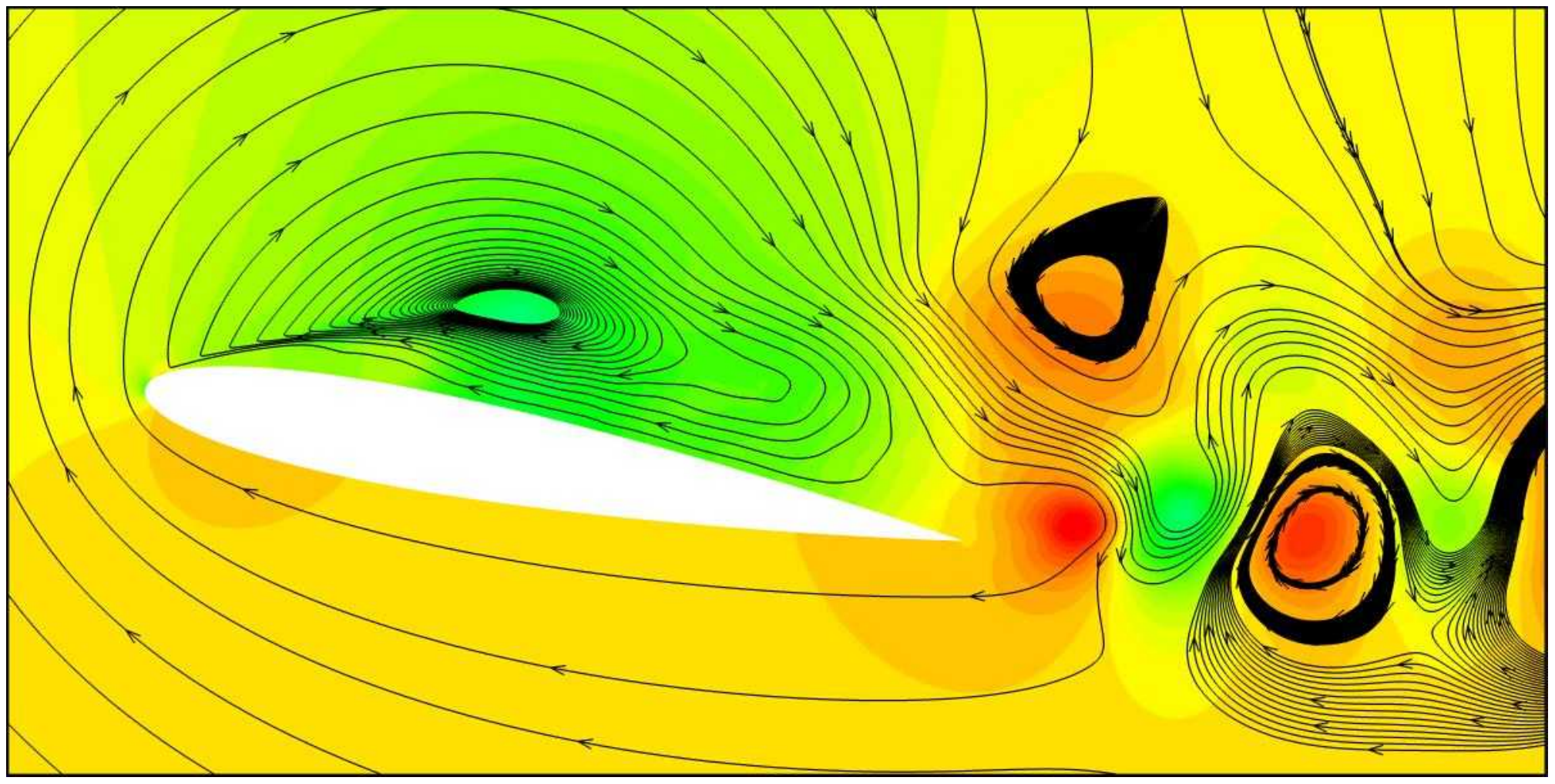}
\textit{$\alpha = 10.5^{\circ}$, LFO 2}
\end{minipage}
\caption{Streamlines patterns superimposed on colour maps of the pressure f\/ield for the POD reconstruction of the oscillating-f\/low using the LFO mode 1 and the LFO mode 2 for the angles of attack of $9.8^{\circ}$--$10.5^{\circ}$.}
\label{POD_LFO1_LFO2_2}
\end{center}
\end{figure}

The POD eigenvalues are then utilized to estimate the cumulative energy content using equation~\ref{POD_xi_t}. F\/igure~\ref{POD_cumulative_energy} shows the cumulative energy plotted versus the number of POD modes used to estimate it. The arrow denotes the direction in which the angle of attack increases. The range of angles of attack is $\alpha=9.25^{\circ}$--$9.8^{\circ}$ on the left-hand side of the f\/igure, and $\alpha=9.8^{\circ}$--$10.5^{\circ}$ on the right-hand side. The angle of attack of $9.8^{\circ}$ is duplicated in both f\/igures because it is the angle at which the POD modes have the fastest convergence to the total energy. That is, the minimum number of POD modes used to attain more than $99\%$ of the total energy. Thus, it is displayed in both f\/igures to compare the convergence of POD modes at other angles of attack. As seen in the f\/igure, at the angle of attack of $9.25^{\circ}$ the cumulative energy converges slowly towards $1$ as the number of POD modes approaches $250$ modes. The cumulative energy converges faster as the angle of attack increases. The fastest cumulative energy convergence is achieved at the angle of attack of $9.8^{\circ}$ as mentioned before. The convergence process of the POD energy slows down as the angle of attack is further increased as seen on the right-hand side of the f\/igure.\newline
The percentage of the energy content of each POD mode, $\xi^{(n)}$, is shown in f\/igure~\ref{POD_energy_percentage}. The energy percentage is estimated from the fraction of each POD eigenvalues, $\lambda^{(n)}$, to the sum of all POD eigenvalues using equation~\ref{POD_xi}. The f\/illed black, red, and blue bars denote the energy percentage of the LFO mode 1, the LFO mode 2, and the HFO mode, respectively. The LFO mode 1 dominates the f\/low-f\/ield in all of the investigated angles of attack. The LFO mode 2 is the second most dominant f\/low mode in the range of angle of attack of $9.4^{\circ} \leq \alpha \leq 10.0^{\circ}$, the third at $\alpha=9.25^{\circ}$, the fourth at $\alpha=10.1^{\circ}$, and the sixth most dominant f\/low mode at the angle of attack of $10.5^{\circ}$. The HFO mode is the second most dominant f\/low mode for the angles of attack $\alpha = 9.25^{\circ}$ and $\alpha \geq 10.1^{\circ}$, and the third most dominant f\/low mode in the range of angle of attack of $9.4^{\circ} \leq \alpha \leq 10.0^{\circ}$. The analysis of the energy content of each POD mode shows that the order of the POD modes does not change in the range of angles of attack of $9.25^{\circ} < \alpha < 10.1^{\circ}$. This is indicative that the range of angles of attack of interest is covered in this study as noted before in~\citet{eljack2018bursting1} and~\citet{eljack2018bursting2}. At the angle of attack of $9.8^{\circ}$, the three dominant POD modes contain more than 65\% of the energy. The remaining energy percentage is distributed among other higher POD modes.\newline
F\/igure~\ref{POD_Vectors} shows the temporal POD modes, $\varphi^{(n)}$, at the angle of attack of $9.7^{\circ}$--$10.0^{\circ}$. The LFO process is fully developed at this range of angles of attack. The left-hand side of the f\/igure shows the LFO mode 1, $\varphi^{(1)}$, denoted by the solid black line, and the LFO mode 2, $\varphi^{(2)}$, denoted by the solid red line. The HFO mode, $\varphi^{(3)}$, is shown on the right-hand side of the f\/igure. The two low-frequency modes are out of phase. The LFO mode 1 leads the LFO mode 2 at all of the investigated angles of attack by $\pi/2$. The HFO mode seems to f\/luctuate more energetically when the LFO mode 2 is at its maximum and minimum values.\newline
The temporal POD modes, $\varphi^{(n)}$, are not scaled with their corresponding eigenvalues, $\lambda^{(n)}$. The temporal POD modes show how a specif\/ic POD mode evolves in time regardless of its amplitude or fraction of energy compared to other POD modes at that instant. Therefore, the temporal POD modes are scaled with their corresponding eigenvalues, and the percentage of the scaled amplitude of each of the temporal POD modes is estimated at each instant in time as indicated by equation~\ref{eq:POD_zeta}. F\/igure~\ref{POD_zeta} shows the percentage of the scaled amplitudes of the oscillating-f\/low in each POD mode as a function of time at the angle of attack of $9.25^{\circ}$--$10.5^{\circ}$. The black, red, and blue lines display the LFO mode 1, the LFO mode 2, and the HFO mode, respectively. The scaled temporal POD modes make more sense since they feature the temporal evolution of the f\/low modes with their real magnitudes.\newline
At the angle of attack of $9.25^{\circ}$, the LFO mode 2 is actually a high-frequency mode. Moreover, it is less energetic than the HFO mode as shown in f\/igure~\ref{POD_energy_percentage}. Thus, the f\/low oscillates at a low-frequency, but the LFO mode 2 is not energetic enough to separate the f\/low. Therefore, the f\/low at this angle of attack remains attached with occasional separations at small amplitudes. As the angle of attack is increased to $9.4^{\circ}$, the LFO mode 1 becomes more pronounced in amplitude and repeats almost regularly. The LFO mode 2 becomes more signif\/icant and peaks at a percentage of more than $20\%$. The HFO mode oscillates at relatively smaller amplitude, but it seems to be more energetic when the LFO mode 1 vanishes and the LFO mode 2 peaks. As discussed before, the LFO mode 2 peaks when the UV of the TCV expands and advects downstream. Thus, it seems that when the UV expands and advects downstream it energies the f\/low oscillation at the trailing-edge and along the aerofoil wake. Hence, the HFO mode f\/luctuates at a relatively higher amplitude when the LFO mode 2 peaks. The magnitude of the maximum or minimum peak of the LFO mode 1 and 2 are interlinked. That is, if the LFO mode 1 peaks at a relatively high amplitude, so will the LFO mode 2, and vice versa.\newline
At the angles of attack of $9.7^{\circ}$--$10.0^{\circ}$, the LFO mode 1 peaks at 75\% and the LFO mode 2 peaks at about 25\%. At these angles, the cycle becomes more regular and repeats periodically with some disturbed cycles. when the periodic cycle of the LFO mode 1 is disturbed, so does that of the LFO mode 2. The LFO mode 2 becomes a high-frequency mode, again, at the angles of attack of $10.1^{\circ}$ and $10.5^{\circ}$ with a percentage comparable to that of the HFO mode. However, the percentage of the energy of the LFO mode 2 increases signif\/icantly when the percentage of the LFO mode 1 increases. The LFO mode 1 and 2 are out of phase by $\pi/2$. It has been consistently reported in the literature that there is a phase difference of $\pi/2$ between the lift and the drag coeff\/icient. This seems to be connected to the phase lag between LFO mode 1 and 2 as will be discussed later.\newline
F\/igures~\ref{POD_LFO1_LFO2_1} and~\ref{POD_LFO1_LFO2_2} show streamlines patterns superimposed on colour maps of the pressure f\/ield of the two most dominant low-frequency POD modes, the LFO mode 1 and the LFO mode 2, for the angles of attack of $9.25^{\circ}$--$10.5^{\circ}$. The left-hand side of the f\/igure displays the LFO mode 1, and the right-hand side of the f\/igure shows the LFO mode 2. The two low-frequency modes were constructed by multiplying their corresponding orthonormal spatial POD modes by the average amplitude of their corresponding POD coeff\/icients, $\overline{|a^{(1)}(t)|}$ and $\overline{|a^{(2)}(t)|}$. The POD analysis extracted only two low-frequency modes, the LFO mode 1 and the LFO mode 2. Whereas, the DMD analysis decomposed the low-frequency oscillating-f\/low into too many low-frequency modes. Furthermore, the low-frequency modes extracted by the POD method are more coherent and consistent compared to those extracted by the DMD method. The POD method objectively recovers coherent motion based on its energy content. Thus, the POD captured eff\/iciently the most energetic two low-frequency modes. On the contrary, the DMD method recovers dynamic modes based on their frequencies and growth rates. Thus, the DMD decomposes the signal of the oscillating-f\/low into too many modes to satisfy the sinusoidal nature of each mode. However, the DMD method provides the f\/low modes and their corresponding frequencies and growth rates. Thus, combining the two methods provided all the information that govern and control the dynamics of the f\/low-f\/ield. The POD method consistently constructed the LFO mode 1 and 2. Thus, description of the spatial evolution of these two modes will be discussed in accordance with the POD results. However, the DMD method provided their frequencies and growth rates. Hence, the frequency and growth of these f\/low modes will be discussed in accordance with the corresponding DMD results.\newline
The LFO mode 1 features the process that generates and sustains the triad of vortices, and the LFO mode 2 features the expansion and advection of the upstream vortex (UV) of the two co-rotating vortices (TCV) as discussed in~\citet{eljack2018bursting2}. The oscillating-f\/low is rotating in the clockwise and the f\/low is fully attached as seen on the left-hand side of the f\/igure. Thus, the UV of the TCV separates the f\/low as it expands and advects downstream. However, the process reverses its direction as the LFO mode 1 and 2 change their sign. At the angles of attack of $9.7^{\circ} \leq \alpha \leq 10.0^{\circ}$, the LFO mode 2 seems to interact directly with and energies the HFO mode. That is, as the UV of the TCV expands and advects downstream, it is attracted by the low-pressure region at the trailing-edge; thus, energies the trailing-edge shedding. At higher angles of attack, the LFO mode 2 breaks down into multiple vortices as it interacts with the trailing-edge instability. However, the LFO mode 2 energies the HFO mode and the trailing-edge shedding when the f\/low is fully attached and when it is fully separated. That is, the interaction takes place and the HFO mode becomes more energetic whenever LFO mode 2 peaks. In total agreement with the previous discussion about the scaled temporal POD modes, and contradicts the conclusion made by~\citet{almutairi2017dynamics} that the tailing-edge shedding energies when the f\/low is separated and dies down when the f\/low is attached.\newline
It is worth noting that, the POD method decomposes the high-frequency oscillating-f\/low into too many modes featuring the oscillating mode along the aerofoil wake. On the contrary, the DMD method effectively recovered the sinusoidal HFO mode. The HFO mode recovered by the POD method is similar to that constructed using the DMD method. However, the HFO mode constructed using the POD method is less coherent compared to its DMD counterpart. Thus, all discussions concerning the shape and coherence of the HFO mode is discussed in accordance with the DMD results.

\subsubsection{Reconstruction of the f\/low-f\/ield}\label{sec:POD_reconstruction}
\begin{figure}
\begin{center}
\begin{minipage}{220pt}
\centering
\includegraphics[width=220pt, trim={0mm 0mm 0mm 0mm}, clip]{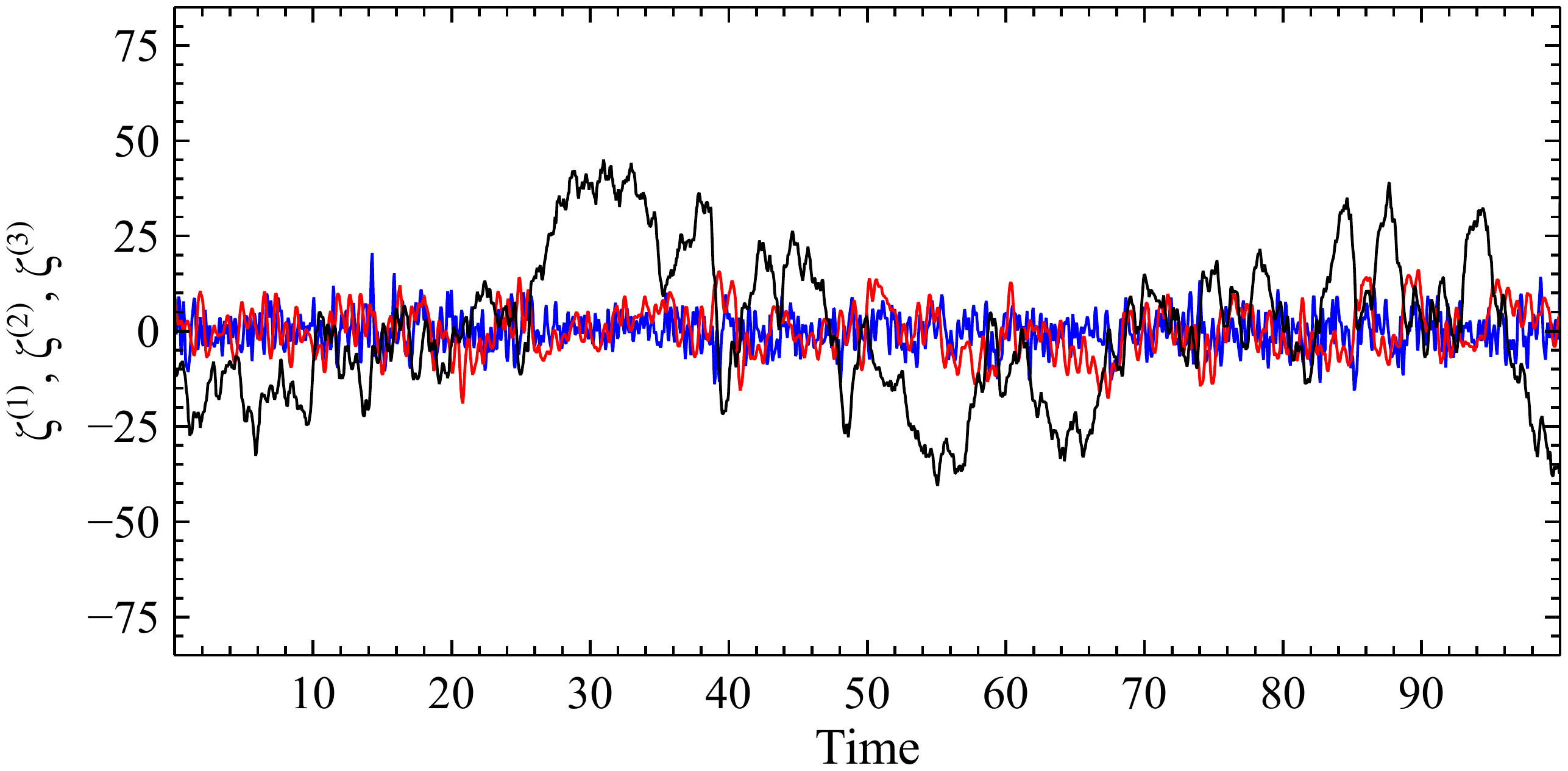}
\textit{$\alpha = 9.25^{\circ}$}
\end{minipage}
\begin{minipage}{220pt}
\centering
\includegraphics[width=220pt, trim={0mm 0mm 0mm 0mm}, clip]{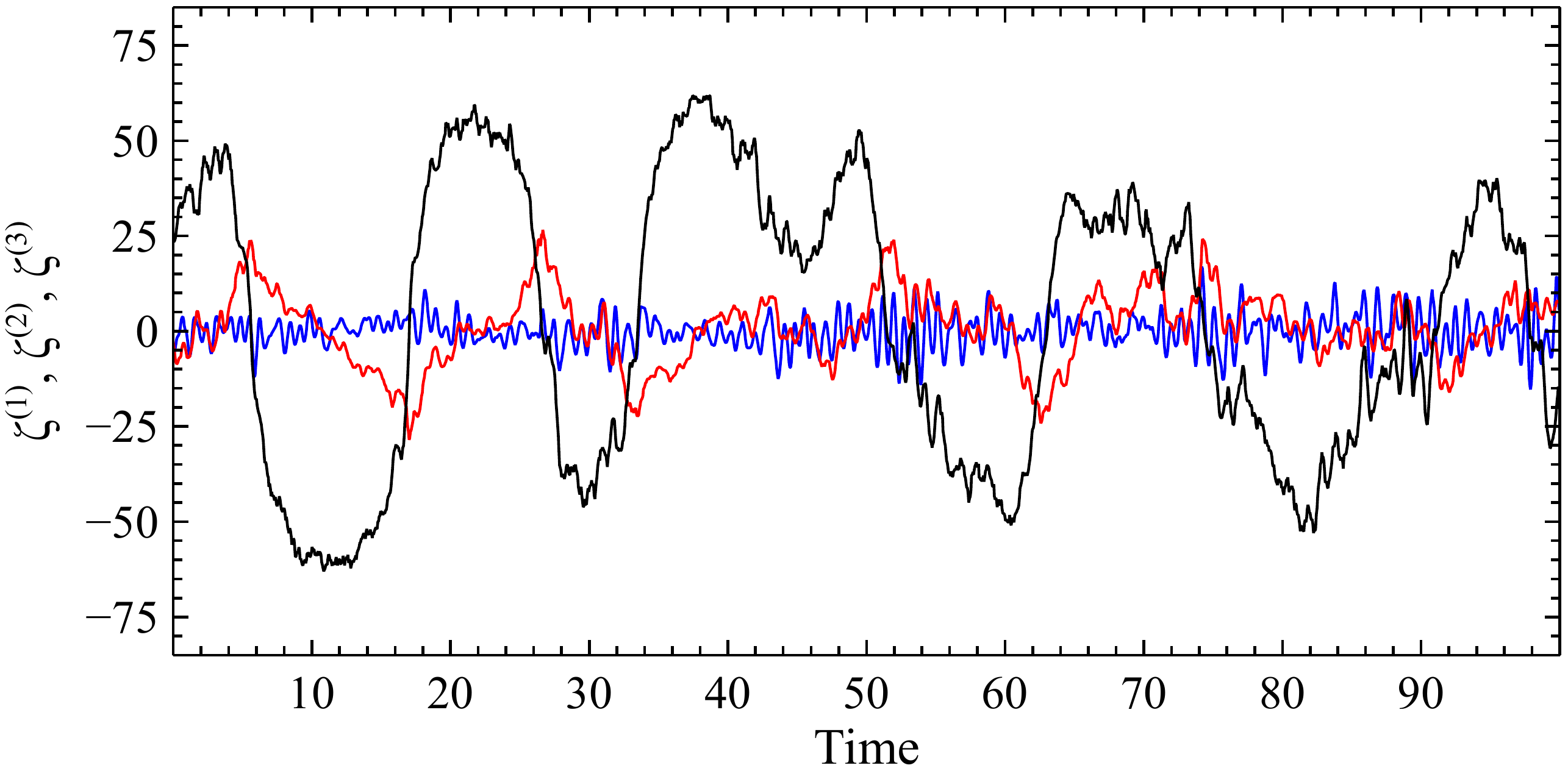}
\textit{$\alpha = 9.4^{\circ}$}
\end{minipage}
\begin{minipage}{220pt}
\centering
\includegraphics[width=220pt, trim={0mm 0mm 0mm 0mm}, clip]{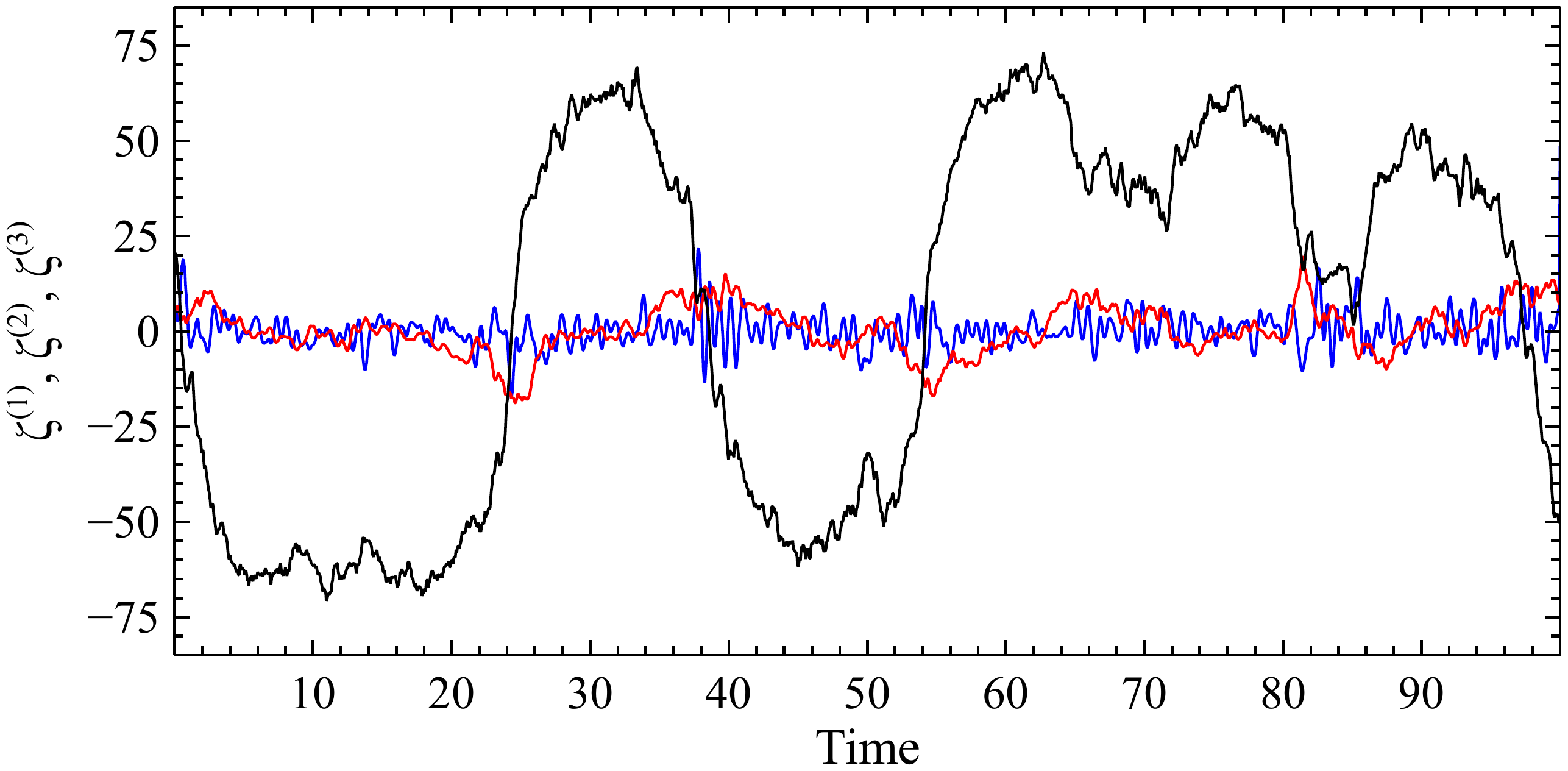}
\textit{$\alpha = 9.5^{\circ}$}
\end{minipage}
\begin{minipage}{220pt}
\centering
\includegraphics[width=220pt, trim={0mm 0mm 0mm 0mm}, clip]{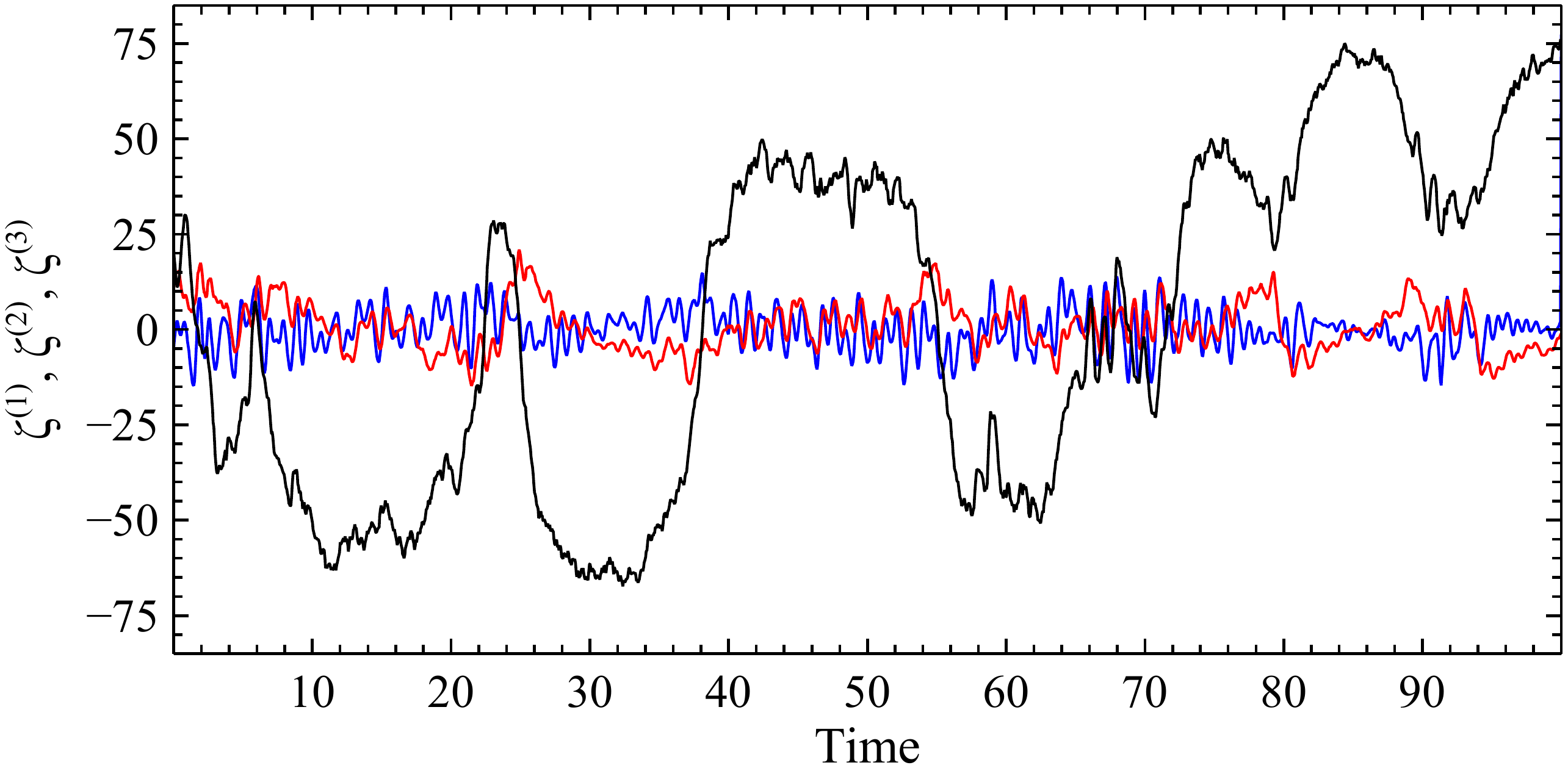}
\textit{$\alpha = 9.6^{\circ}$}
\end{minipage}
\begin{minipage}{220pt}
\centering
\includegraphics[width=220pt, trim={0mm 0mm 0mm 0mm}, clip]{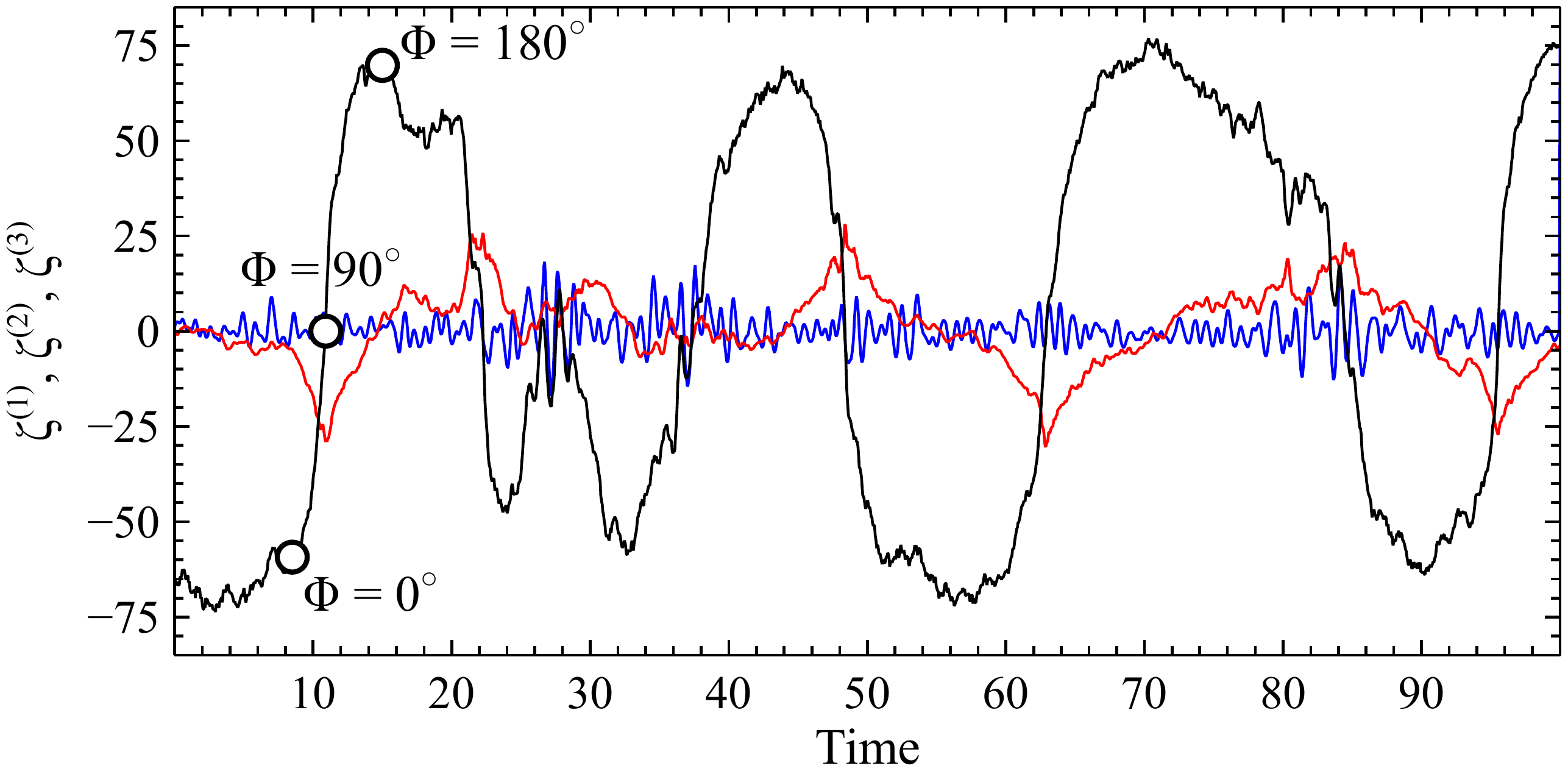}
\textit{$\alpha = 9.7^{\circ}$}
\end{minipage}
\begin{minipage}{220pt}
\centering
\includegraphics[width=220pt, trim={0mm 0mm 0mm 0mm}, clip]{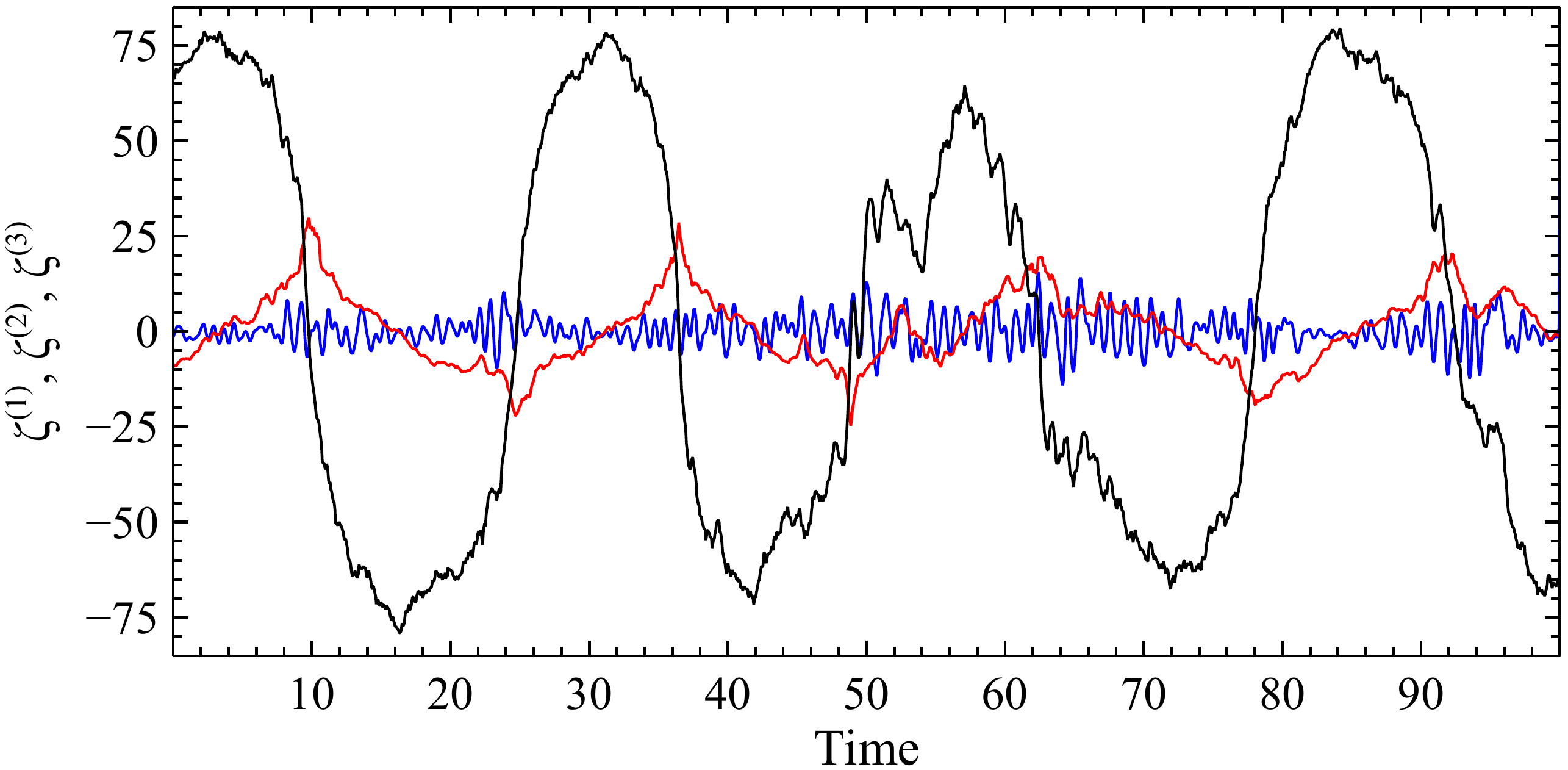}
\textit{$\alpha = 9.8^{\circ}$}
\end{minipage}
\begin{minipage}{220pt}
\centering
\includegraphics[width=220pt, trim={0mm 0mm 0mm 0mm}, clip]{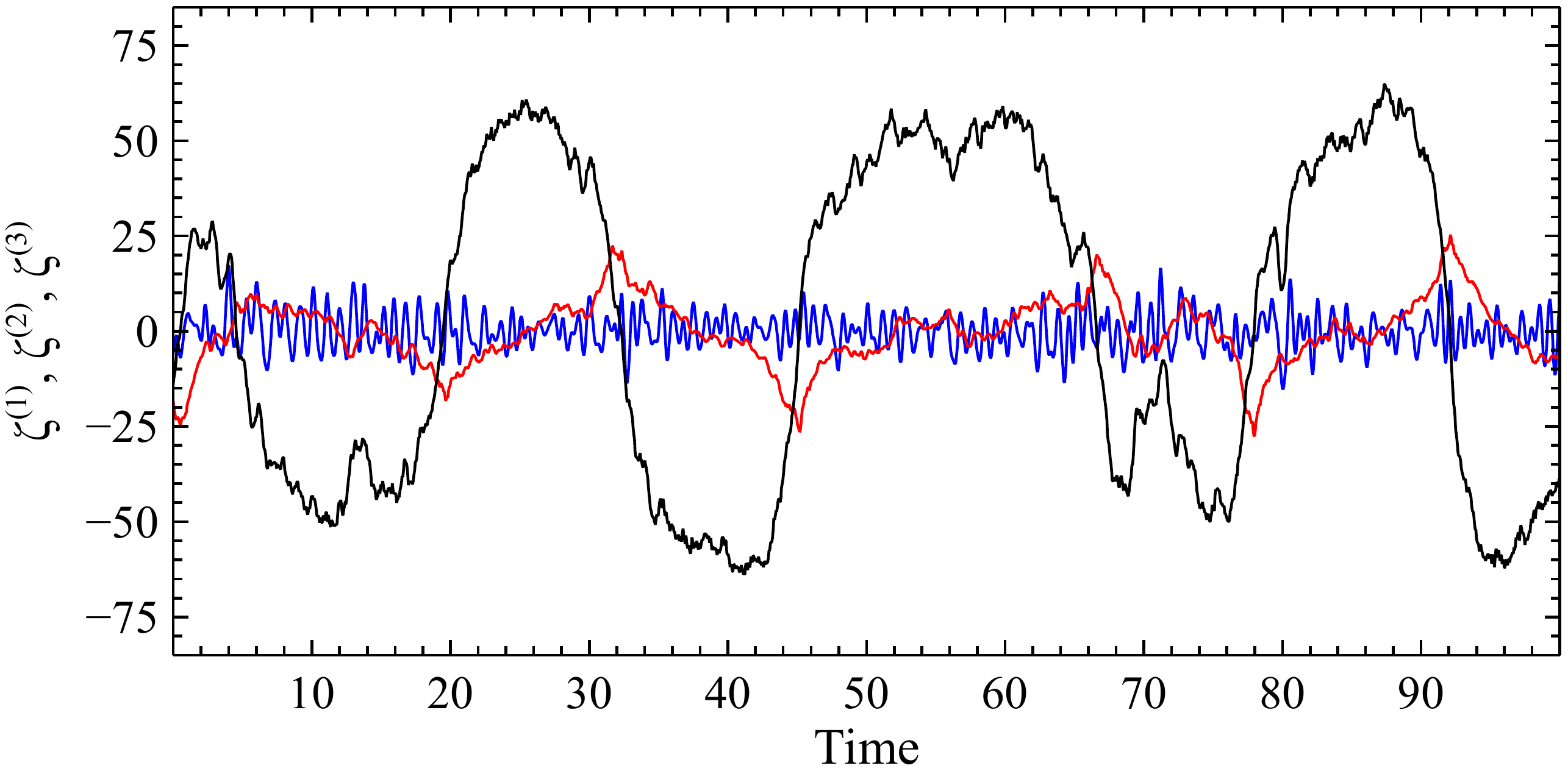}
\textit{$\alpha = 9.9^{\circ}$}
\end{minipage}
\begin{minipage}{220pt}
\centering
\includegraphics[width=220pt, trim={0mm 0mm 0mm 0mm}, clip]{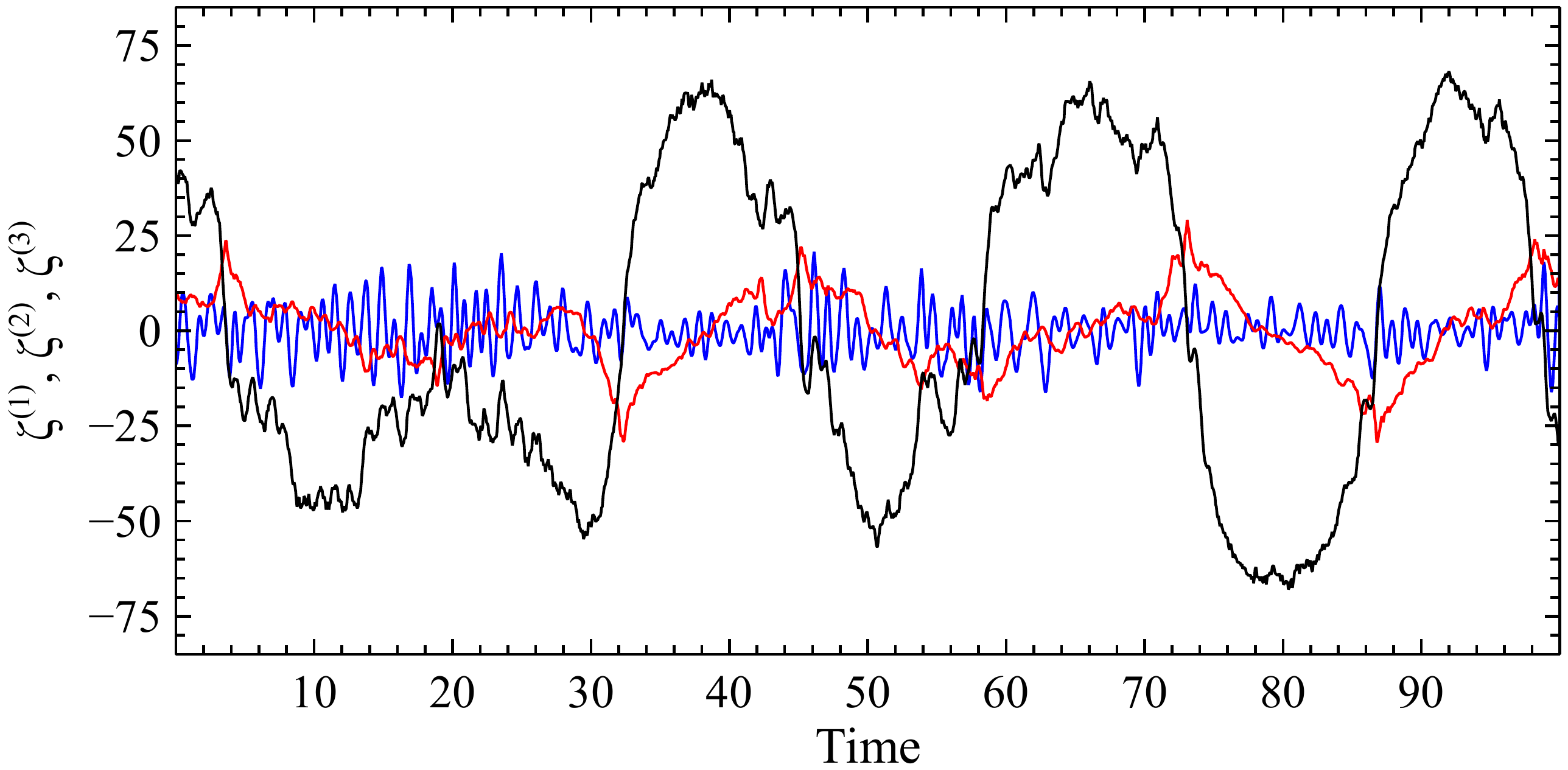}
\textit{$\alpha = 10.0^{\circ}$}
\end{minipage}
\begin{minipage}{220pt}
\centering
\includegraphics[width=220pt, trim={0mm 0mm 0mm 0mm}, clip]{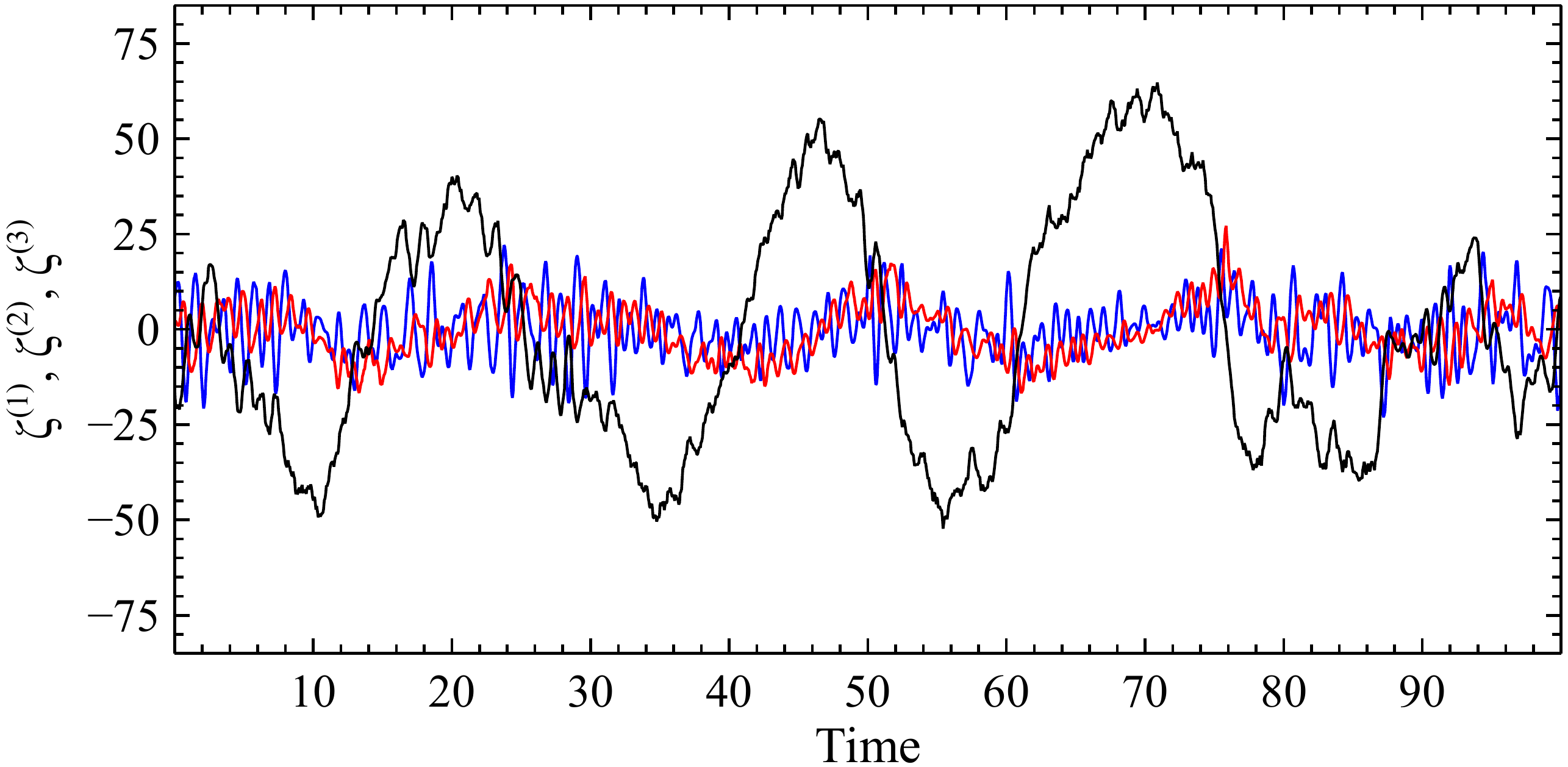}
\textit{$\alpha = 10.1^{\circ}$}
\end{minipage}
\begin{minipage}{220pt}
\centering
\includegraphics[width=220pt, trim={0mm 0mm 0mm 0mm}, clip]{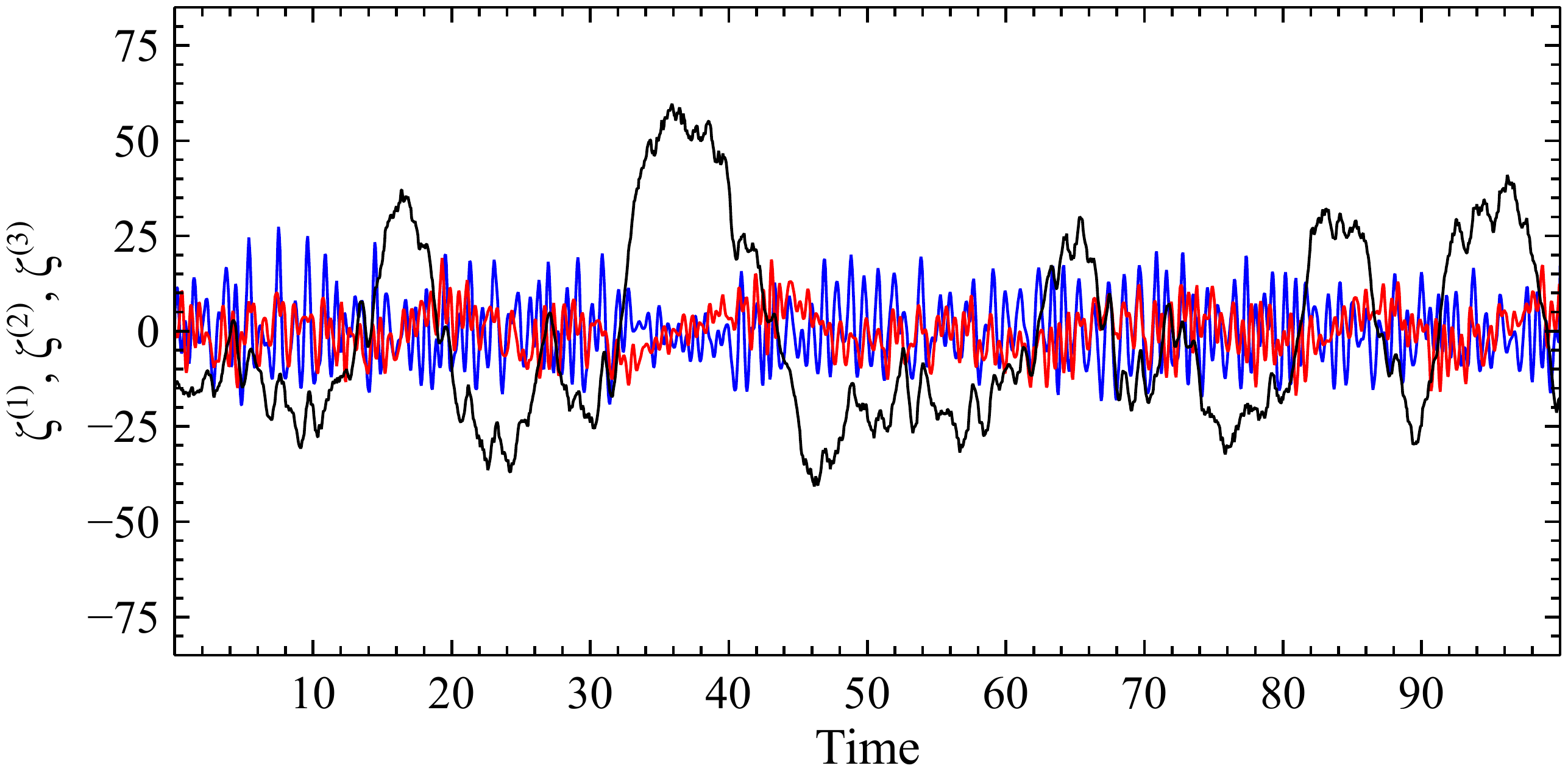}
\textit{$\alpha = 10.5^{\circ}$}
\end{minipage}
\caption{The percentage of the scaled amplitudes of the oscillating-f\/low in each POD mode as a function of time for the angles of attack of $9.25^{\circ}$--$10.5^{\circ}$. The black line displays the LFO mode 1, the red line indicates the LFO mode 2, and the blue line denotes the HFO mode.}
\label{POD_zeta}
\end{center}
\end{figure}

\begin{figure}
\begin{center}
\begin{minipage}{220pt}
\centering
\includegraphics[width=220pt, trim={0mm 0mm 0mm 0mm}, clip]{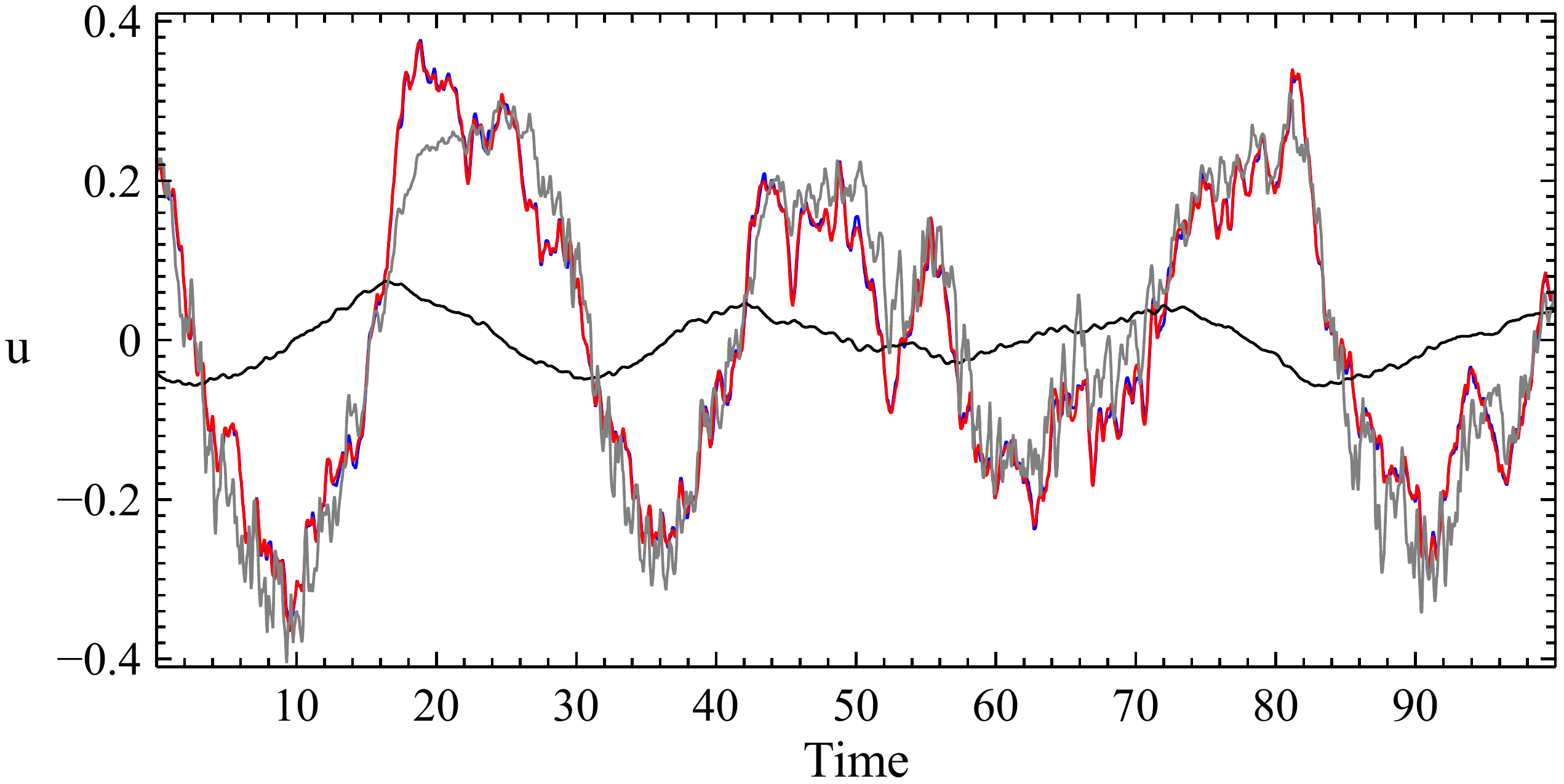}
\textit{(a) The streamwise velocity at the leading-edge}
\end{minipage}
\medskip
\begin{minipage}{220pt}
\centering
\includegraphics[width=220pt, trim={0mm 0mm 0mm 0mm}, clip]{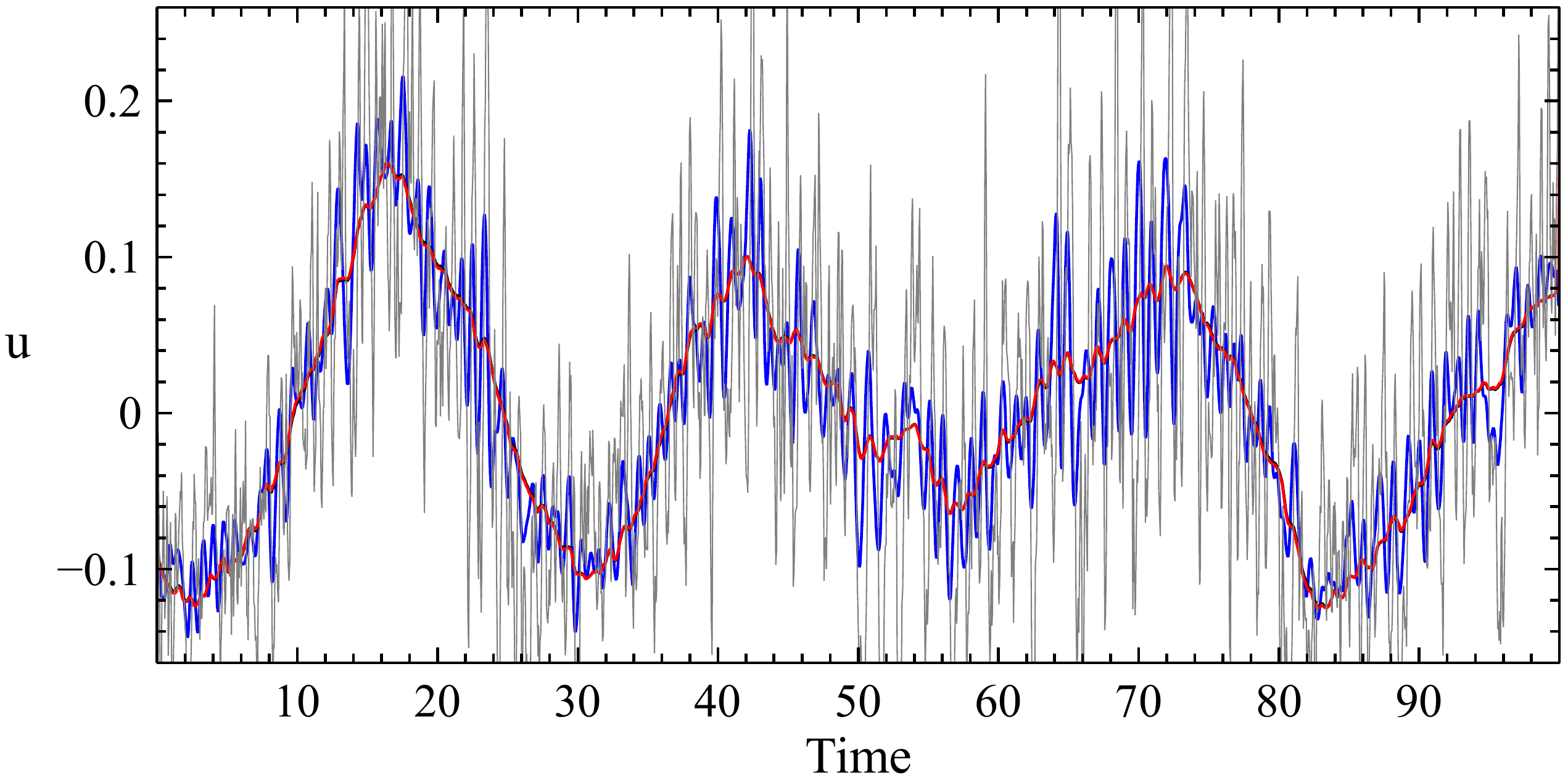}
\textit{(b) The streamwise velocity at the trailing-edge}
\end{minipage}
\medskip
\begin{minipage}{220pt}
\centering
\includegraphics[width=220pt, trim={0mm 0mm 0mm 0mm}, clip]{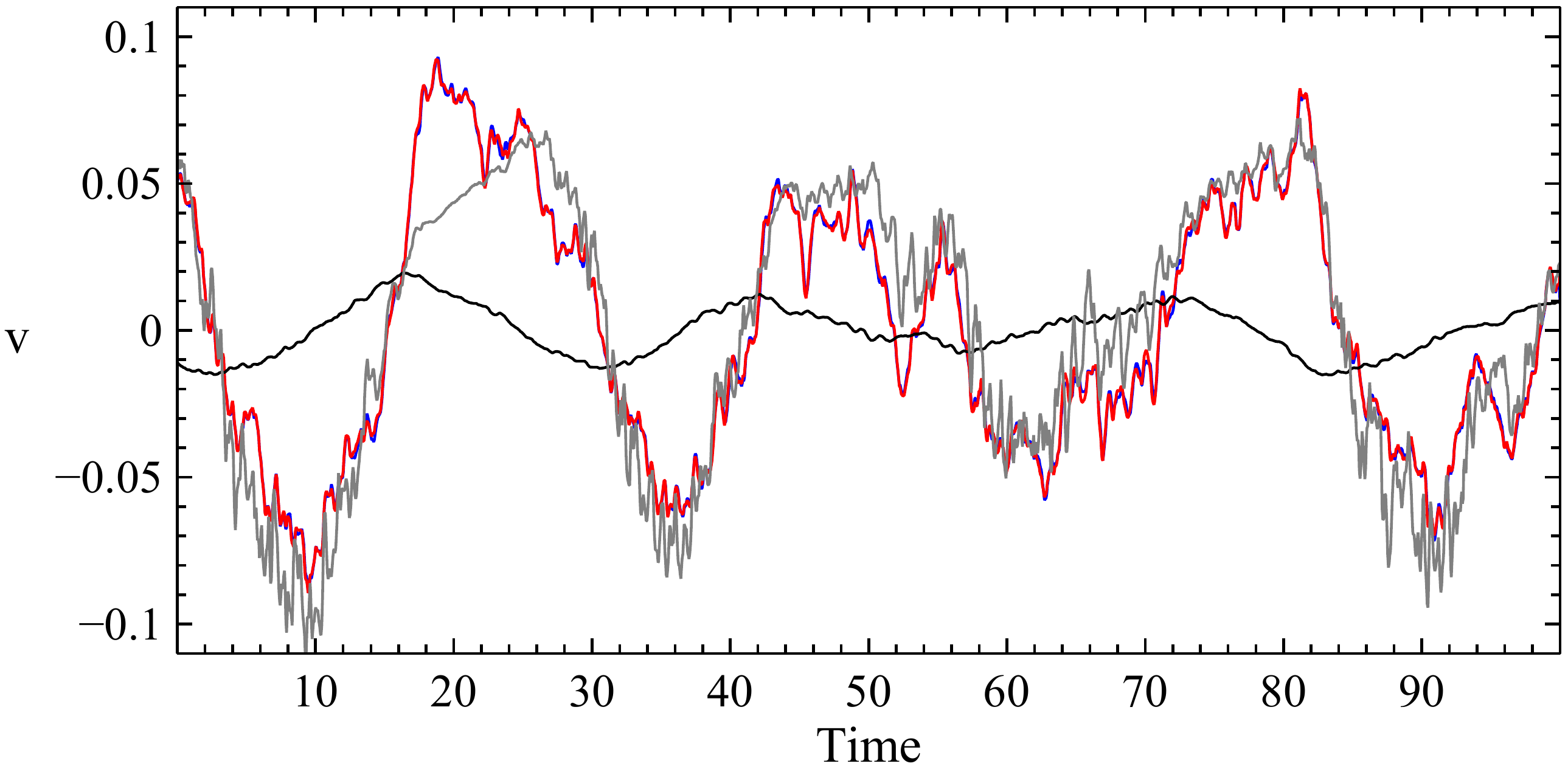}
\textit{(c) The wall-normal velocity at the leading-edge}
\end{minipage}
\medskip
\begin{minipage}{220pt}
\centering
\includegraphics[width=220pt, trim={0mm 0mm 0mm 0mm}, clip]{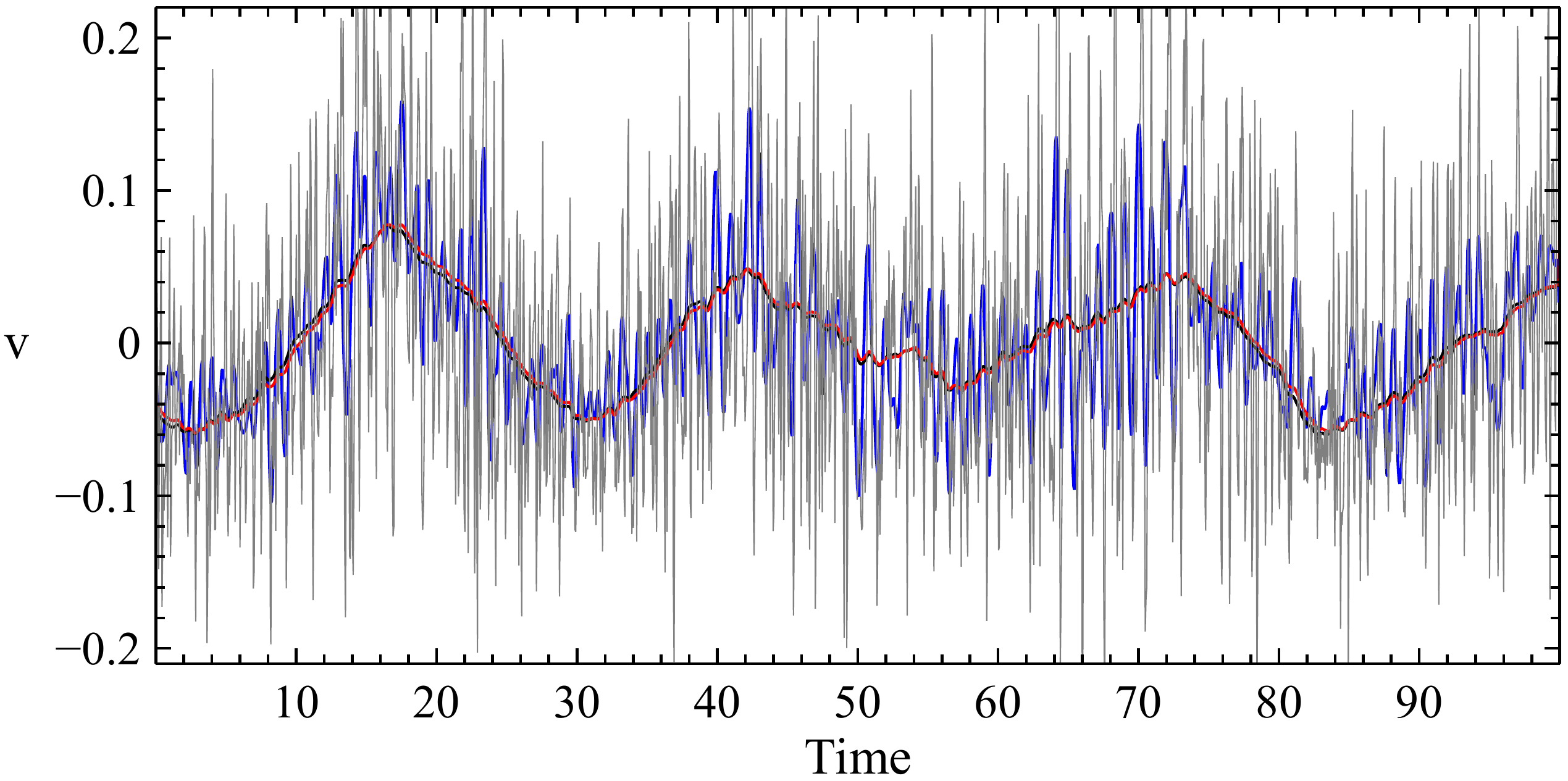}
\textit{(d) The wall-normal velocity at the trailing-edge}
\end{minipage}
\medskip
\begin{minipage}{220pt}
\centering
\includegraphics[width=220pt, trim={0mm 0mm 0mm 0mm}, clip]{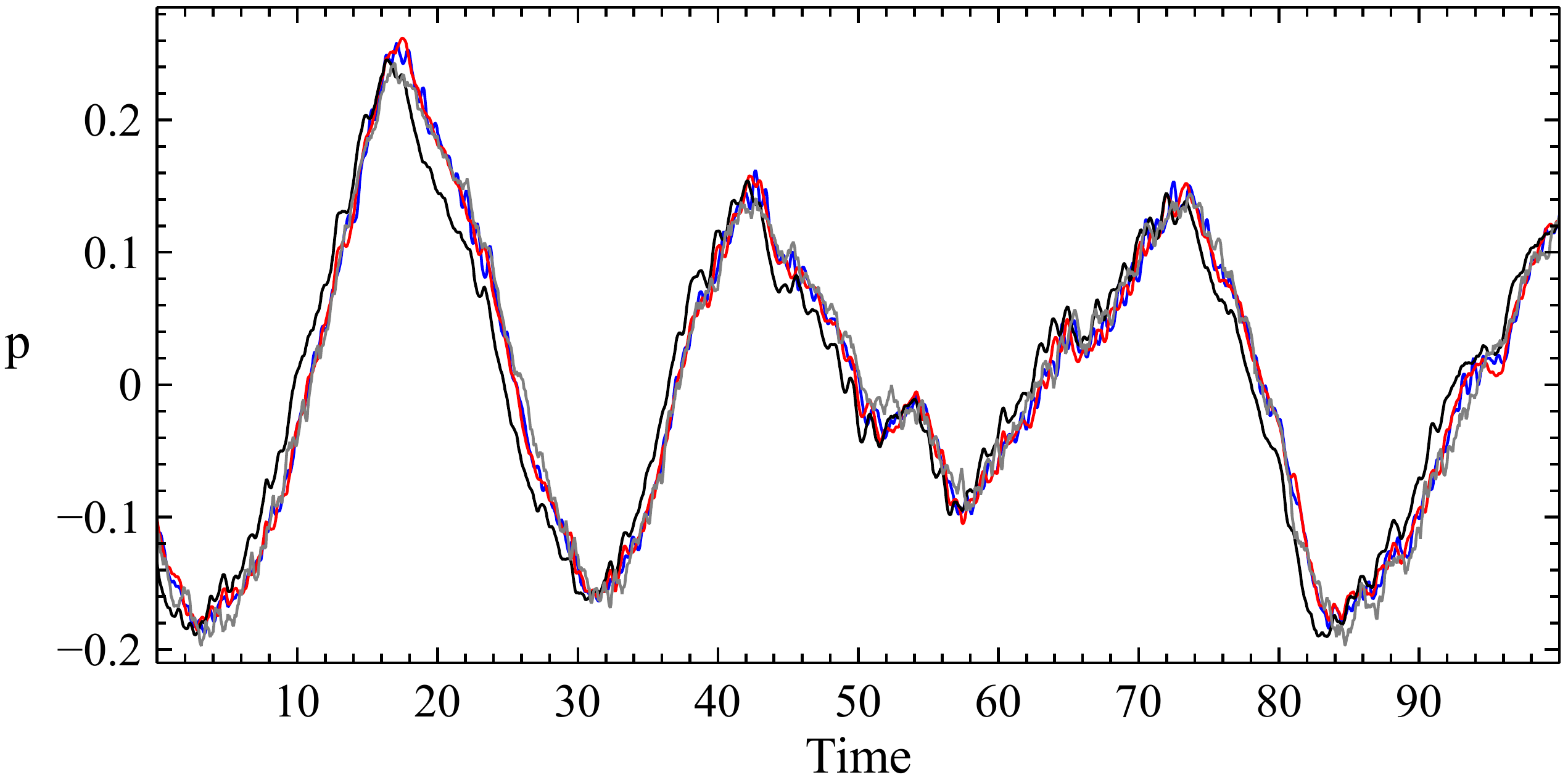}
\textit{(e) The pressure at the leading-edge}
\end{minipage}
\begin{minipage}{220pt}
\centering
\includegraphics[width=220pt, trim={0mm 0mm 0mm 0mm}, clip]{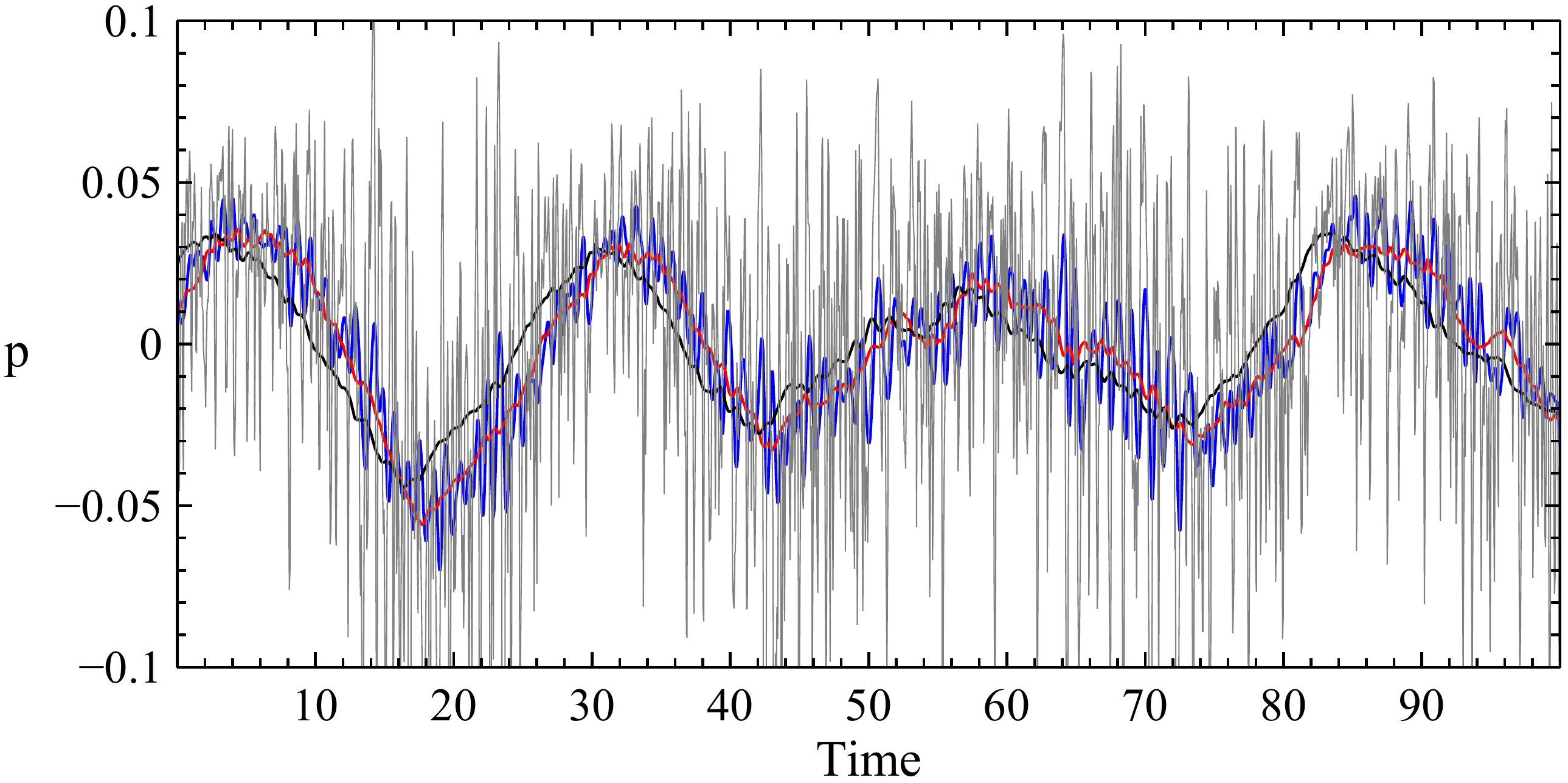}
\textit{(f) The pressure at the trailing-edge}
\end{minipage}
\caption{Signals of the LES data of the streamwise velocity, wall-normal velocity, and the pressure probed in the vicinity of the aerofoil leading and trailing edges compared to their corresponding POD reconstruction of the data at the angle of attack of $9.8^{\circ}$. Grey solid line: LES data; black solid line: reconstructed data using the LFO mode 1; red solid line: reconstructed data using the LFO mode 1 and the LFO mode 2; and the blue solid line: reconstructed data using the LFO mode 1, the LFO mode 2, and the HFO mode.}
\label{POD_rec_probes}
\end{center}
\end{figure}

\begin{figure}
\begin{center}
\begin{minipage}{220pt}
\centering
\includegraphics[width=220pt, trim={0mm 0mm 0mm 0mm}, clip]{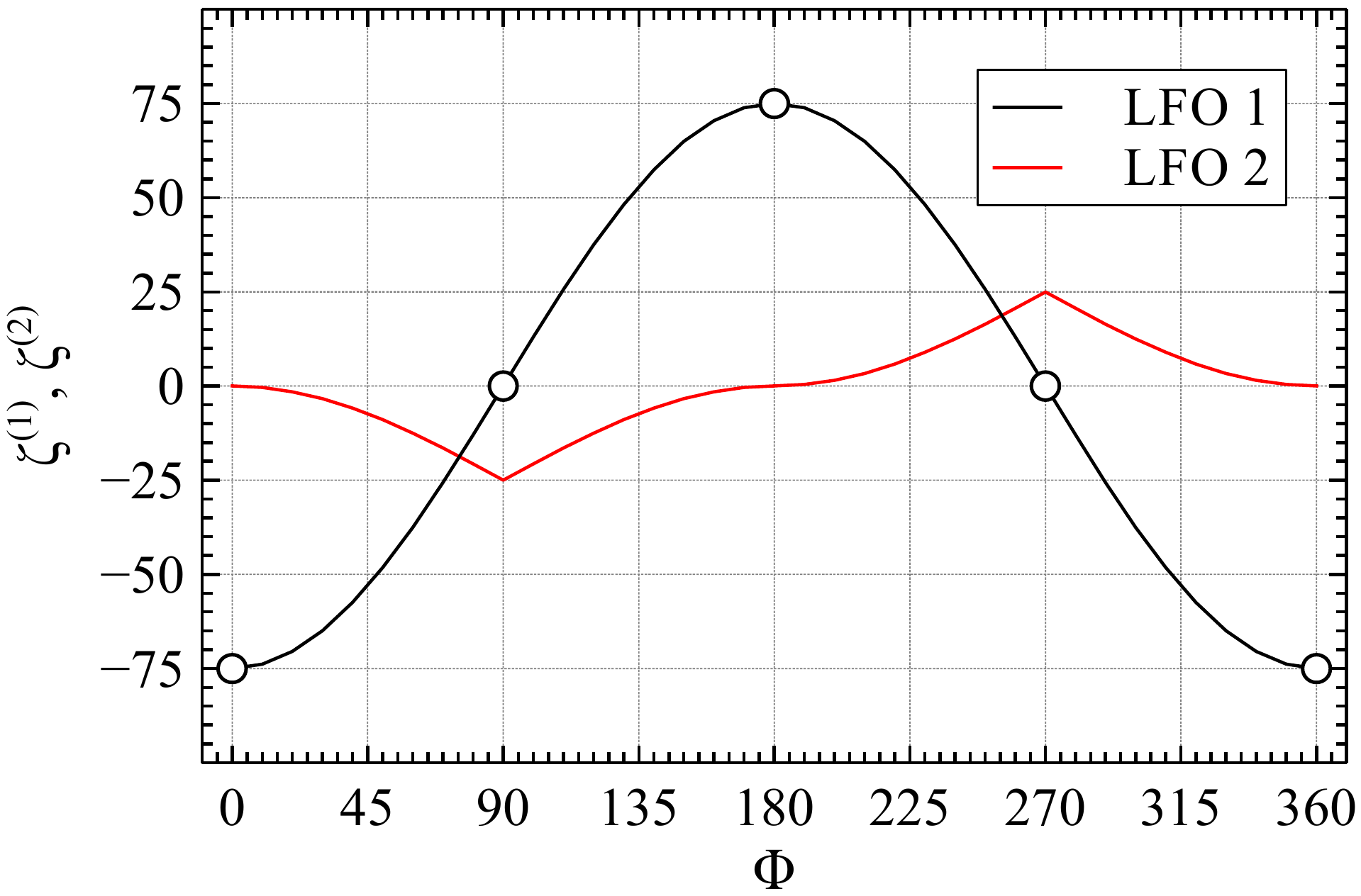}
\textit{(a) Ideal cycle of the LFO mode 1 and 2}
\end{minipage}
\begin{minipage}{220pt}
\centering
\includegraphics[width=220pt, trim={0mm 0mm 0mm 0mm}, clip]{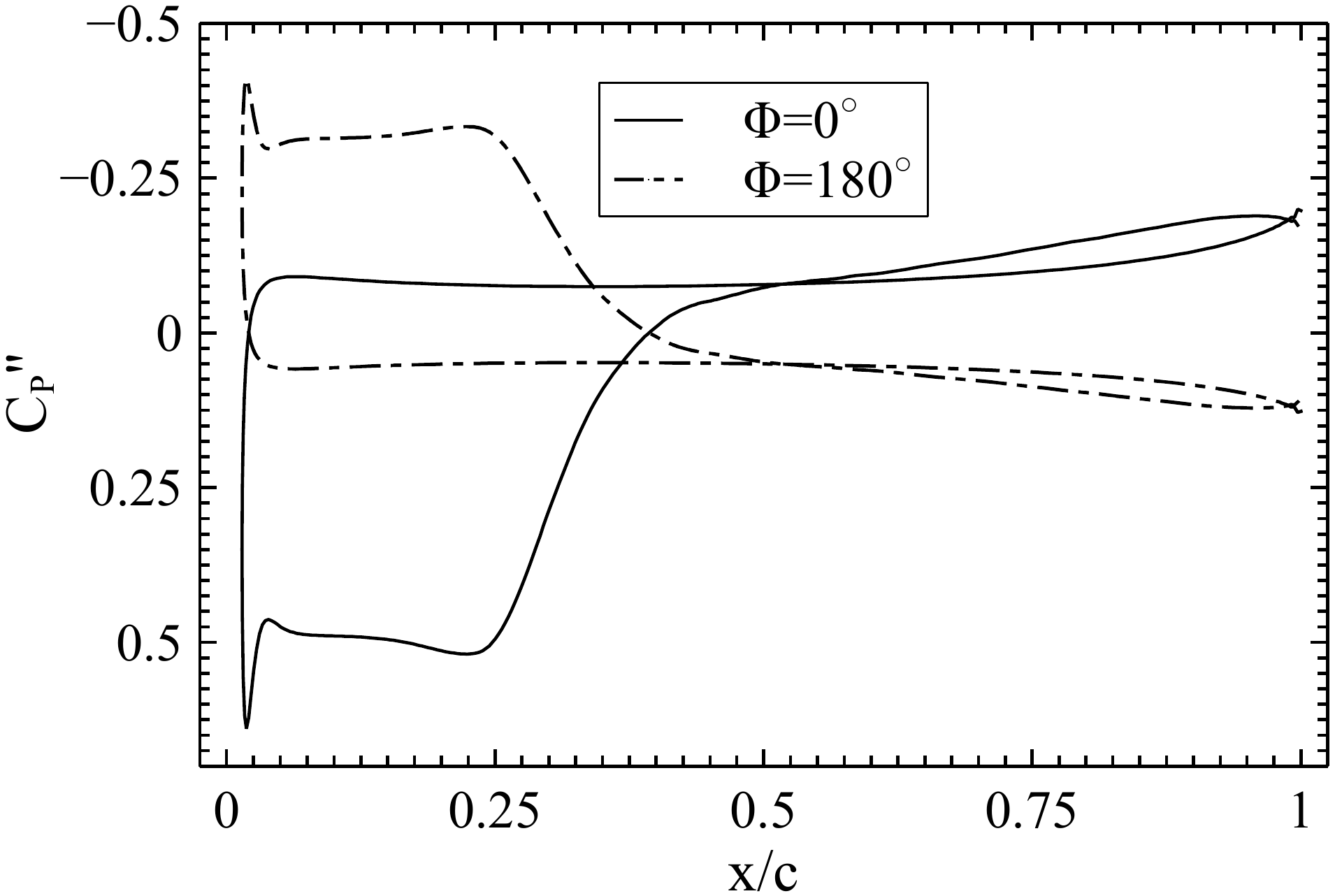}
\textit{(b) The LFO mode 1}
\end{minipage}
\begin{minipage}{220pt}
\centering
\includegraphics[width=220pt, trim={0mm 0mm 0mm 0mm}, clip]{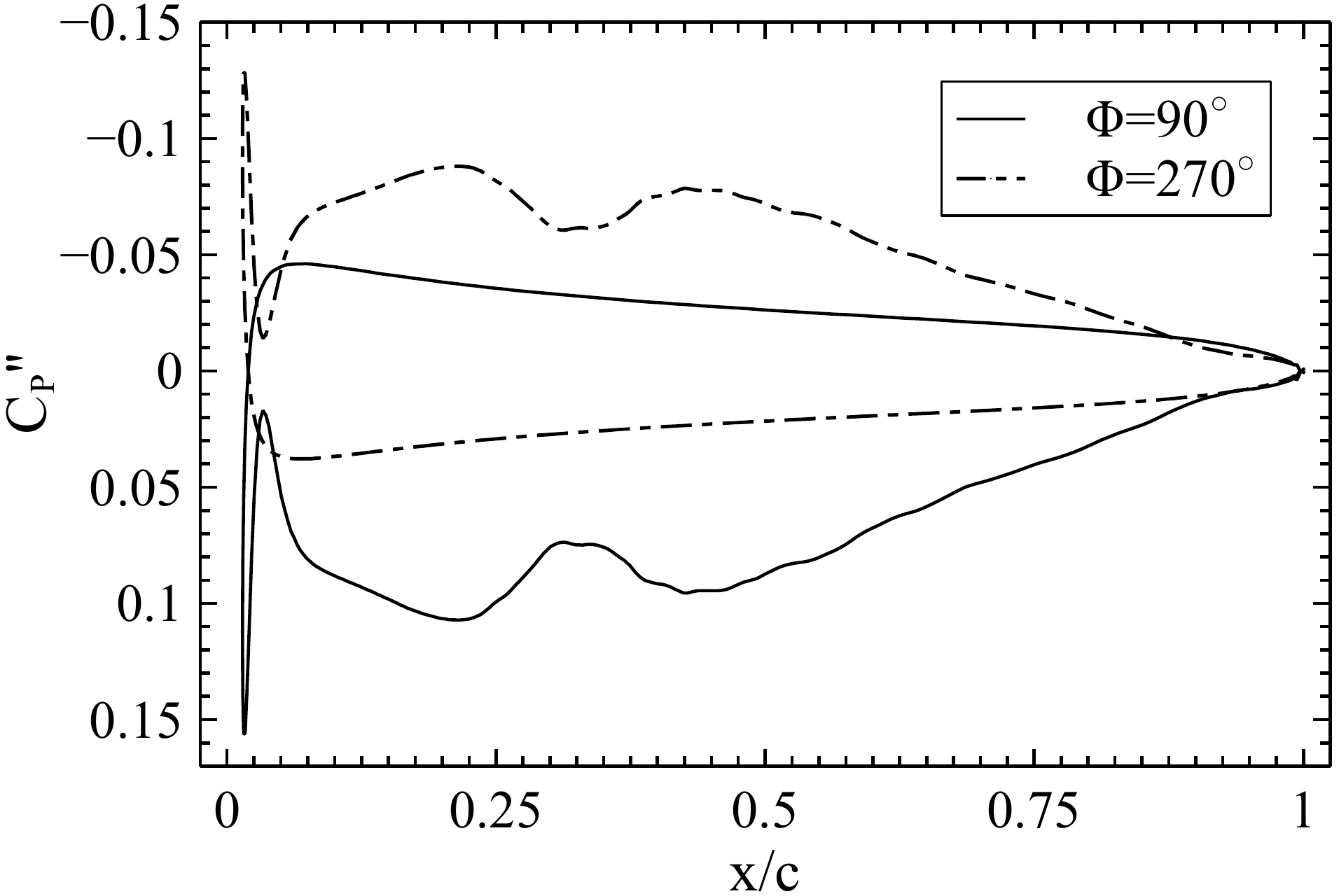}
\textit{(c) The LFO mode 2}
\end{minipage}
\begin{minipage}{220pt}
\centering
\includegraphics[width=220pt, trim={0mm 0mm 0mm 0mm}, clip]{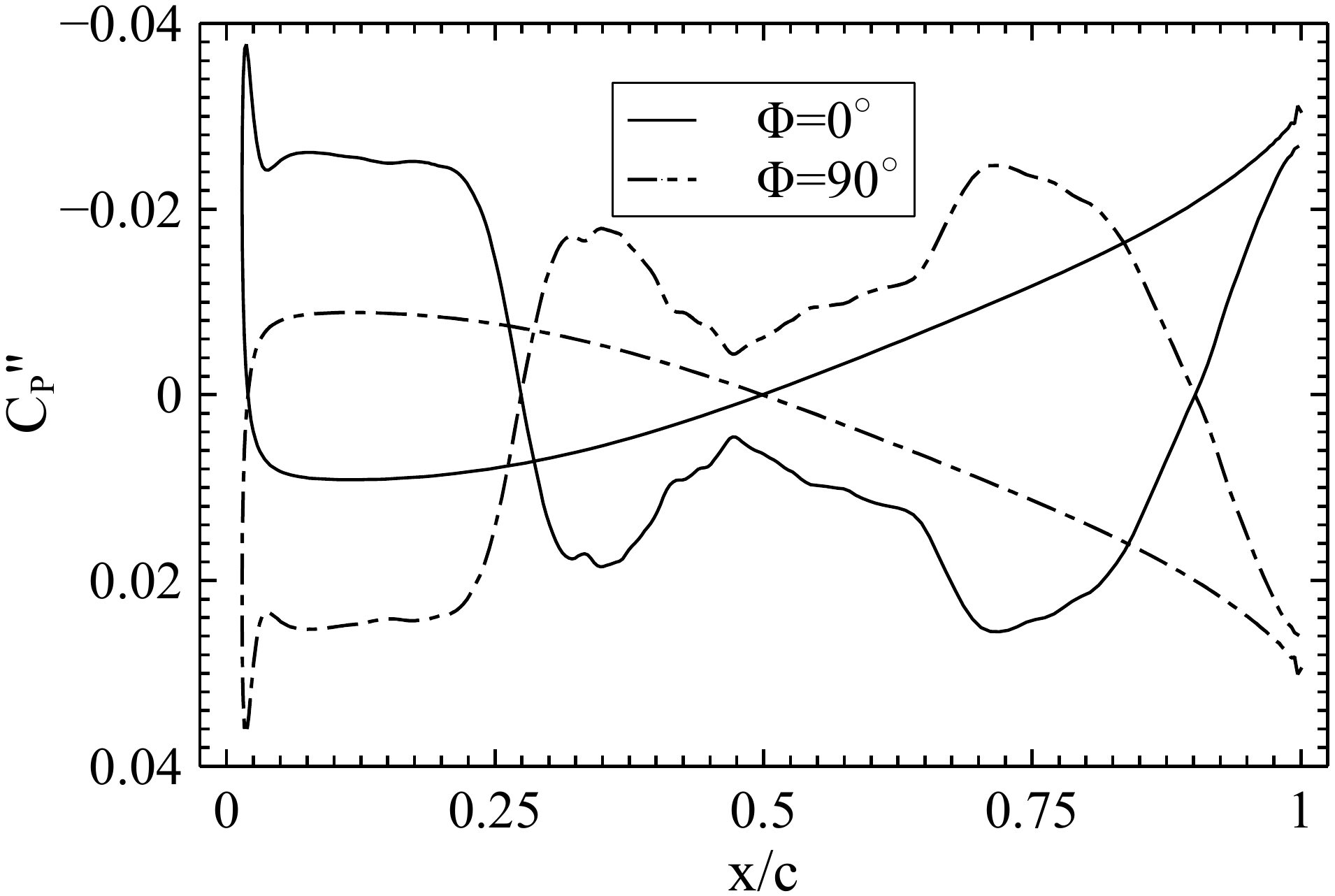}
\textit{(d) The HFO mode}
\end{minipage}
\caption{POD reconstruction of the oscillating-pressure-coeff\/icient using the LFO mode 1, the LFO mode 2, and the HFO mode at the phase angles of $\Phi=0^{\circ}$, $90^{\circ}$, $180^{\circ}$, and $270^{\circ}$ at the angle of attack of $9.8^{\circ}$.}
\label{POD_rec_Cp}
\end{center}
\end{figure}

\begin{figure}
\begin{center}
\begin{minipage}{220pt}
\centering
\includegraphics[width=220pt, trim={0mm 0mm 0mm 0mm}, clip]{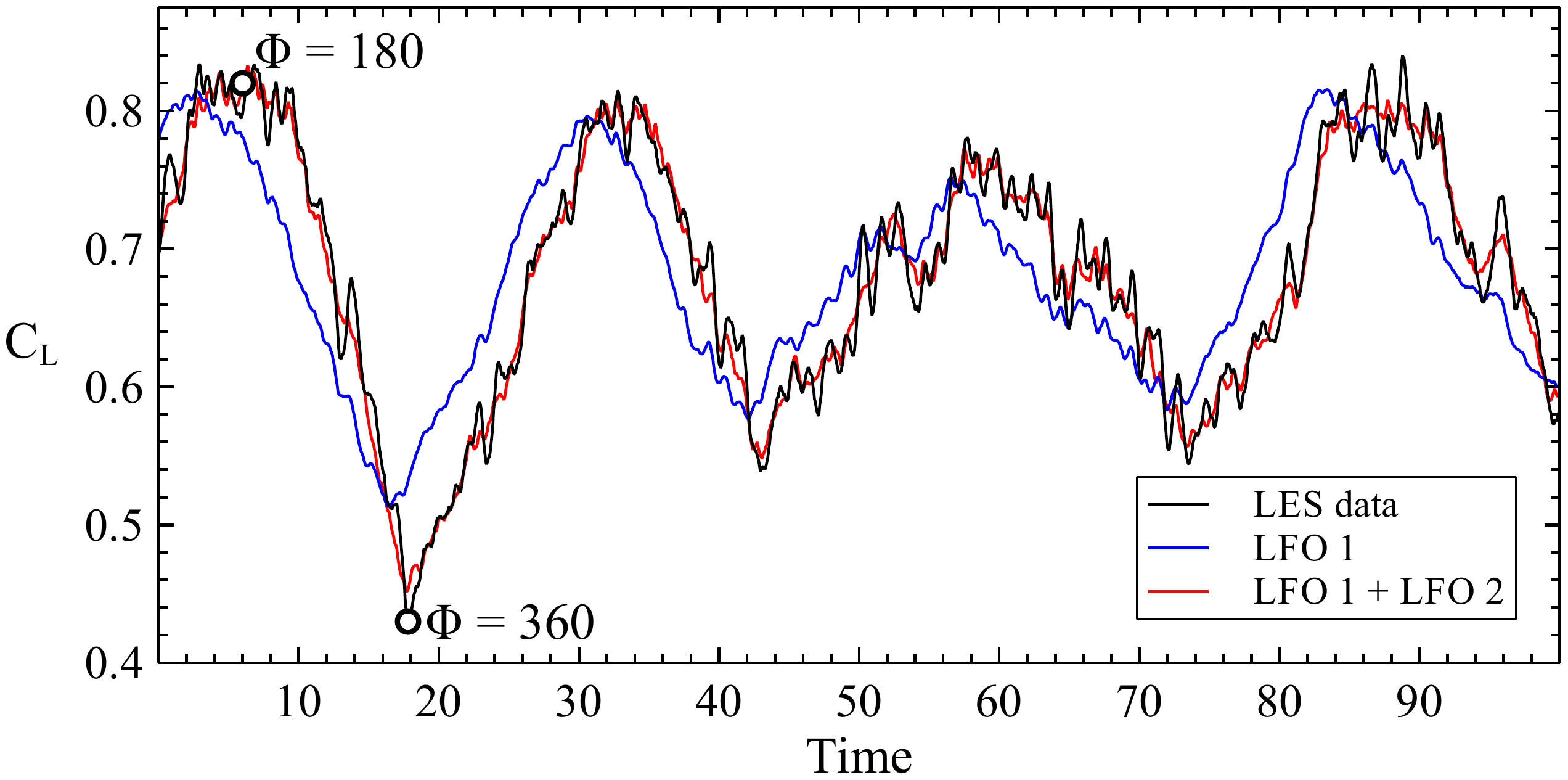}
\textit{(a) The lift coefficient}
\end{minipage}
\begin{minipage}{220pt}
\centering
\includegraphics[width=220pt, trim={0mm 0mm 0mm 0mm}, clip]{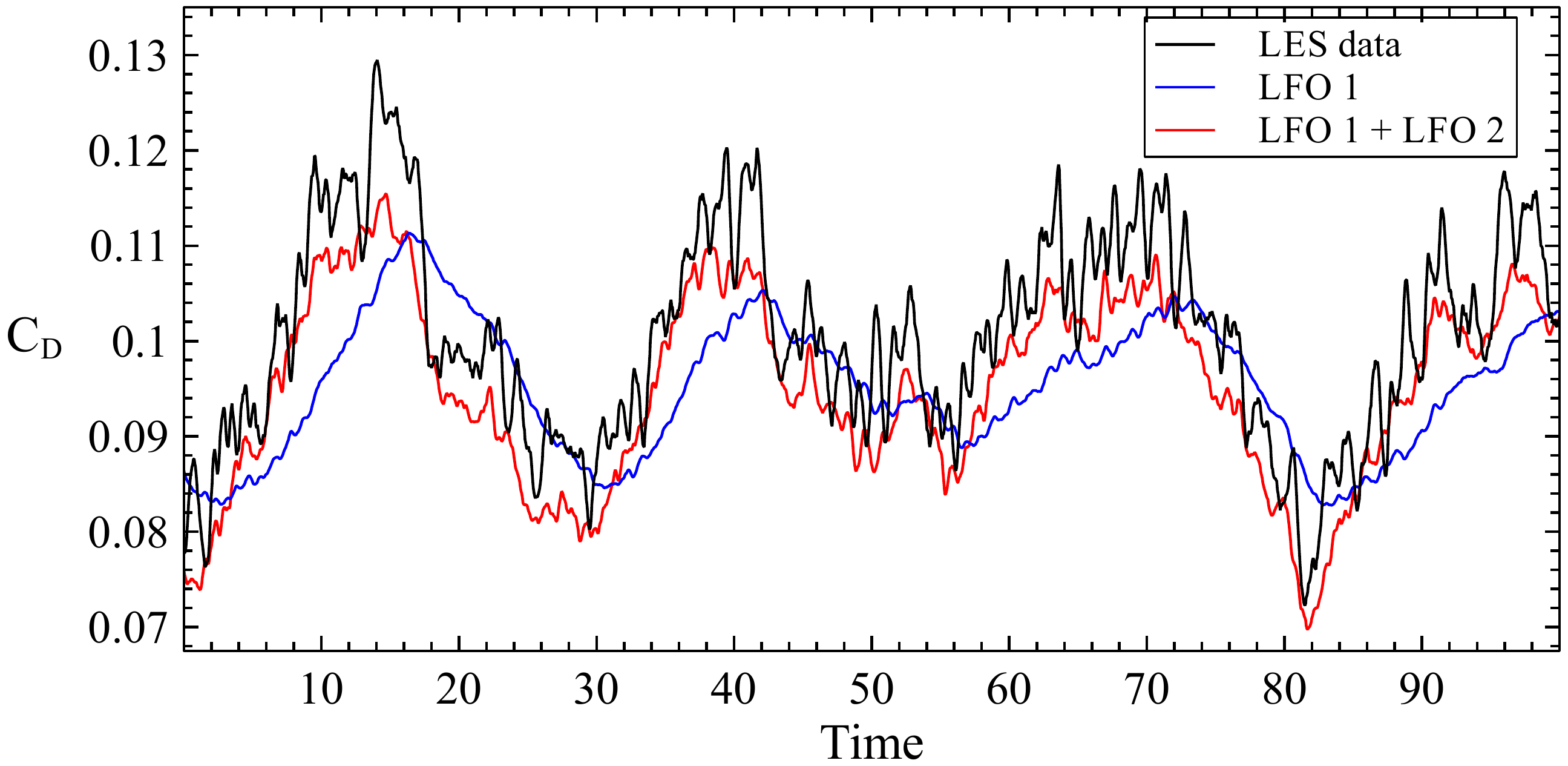}
\textit{(b) The pressure-drag coefficient}
\end{minipage}
\begin{minipage}{220pt}
\centering
\includegraphics[width=220pt, trim={0mm 0mm 0mm 0mm}, clip]{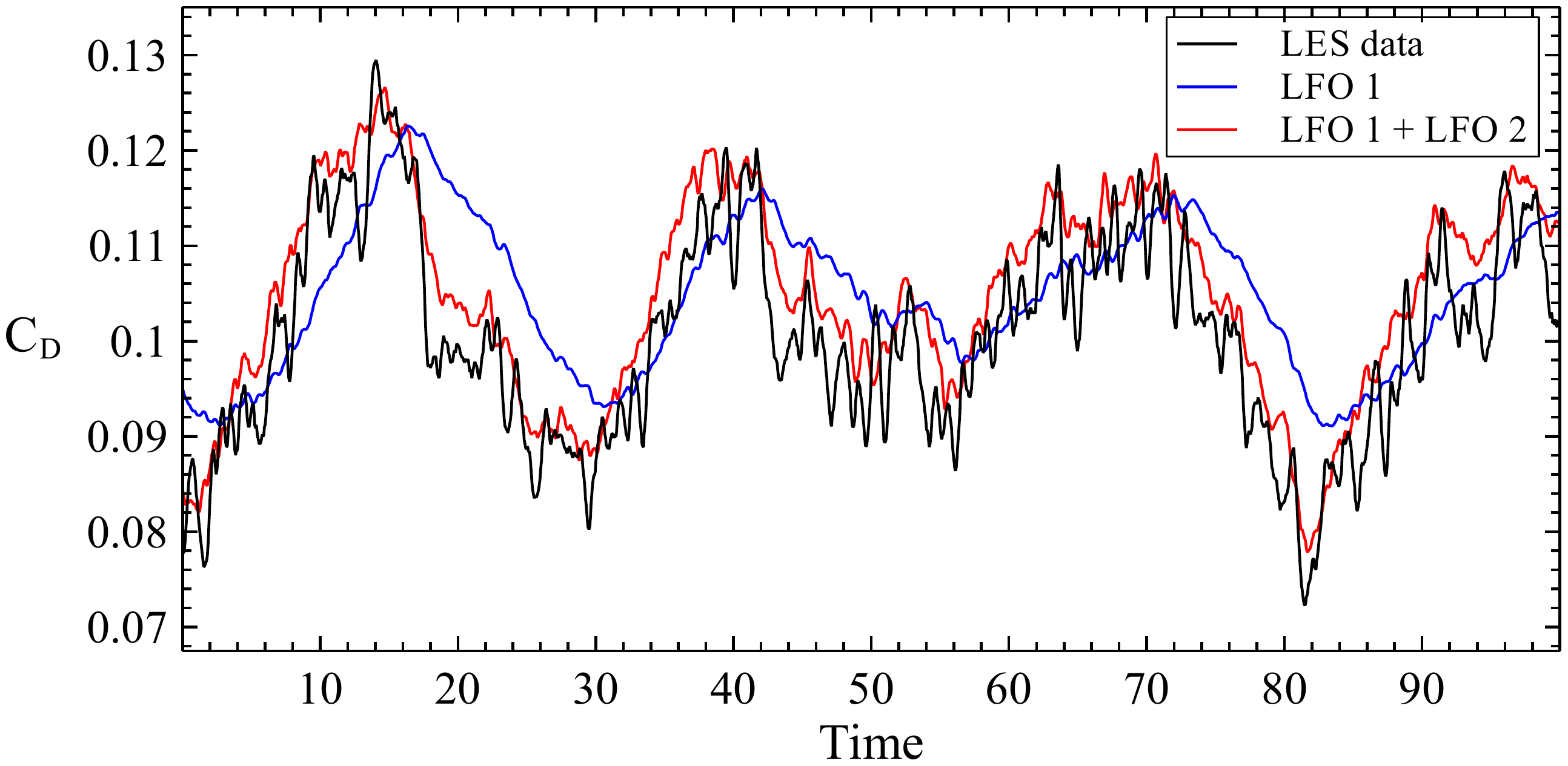}
\textit{(c) The drag coefficient}
\end{minipage}
\begin{minipage}{220pt}
\centering
\includegraphics[width=220pt, trim={0mm 0mm 0mm 0mm}, clip]{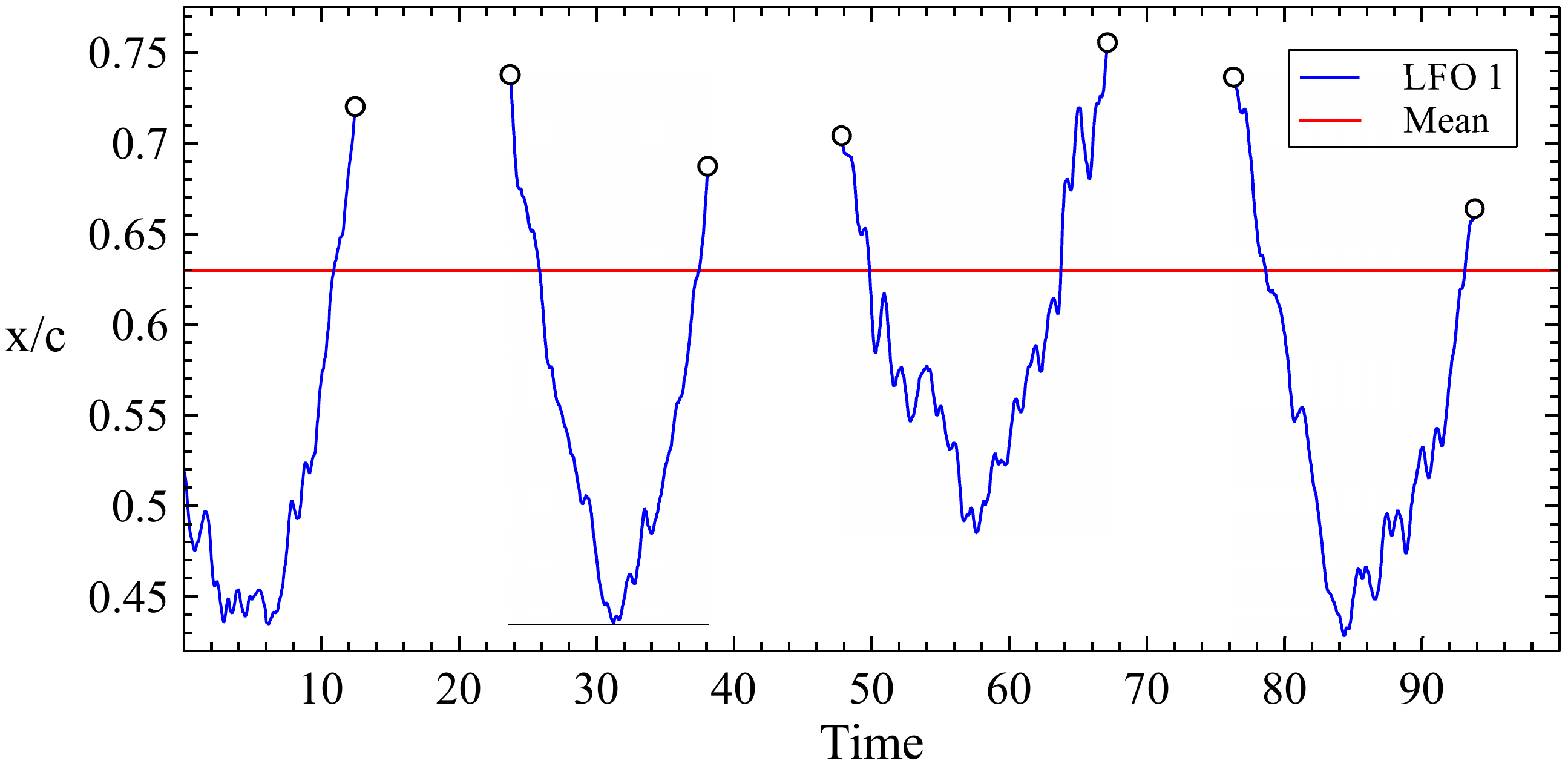}
\textit{(d) The reattachment location}
\end{minipage}
\caption{POD reconstruction of the lift coeff\/icient, the pressure-drag coeff\/icient, the drag coeff\/icient, and the reattachment location of the shear layer using the LFO mode 1 and the LFO mode 2 at the angle of attack of $9.8^{\circ}$.}
\label{POD_rec_cl_cd}
\end{center}
\end{figure}

\begin{figure}
\begin{center}
\begin{minipage}{145pt}
\centering
\includegraphics[width=145pt, trim={0mm 0mm 0mm 0mm}, clip]{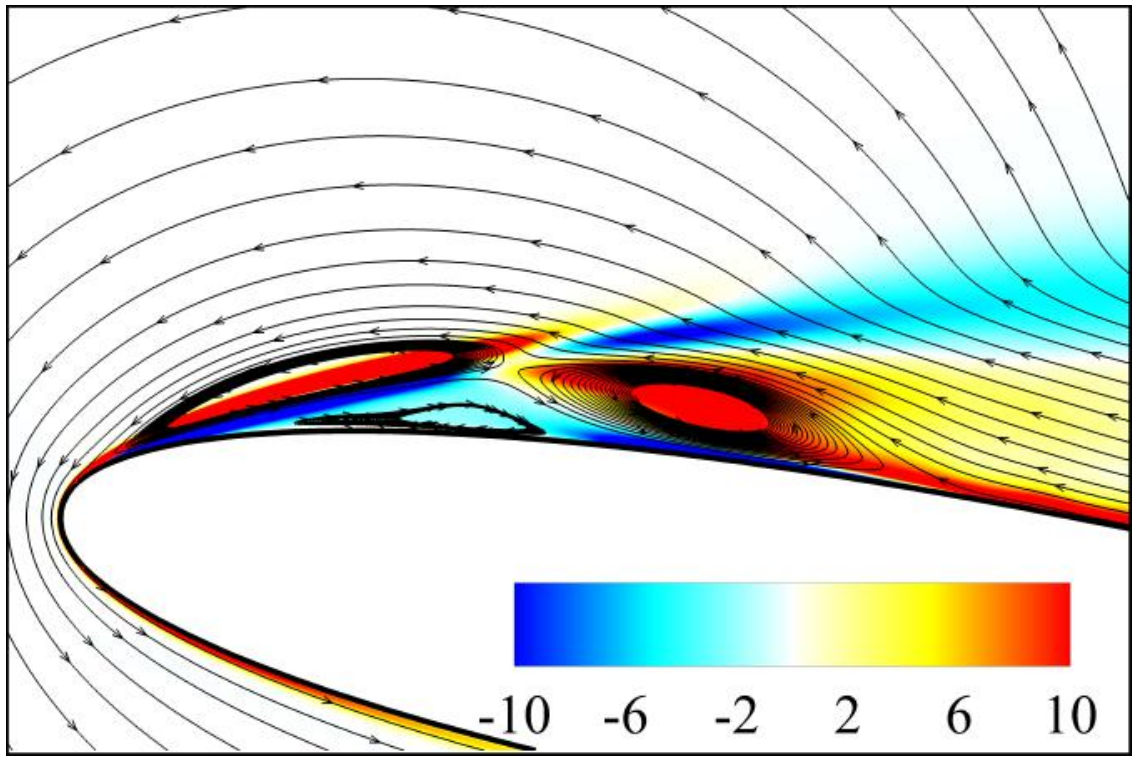}
\textit{time=8.5, $\Phi = 0^{\circ}$}
\end{minipage}
\begin{minipage}{145pt}
\centering
\includegraphics[width=145pt, trim={0mm 0mm 0mm 0mm}, clip]{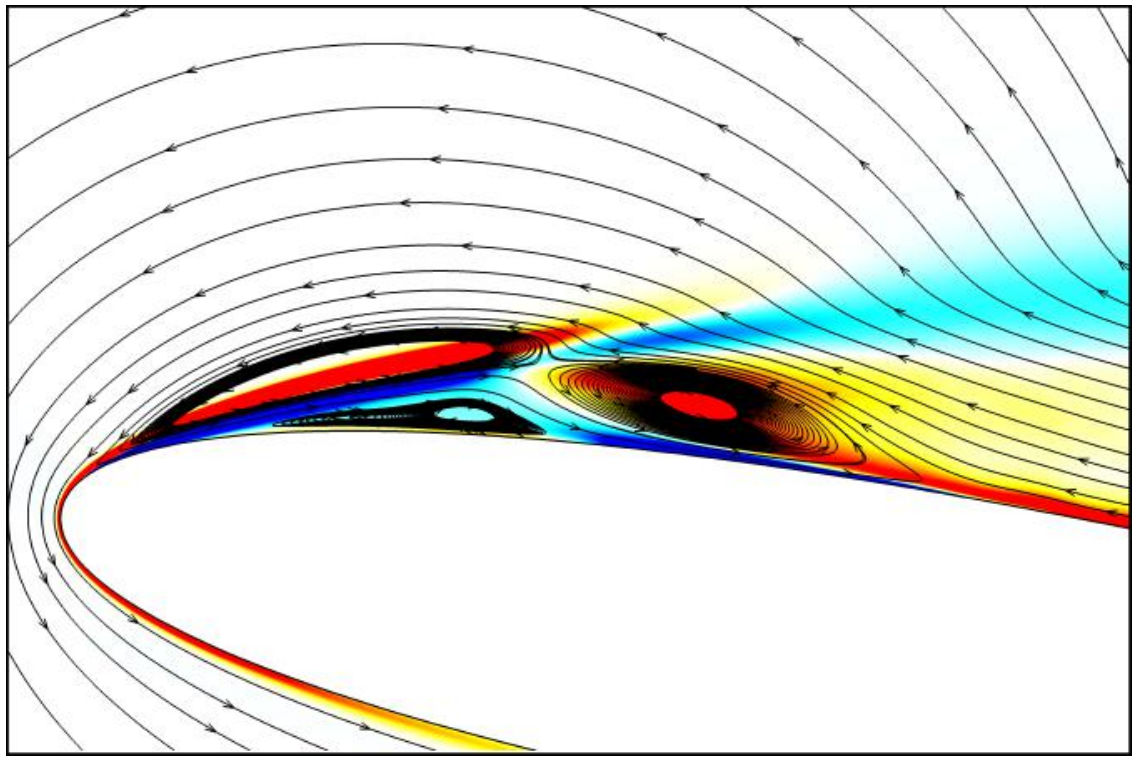}
\textit{time=8.75}
\end{minipage}
\begin{minipage}{145pt}
\centering
\includegraphics[width=145pt, trim={0mm 0mm 0mm 0mm}, clip]{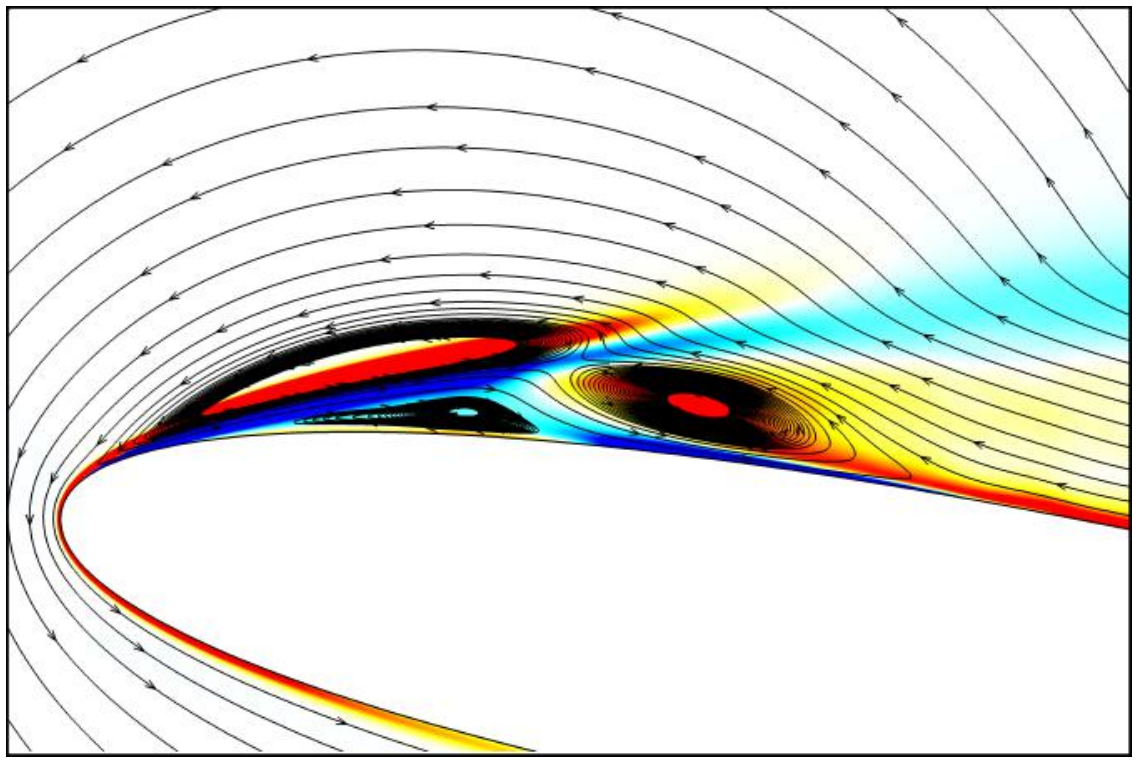}
\textit{time=9.0}
\end{minipage}
\begin{minipage}{145pt}
\centering
\includegraphics[width=145pt, trim={0mm 0mm 0mm 0mm}, clip]{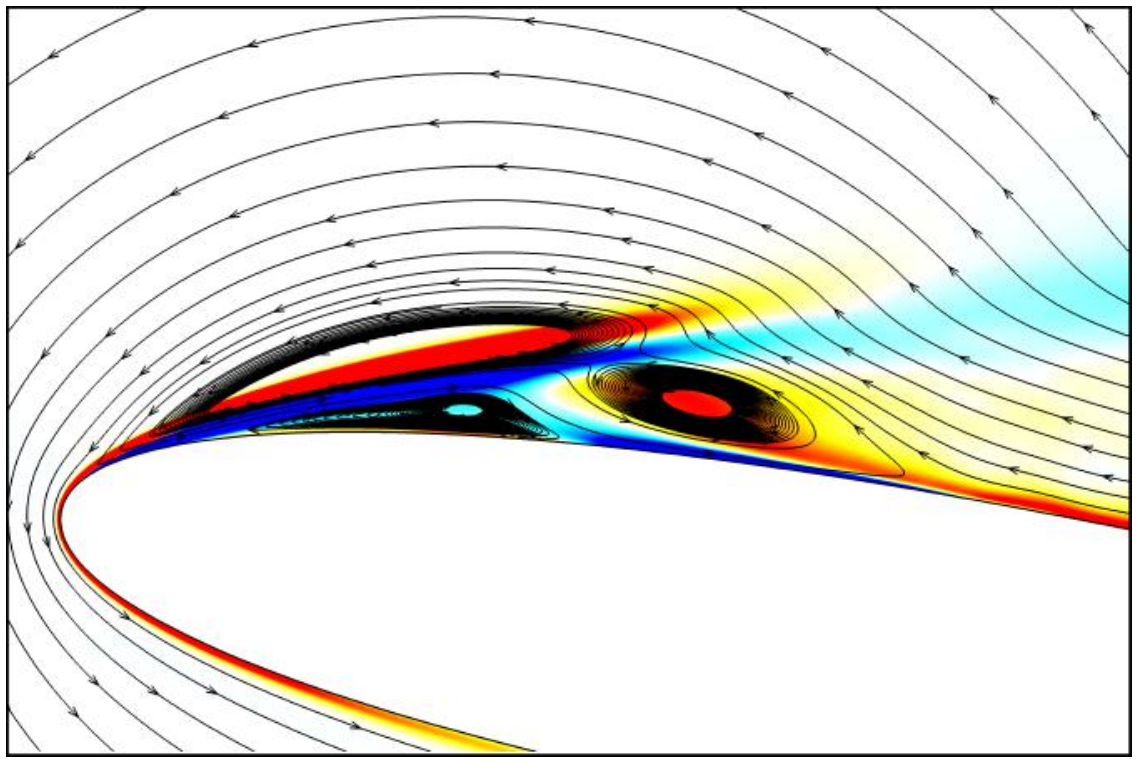}
\textit{time=9.25}
\end{minipage}
\begin{minipage}{145pt}
\centering
\includegraphics[width=145pt, trim={0mm 0mm 0mm 0mm}, clip]{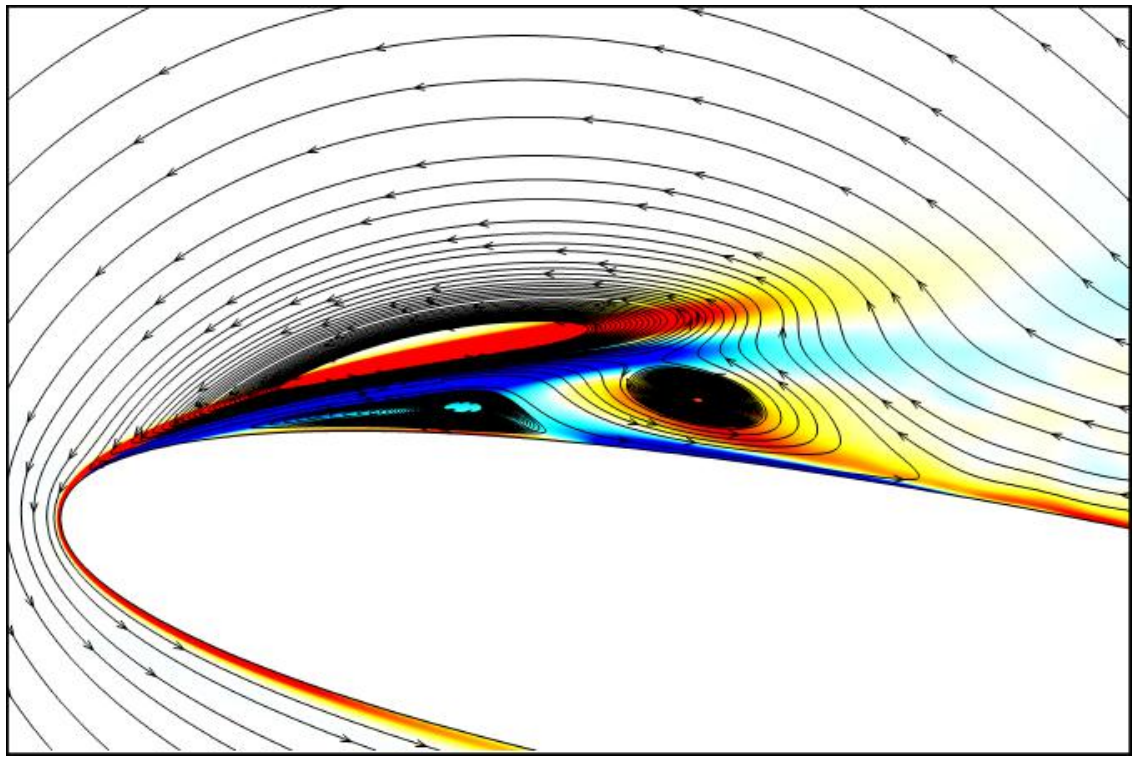}
\textit{time=9.5}
\end{minipage}
\begin{minipage}{145pt}
\centering
\includegraphics[width=145pt, trim={0mm 0mm 0mm 0mm}, clip]{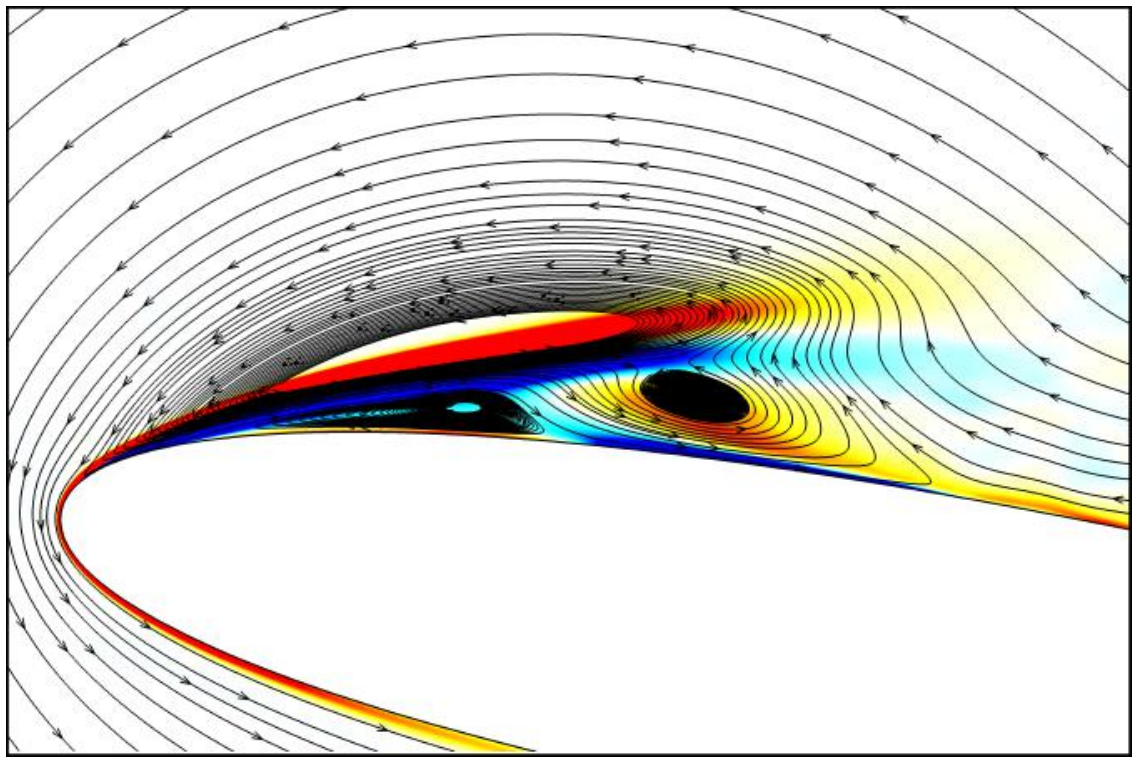}
\textit{time=9.75}
\end{minipage}
\begin{minipage}{145pt}
\centering
\includegraphics[width=145pt, trim={0mm 0mm 0mm 0mm}, clip]{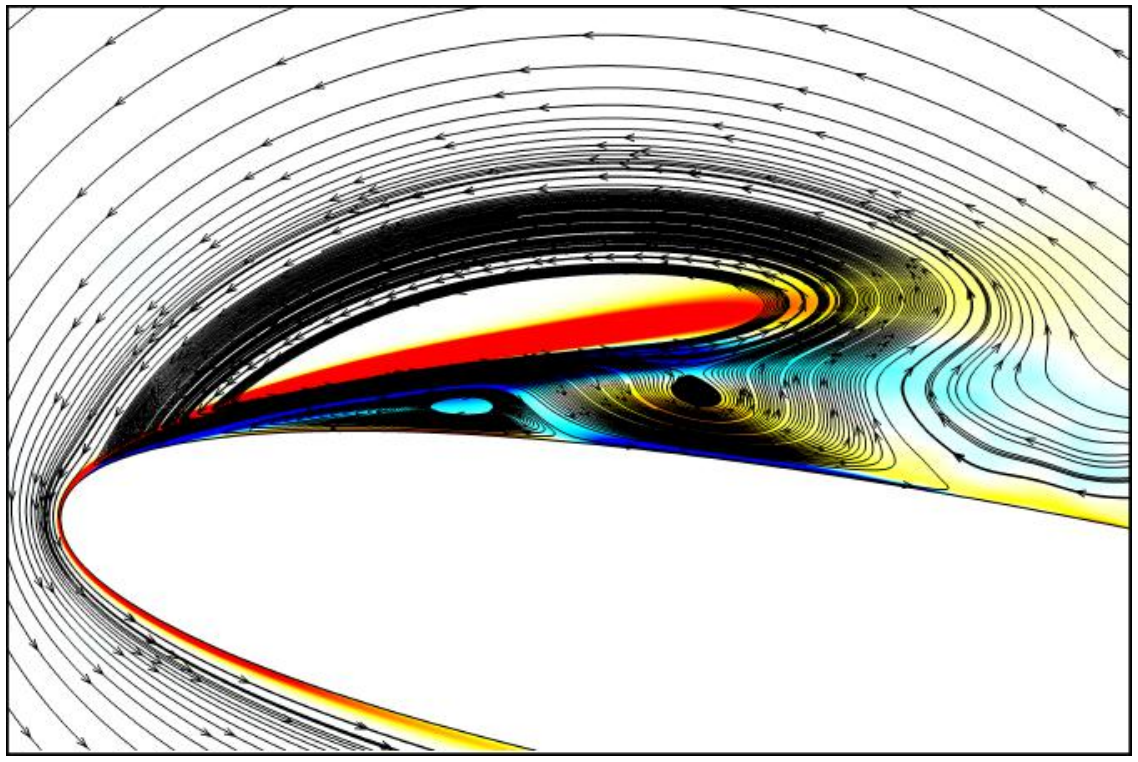}
\textit{time=10.0}
\end{minipage}
\begin{minipage}{145pt}
\centering
\includegraphics[width=145pt, trim={0mm 0mm 0mm 0mm}, clip]{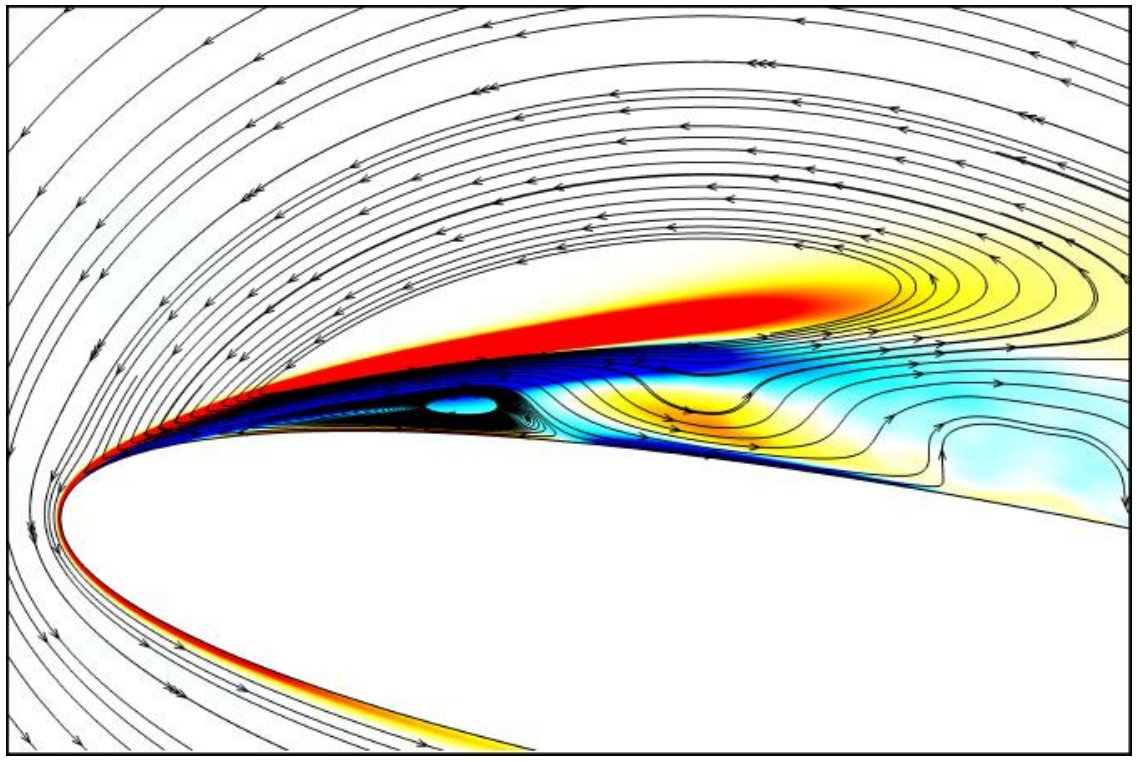}
\textit{time=10.25}
\end{minipage}
\begin{minipage}{145pt}
\centering
\includegraphics[width=145pt, trim={0mm 0mm 0mm 0mm}, clip]{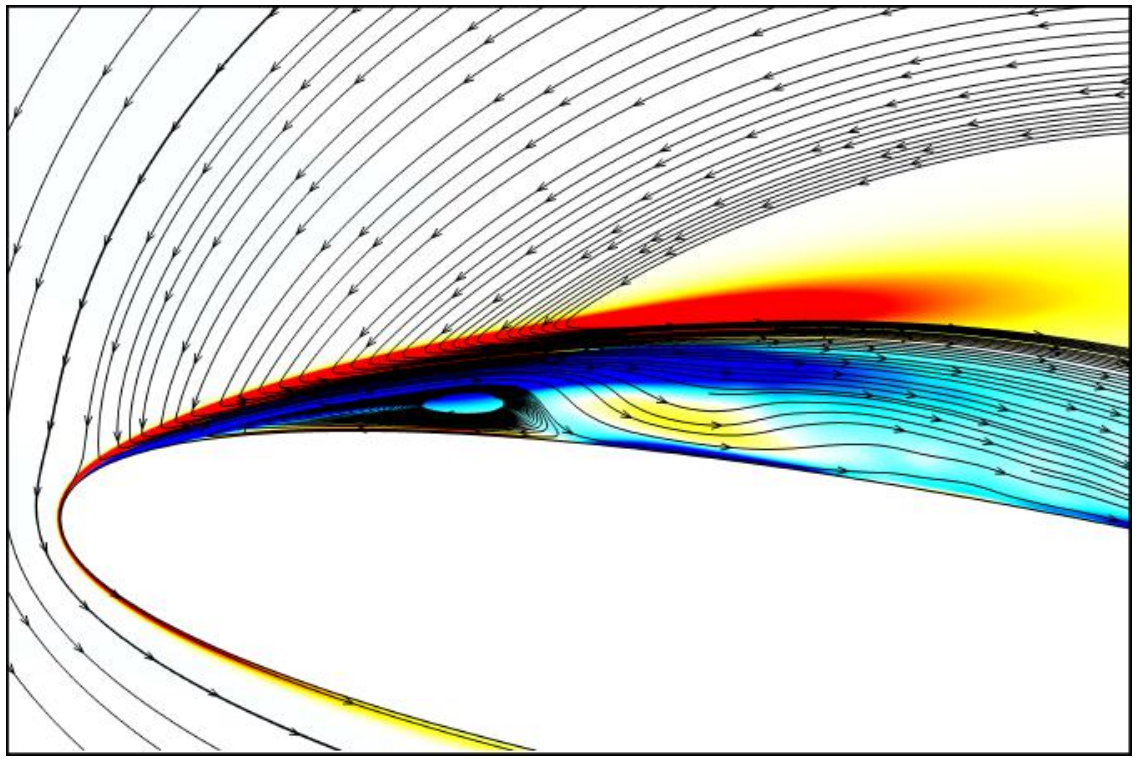}
\textit{time=11.0, $\Phi = 90^{\circ}$}
\end{minipage}
\begin{minipage}{145pt}
\centering
\includegraphics[width=145pt, trim={0mm 0mm 0mm 0mm}, clip]{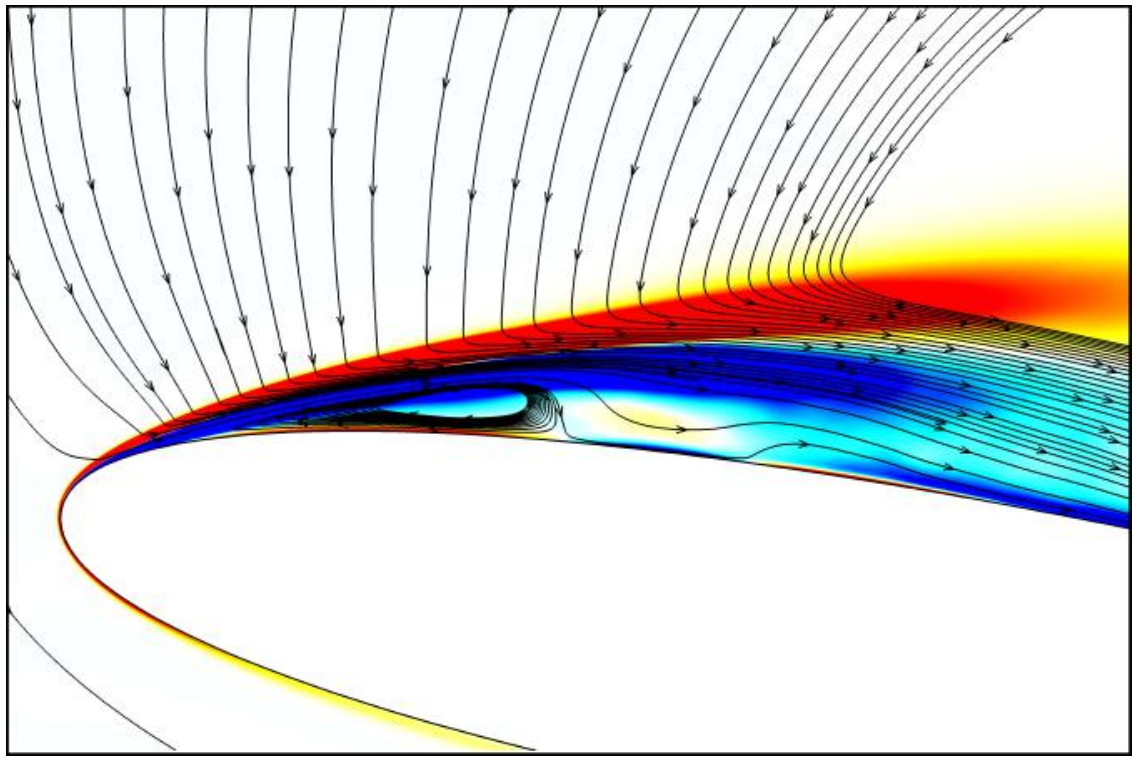}
\textit{time=11.75}
\end{minipage}
\begin{minipage}{145pt}
\centering
\includegraphics[width=145pt, trim={0mm 0mm 0mm 0mm}, clip]{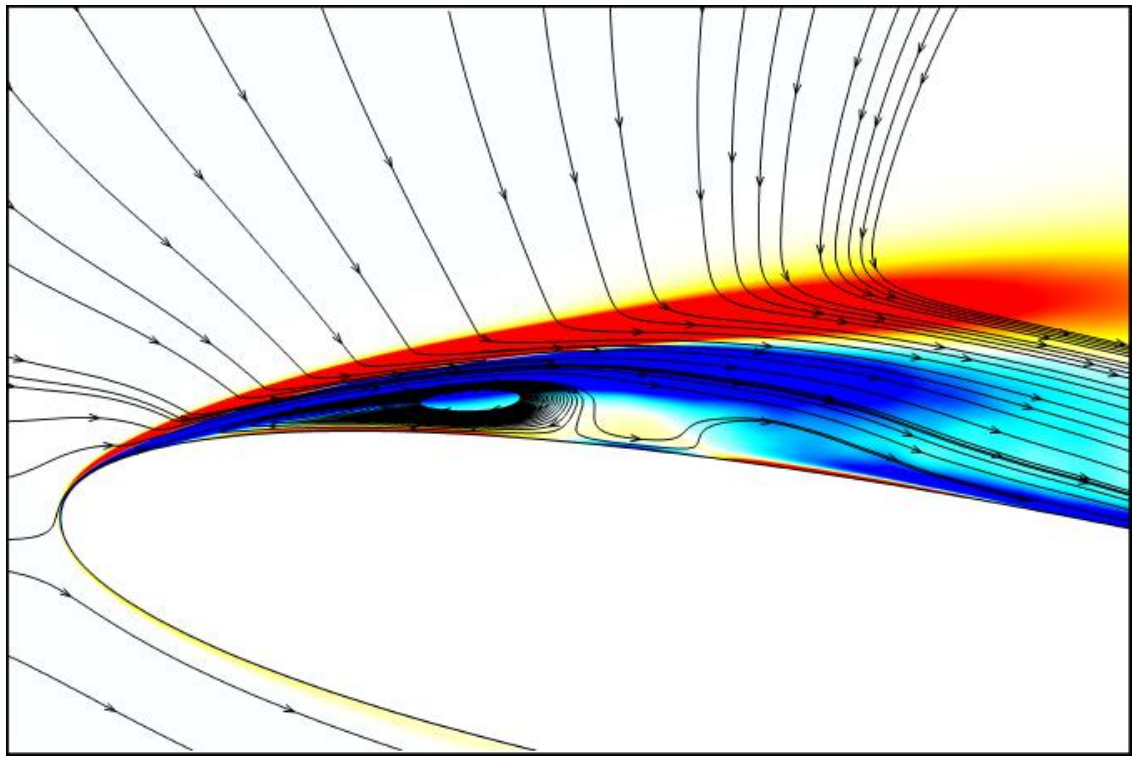}
\textit{time=12.25}
\end{minipage}
\begin{minipage}{145pt}
\centering
\includegraphics[width=145pt, trim={0mm 0mm 0mm 0mm}, clip]{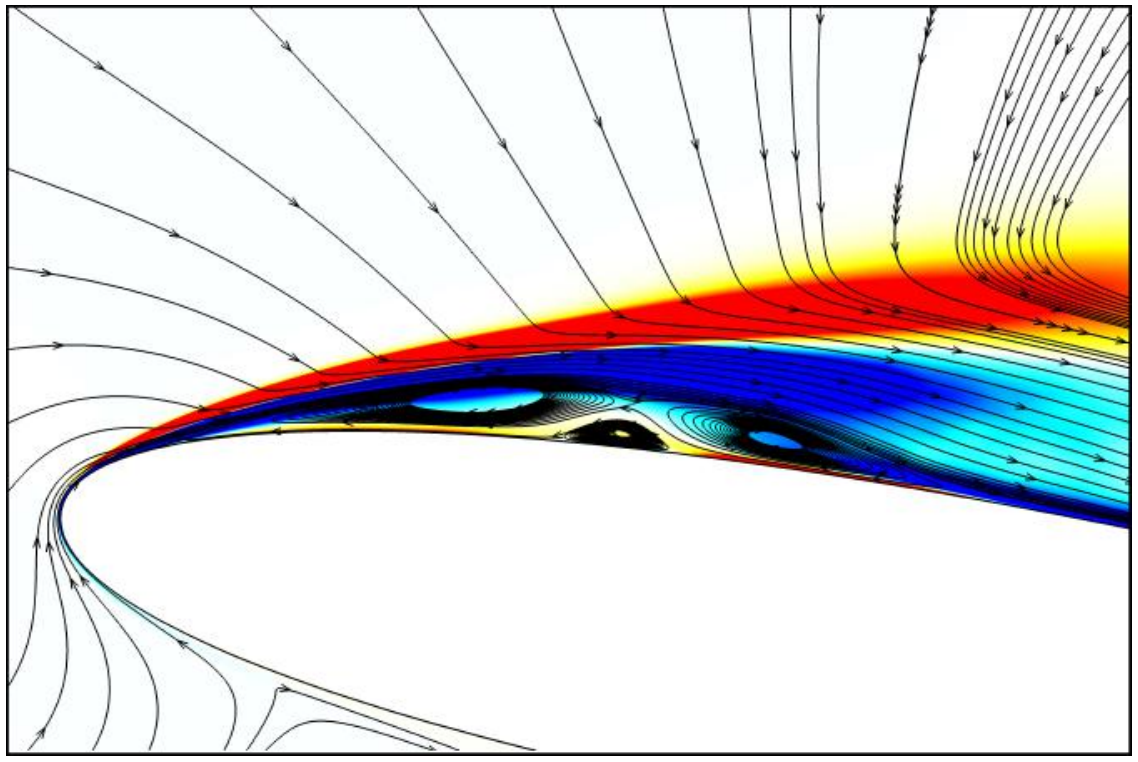}
\textit{time=12.5}
\end{minipage}
\begin{minipage}{145pt}
\centering
\includegraphics[width=145pt, trim={0mm 0mm 0mm 0mm}, clip]{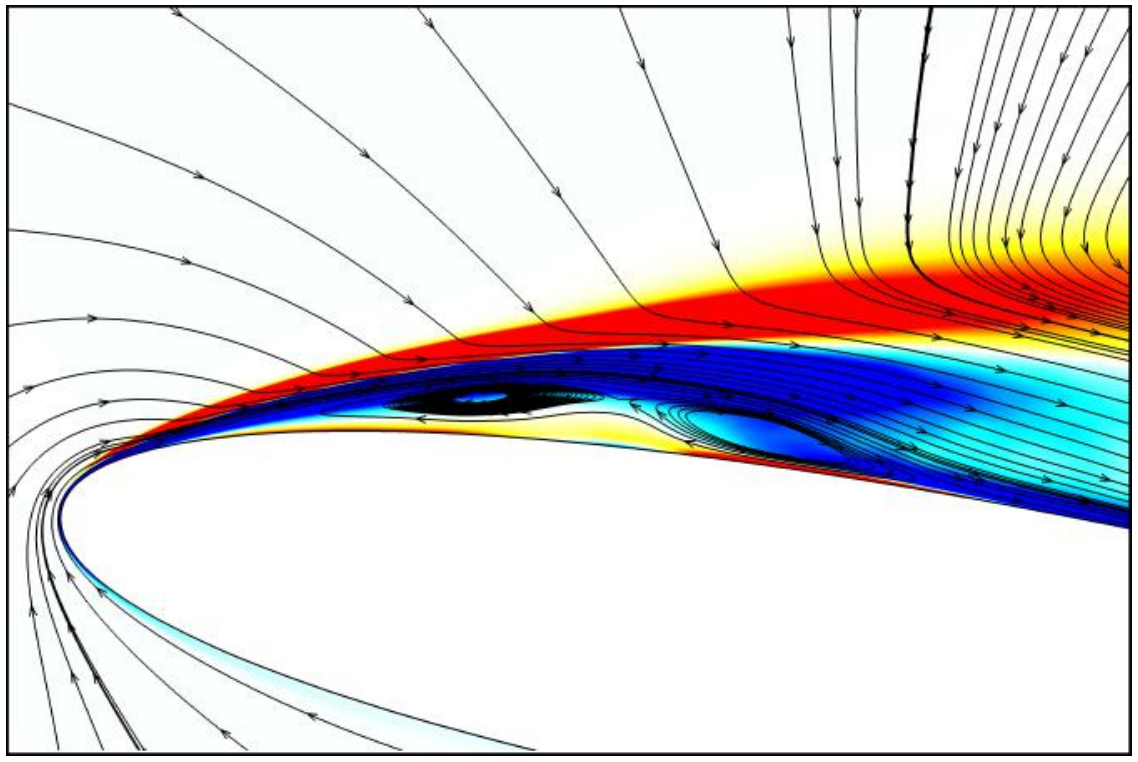}
\textit{time=12.75}
\end{minipage}
\begin{minipage}{145pt}
\centering
\includegraphics[width=145pt, trim={0mm 0mm 0mm 0mm}, clip]{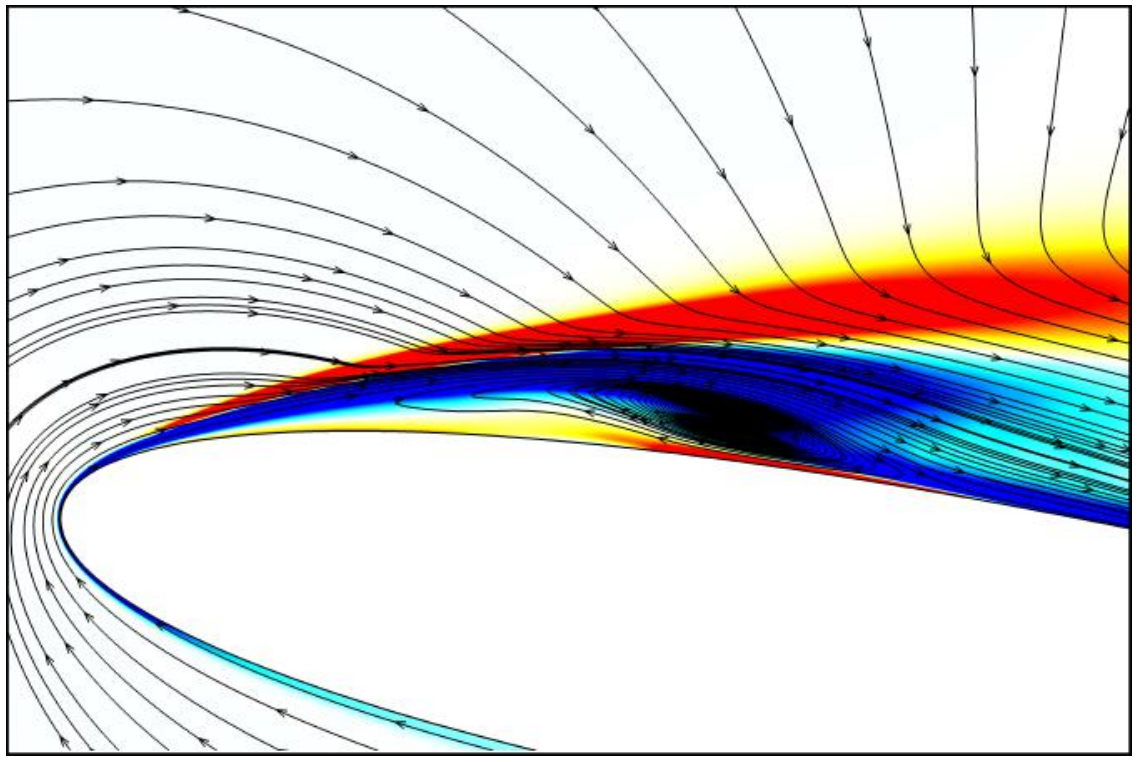}
\textit{time=13.5}
\end{minipage}
\begin{minipage}{145pt}
\centering
\includegraphics[width=145pt, trim={0mm 0mm 0mm 0mm}, clip]{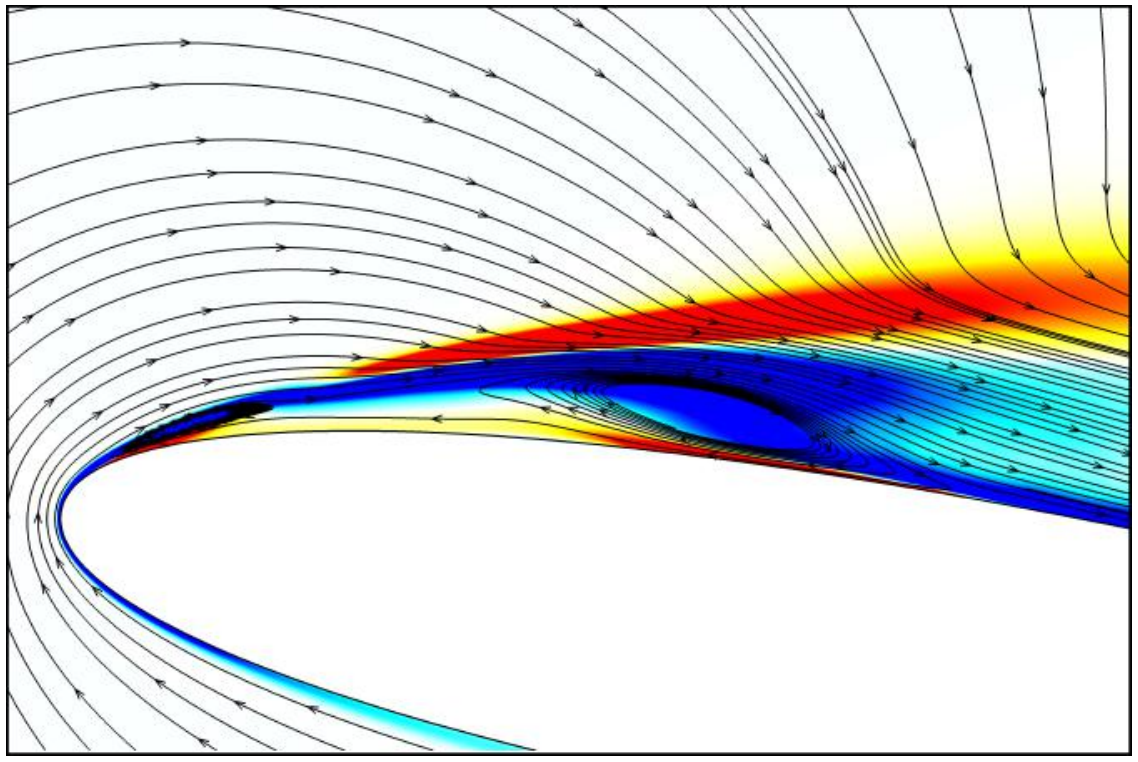}
\textit{time=14.0}
\end{minipage}
\begin{minipage}{145pt}
\centering
\includegraphics[width=145pt, trim={0mm 0mm 0mm 0mm}, clip]{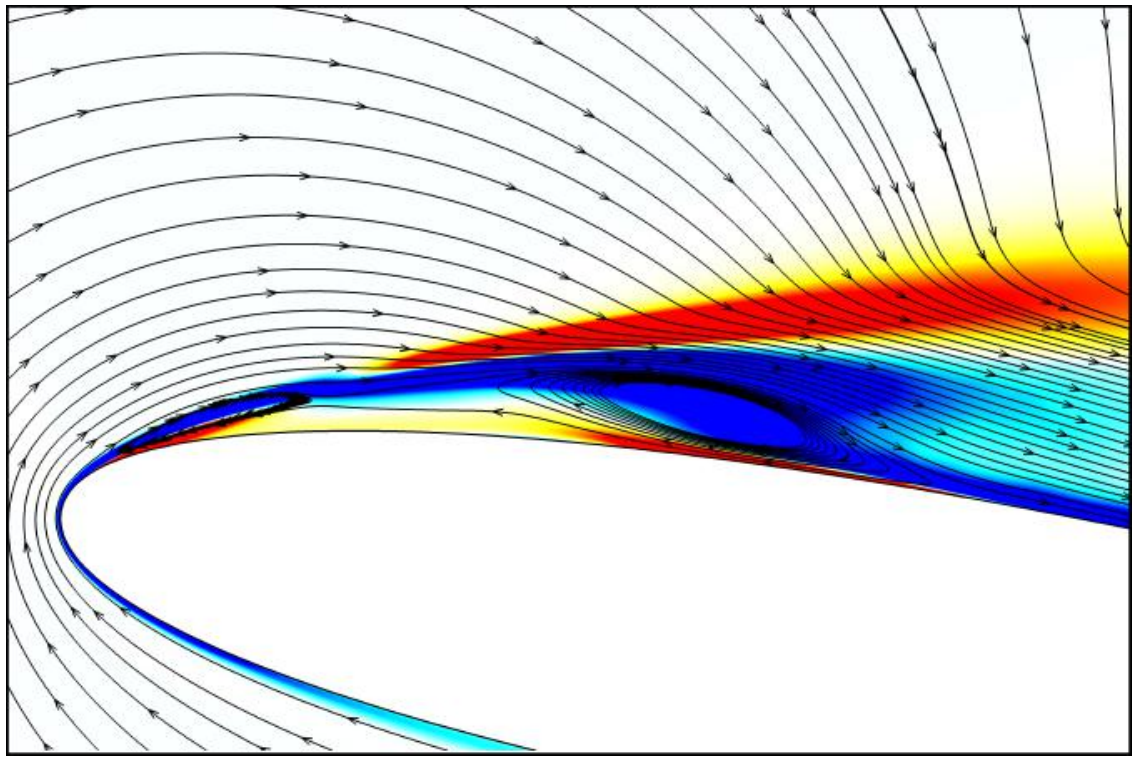}
\textit{time=14.5}
\end{minipage}
\begin{minipage}{145pt}
\centering
\includegraphics[width=145pt, trim={0mm 0mm 0mm 0mm}, clip]{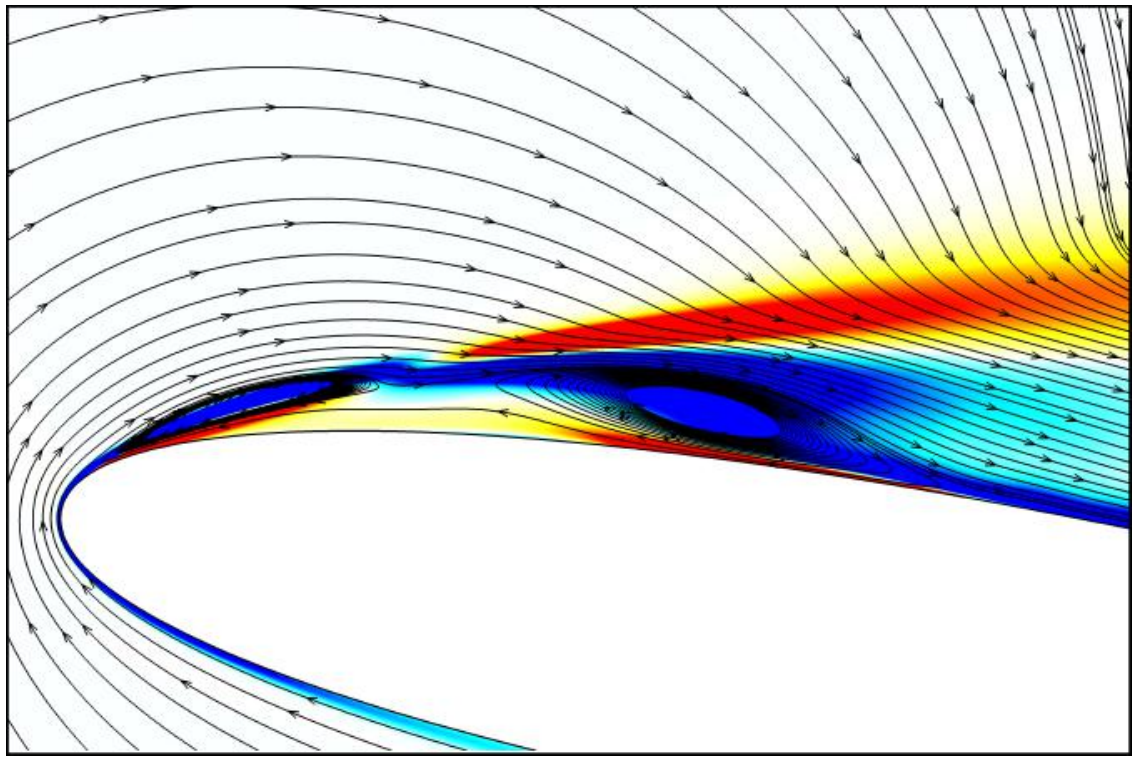}
\textit{time=14.75}
\end{minipage}
\begin{minipage}{145pt}
\centering
\includegraphics[width=145pt, trim={0mm 0mm 0mm 0mm}, clip]{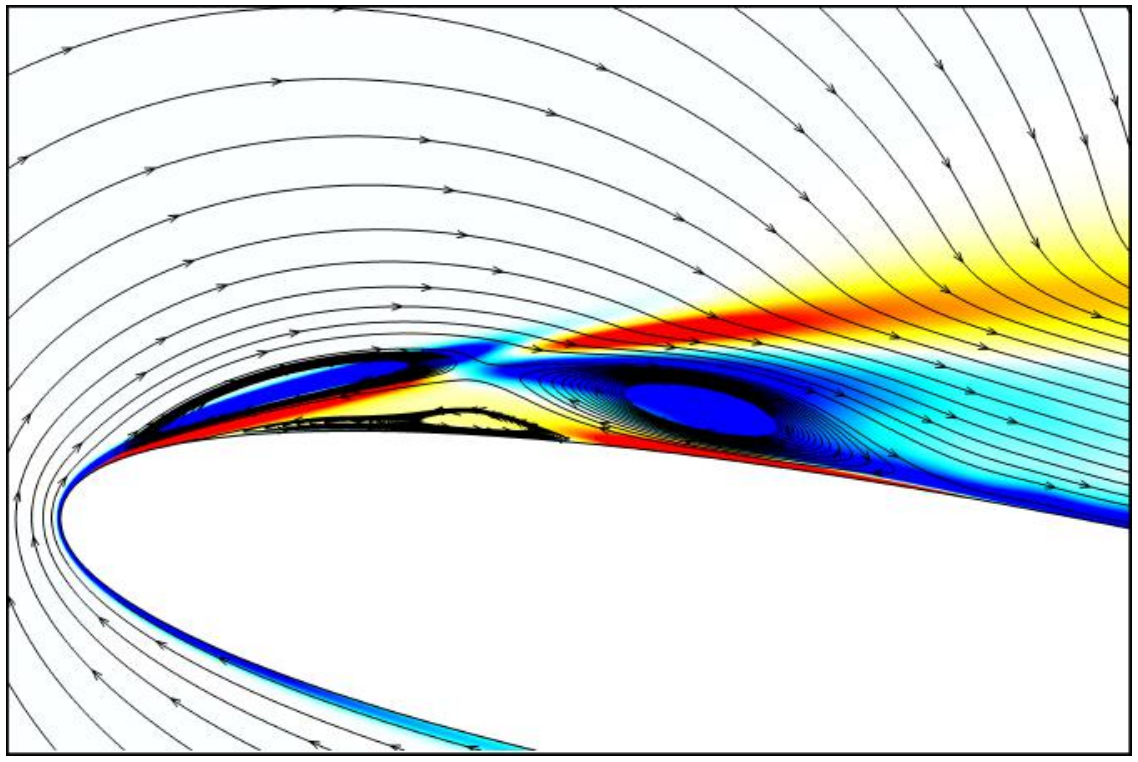}
\textit{time=15.0, $\Phi = 180^{\circ}$}
\end{minipage}
\caption{Streamlines patterns superimposed on colour maps of the spanwise vorticity $\omega_z$ for the POD reconstruction of the oscillating-f\/low using the LFO mode 1, the LFO mode 2, and the HFO mode for the angle of attack of $9.7^{\circ}$. The f\/low-f\/ield is attaching.}
\label{POD_rec_TCV_970}
\end{center}
\end{figure}

\begin{figure}
\begin{center}
\begin{minipage}{220pt}
\centering
\includegraphics[width=220pt, trim={0mm 0mm 0mm 0mm}, clip]{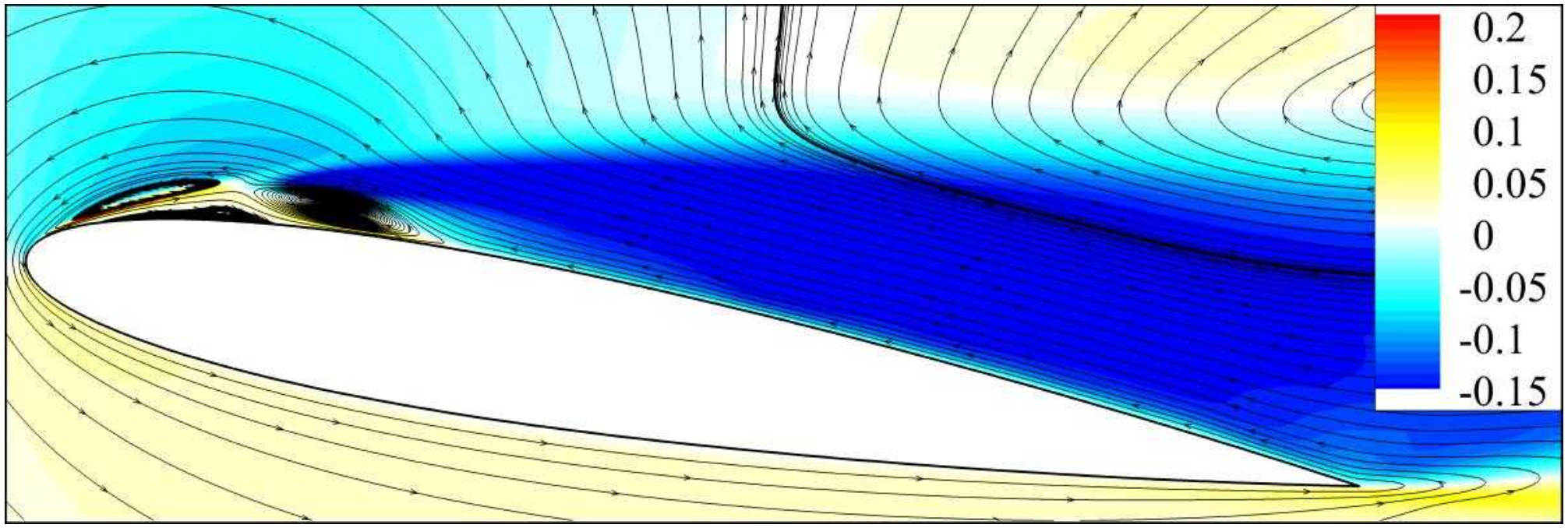}
\textit{time=8.5, $\Phi = 0^{\circ}$}
\end{minipage}
\medskip
\begin{minipage}{220pt}
\centering
\includegraphics[width=220pt, trim={0mm 0mm 0mm 0mm}, clip]{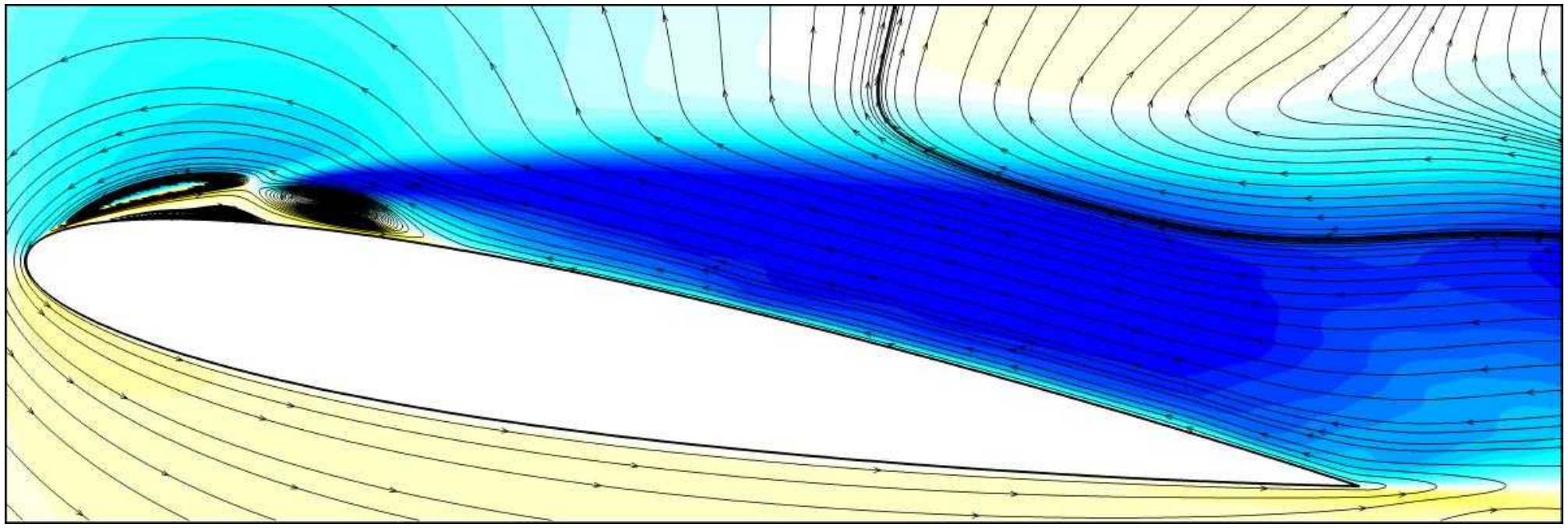}
\textit{time=8.75}
\end{minipage}
\medskip
\begin{minipage}{220pt}
\centering
\includegraphics[width=220pt, trim={0mm 0mm 0mm 0mm}, clip]{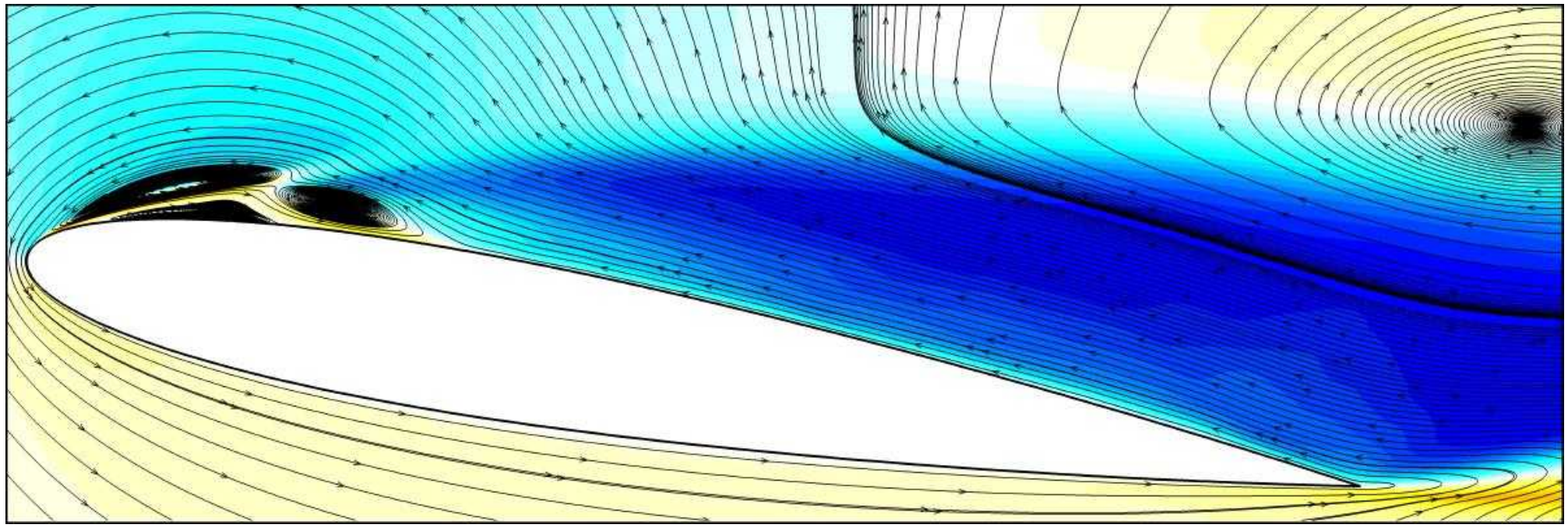}
\textit{time=9.0}
\end{minipage}
\medskip
\begin{minipage}{220pt}
\centering
\includegraphics[width=220pt, trim={0mm 0mm 0mm 0mm}, clip]{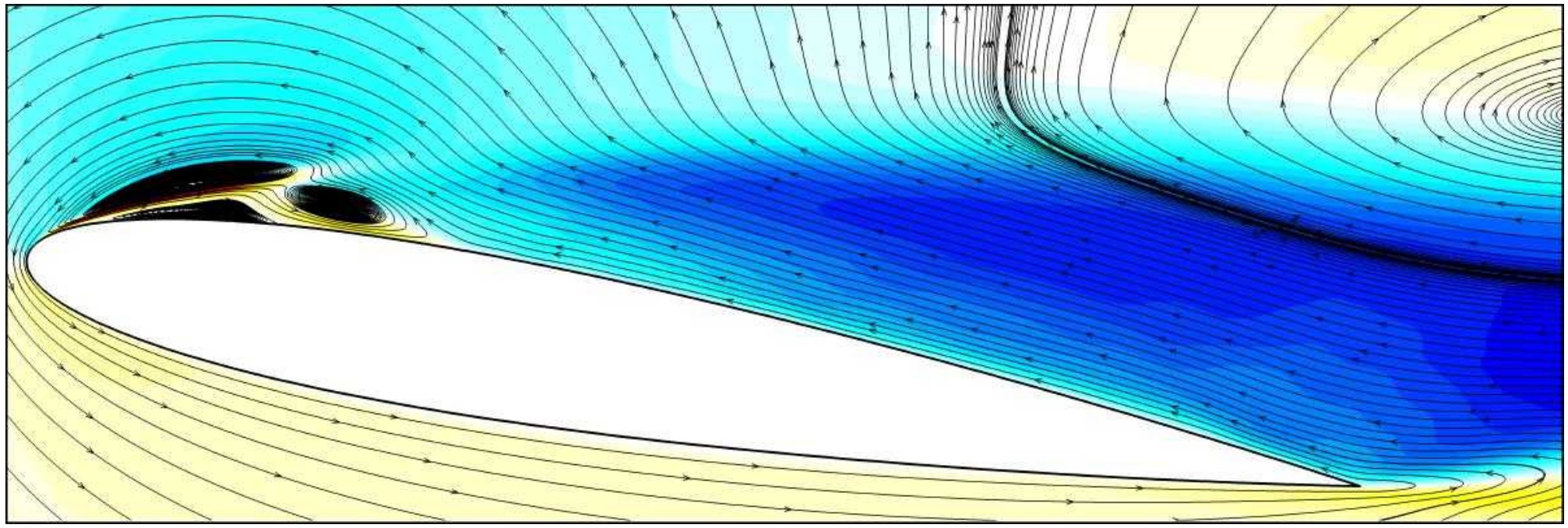}
\textit{time=9.25}
\end{minipage}
\medskip
\begin{minipage}{220pt}
\centering
\includegraphics[width=220pt, trim={0mm 0mm 0mm 0mm}, clip]{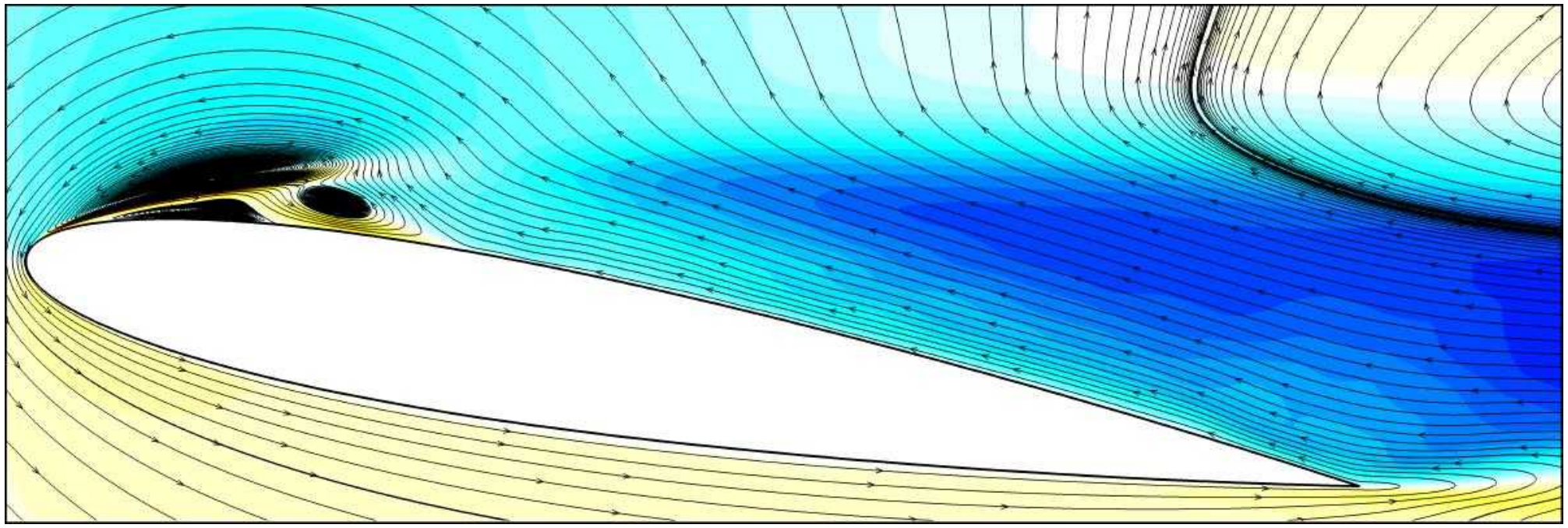}
\textit{time=9.5}
\end{minipage}
\medskip
\begin{minipage}{220pt}
\centering
\includegraphics[width=220pt, trim={0mm 0mm 0mm 0mm}, clip]{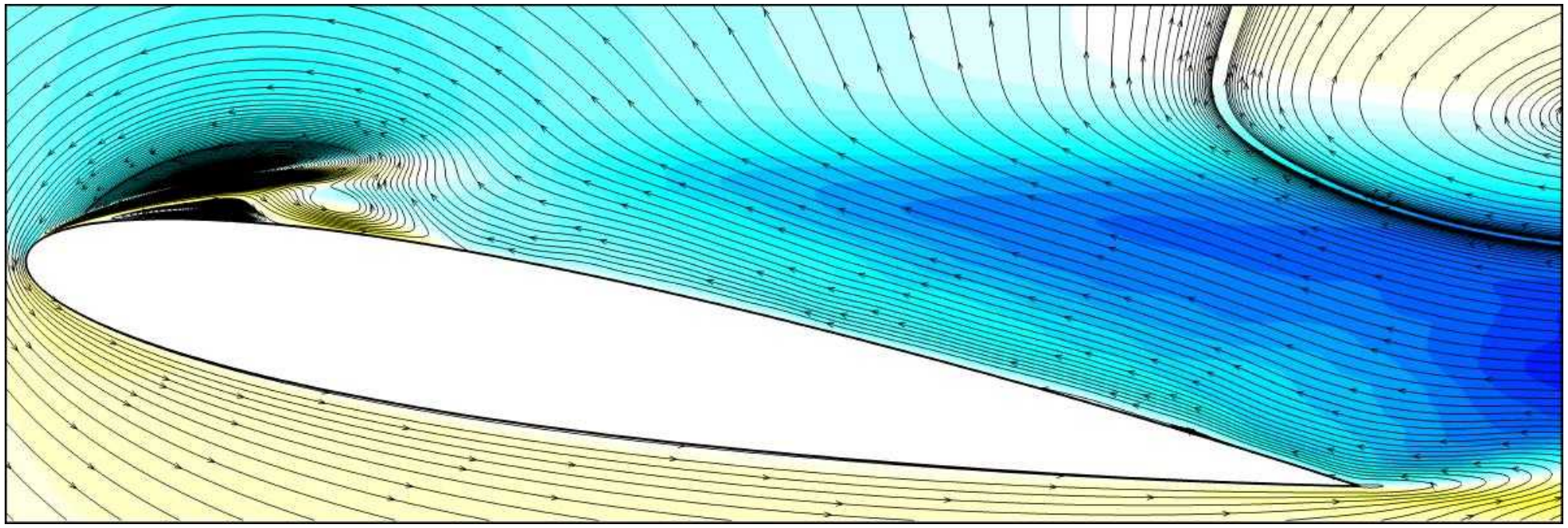}
\textit{time=9.75}
\end{minipage}
\medskip
\begin{minipage}{220pt}
\centering
\includegraphics[width=220pt, trim={0mm 0mm 0mm 0mm}, clip]{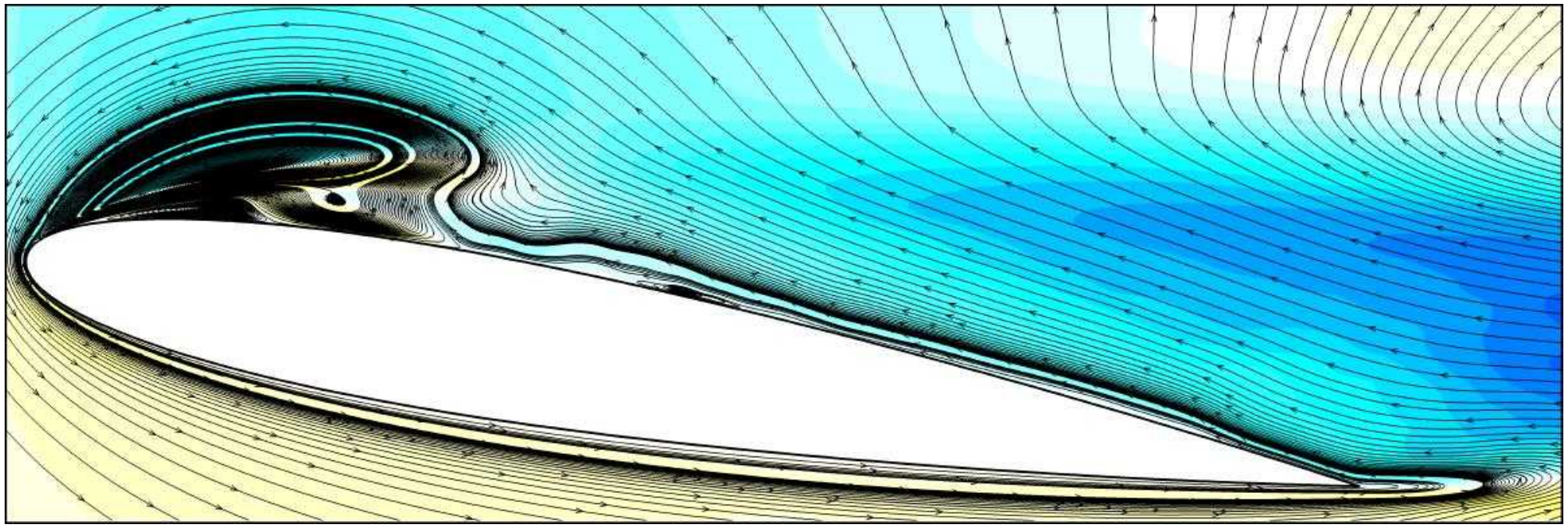}
\textit{time=10.0}
\end{minipage}
\medskip
\begin{minipage}{220pt}
\centering
\includegraphics[width=220pt, trim={0mm 0mm 0mm 0mm}, clip]{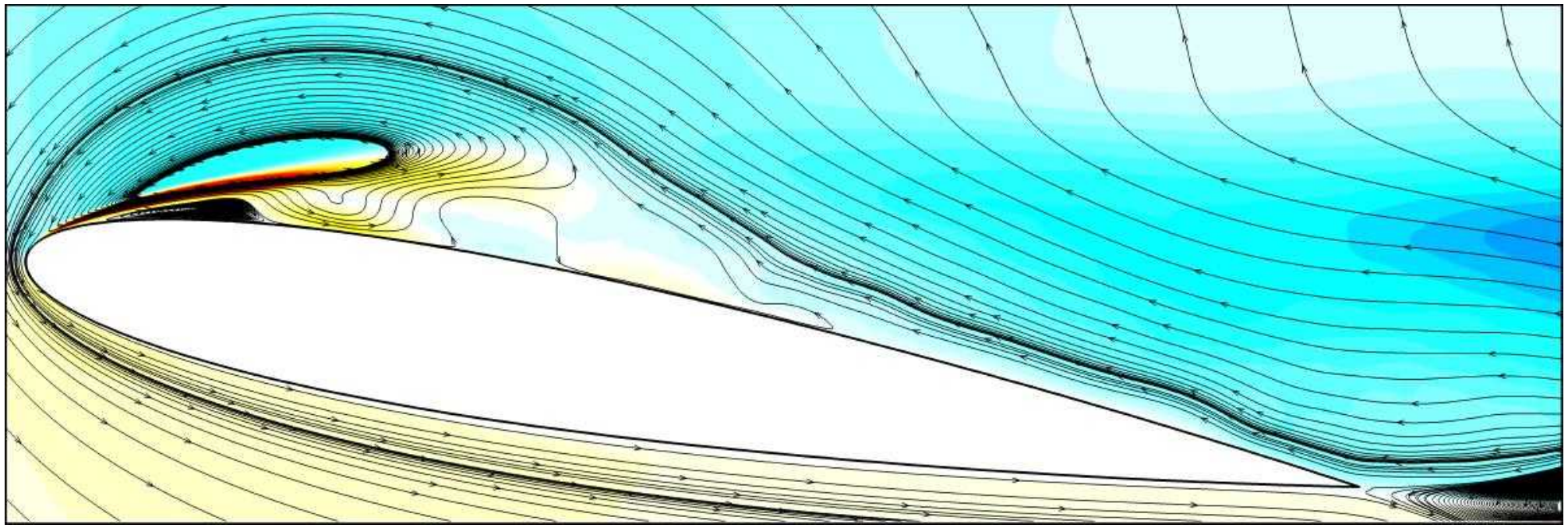}
\textit{time=10.25}
\end{minipage}
\medskip
\begin{minipage}{220pt}
\centering
\includegraphics[width=220pt, trim={0mm 0mm 0mm 0mm}, clip]{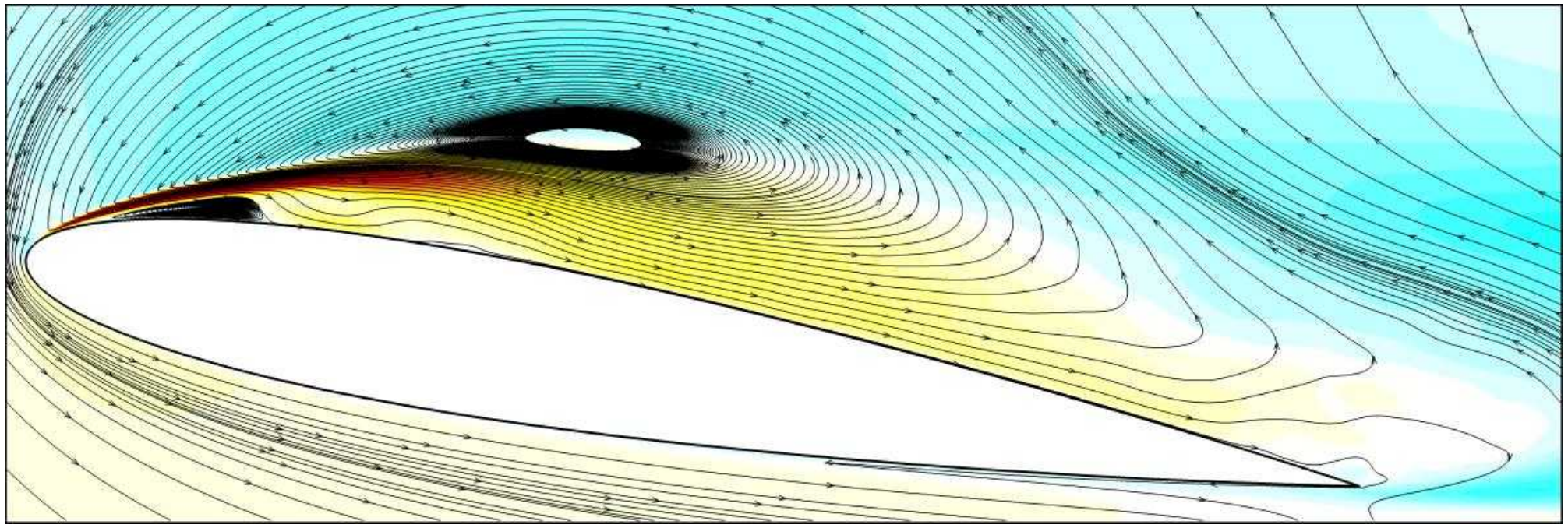}
\textit{time=11.0, $\Phi = 90^{\circ}$}
\end{minipage}
\medskip
\begin{minipage}{220pt}
\centering
\includegraphics[width=220pt, trim={0mm 0mm 0mm 0mm}, clip]{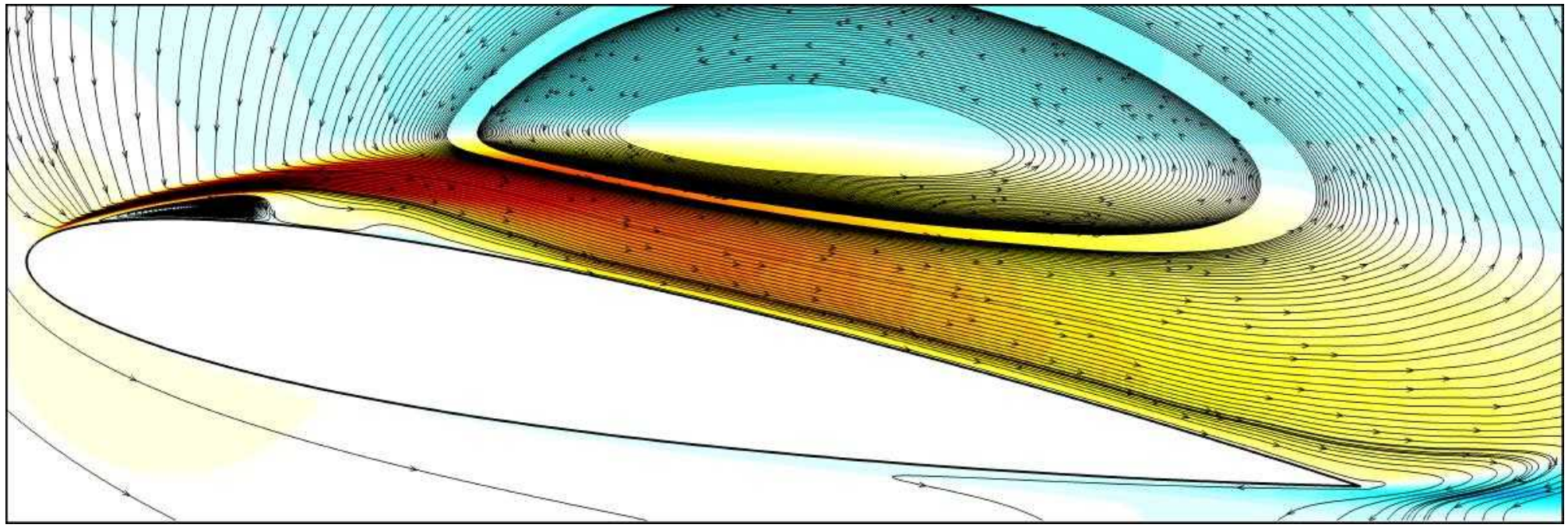}
\textit{time=11.75}
\end{minipage}
\medskip
\begin{minipage}{220pt}
\centering
\includegraphics[width=220pt, trim={0mm 0mm 0mm 0mm}, clip]{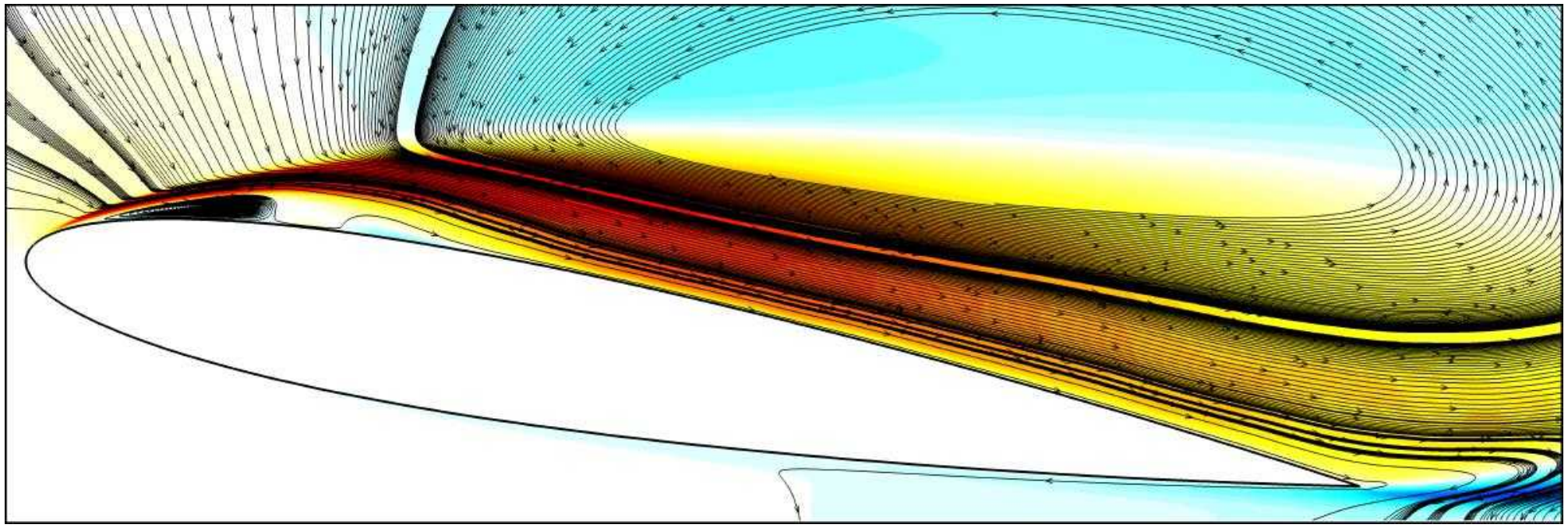}
\textit{time=12.25}
\end{minipage}
\begin{minipage}{220pt}
\centering
\includegraphics[width=220pt, trim={0mm 0mm 0mm 0mm}, clip]{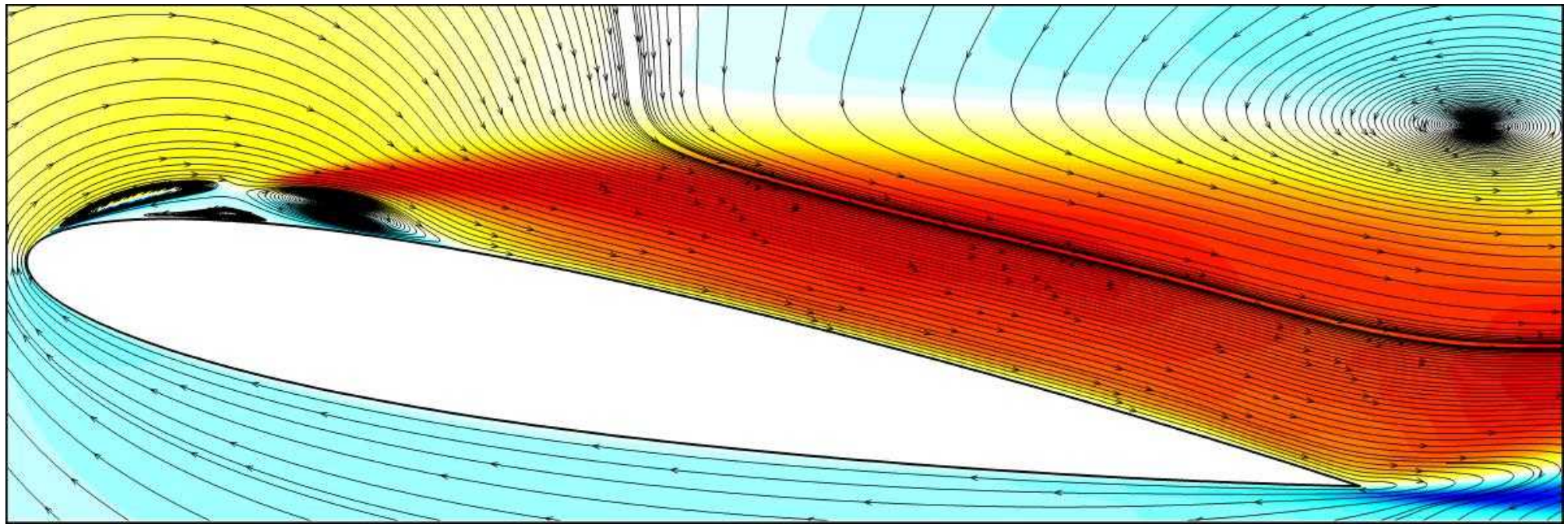}
\textit{time=15.0, $\Phi = 180^{\circ}$}
\end{minipage}
\caption{Streamlines patterns superimposed on colour maps of the streamwise velocity component for the POD reconstruction of the oscillating-f\/low using the LFO mode 1, the LFO mode 2, and the HFO mode for the angle of attack of $9.7^{\circ}$. The f\/low-f\/ield is attaching.}
\label{POD_rec_970}
\end{center}
\end{figure}

\begin{figure}
\begin{center}
\begin{minipage}{220pt}
\centering
\includegraphics[width=220pt, trim={0mm 0mm 0mm 0mm}, clip]{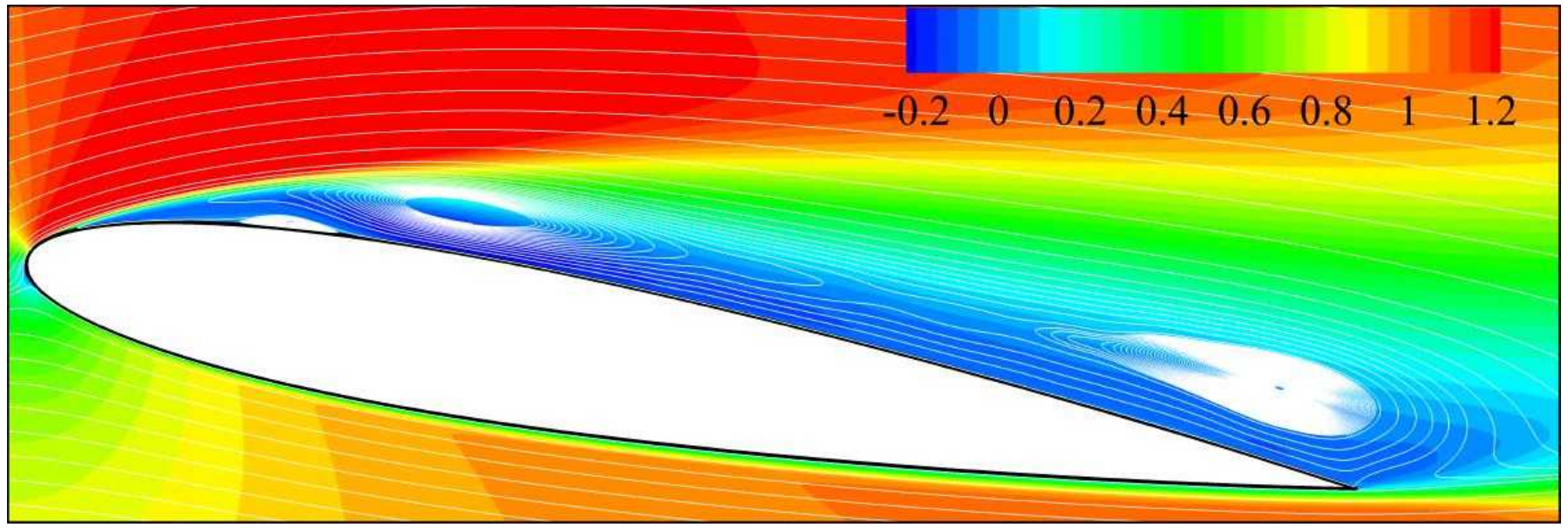}
\textit{time=8.5, $\Phi = 0^{\circ}$}
\end{minipage}
\medskip
\begin{minipage}{220pt}
\centering
\includegraphics[width=220pt, trim={0mm 0mm 0mm 0mm}, clip]{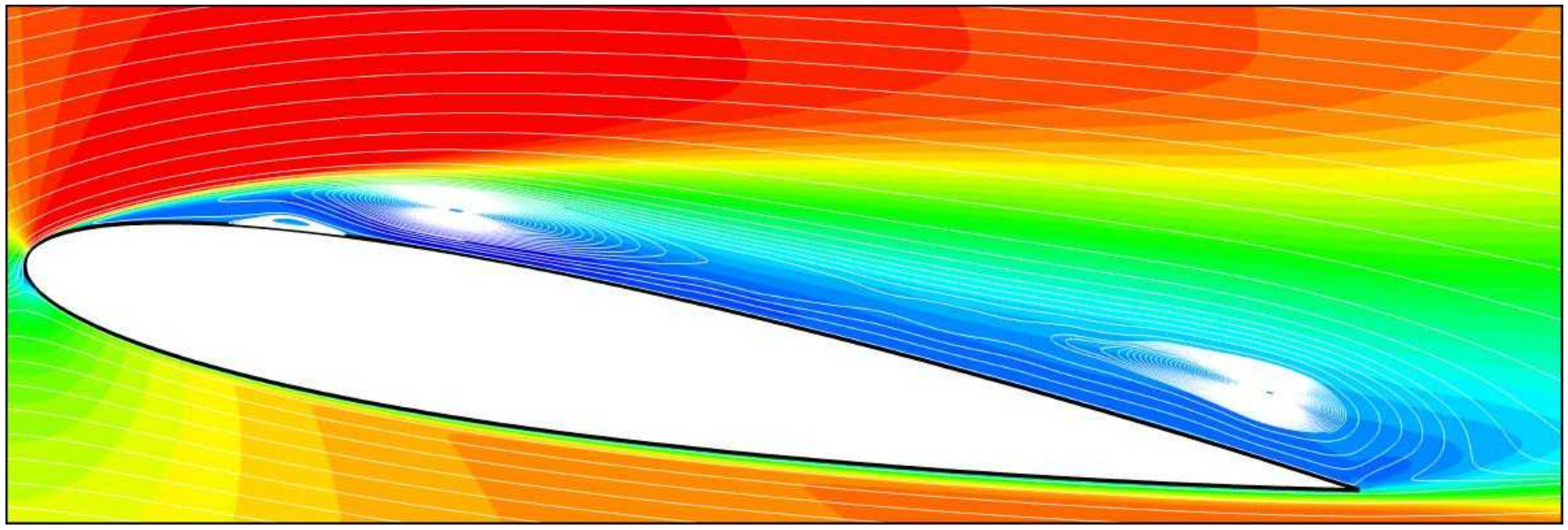}
\textit{time=8.75}
\end{minipage}
\medskip
\begin{minipage}{220pt}
\centering
\includegraphics[width=220pt, trim={0mm 0mm 0mm 0mm}, clip]{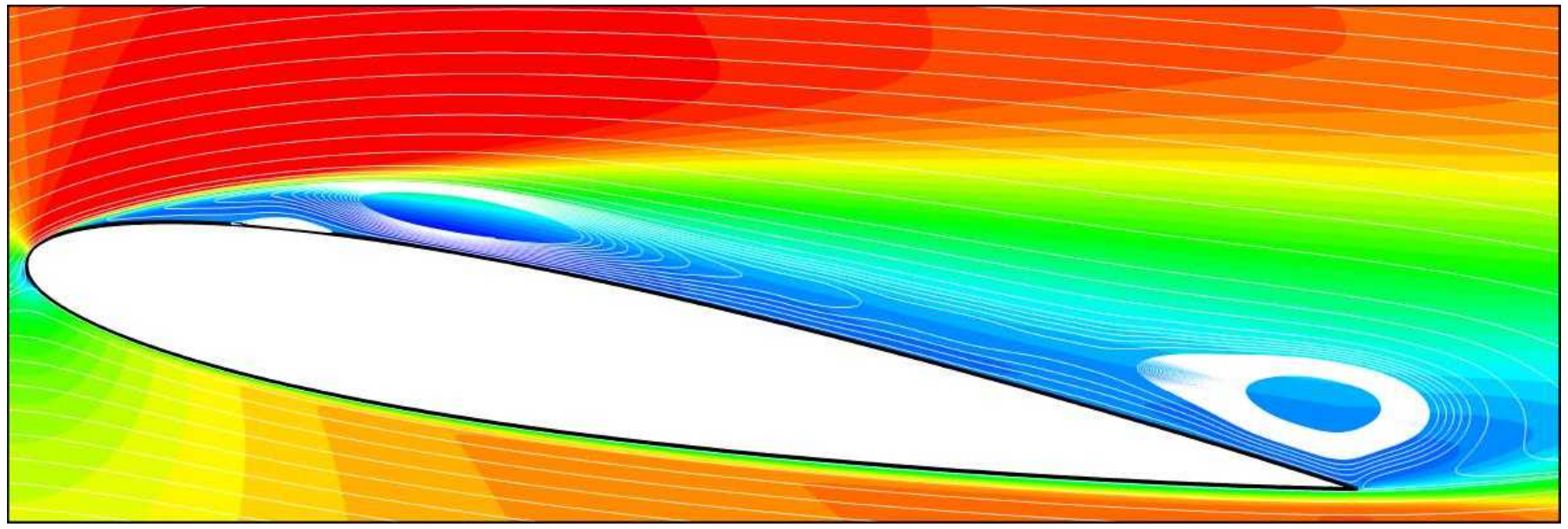}
\textit{time=9.0}
\end{minipage}
\medskip
\begin{minipage}{220pt}
\centering
\includegraphics[width=220pt, trim={0mm 0mm 0mm 0mm}, clip]{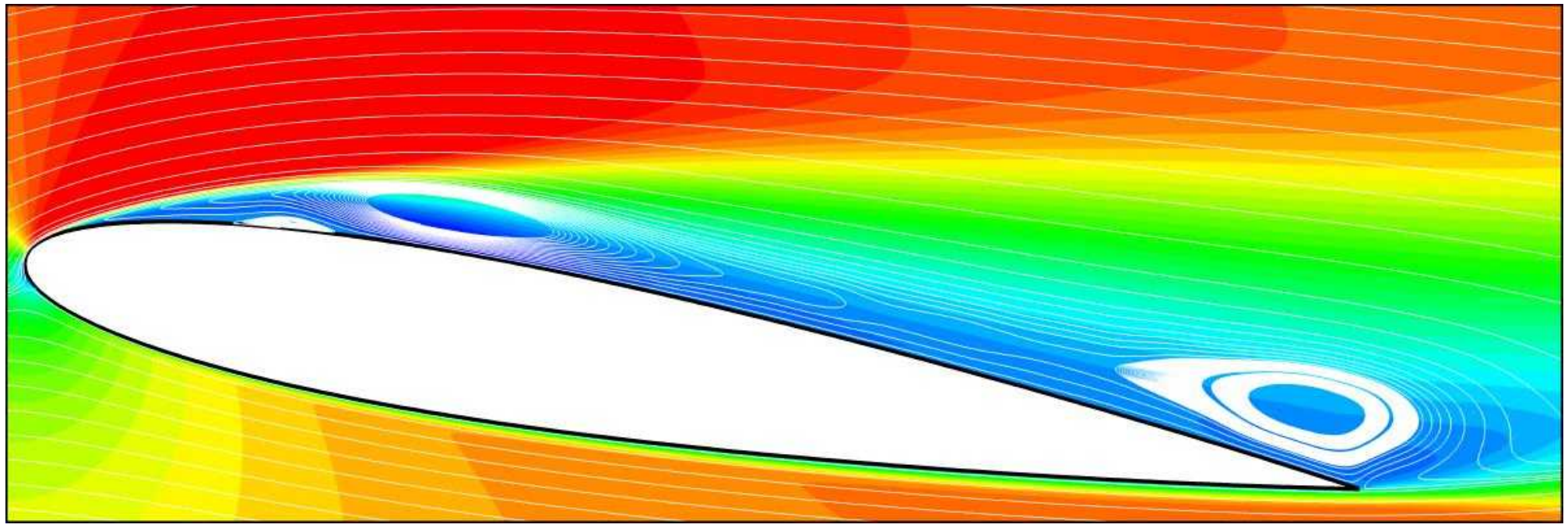}
\textit{time=9.25}
\end{minipage}
\medskip
\begin{minipage}{220pt}
\centering
\includegraphics[width=220pt, trim={0mm 0mm 0mm 0mm}, clip]{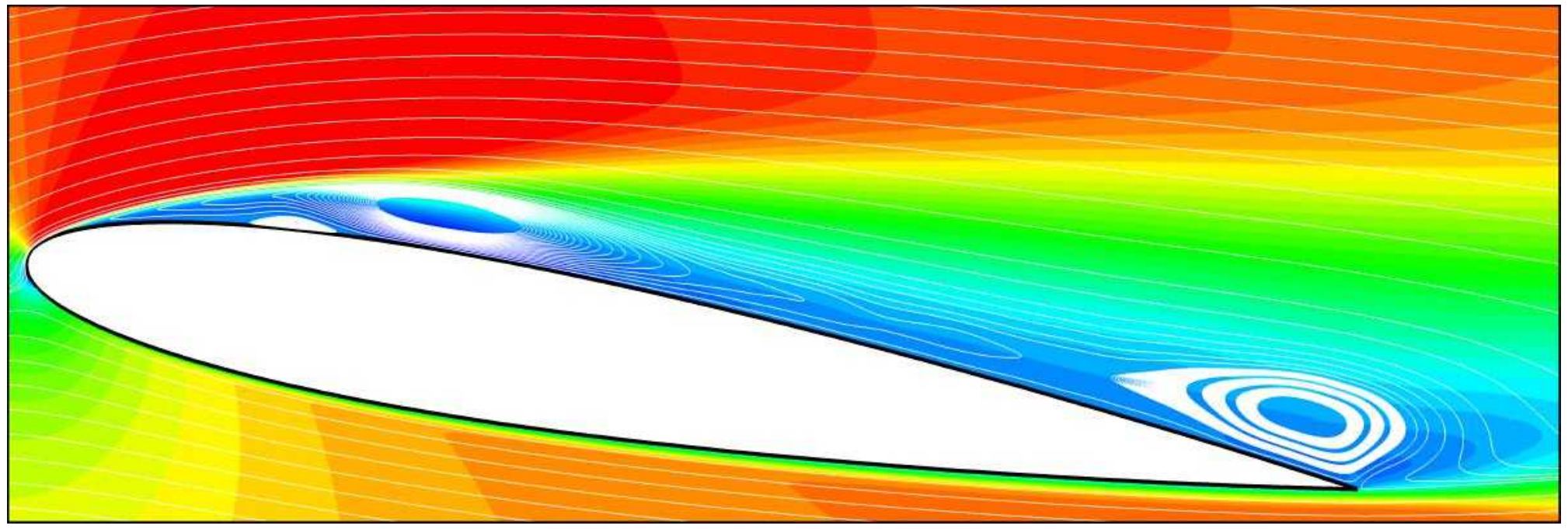}
\textit{time=9.5}
\end{minipage}
\medskip
\begin{minipage}{220pt}
\centering
\includegraphics[width=220pt, trim={0mm 0mm 0mm 0mm}, clip]{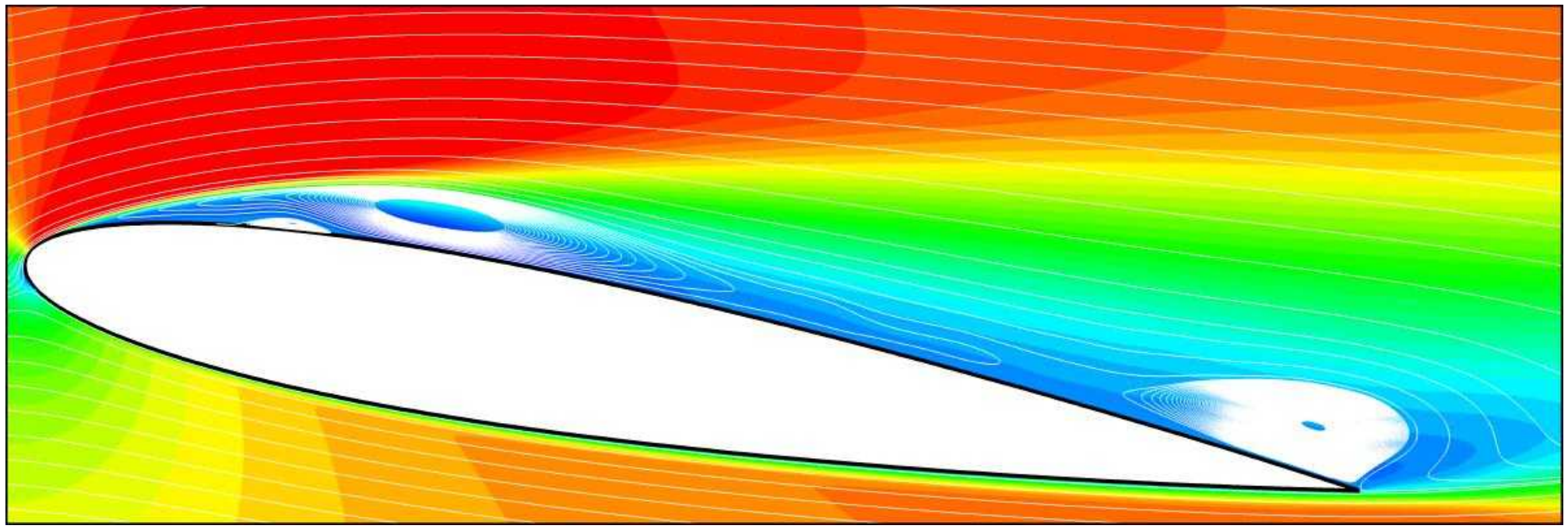}
\textit{time=9.75}
\end{minipage}
\medskip
\begin{minipage}{220pt}
\centering
\includegraphics[width=220pt, trim={0mm 0mm 0mm 0mm}, clip]{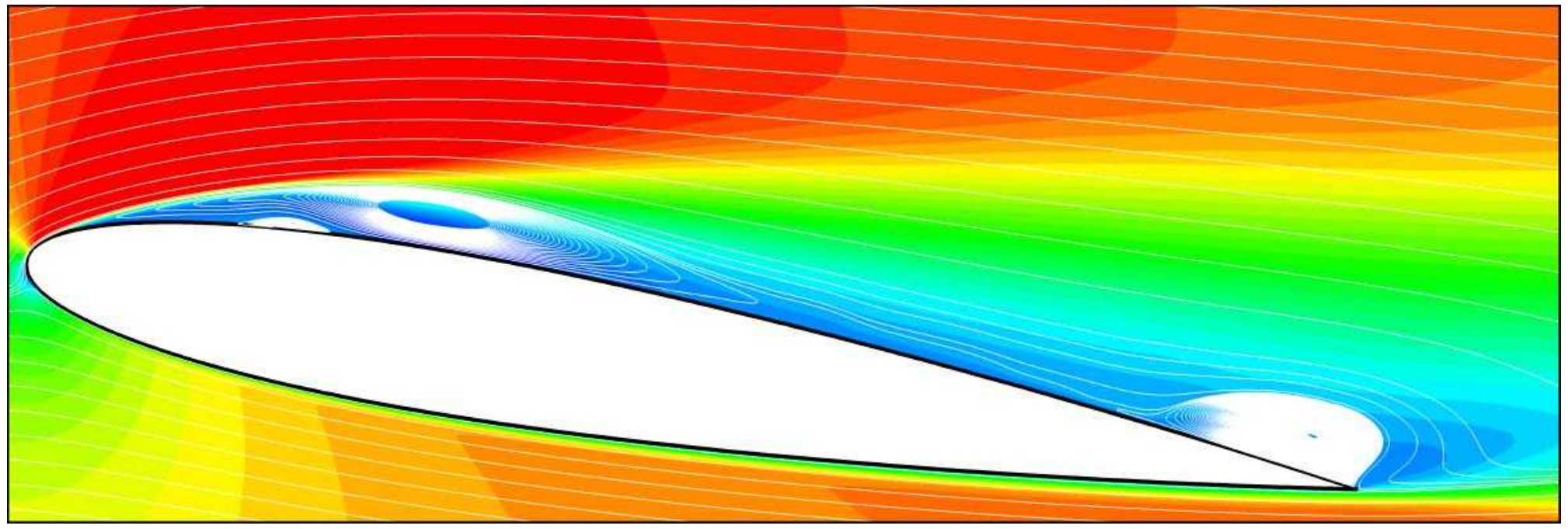}
\textit{time=10.0}
\end{minipage}
\medskip
\begin{minipage}{220pt}
\centering
\includegraphics[width=220pt, trim={0mm 0mm 0mm 0mm}, clip]{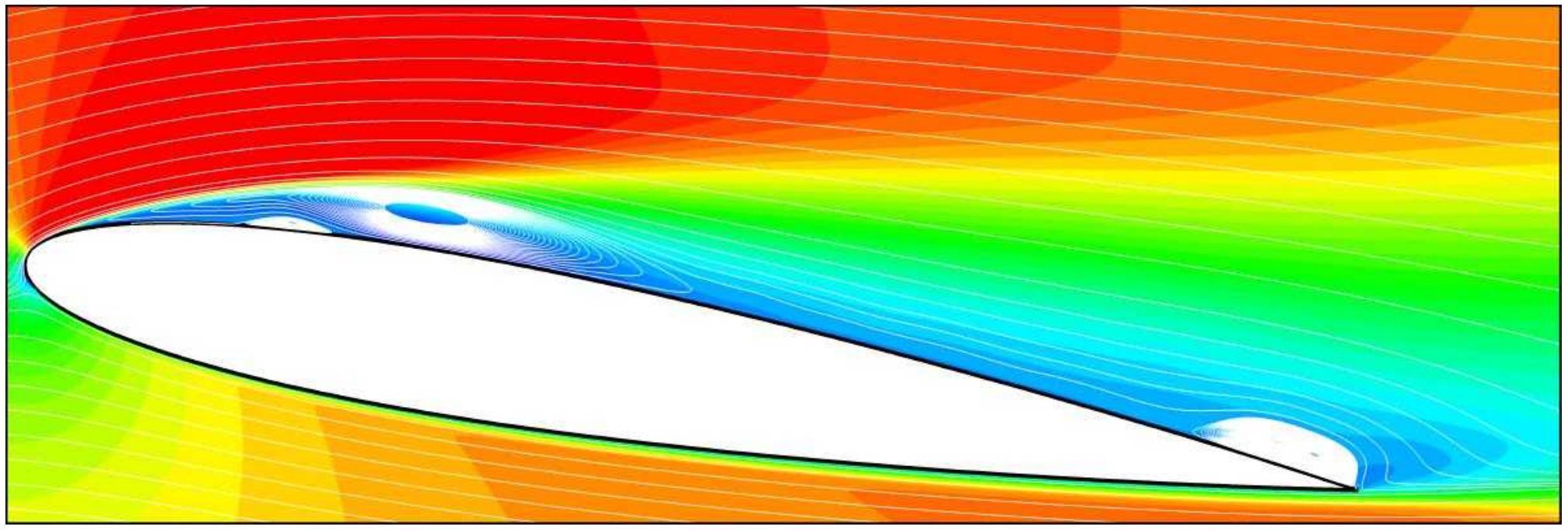}
\textit{time=10.25}
\end{minipage}
\medskip
\begin{minipage}{220pt}
\centering
\includegraphics[width=220pt, trim={0mm 0mm 0mm 0mm}, clip]{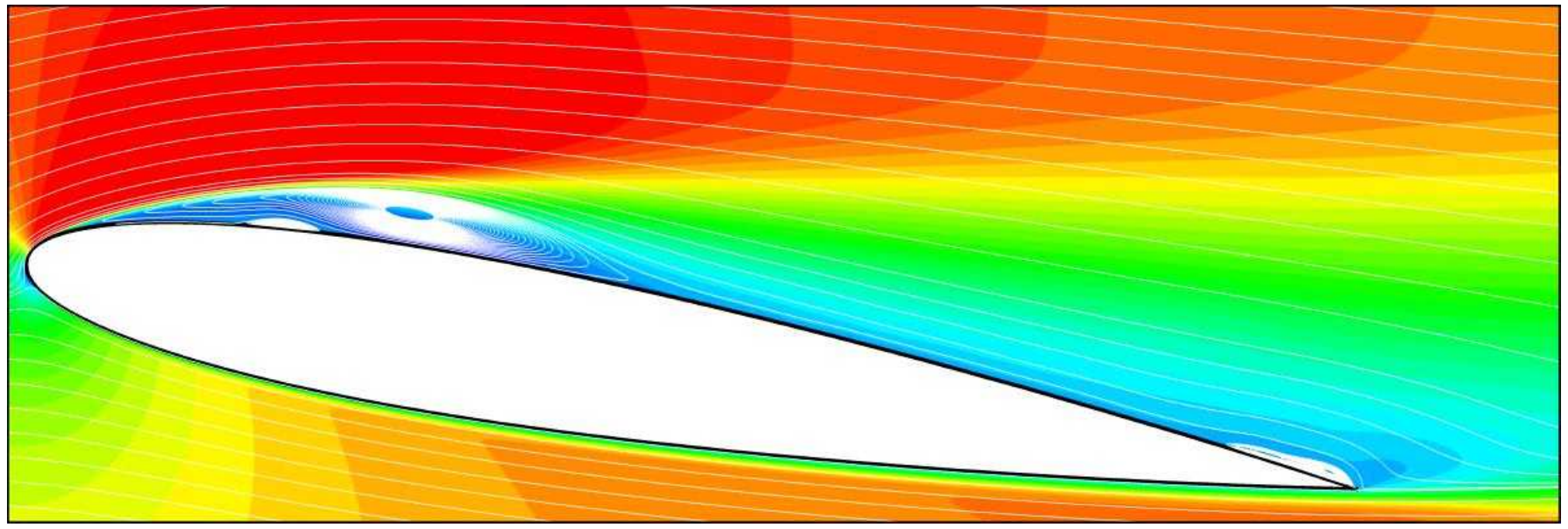}
\textit{time=11.0, $\Phi = 90^{\circ}$}
\end{minipage}
\medskip
\begin{minipage}{220pt}
\centering
\includegraphics[width=220pt, trim={0mm 0mm 0mm 0mm}, clip]{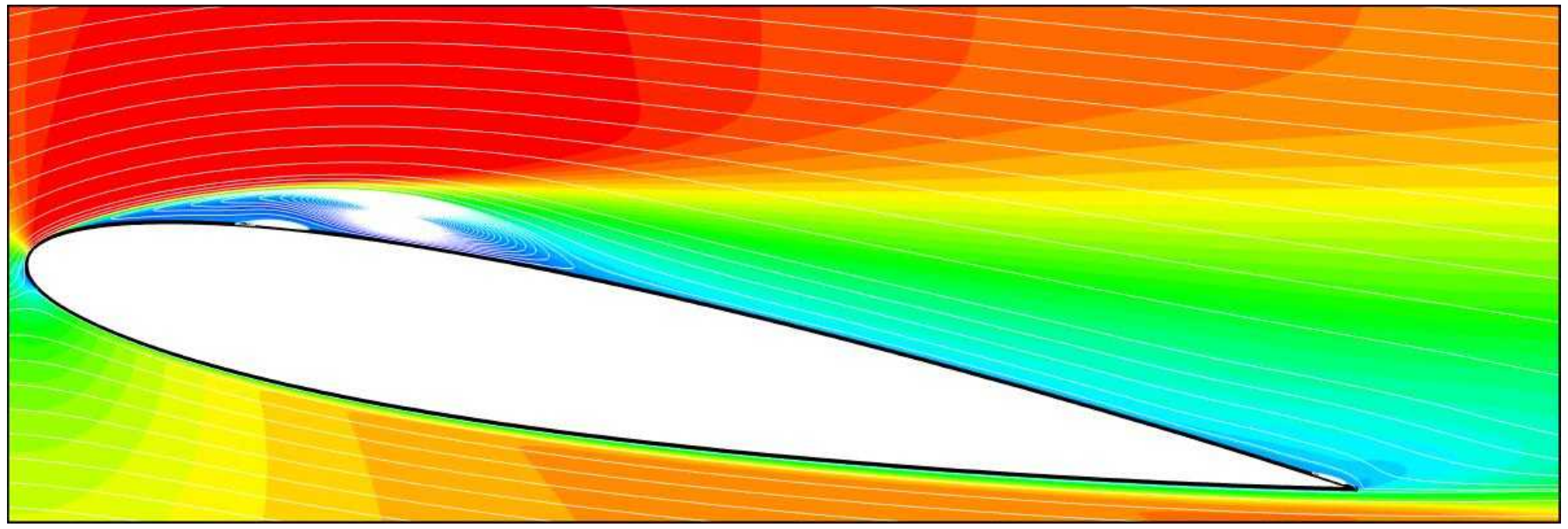}
\textit{time=11.75}
\end{minipage}
\medskip
\begin{minipage}{220pt}
\centering
\includegraphics[width=220pt, trim={0mm 0mm 0mm 0mm}, clip]{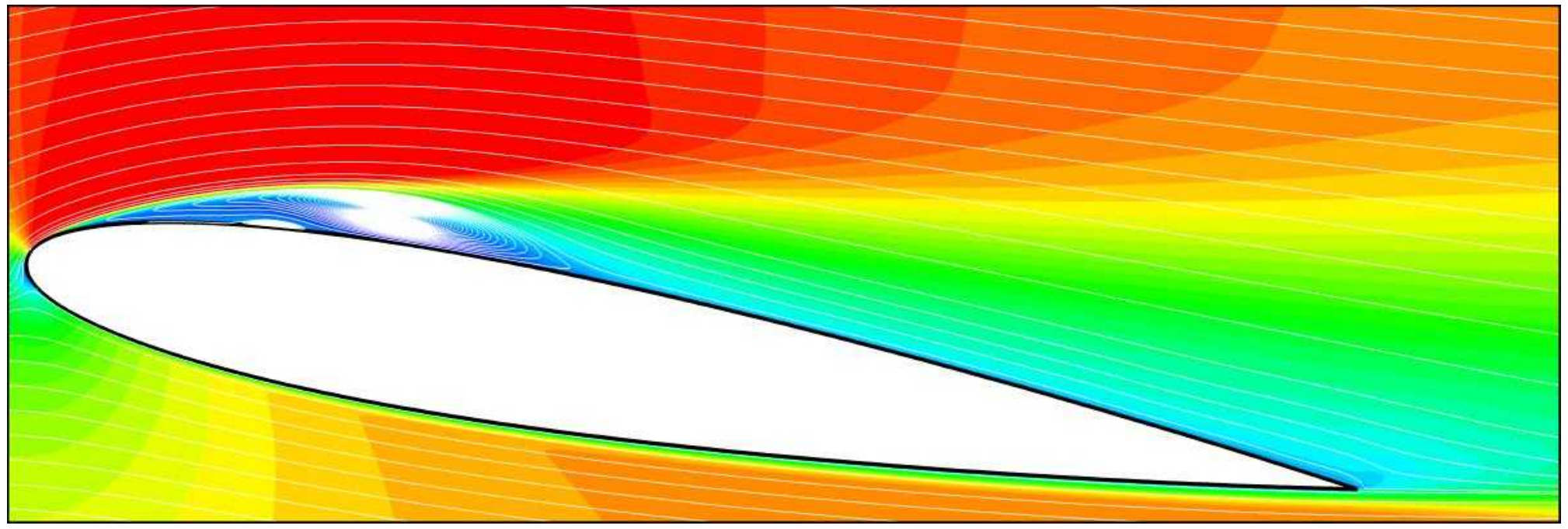}
\textit{time=12.25}
\end{minipage}
\begin{minipage}{220pt}
\centering
\includegraphics[width=220pt, trim={0mm 0mm 0mm 0mm}, clip]{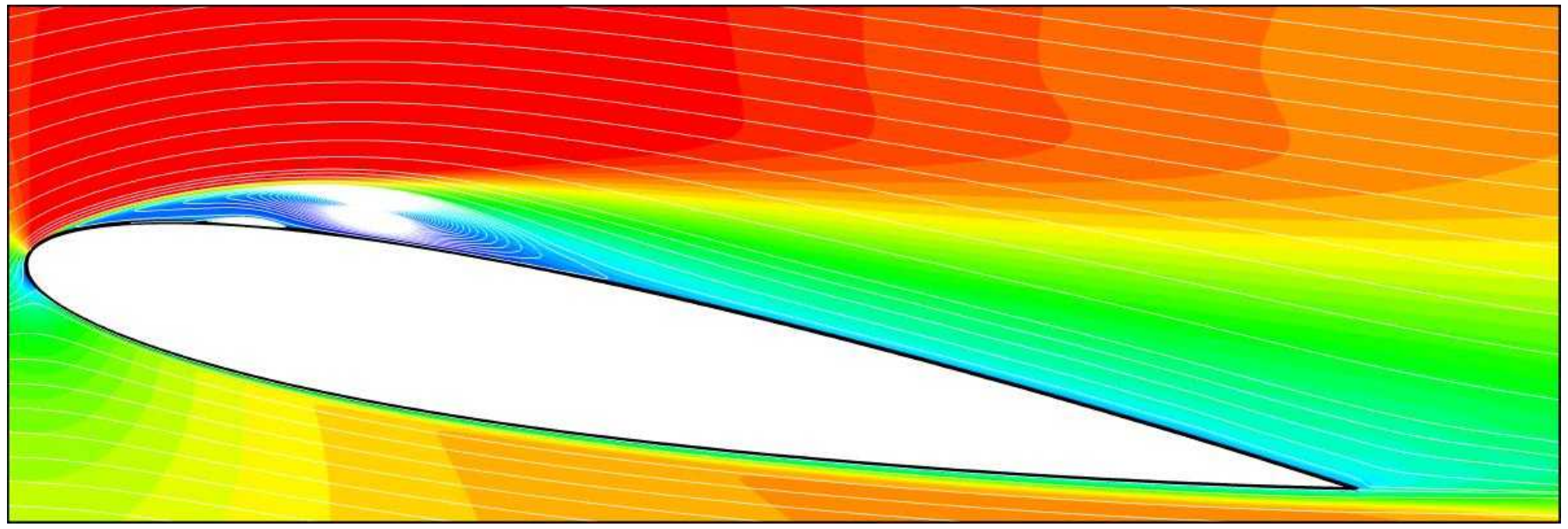}
\textit{time=15.0, $\Phi = 180^{\circ}$}
\end{minipage}
\caption{Streamlines patterns superimposed on colour maps of the streamwise velocity component for the POD reconstruction of the instantaneous f\/low-f\/ield using the LFO mode 1, the LFO mode 2, and the HFO mode for the angle of attack of $9.7^{\circ}$. The f\/low-f\/ield is attaching.}
\label{POD_rec_instant_970}
\end{center}
\end{figure}

\begin{figure}
\begin{center}
\begin{minipage}{145pt}
\centering
\includegraphics[width=145pt, trim={0mm 0mm 0mm 0mm}, clip]{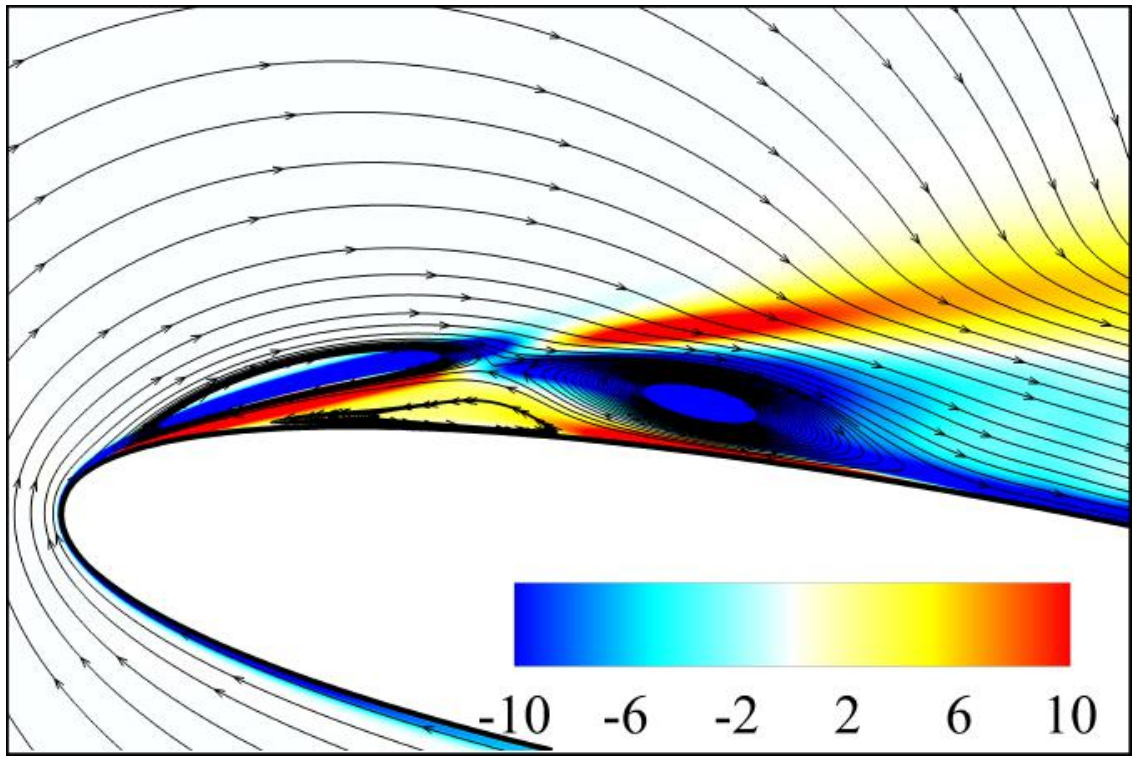}
\textit{time=6.0, $\Phi = 180^{\circ}$}
\end{minipage}
\begin{minipage}{145pt}
\centering
\includegraphics[width=145pt, trim={0mm 0mm 0mm 0mm}, clip]{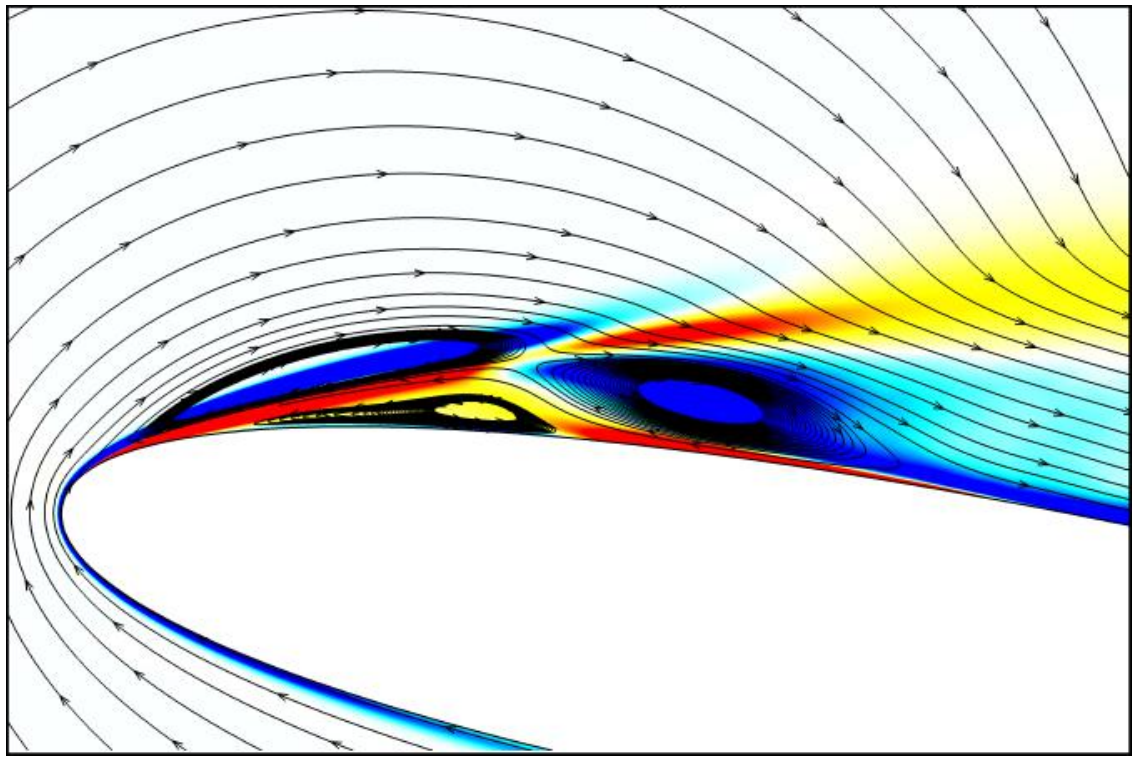}
\textit{time=6.25}
\end{minipage}
\begin{minipage}{145pt}
\centering
\includegraphics[width=145pt, trim={0mm 0mm 0mm 0mm}, clip]{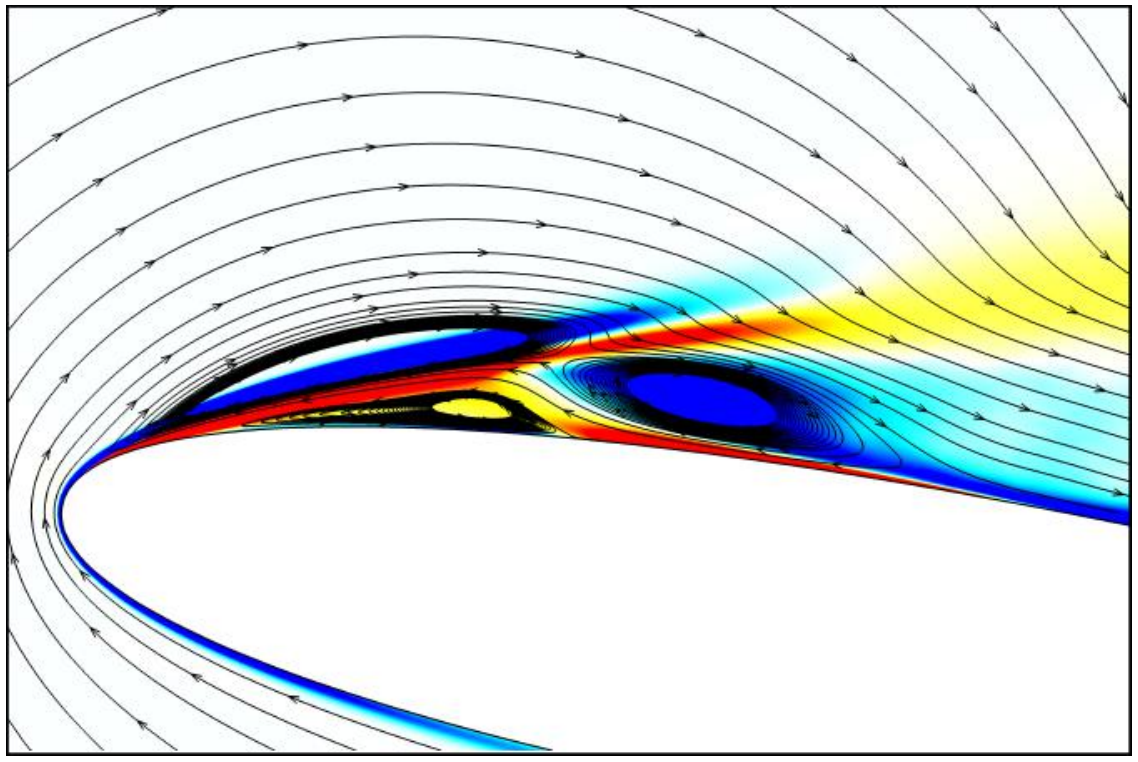}
\textit{time=6.5}
\end{minipage}
\begin{minipage}{145pt}
\centering
\includegraphics[width=145pt, trim={0mm 0mm 0mm 0mm}, clip]{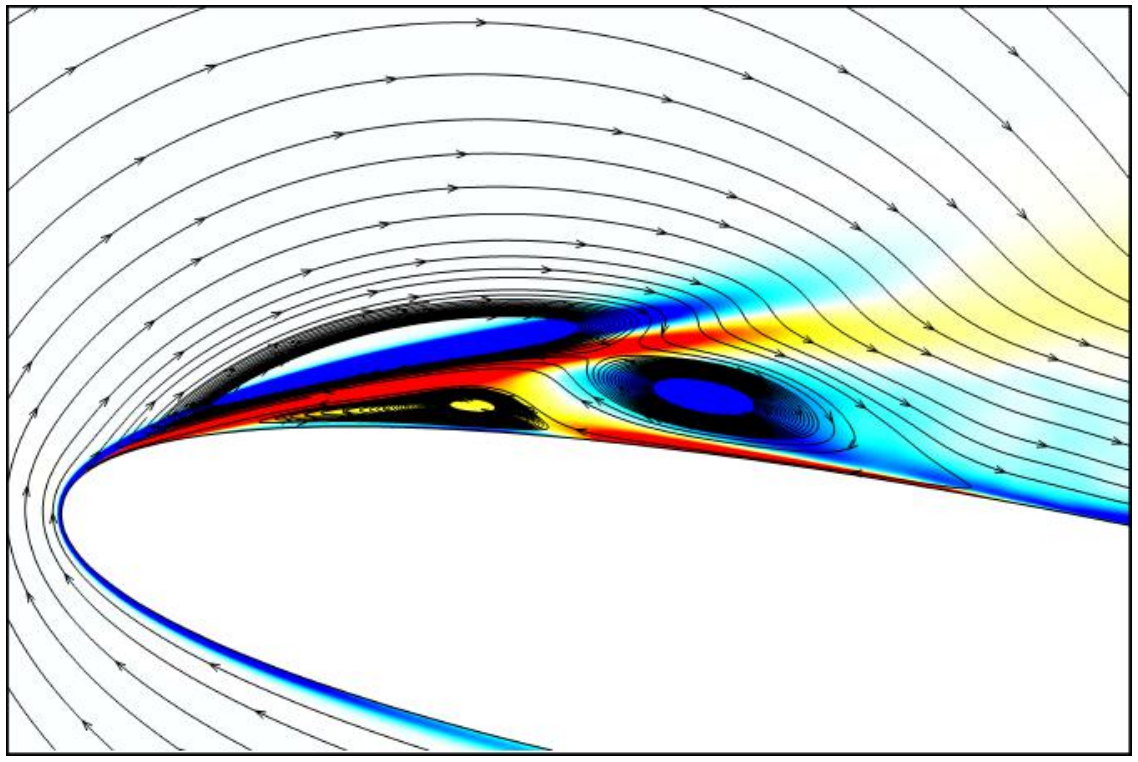}
\textit{time=7.5}
\end{minipage}
\begin{minipage}{145pt}
\centering
\includegraphics[width=145pt, trim={0mm 0mm 0mm 0mm}, clip]{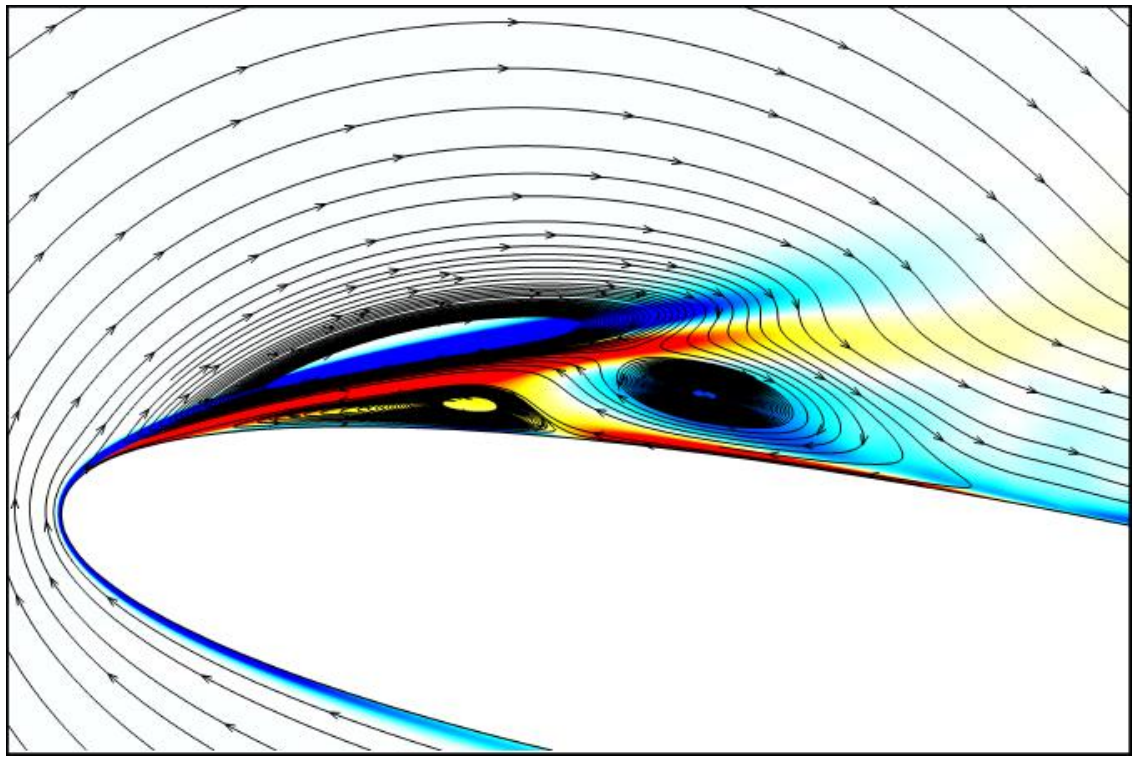}
\textit{time=8.0}
\end{minipage}
\begin{minipage}{145pt}
\centering
\includegraphics[width=145pt, trim={0mm 0mm 0mm 0mm}, clip]{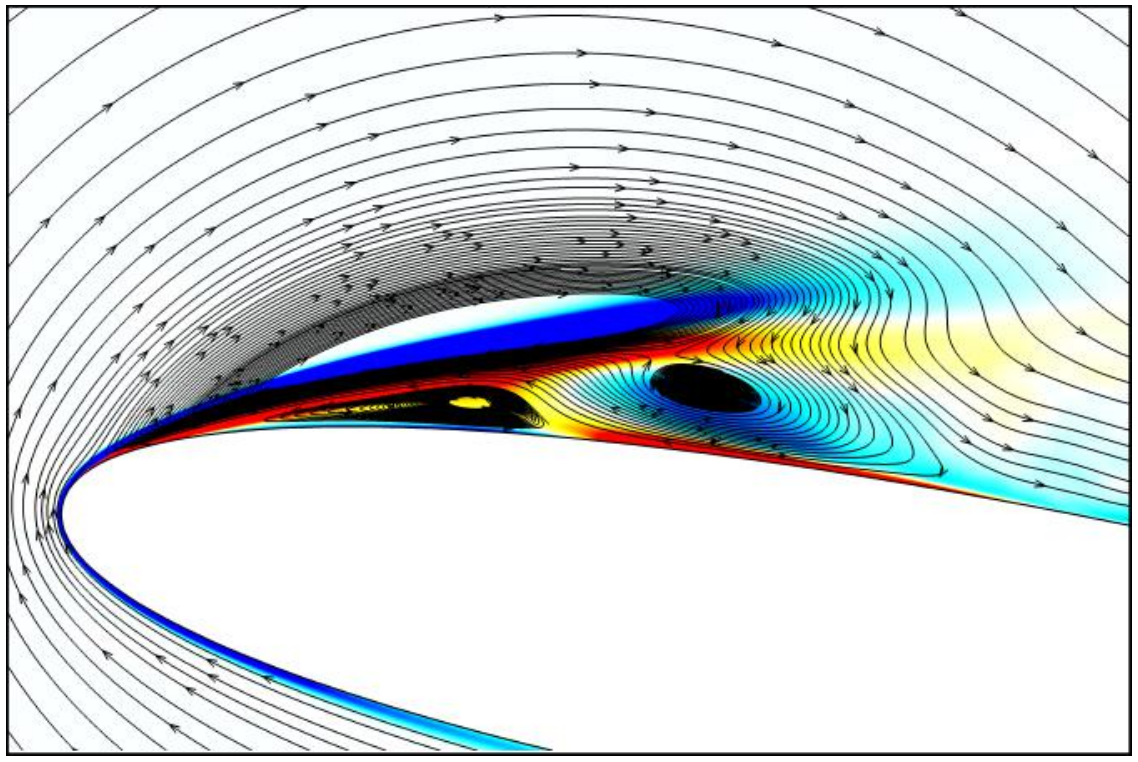}
\textit{time=8.5}
\end{minipage}
\begin{minipage}{145pt}
\centering
\includegraphics[width=145pt, trim={0mm 0mm 0mm 0mm}, clip]{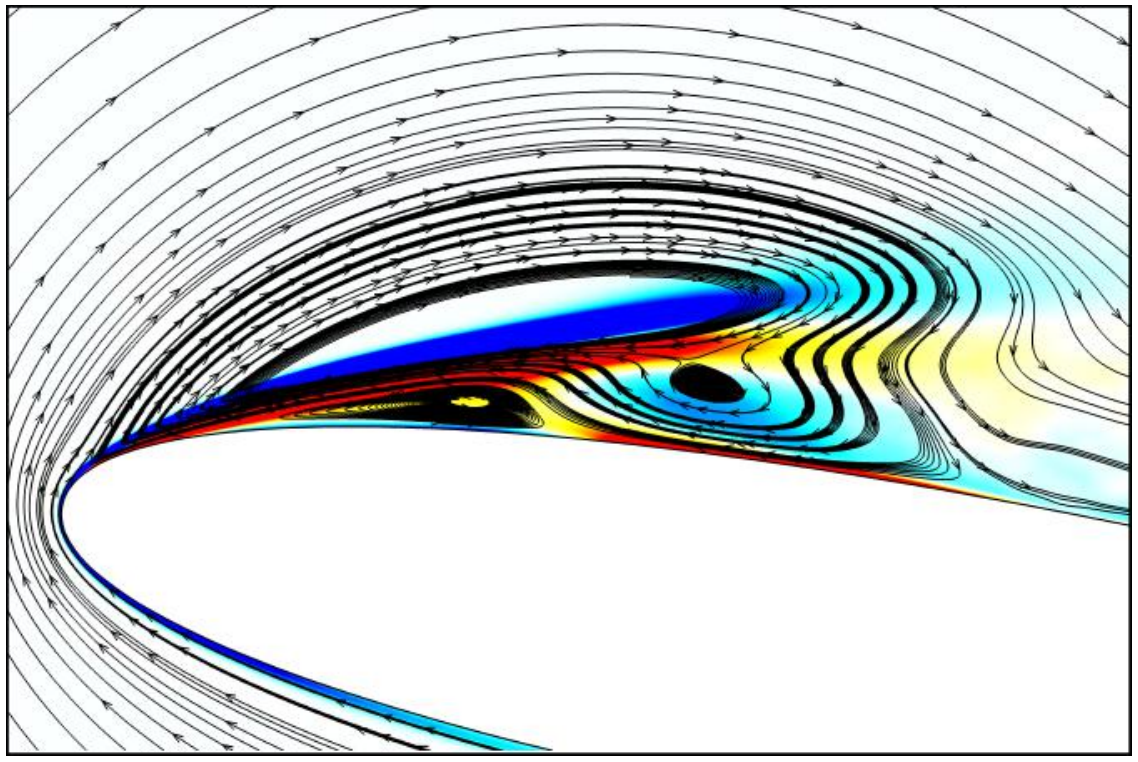}
\textit{time=9.0}
\end{minipage}
\begin{minipage}{145pt}
\centering
\includegraphics[width=145pt, trim={0mm 0mm 0mm 0mm}, clip]{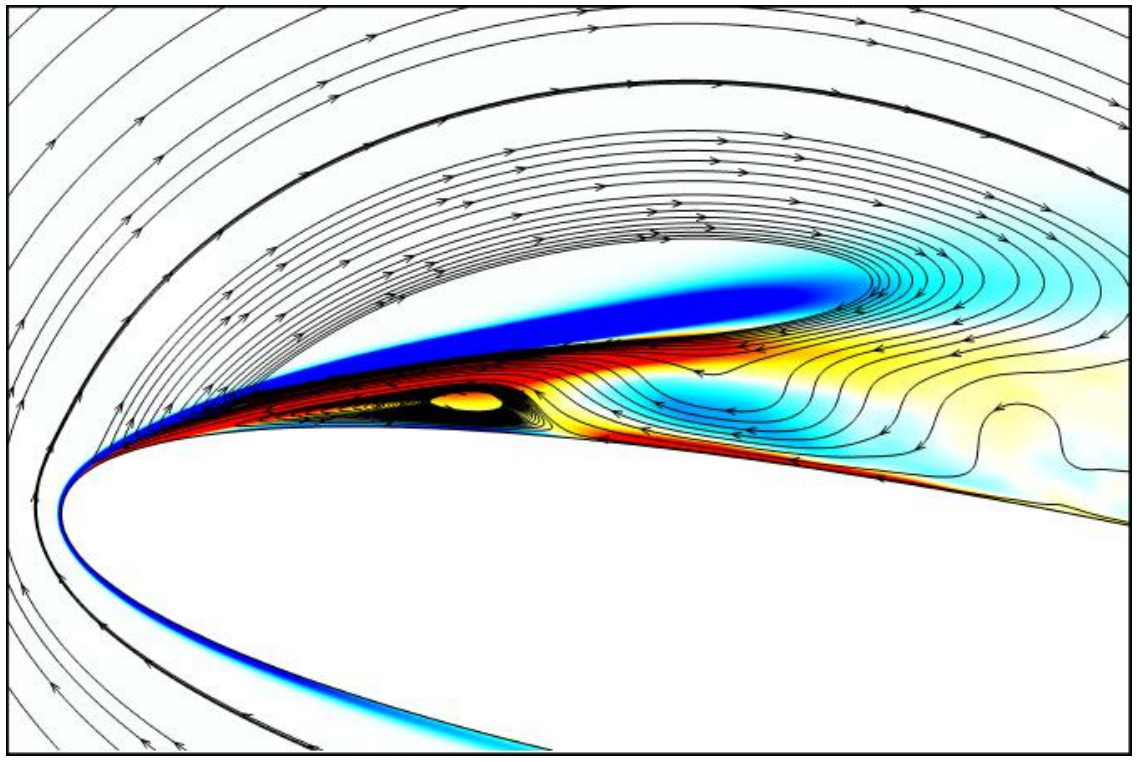}
\textit{time=9.25}
\end{minipage}
\begin{minipage}{145pt}
\centering
\includegraphics[width=145pt, trim={0mm 0mm 0mm 0mm}, clip]{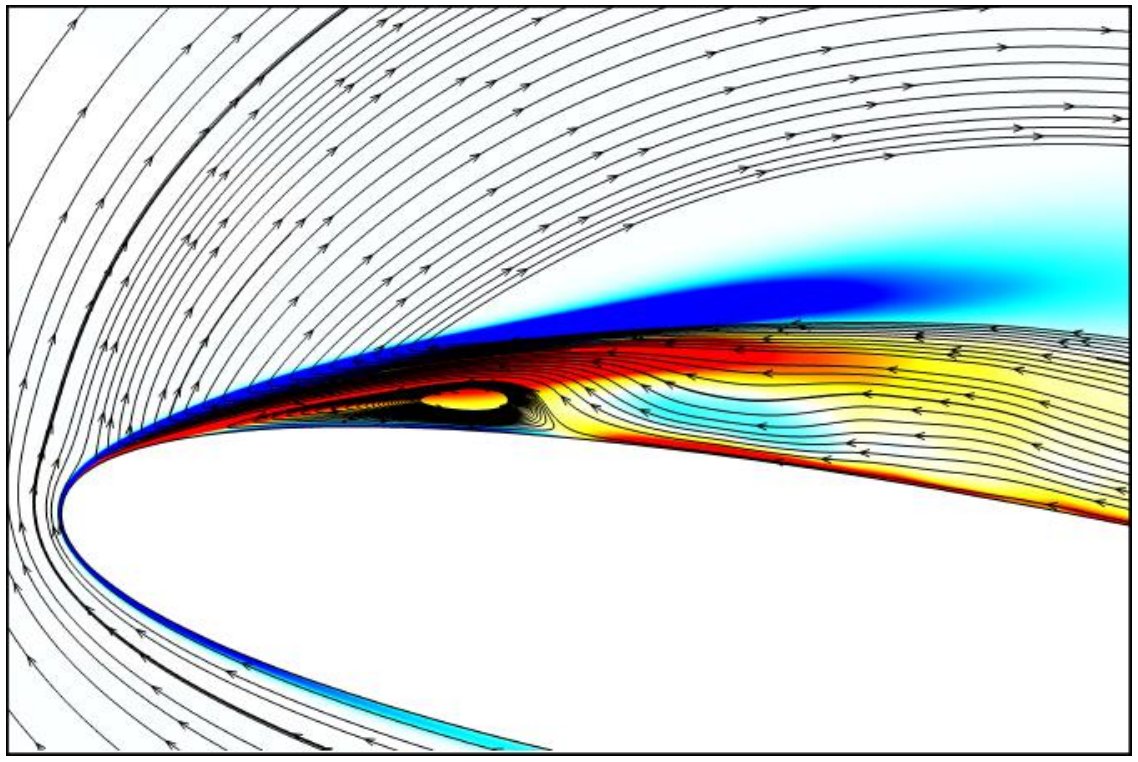}
\textit{time=9.75, $\Phi = 270^{\circ}$}
\end{minipage}
\begin{minipage}{145pt}
\centering
\includegraphics[width=145pt, trim={0mm 0mm 0mm 0mm}, clip]{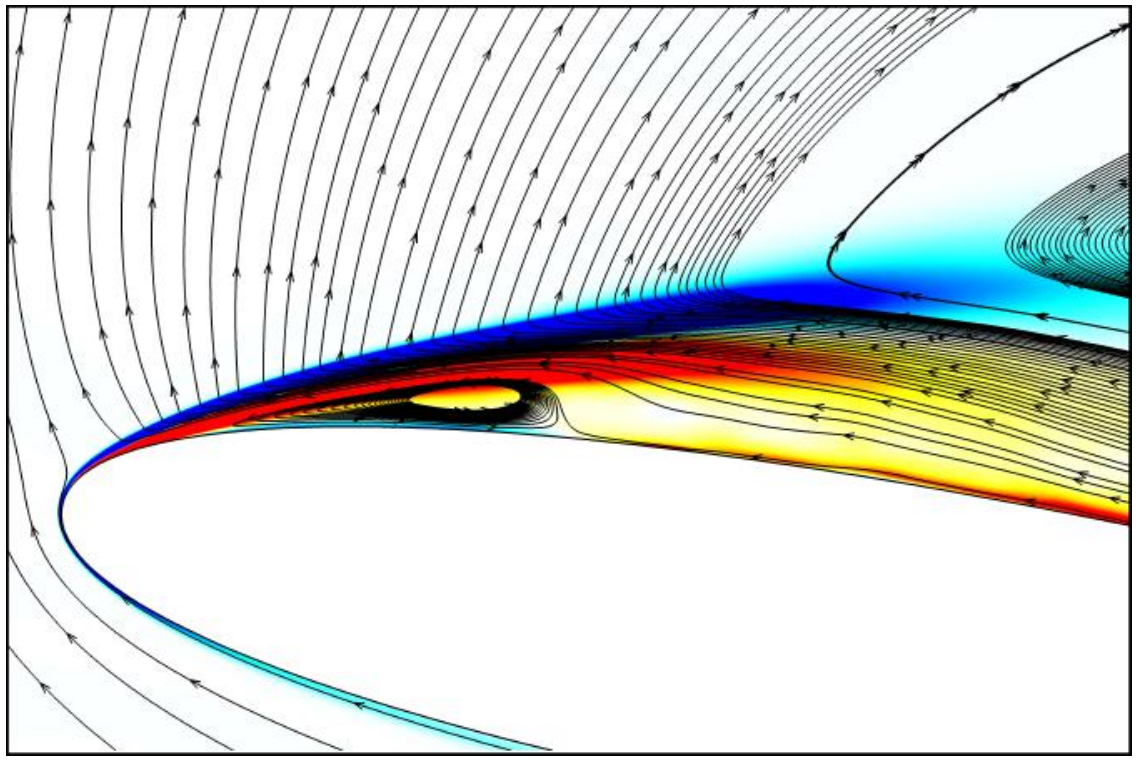}
\textit{time=10.75}
\end{minipage}
\begin{minipage}{145pt}
\centering
\includegraphics[width=145pt, trim={0mm 0mm 0mm 0mm}, clip]{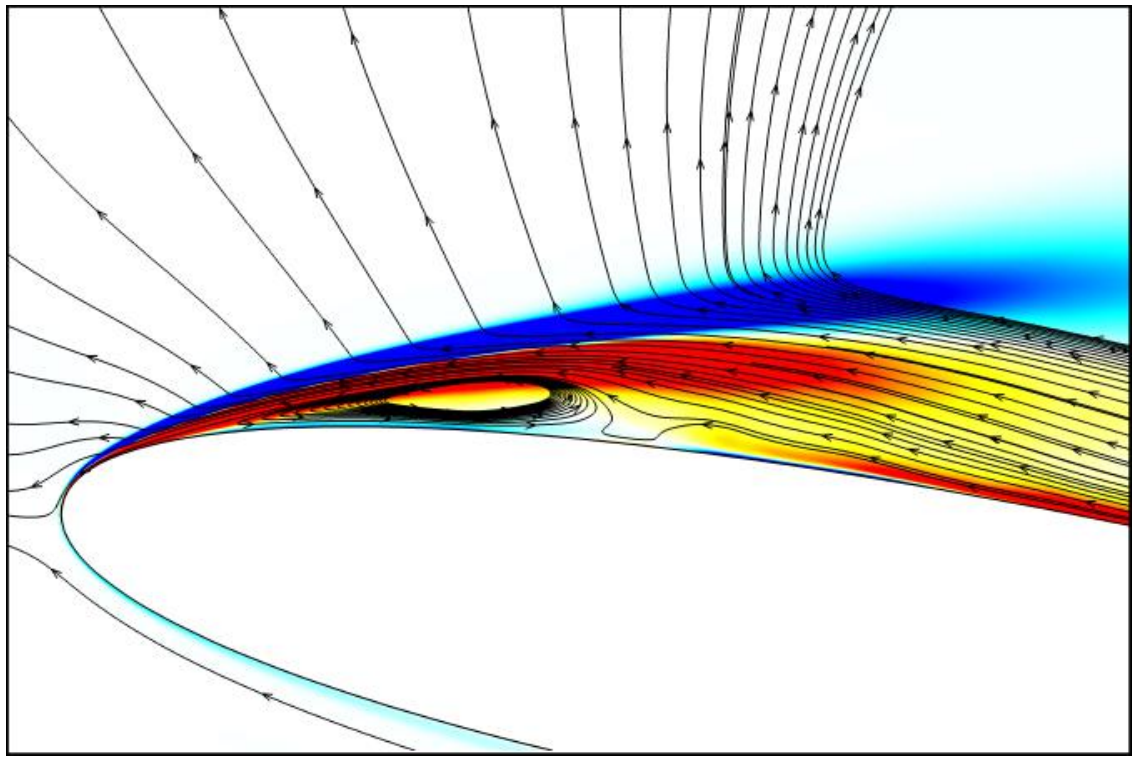}
\textit{time=11.75}
\end{minipage}
\begin{minipage}{145pt}
\centering
\includegraphics[width=145pt, trim={0mm 0mm 0mm 0mm}, clip]{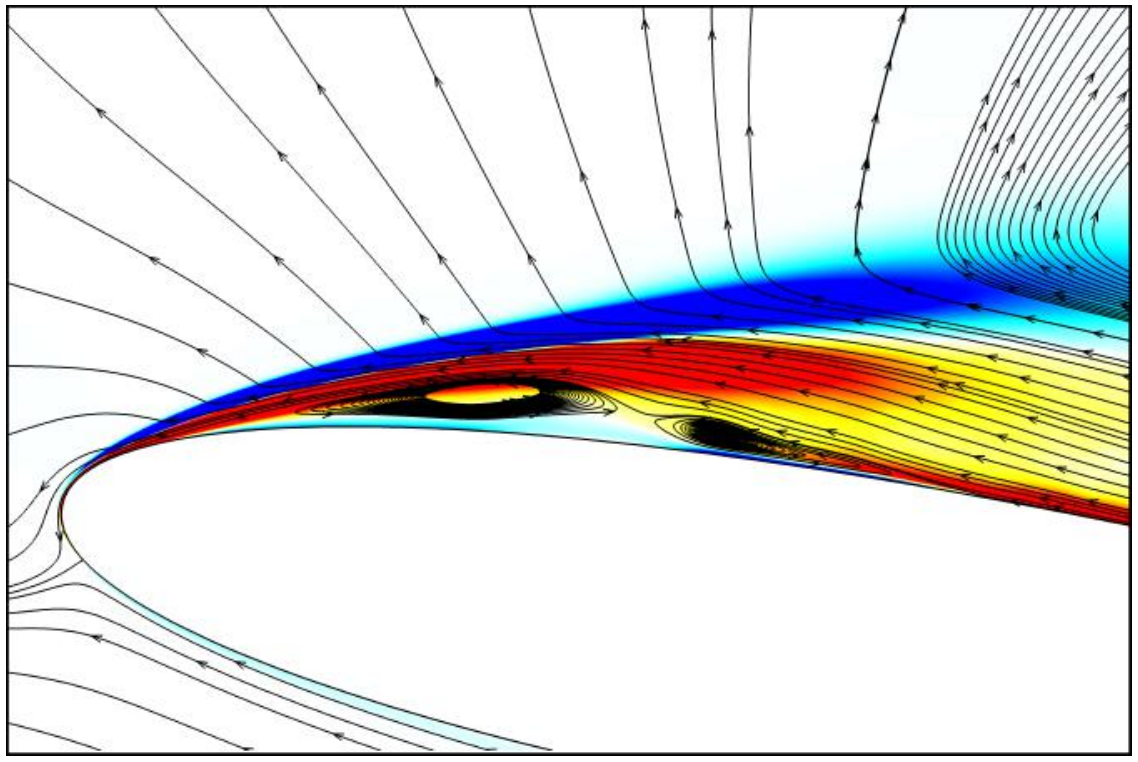}
\textit{time=12.0}
\end{minipage}
\begin{minipage}{145pt}
\centering
\includegraphics[width=145pt, trim={0mm 0mm 0mm 0mm}, clip]{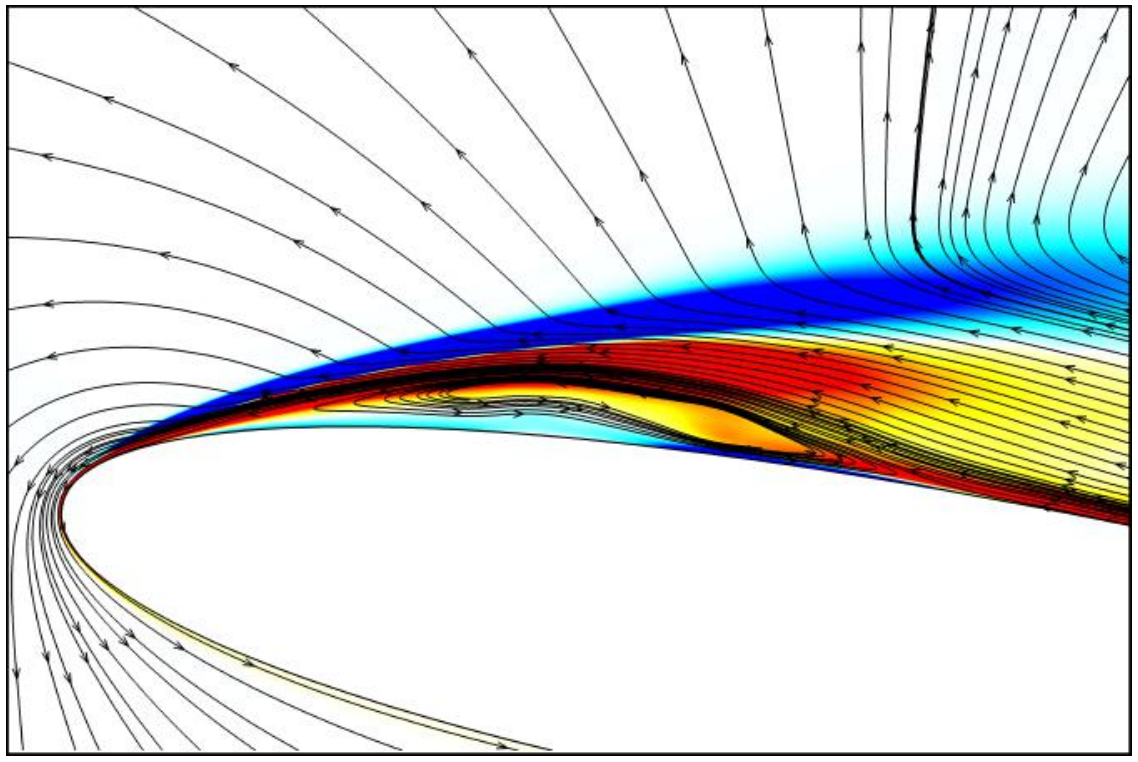}
\textit{time=12.5}
\end{minipage}
\begin{minipage}{145pt}
\centering
\includegraphics[width=145pt, trim={0mm 0mm 0mm 0mm}, clip]{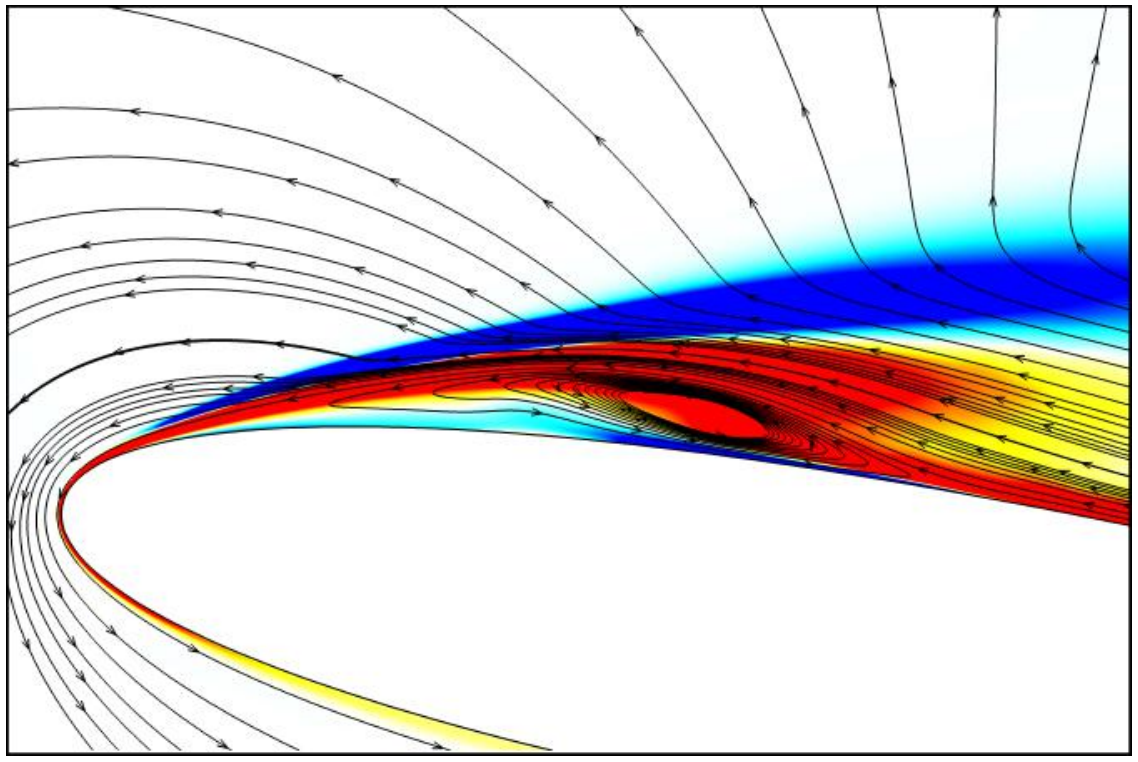}
\textit{time=14.0}
\end{minipage}
\begin{minipage}{145pt}
\centering
\includegraphics[width=145pt, trim={0mm 0mm 0mm 0mm}, clip]{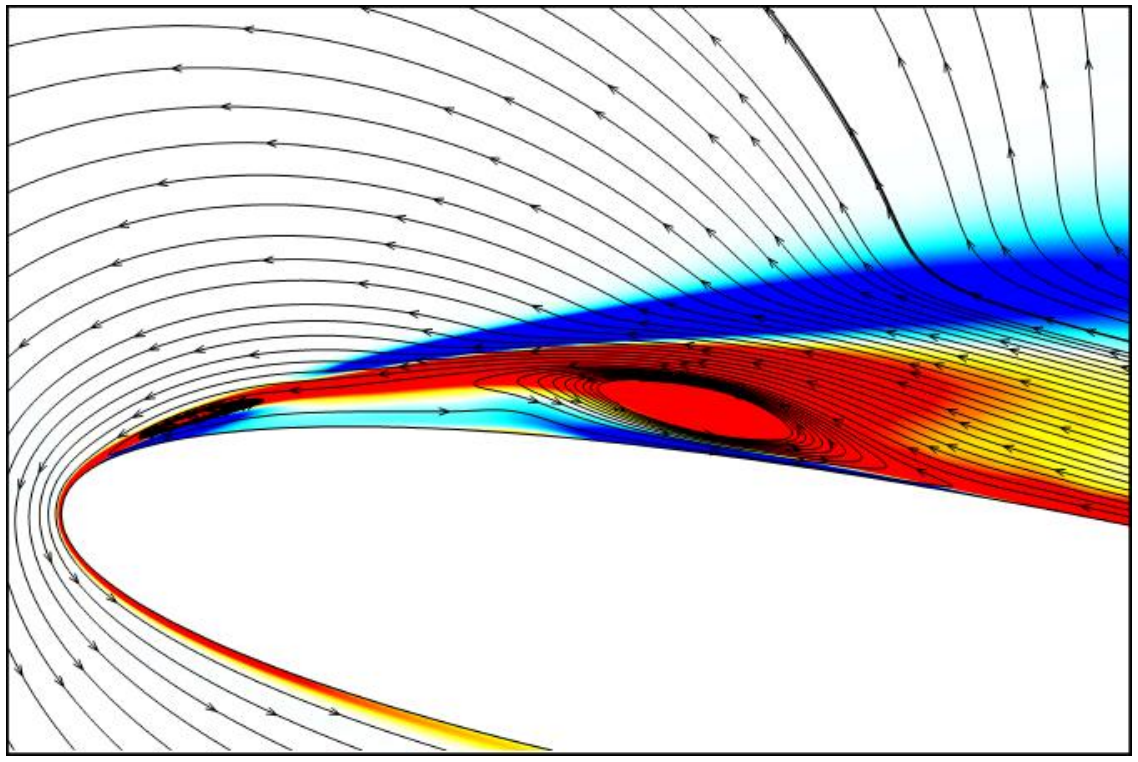}
\textit{time=15.25}
\end{minipage}
\begin{minipage}{145pt}
\centering
\includegraphics[width=145pt, trim={0mm 0mm 0mm 0mm}, clip]{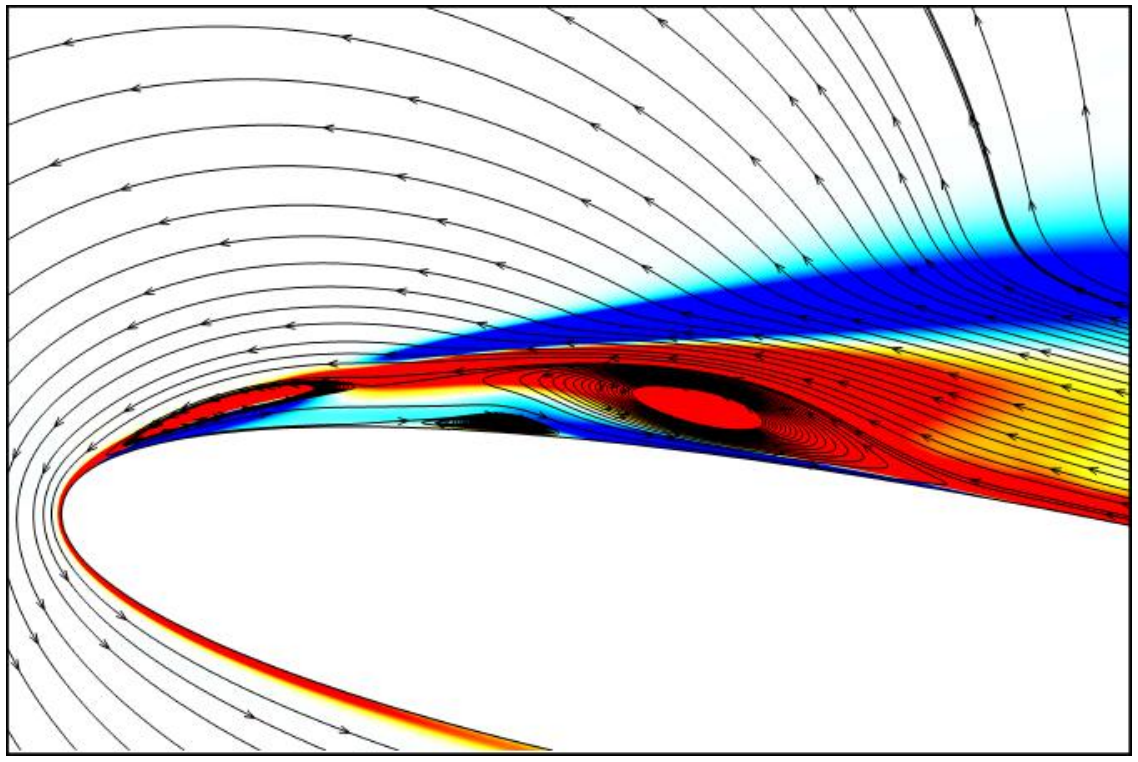}
\textit{time=16.25}
\end{minipage}
\begin{minipage}{145pt}
\centering
\includegraphics[width=145pt, trim={0mm 0mm 0mm 0mm}, clip]{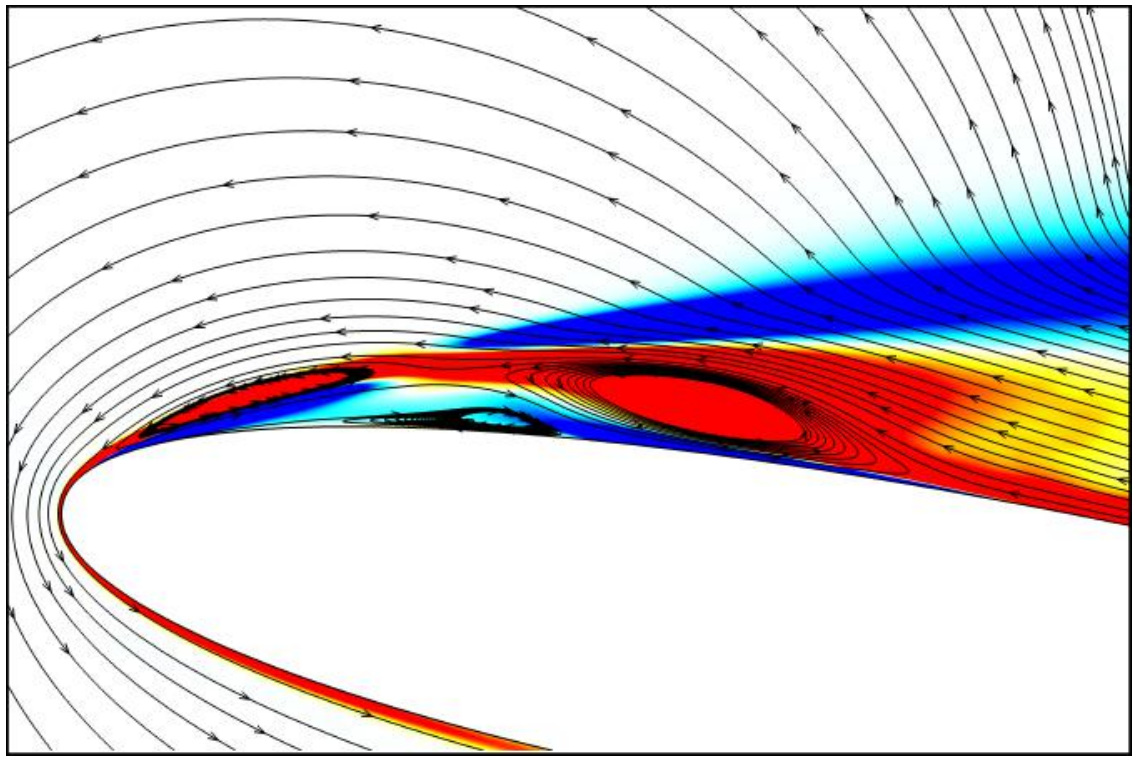}
\textit{time=16.75}
\end{minipage}
\begin{minipage}{145pt}
\centering
\includegraphics[width=145pt, trim={0mm 0mm 0mm 0mm}, clip]{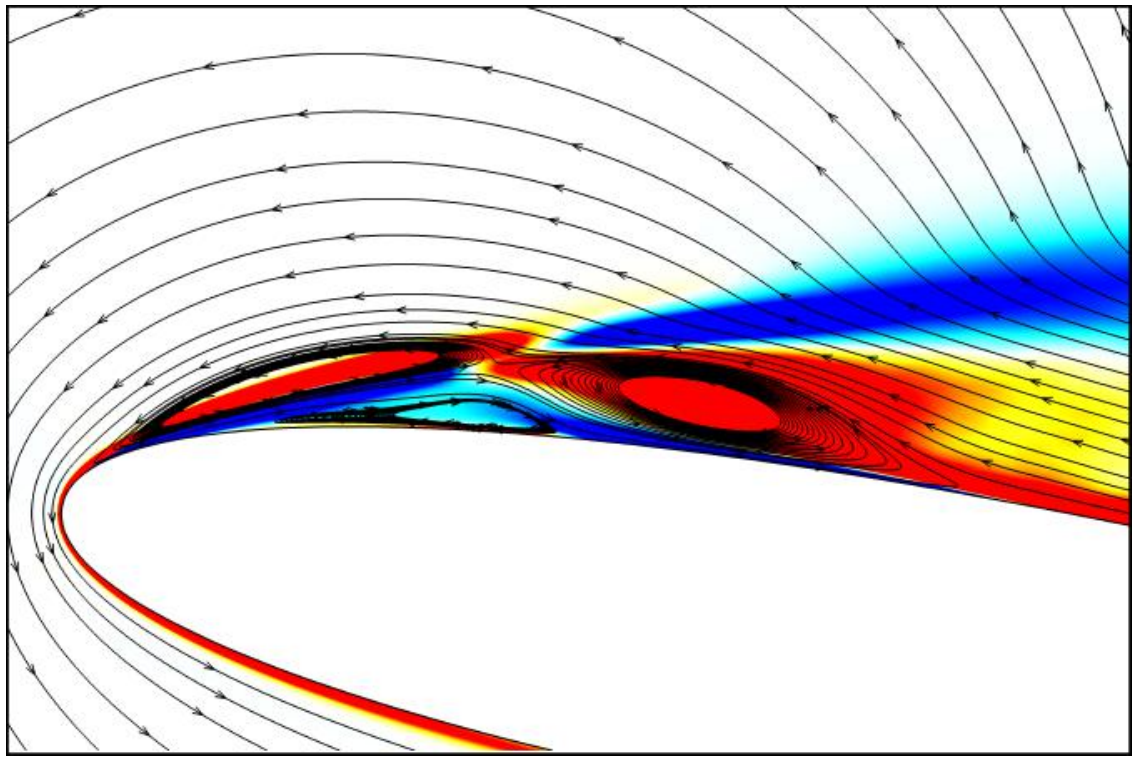}
\textit{time=17.5, $\Phi = 360^{\circ}$}
\end{minipage}
\caption{Streamlines patterns superimposed on colour maps of the spanwise vorticity $\omega_z$ for the POD reconstruction of the oscillating-f\/low using the LFO mode 1, the LFO mode 2, and the HFO mode for the angle of attack of $9.8^{\circ}$. The f\/low-f\/ield is separating.}
\label{POD_rec_TCV_980}
\end{center}
\end{figure}

\begin{figure}
\begin{center}
\begin{minipage}{220pt}
\centering
\includegraphics[width=220pt, trim={0mm 0mm 0mm 0mm}, clip]{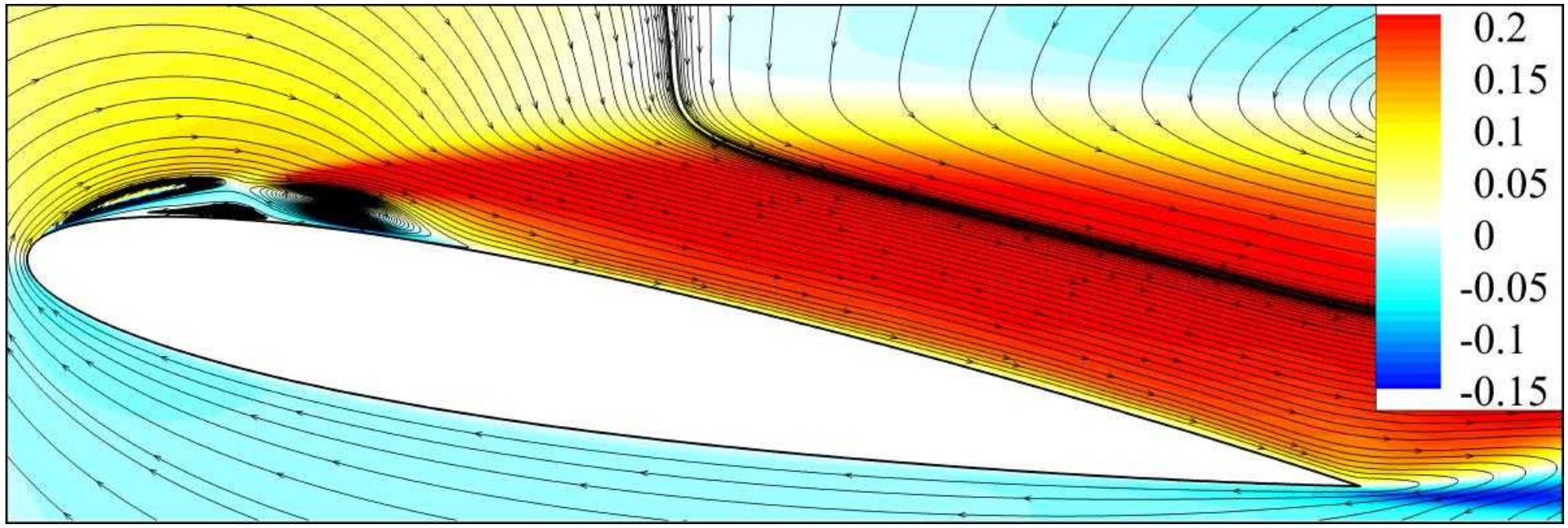}
\textit{time=6.0, $\Phi = 180^{\circ}$}
\end{minipage}
\medskip
\begin{minipage}{220pt}
\centering
\includegraphics[width=220pt, trim={0mm 0mm 0mm 0mm}, clip]{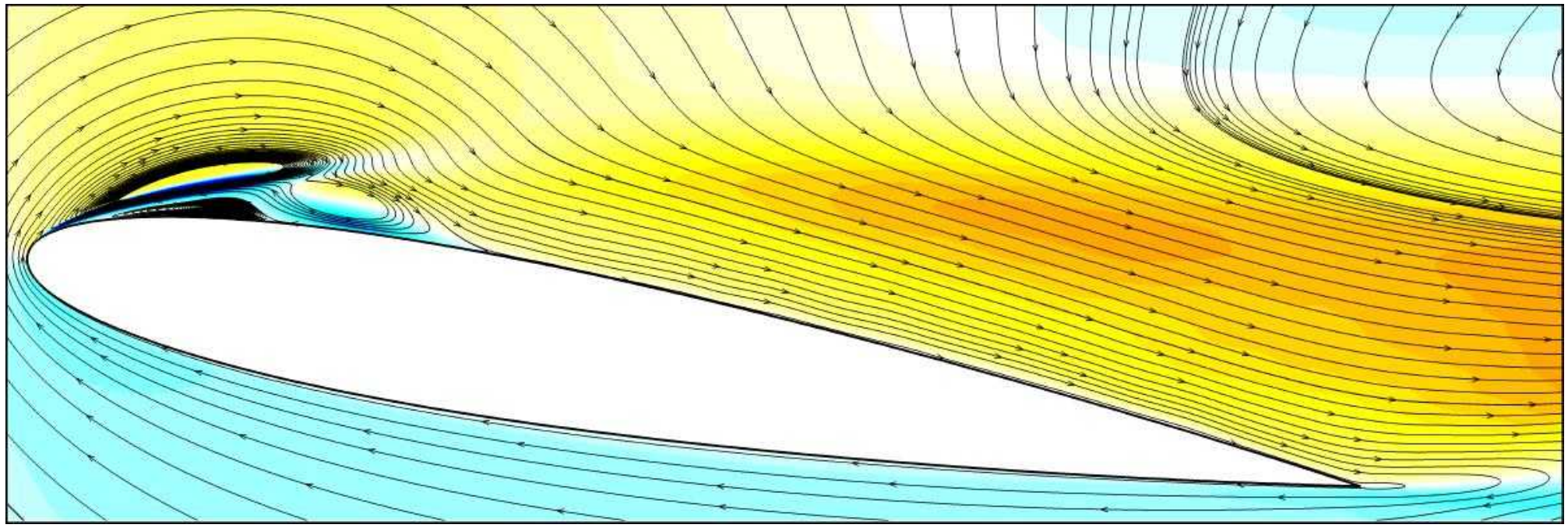}
\textit{time=8.0}
\end{minipage}
\medskip
\begin{minipage}{220pt}
\centering
\includegraphics[width=220pt, trim={0mm 0mm 0mm 0mm}, clip]{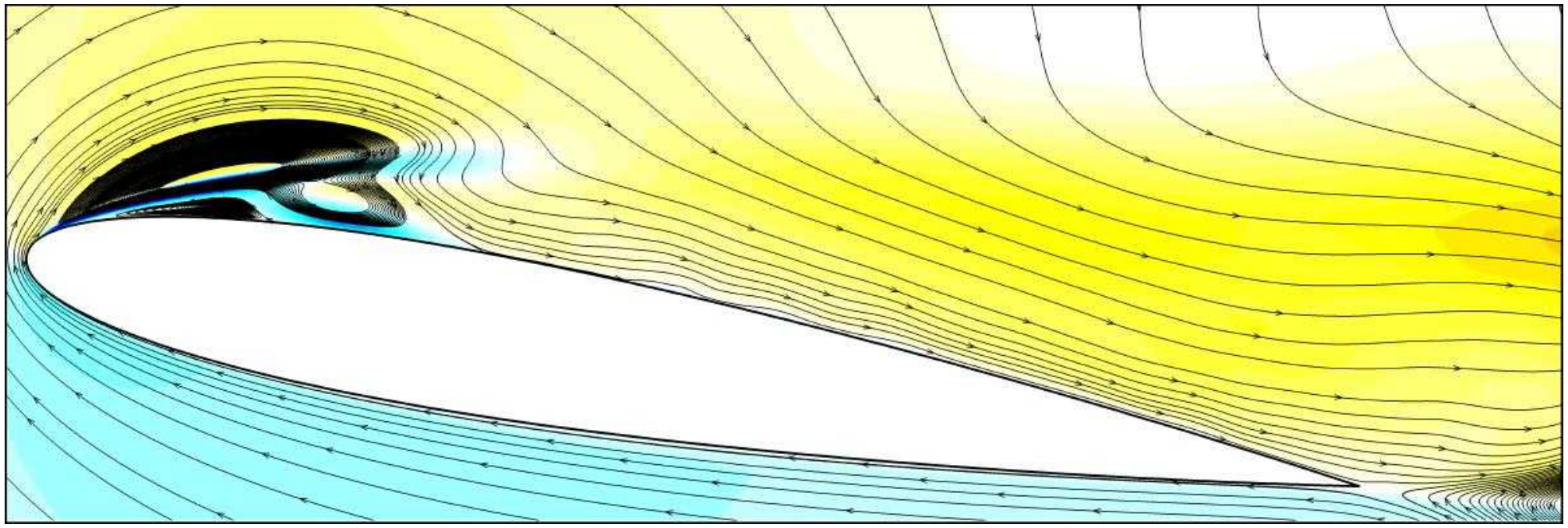}
\textit{time=9.75, $\Phi = 270^{\circ}$}
\end{minipage}
\medskip
\begin{minipage}{220pt}
\centering
\includegraphics[width=220pt, trim={0mm 0mm 0mm 0mm}, clip]{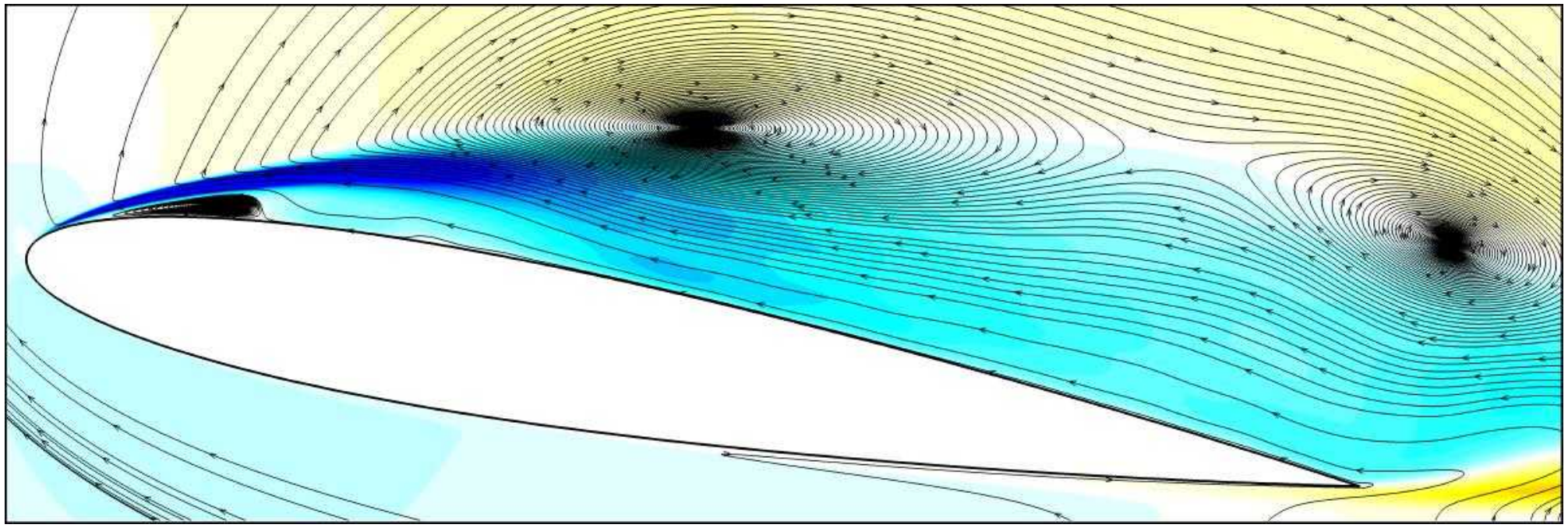}
\textit{time=10.75}
\end{minipage}
\medskip
\begin{minipage}{220pt}
\centering
\includegraphics[width=220pt, trim={0mm 0mm 0mm 0mm}, clip]{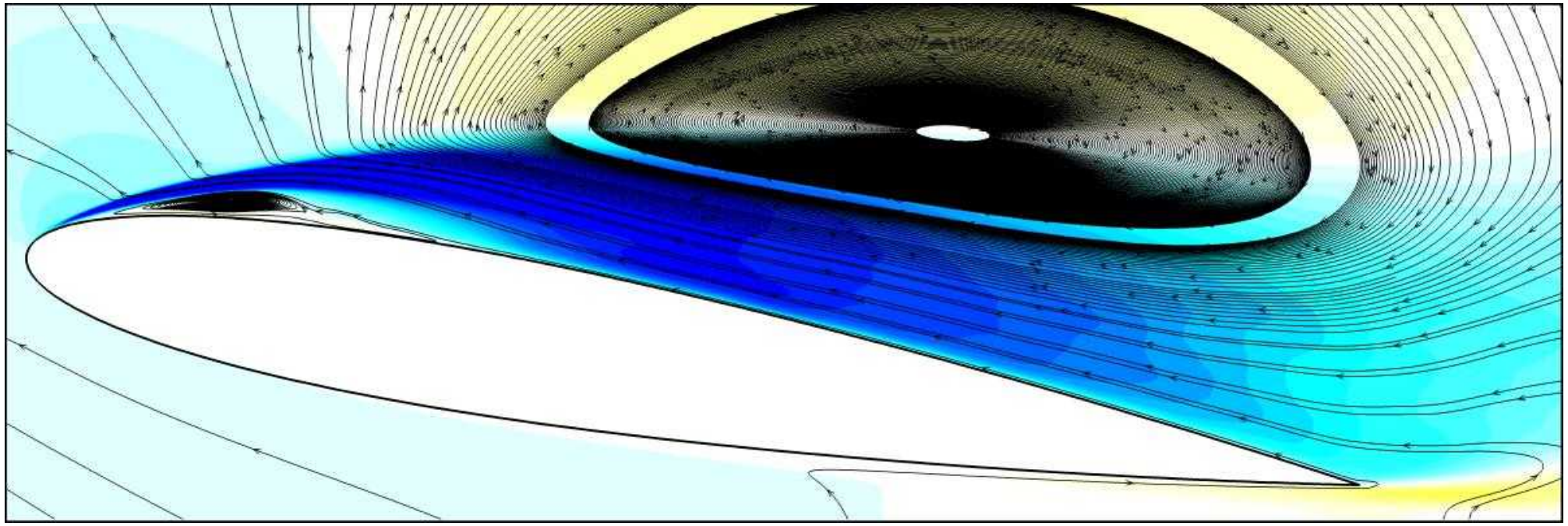}
\textit{time=11.75}
\end{minipage}
\medskip
\begin{minipage}{220pt}
\centering
\includegraphics[width=220pt, trim={0mm 0mm 0mm 0mm}, clip]{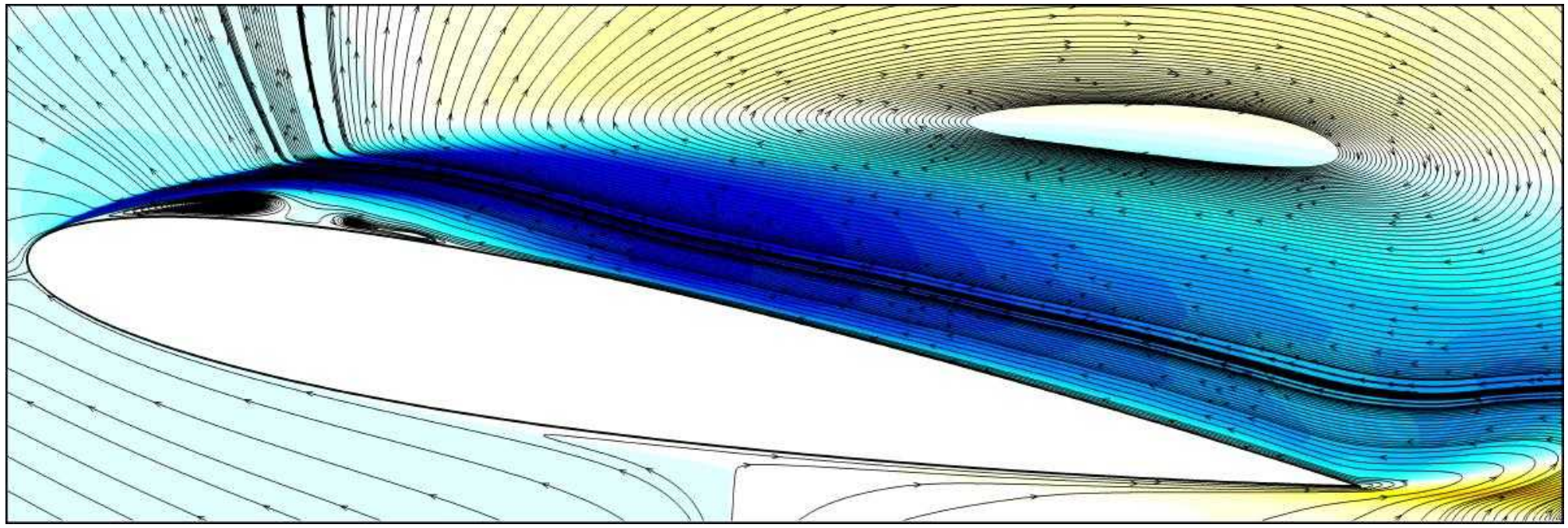}
\textit{time=12.0}
\end{minipage}
\medskip
\begin{minipage}{220pt}
\centering
\includegraphics[width=220pt, trim={0mm 0mm 0mm 0mm}, clip]{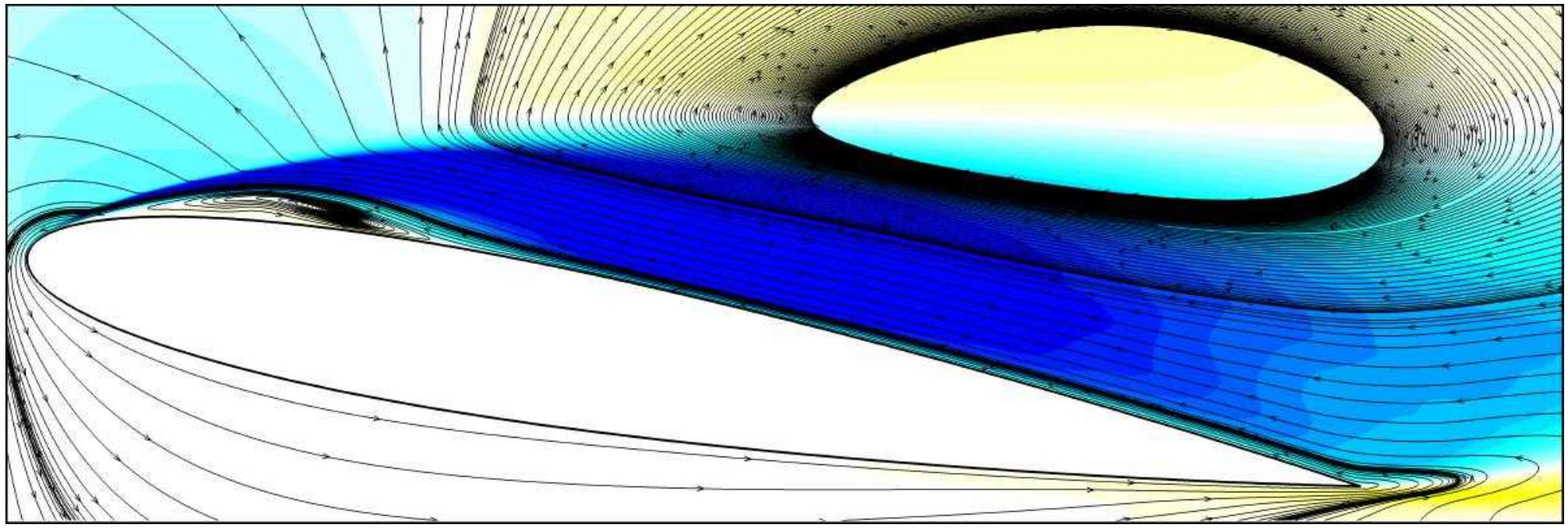}
\textit{time=12.5}
\end{minipage}
\medskip
\begin{minipage}{220pt}
\centering
\includegraphics[width=220pt, trim={0mm 0mm 0mm 0mm}, clip]{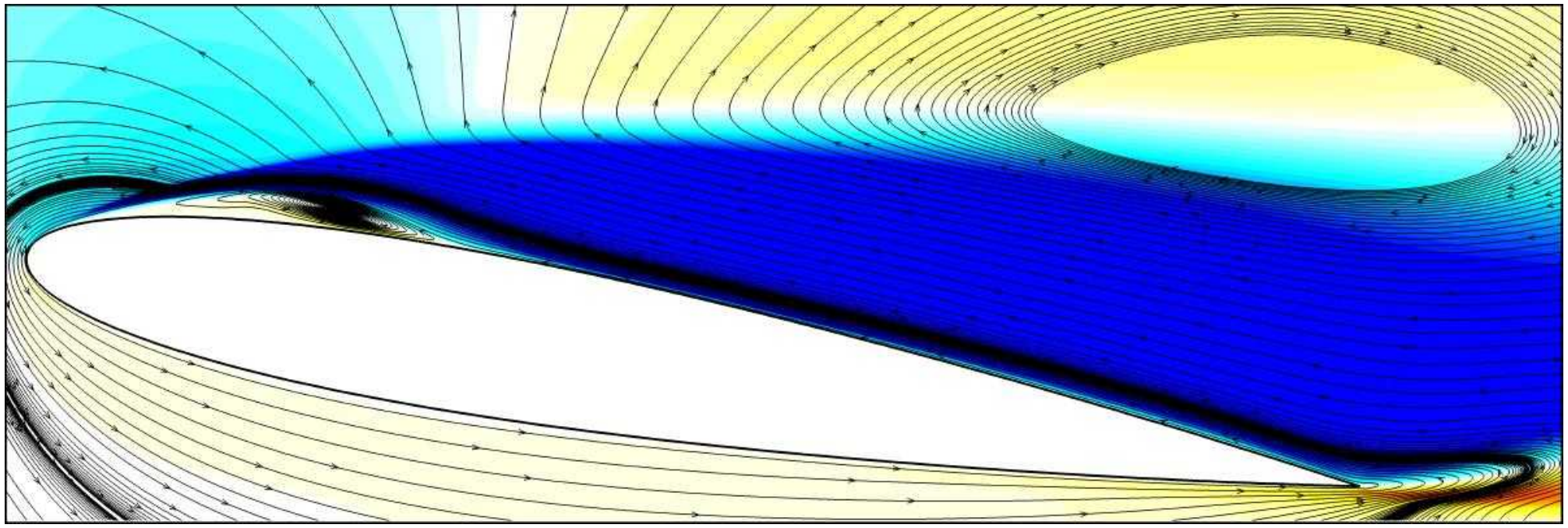}
\textit{time=14.0}
\end{minipage}
\medskip
\begin{minipage}{220pt}
\centering
\includegraphics[width=220pt, trim={0mm 0mm 0mm 0mm}, clip]{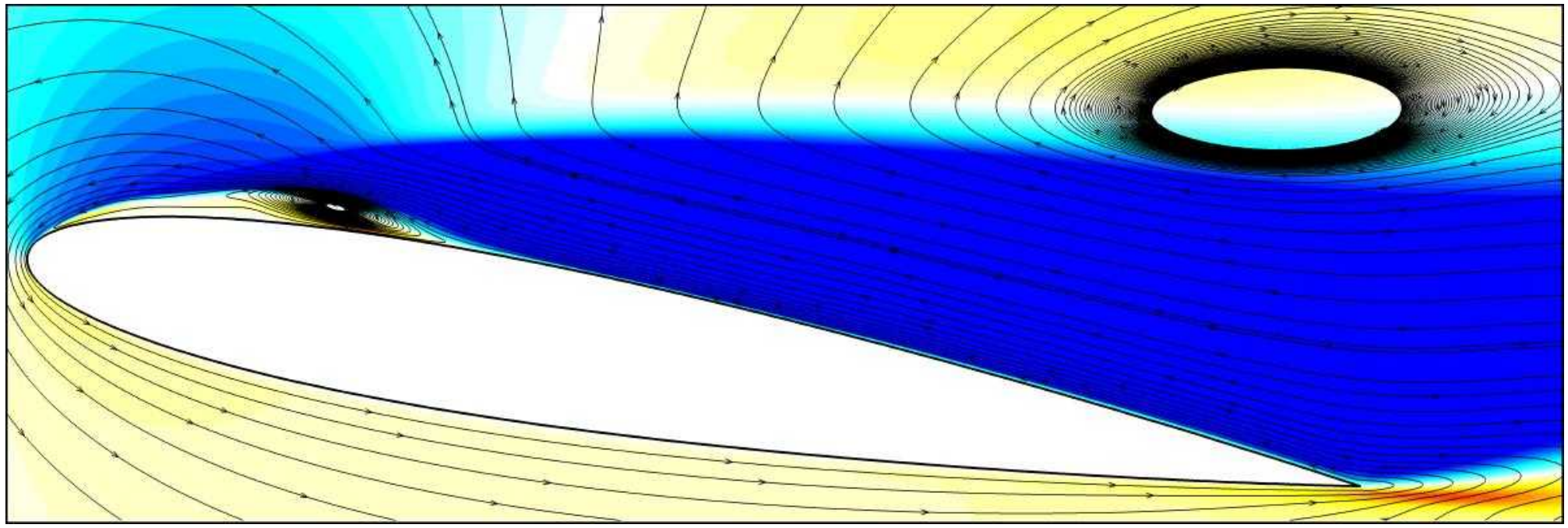}
\textit{time=15.25}
\end{minipage}
\medskip
\begin{minipage}{220pt}
\centering
\includegraphics[width=220pt, trim={0mm 0mm 0mm 0mm}, clip]{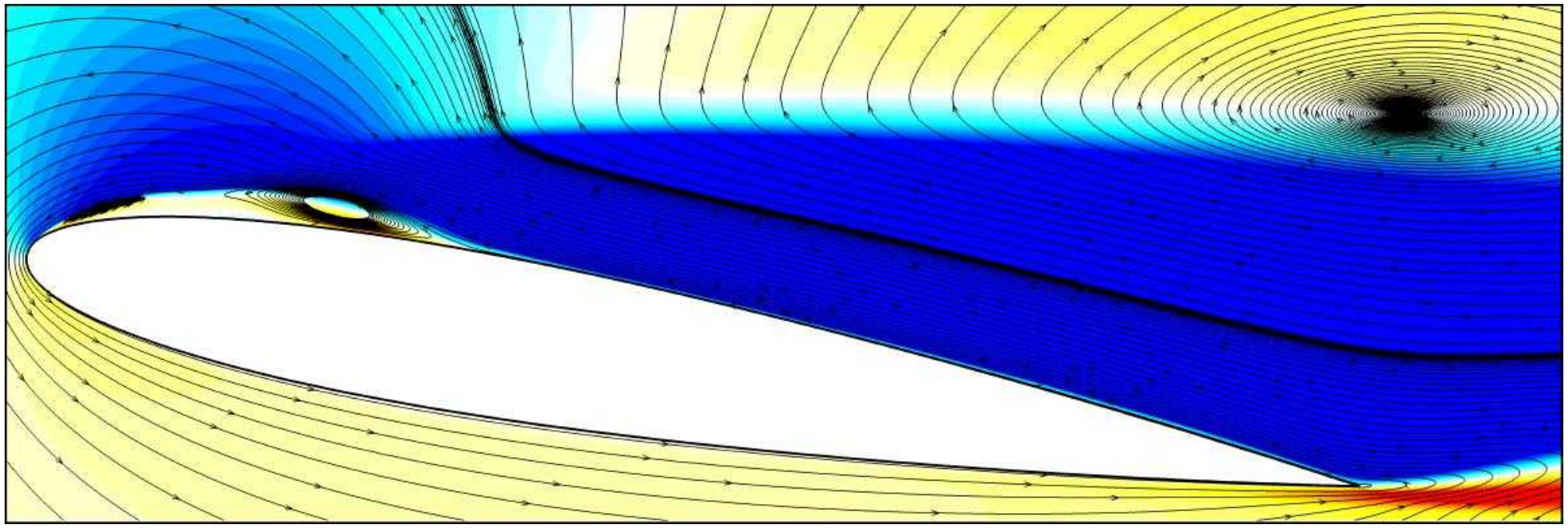}
\textit{time=16.25}
\end{minipage}
\medskip
\begin{minipage}{220pt}
\centering
\includegraphics[width=220pt, trim={0mm 0mm 0mm 0mm}, clip]{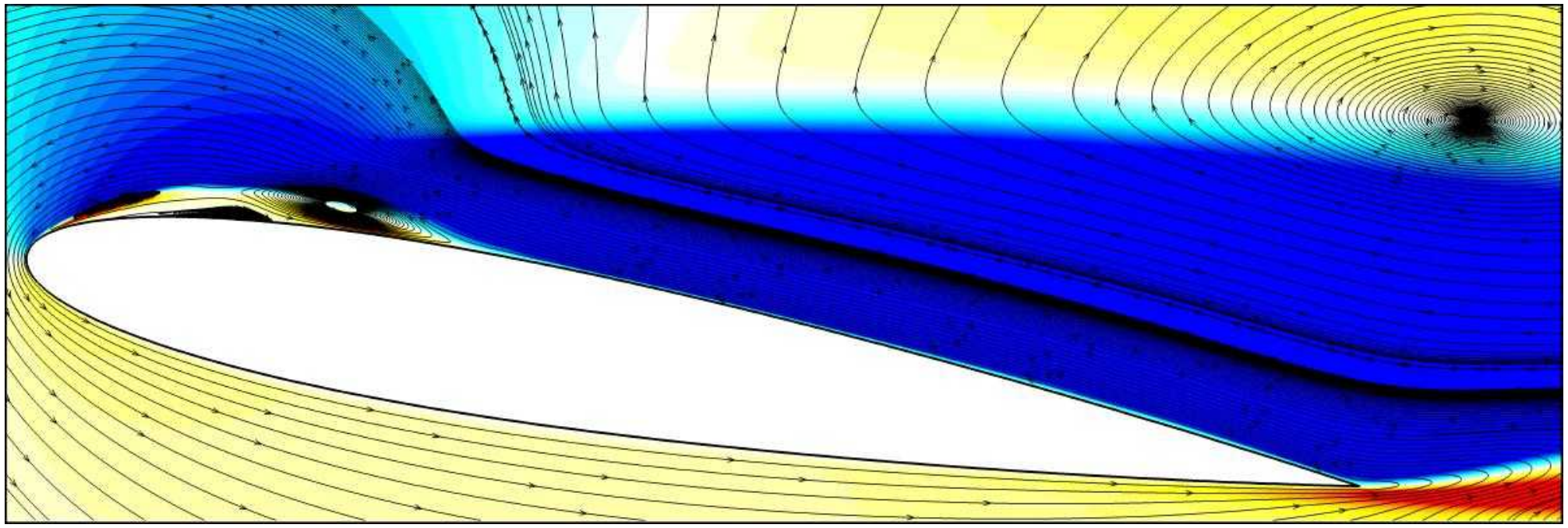}
\textit{time=16.75}
\end{minipage}
\begin{minipage}{220pt}
\centering
\includegraphics[width=220pt, trim={0mm 0mm 0mm 0mm}, clip]{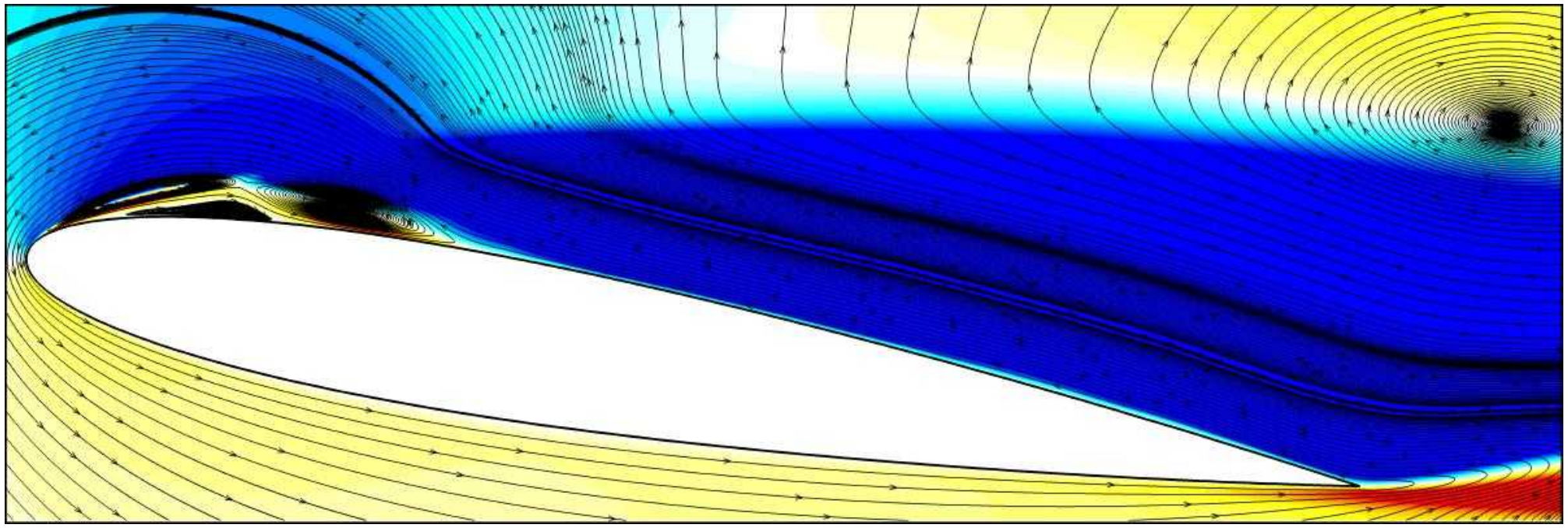}
\textit{time=17.5, $\Phi = 360^{\circ}$}
\end{minipage}
\caption{Streamlines patterns superimposed on colour maps of the streamwise velocity component for the POD reconstruction of the oscillating-f\/low using the LFO mode 1, the LFO mode 2, and the HFO mode for the angle of attack of $9.8^{\circ}$. The f\/low-f\/ield is separating.}
\label{POD_rec_980}
\end{center}
\end{figure}

F\/igures~\ref{POD_LFO1_LFO2_1} and~\ref{POD_LFO1_LFO2_2} show an ``average'' spatial distribution of the LFO mode 1 and 2 without any description of the temporal evolution of these modes. The orthonormal POD spatial modes and the POD coeff\/icients can be used to reconstruct the original f\/low-f\/ield using any subset of the POD modes. The f\/low-f\/ield was reconstructed and probed at selected locations. Comparison of the reconstructed signals probed at different locations shows that the streamwise velocity, the wall-normal velocity, and the pressure have interesting behaviour at the leading and the trailing edges of the aerofoil. F\/igure~\ref{POD_rec_probes} shows comparisons of the original LES data and the reconstructed POD data using the LFO mode 1, the LFO mode 2, and the HFO mode for the streamwise velocity, the wall-normal velocity, and the pressure in the vicinity of the aerofoil leading and trailing edges at the angle of attack of $9.8^{\circ}$. The leading-edge probe was chosen to be in a location inside the upstream vortex of the two co-rotating vortices. The trailing-edge probe was chosen to be located at about $0.4$ chords downstream the trailing-edge. The grey solid line indicates the original LES data; whereas, the black, red, and blue solid lines display the reconstructed data using the LFO mode 1; the LFO mode 1 and the LFO mode 2; and the LFO mode 1, the LFO mode 2, and the HFO mode, respectively.\newline
The left and right-hand sides of the f\/igure show signals of the f\/low variables in the vicinity of the aerofoil leading-edge and downstream the trailing-edge, respectively. The three most dominant POD modes; the LFO mode 1, the LFO mode 2, and the HFO mode; reconstructed the oscillating-f\/low favourably. However, these three dominant f\/low modes contain a maximum of $65\%$ of the energy of the f\/low in all of the investigated angles of attack. Thus, at least $45\%$ of the energy content is not accounted for by these three dominant modes. It is noted that these three dominant f\/low modes represent the oscillating-f\/low and not the ``turbulent-f\/luctuating'' part of the f\/low. Thus, the remaining $45\%$ of energy content is ``turbulent-f\/luctuating''; therefore, it is not ref\/lected by these three modes.\newline
The leading-edge probe shows that the velocity components and the pressure are reconstructed mostly by the LFO mode 1 and the LFO mode 2; thus, the HFO mode does not contribute much to any of the f\/low variables. This is indicative that, the HFO mode does not inf\/luence, directly, the f\/low at this location. Furthermore, the original data of the velocity components are mostly recovered by the LFO mode 2. While the LFO mode 1 has a little and uniform effect, it is the LFO mode 2 that shaped the signal to its original LES form. This is indicative that the velocity components in this location are mostly inf\/luenced by the expansion and advection of the UV of the TCV rather than being inf\/luenced by the LFO mode 1. The pressure f\/ield is mostly recovered by the LFO mode 1 in this location with very little and limited effect of the LFO mode 2 and the HFO mode.
The trailing-edge probes show that the f\/low variables are overwhelmed by the HFO mode and its subharmonic. However, the low-frequency pattern exists in all of the f\/low variables at this location. Thus, the velocity components are inf\/luenced exclusively by the LFO mode 1 and the HFO mode. Whereas, the pressure at this location is inf\/luenced by the LFO mode 1 and the HFO mode in addition to a limited effect of the LFO mode 2.\newline
As the original oscillating-f\/low is reconstructed using the three most dominant POD modes, the aerodynamic forces could be reconstructed as well. The pressure coeff\/icient was estimated utilizing the pressure f\/ield reconstructed using the LFO mode 1, the LFO mode 2, and the HFO mode. The obtained pressure coeff\/icient was then duly integrated around the aerofoil to obtain the reconstructed lift and pressure-drag coeff\/icients. The reconstructed velocity components were also used to estimate the reconstructed skin-friction coeff\/icient and the reconstructed reattachment location of the shear layer.\newline
F\/igure~\ref{POD_rec_Cp}a shows an ideal sinusoidal cycle of the LFO mode 1 and 2. The ideal cycle for the LFO mode 1 and the LFO mode 2 mimics the cycle of the lift and the drag coeff\/icients, respectively. The negative sign of the percentage of the two modes indicates that the f\/low switches its direction rather than the percentage being negative. That is, when the percentage is positive, the oscillating f\/low is rotating in the clockwise direction and the f\/luctuating lift coeff\/icient is positive, and vice versa. The ideal cycle has a minimum of $(-)$ $75\%$ of the LFO mode 1 at the phase angle of $0^{\circ}$. The percentage of the LFO mode 1 then decreases until it vanishes at the phase angle of $90^{\circ}$. After that, the percentage of the LFO mode 1 becomes positive and increases until it peaks at $75\%$ at the phase angle of $180^{\circ}$. The percentage decreases, again, until it becomes zero at the phase angle of $270^{\circ}$. The LFO mode 1 and the LFO mode 2 are out of phase by $\pi/2$, and the LFO mode 2 peaks at a maximum percentage of one-third that of the LFO mode 1.\newline
Reconstruction of the oscillating-pressure-coeff\/icient using the LFO mode 1, the LFO mode 2, and the HFO mode at the phase angles of $\Phi=0^{\circ}$, $90^{\circ}$, $180^{\circ}$, and $270^{\circ}$ at the angle of attack of $9.8^{\circ}$ are shown in f\/igure~\ref{POD_rec_Cp}. The LFO mode 1 and the LFO mode 2 vanishes at certain phase angles as seen in f\/igure~\ref{POD_rec_Cp}a. Thus, the reconstructed oscillating-pressure-coeff\/icient using the LFO mode 1; and the LFO mode 2 is zero at the phase angles of $\Phi=90^{\circ}$, $270^{\circ}$; and $0^{\circ}$, $180^{\circ}$, respectively. Generally speaking, the lift and the drag coeff\/icients oscillates in accordance with the LFO mode 1 and the LFO mode 2, respectively. Furthermore, when the oscillating-pressure-coeff\/icient is positive, it decreases the oscillating-lift-coeff\/icient and vice versa. The dramatic changes in the reconstructed oscillating-pressure-coeff\/icient take place on the suction surface of the aerofoil, as seen in the f\/igure. It is noted that the reconstructed oscillating-pressure-coeff\/icient using the LFO mode 1 and the LFO mode 2 mostly inf\/luences the lift coeff\/icient and the drag coeff\/icient, respectively. The reconstructed oscillating-pressure-coeff\/icient using the HFO mode preserves its shape and switches its direction every $\pi/2$. The effect of the HFO mode on both the oscillating-lift and the oscillating-drag is not signif\/icant.\newline
F\/igure~\ref{POD_rec_cl_cd} compares the lift, the pressure-drag, the drag, and the location of the reattachment of the shear layer reconstructed using the LFO mode 1 and the LFO mode 2, to these estimated from the LES data at the angle of attack of $9.8^{\circ}$. The LFO mode 1 and 2 reconstruct the aerodynamic forces favourably as seen in the f\/igure. The location of the reattachment of the shear layer oscillates up and downstream that of the mean bubble length. The discontinuities in the reattachment location are due to the fact that the f\/low becomes fully separated and it does not reattach at all. The location of the reattachment was reconstructed using the LFO mode 1 only, and cannot be compared to that estimated from the LES data. As mentioned before, the LFO cycle is very disturbed at this low Reynolds number. Therefore, the instantaneous skin-friction f\/luctuates several times above and below zero around the point at which it crosses the $x$-axis. Thus, estimating the reattachment location of the shear layer, instantaneously, is not feasible.\newline
The phase-averaged oscillating-f\/low presented in~\citet{eljack2018bursting2} was limited to $36$ phase angles. The POD eigenvalues and eigenfunctions were used to reconstruct the f\/low-f\/ield. The reconstructed f\/low-f\/ield resembles that of the original f\/low-f\/ield. As the POD objectively recovers the most dominant f\/low features, the LFO is captured using the two low-frequency modes and the HFO mode. Although these three POD modes account for about $60\%$ of the total energy in the oscillating-f\/low, they represent the original oscillating-f\/low favourably as seen in f\/igure~\ref{POD_rec_probes}. The rest of the POD modes are at high-frequency; thus, they do not contribute to the oscillating-f\/low and the LFO phenomenon but rather contribute to the ``turbulent-f\/luctuations'' part of the f\/low-f\/ield.\newline
F\/igures~\ref{POD_rec_TCV_970} and~\ref{POD_rec_970} shows the reconstructed oscillating-f\/low using the LFO mode 1, the LFO mode 2, and the HFO mode at the angle of attack of $9.7^{\circ}$. The starting point in time for the reconstructed f\/low at f\/low time $=8.5$ is shown in f\/igure~\ref{POD_zeta}, $\alpha = 9.7^{\circ}$, and indicated by the black circle and $\Phi = 0^{\circ}$. The f\/low is fully separated at zero phase angle, the oscillating-f\/low direction is anti-clockwise, and the triad of vortices are present and in their most coherent state. At the f\/low time of $8.75$ the upstream vortex of the triad of vortices pops up above the downstream vortex of the triad of vortices. At the f\/low time of $9.0$ and $9.25$ the upstream vortex pops a little more above the UV of the TCV. At the f\/low time of $9.5$, the UV starts to slide above the DV of the TCV. The downstream vortex starts to merge with the upstream vortex and energies it at the f\/low time of $9.75$. While the downstream vortex merges with UV, the latter continue to expand until the two vortices form one vortex at the f\/low time of $10.25$. At the f\/low time of $11.0$ and the phase angle of $90^{\circ}$ the recently formed vortex expands abruptly. It is interesting to note that the LFO mode 2 peaks at this instant in time as shown in f\/igure~\ref{POD_zeta}, $\alpha = 9.7^{\circ}$. After that, the recently formed vortex continue to expand and starts to change the direction of the oscillating-f\/low. The secondary vortex seems to be present and intact during this process. At the f\/low time of $13.5$ the secondary vortex vanishes and a new downstream vortex forms. As the oscillating-f\/low continue to change its direction of rotation, a new upstream vortex starts to form at the f\/low time of $14.0$. It is interesting to note how the vorticity changes sign at this instant in time, and how it creates the new UV of the TCV which is rotating in the clockwise direction. The newly formed upstream vortex grows in size and strength at the f\/low time of $14.5$ and $14.75$. The f\/low is fully attached and the LFO mode 1 at its maximum amplitude at the f\/low time of $15.0$ and the phase angle of $180^{\circ}$. The oscillating-f\/low direction is clockwise, and the triad of vortices are present and in their most coherent state. The reconstructed oscillating-f\/low using the LFO mode 1, the LFO mode 2, and the HFO mode is added to the mean f\/low to obtain the reconstructed instantaneous f\/low-f\/ield. F\/igure~\ref{POD_rec_instant_970} shows the reconstructed instantaneous f\/low-f\/ield as it attaches at the angle of attack of $9.7^{\circ}$.\newline
Reconstruction of the f\/low-f\/ield at the angle of attack of $9.7^{\circ}$ provided detailed description of how the f\/low attaches. When the cycle of the LFO proceeds to phase angles $\Phi>180^{\circ}$, the process reverses its direction and the fully attached f\/low starts to separate. Thus, a similar sequence of events takes place, as shown in f\/igures~\ref{POD_rec_TCV_980} and~\ref{POD_rec_980}. The f\/low-f\/ield was reconstructed at the angle of attack of $9.8^{\circ}$. The starting point in time for the reconstructed f\/low is shown in the plot of the lift coeff\/icient in f\/igure~\ref{POD_rec_cl_cd}a and indicated by the black circle and $\Phi = 180^{\circ}$. The f\/low is fully attached and the LFO mode 1 is at its maximum amplitude. The triad of vortices is present and at their most coherent state. The f\/low evolves in a similar fashion to that of the attaching f\/low discussed in the previous paragraph until the abrupt expansion of the vortex at the f\/low time of $9.75$ and the phase angle of $270^{\circ}$. After that, the oscillating-f\/low changes its direction of rotation from the clockwise to the anti-clockwise direction. The triad of vortices continues to deform as the oscillating-f\/low continues to switch its direction globally until the triad of vortices arrives at their most coherent state at the f\/low time of $17.5$ and the phase angle of $360^{\circ}$. F\/inally, the oscillating-f\/low direction is anti-clockwise, and the f\/low is fully separated. Consequently, the lift coeff\/icient drops to its minimum value as seen in f\/igure~\ref{POD_rec_cl_cd}a. The reconstructed instantaneous f\/low-f\/ield at the angle of attack of $9.8^{\circ}$ is shown in f\/igure~\ref{POD_rec_instant_980}. The evolution of the oscillating-f\/low is visualised in a step by step snapshots that reveals how the f\/low attaches and separates. Thus, a detailed description of the LFO phenomenon is given and the underlying mechanism is discussed in detail.

\subsection{The phase lag between the lift, the drag, and the reattachment location}
\begin{figure}
\begin{center}
\begin{minipage}{220pt}
\centering
\includegraphics[width=220pt, trim={0mm 0mm 0mm 0mm}, clip]{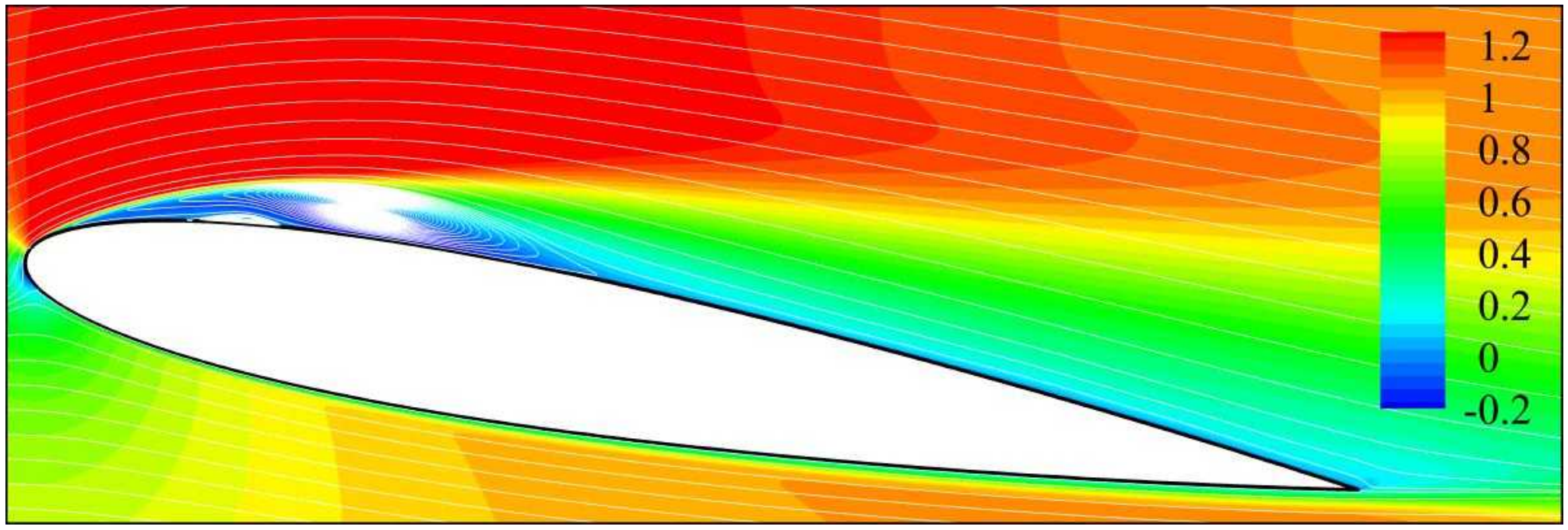}
\textit{time=6.0, $\Phi = 180^{\circ}$}
\end{minipage}
\medskip
\begin{minipage}{220pt}
\centering
\includegraphics[width=220pt, trim={0mm 0mm 0mm 0mm}, clip]{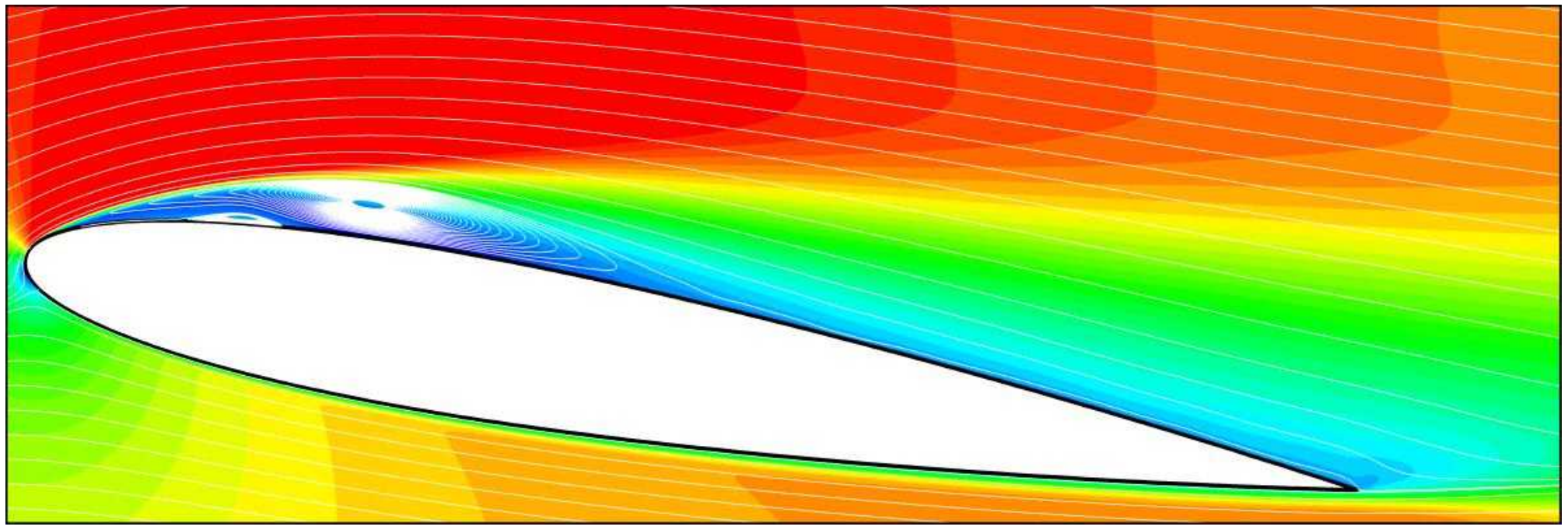}
\textit{time=8.0}
\end{minipage}
\medskip
\begin{minipage}{220pt}
\centering
\includegraphics[width=220pt, trim={0mm 0mm 0mm 0mm}, clip]{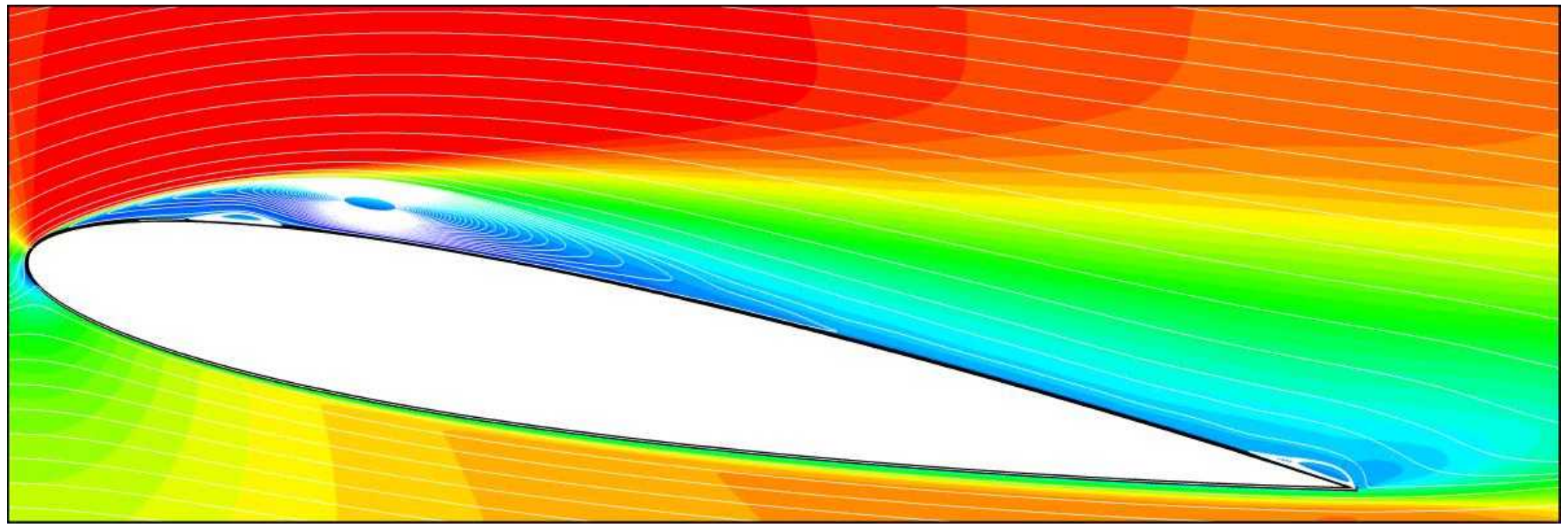}
\textit{time=9.75, $\Phi = 270^{\circ}$}
\end{minipage}
\medskip
\begin{minipage}{220pt}
\centering
\includegraphics[width=220pt, trim={0mm 0mm 0mm 0mm}, clip]{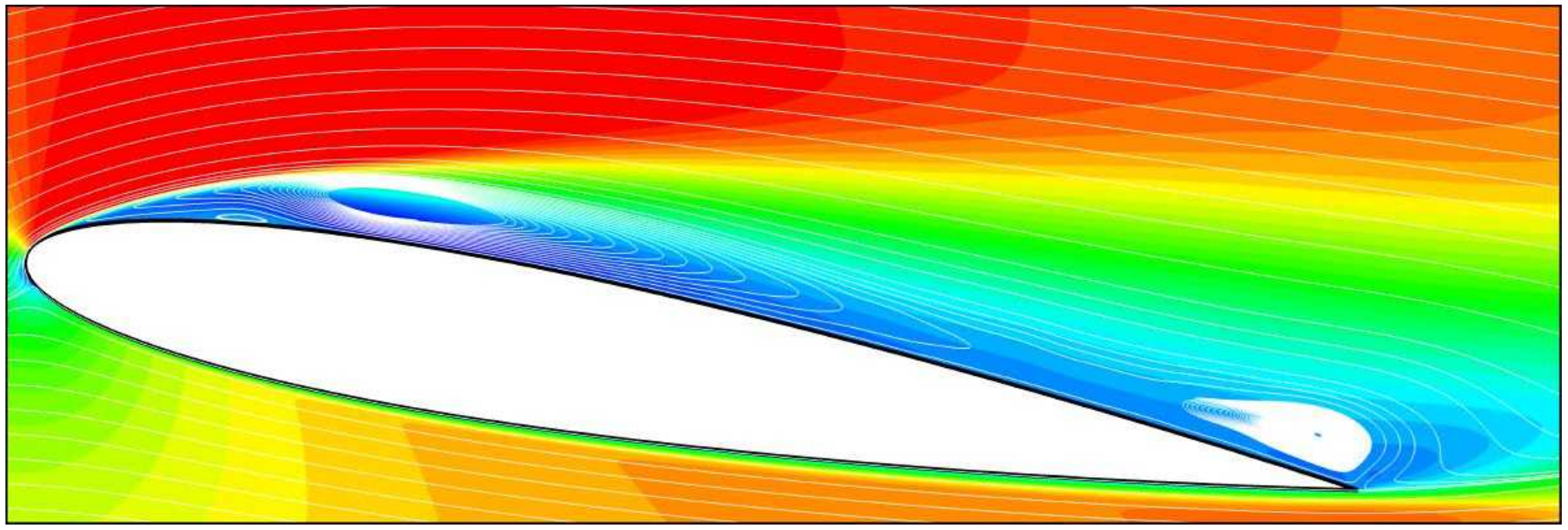}
\textit{time=10.75}
\end{minipage}
\medskip
\begin{minipage}{220pt}
\centering
\includegraphics[width=220pt, trim={0mm 0mm 0mm 0mm}, clip]{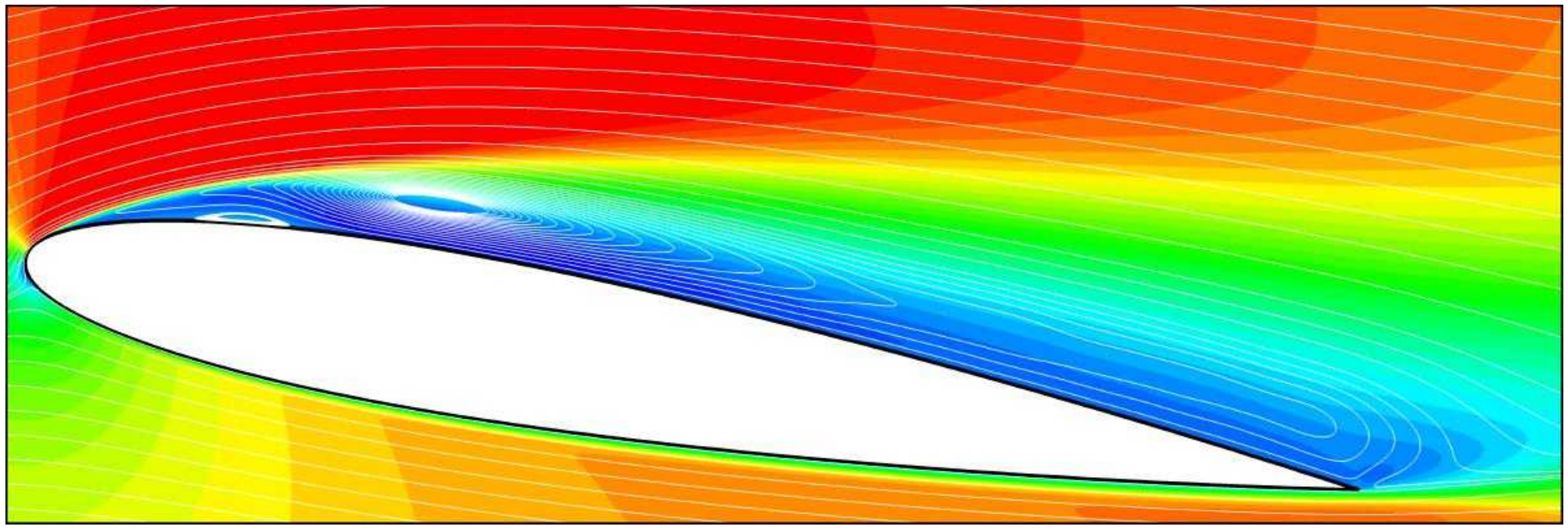}
\textit{time=11.75}
\end{minipage}
\medskip
\begin{minipage}{220pt}
\centering
\includegraphics[width=220pt, trim={0mm 0mm 0mm 0mm}, clip]{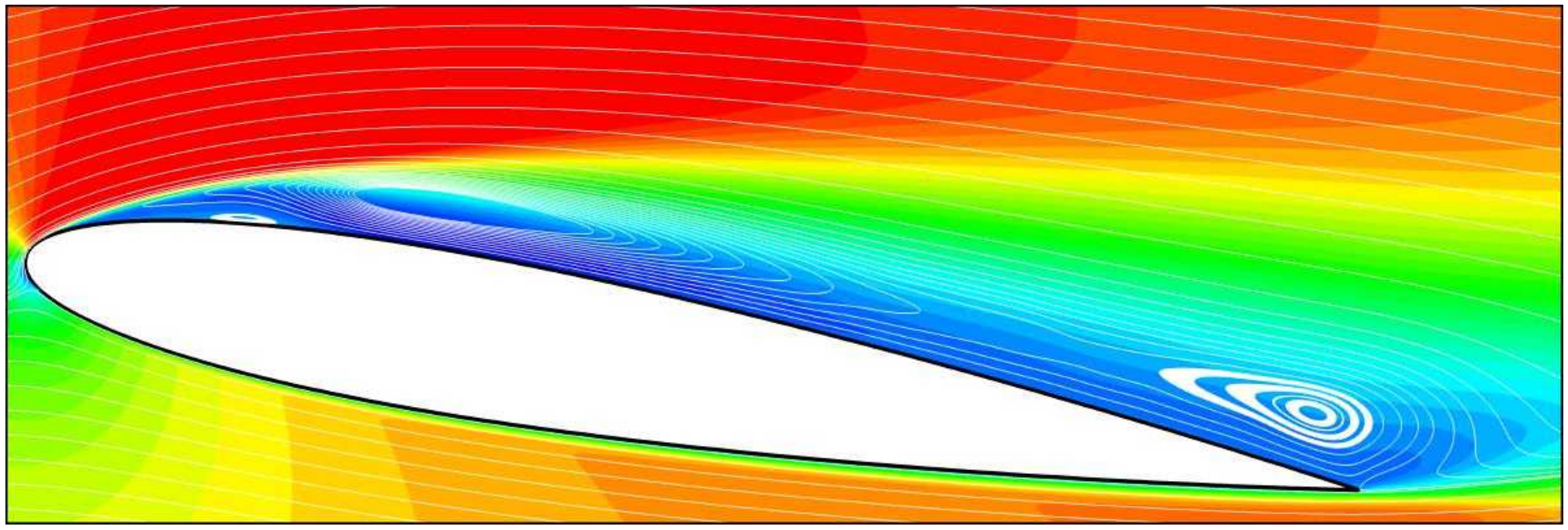}
\textit{time=12.0}
\end{minipage}
\medskip
\begin{minipage}{220pt}
\centering
\includegraphics[width=220pt, trim={0mm 0mm 0mm 0mm}, clip]{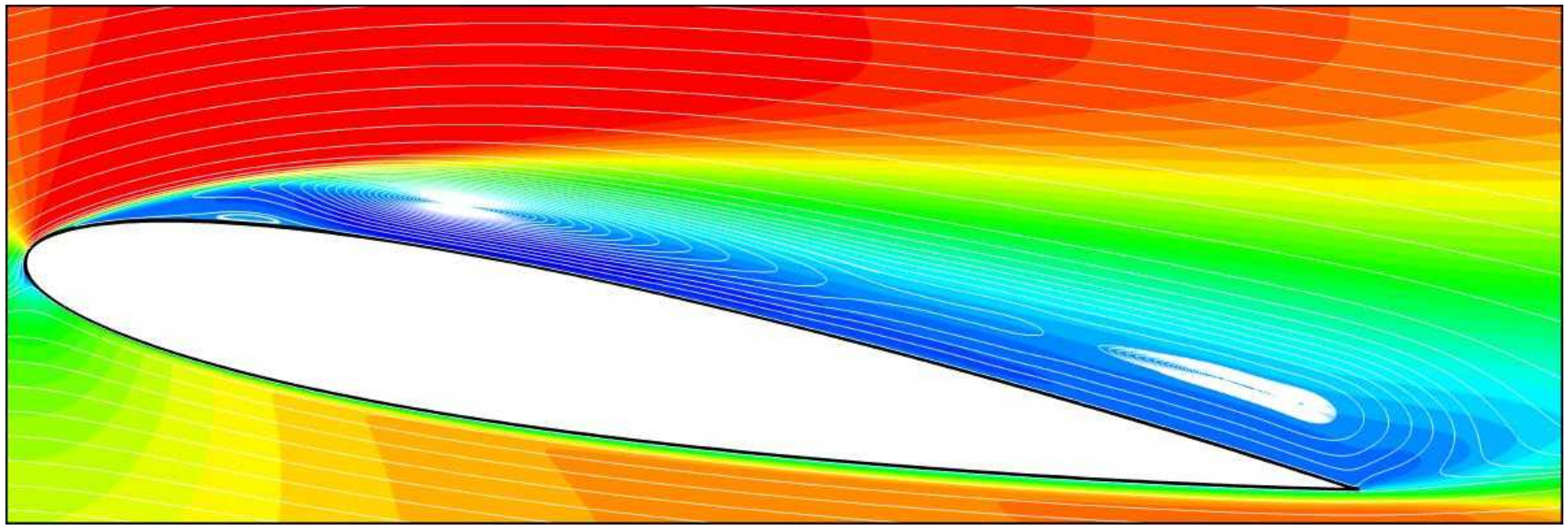}
\textit{time=12.5}
\end{minipage}
\medskip
\begin{minipage}{220pt}
\centering
\includegraphics[width=220pt, trim={0mm 0mm 0mm 0mm}, clip]{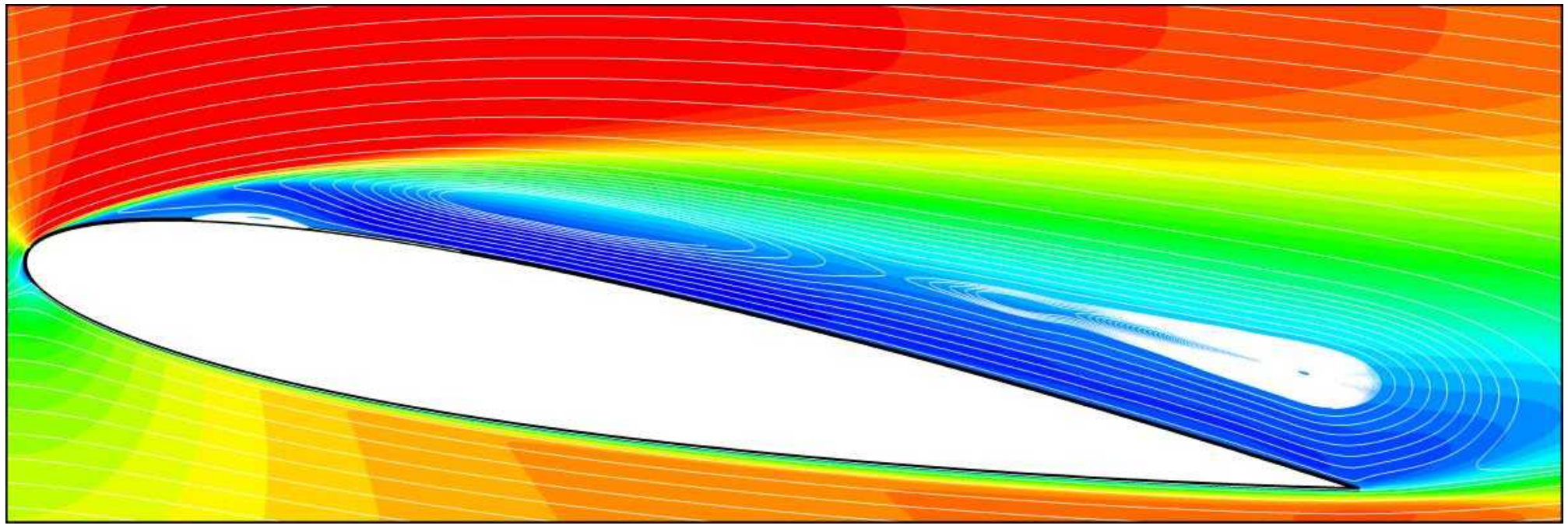}
\textit{time=14.0}
\end{minipage}
\medskip
\begin{minipage}{220pt}
\centering
\includegraphics[width=220pt, trim={0mm 0mm 0mm 0mm}, clip]{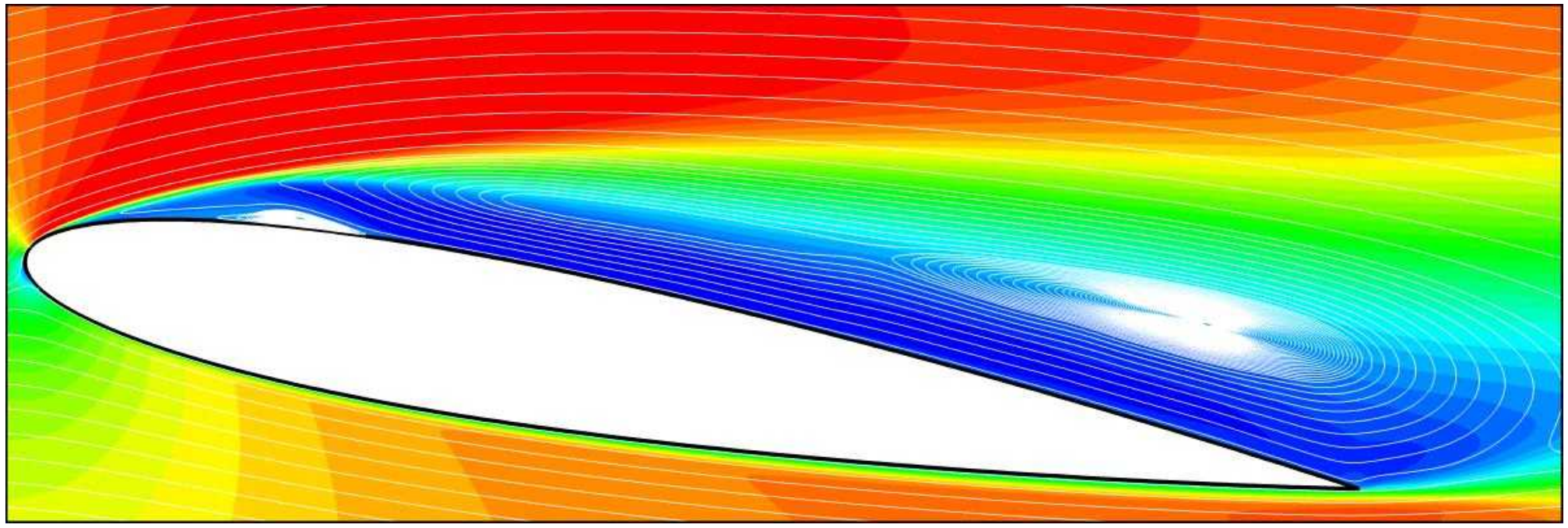}
\textit{time=15.25}
\end{minipage}
\medskip
\begin{minipage}{220pt}
\centering
\includegraphics[width=220pt, trim={0mm 0mm 0mm 0mm}, clip]{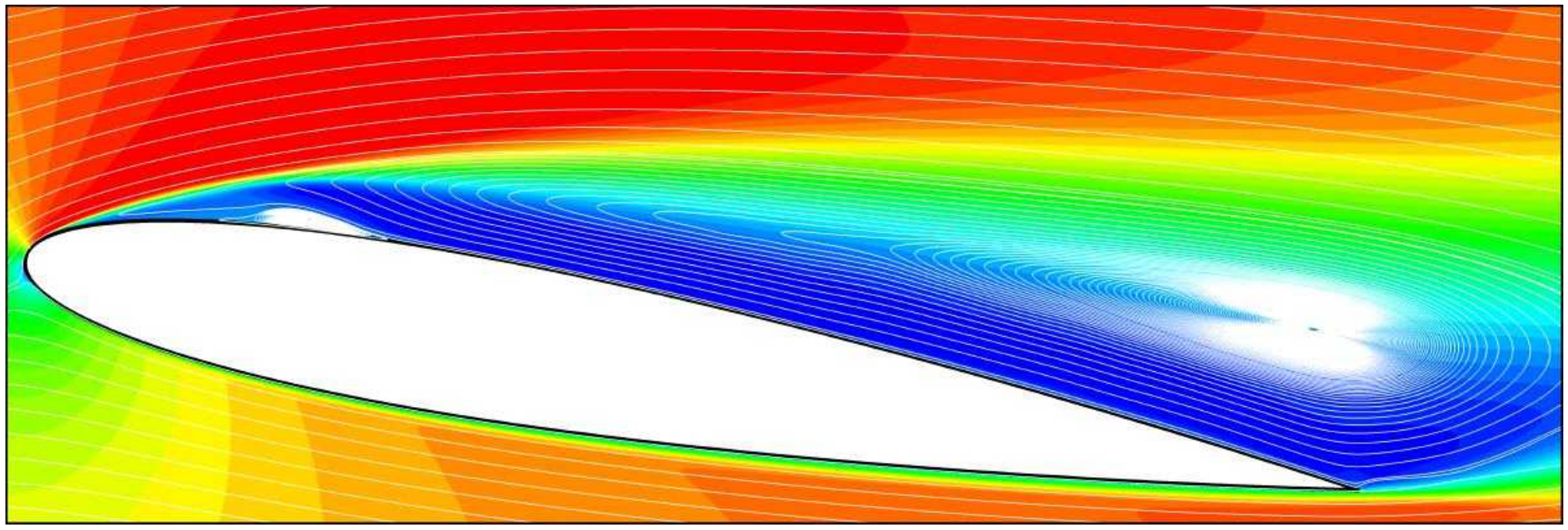}
\textit{time=16.25}
\end{minipage}
\medskip
\begin{minipage}{220pt}
\centering
\includegraphics[width=220pt, trim={0mm 0mm 0mm 0mm}, clip]{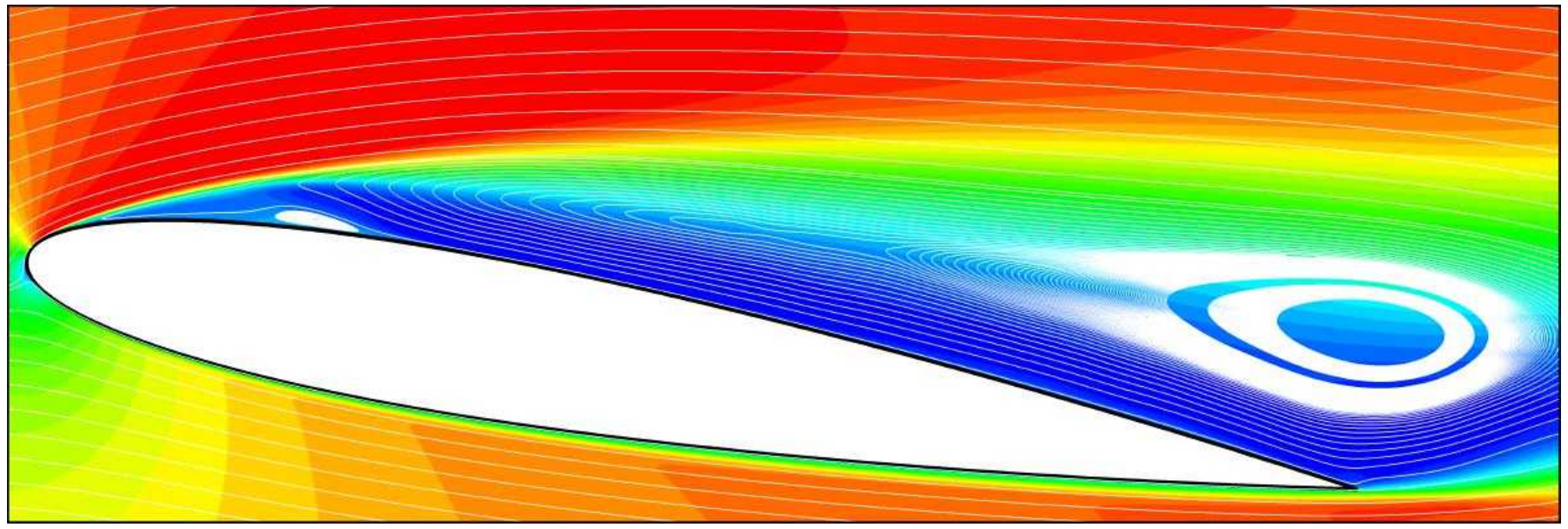}
\textit{time=16.75}
\end{minipage}
\begin{minipage}{220pt}
\centering
\includegraphics[width=220pt, trim={0mm 0mm 0mm 0mm}, clip]{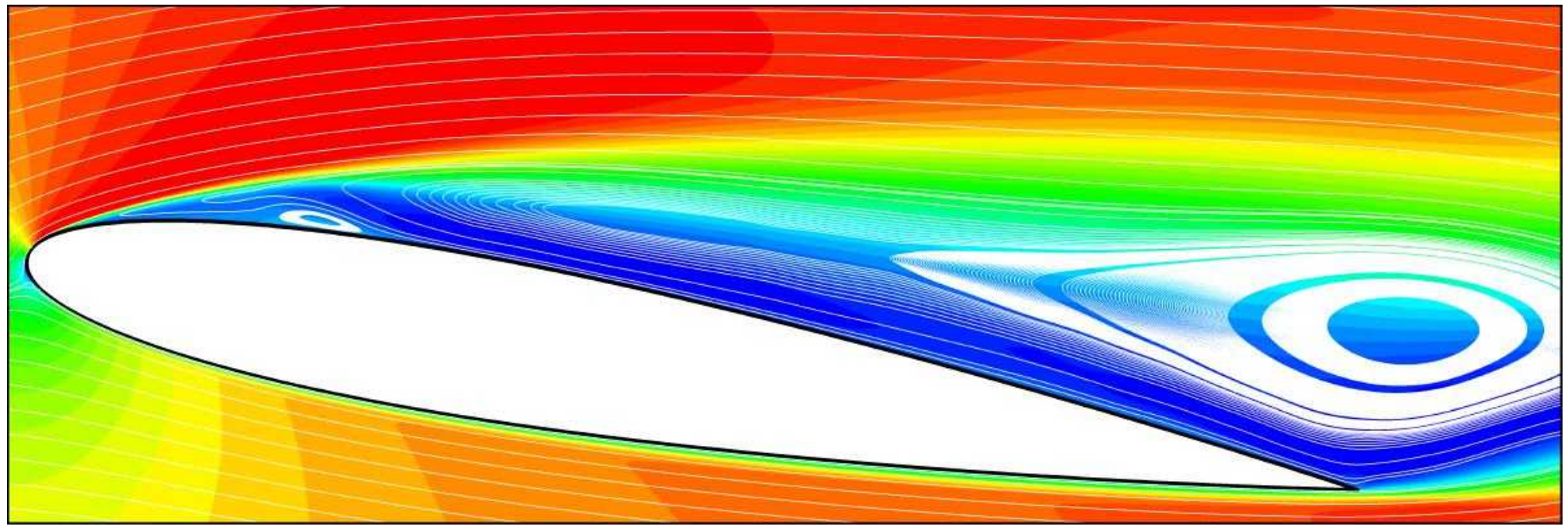}
\textit{time=17.5, $\Phi = 360^{\circ}$}
\end{minipage}
\caption{Streamlines patterns superimposed on colour maps of the streamwise velocity component for the POD reconstruction of the instantaneous f\/low-f\/ield using the LFO mode 1, the LFO mode 2, and the HFO mode for the angle of attack of $9.8^{\circ}$. The f\/low-f\/ield is separating.}
\label{POD_rec_instant_980}
\end{center}
\end{figure}

Time histories of the aerodynamic coeff\/icients provided by~\citet{almutairi2015large} and~\citet{eljack2017high} show that the lift coeff\/icient leads the drag coeff\/icient by a phase of $\pi/2$. The LFO mode 1 features the process that creates and sustains the triad of vortices and adds energy to the upstream vortex of the two counter-rotating vortices. The LFO mode 2 on the suction surface of the aerofoil features the expansion and advection of the UV of the TCV as mentioned before. The life cycle of the triad of vortices is perfectly synchronised with the two low-frequency modes as discussed in $\S$~\ref{sec:POD_reconstruction}. The lift coeff\/icient follows the dynamics of the LFO mode 1 while the drag coeff\/icient follows the LFO mode 2. F\/igure~\ref{POD_zeta} show that the LFO mode 1 leads the LFO mode 2 by a phase of $\pi/2$. Furthermore, f\/igures~\ref{POD_rec_Cp} and~\ref{POD_rec_cl_cd} show that the oscillation of the pressure coeff\/icient reconstructed using LFO mode 1 contribute mostly to the lift coeff\/icient. Whereas, the reconstructed oscillating-pressure-coeff\/icient using the LFO mode 2 contributes mostly to oscillations in the drag coeff\/icient. Thus, the lift coeff\/icient leads the drag coeff\/icient by a phase angle of $\pi/2$.\newline
F\/igure~\ref{POD_rec_cl_cd} shows that the reattachment location of the shear layer oscillates along the aerofoil chord up and downstream the location of the mean bubble size. The oscillation mimics that of the LFO mode 1 and the lift coeff\/icient with a phase difference of $\pi$. However, the location of the reattachment of the shear layer exhibits discontinuities as a result of the f\/low being fully separated and the reattachment location on the aerofoil surface is not def\/ined. F\/igure~\ref{POD_rec_instant_980} shows that at the phase angle of $\Phi=180^{\circ}$, the f\/low is fully attached. Thus, the reattachment location is at its closest point to the leading-edge and has a minimum on the plot; whereas, the lift coeff\/icient has a maximum when the f\/low is fully attached. At the phase angle of $\Phi=90^{\circ}$ and $\Phi=270^{\circ}$, the f\/low is partially separated. Thus, the lift coeff\/icient is at its mean value and the reattachment location takes place around the mean bubble length. At the phase angles $\Phi < 90^{\circ}$ and $\Phi > 270^{\circ}$, the reattachment location moves downstream the mean bubble length until it becomes undef\/ined as the phase angle approaches $0^{\circ}$ or $360^{\circ}$, as seen in f\/igure~\ref{POD_rec_instant_980}.\newline
~\citet{ansell2015characterization} carried out wind-tunnel measurements to characterize the LFO present in the f\/low-f\/ield around a NACA-0012 aerofoil with a horn-ice shape. The authors integrated the pressure coeff\/icient around the aerofoil at mid-span location to obtain the lift coeff\/icient. They correlated the lift coeff\/icient and the location of the shear layer reattachment. They found that the elongation of the LSB takes place at the maximum lift and the shrinkage of LSB presents when the lift coeff\/icient decreases. The authors reported that at the low-frequency mode, the lift coeff\/icient leads the reattachment location of the shear layer by a phase of $\pi/2$. The discrepancy is due to the fact that~\citet{ansell2015characterization} investigated the LFO around an aerofoil with an iced leading-edge, and the LFO phenomenon could be governed by a different underlying mechanism.

\subsection{The dynamics of the f\/low}
The f\/low-f\/ield around the NACA-0012 aerofoil exhibits an instability at the trailing-edge that drives alternating vortices at a high frequency, the HFO mode. The stagnation point in the vicinity of the leading-edge pitches up and down in accordance with the dynamics of the alternating vortices. Thus, a global f\/low oscillation around the aerofoil initiates at low-frequencies, the LFO mode 1 and the LFO mode 2. Extensive investigation of the various f\/low modes and their interaction among each other revealed that the LFO mode 1, the LFO mode 2, and the HFO mode are the most important and dominant f\/low modes that govern the dynamics of the f\/low at the onset of stall. The LFO mode 1 of the streamwise velocity and the wall-normal velocity is the globally oscillating-f\/low around the aerofoil. The LFO mode 1 of the pressure f\/ield is the oscillating-pressure along the aerofoil chord. Thus, the LFO mode 1 features the process that creates and sustains the triad of vortices and adds energy to the upstream vortex (UV) of the two counter-rotating vortices (TCV). The LFO mode 2 of the wall-normal velocity features the ejection of the UV of the TCV. The HFO mode of the wall-normal velocity is the oscillating-wall-normal-velocity along the aerofoil wake. At relatively low angles of attack, the amplitude of the LFO mode 1 and 2 is insignif\/icant; however, it increases as the angle of attack increases. The LFO mode 1, the LFO mode 2, and the HFO mode are interlinked, and mutually strengthen each other until the LFO mode 1 and 2 overtake the HFO mode at the inception of a stall. Consequently, the aerodynamic coeff\/icients start to oscillate at low-frequency. The LFO mode 1 dominates the streamwise velocity, the wall-normal velocity, and the pressure f\/ield in the angles of attack of $9.7^{\circ} \leq \alpha \leq 10.0^{\circ}$. Thus, the LFO is fully developed in this range of angles of attack. The LFO loose momentum as the angle of attack is further increased, and the HFO mode dominates the f\/low-f\/ield.

\section*{Conclusions}
The objective of the present work was to carry out a detailed f\/low dissection and shed some light on the dynamics of the f\/low-f\/ield around a NACA-0012 aerofoil at a Reynolds number of $5\times10^4$, Mach number of $0.4$, and at various angles of attack around the onset of stall. The Dynamic Mode Decomposition (DMD) and the Proper Orthogonal Decomposition (POD) methods were applied to data sets sampled on the $x$--$y$ plane including the velocity components, the pressure, and the aerodynamic coeff\/icients. The data sets span four low-frequency cycles and were locally-time-averaged every $50$ time-steps and ensemble-averaged in the spanwise direction on the f\/ly before they were recorded. Three distinct dominant f\/low modes are identif\/ied by the DMD and the POD:
\begin{enumerate}
  \item A globally oscillating f\/low mode at a low-frequency (LFO mode 1). The LFO mode 1 of the streamwise and the wall-normal velocity components is the globally oscillating-f\/low around the aerofoil. The LFO mode 1 of the pressure f\/ield is the oscillating-pressure along the aerofoil chord. Thus, the LFO mode 1 features the process that creates and sustains the triad of vortices and adds energy to the upstream vortex (UV) of the two counter-rotating vortices (TCV).
  \item A locally oscillating f\/low mode on the suction surface of the aerofoil at a low-frequency (LFO mode 2) presenting the ejection of the UV of the TCV.
  \item A locally oscillating f\/low mode along the wake of the aerofoil at a high-frequency (HFO mode) highlighting the oscillating-wall-normal-velocity along the aerofoil wake.
\end{enumerate}
The life-cycle of the triad of vortices is perfectly synchronised with the LFO mode 1 and 2. Time histories of the lift and the drag coeff\/icients mimic the temporal evolution of the LFO mode 1 and 2, respectively. The POD temporal modes show that the LFO mode 1 leads the LFO mode 2 by a phase of $\pi/2$. This explains the previously reported observations that the lift coeff\/icient leads the drag coeff\/icient by a phase of $\pi/2$. The reattachment location of the shear layer oscillates along the aerofoil chord in harmony with the lift coeff\/icient and the LFO mode 1 with a phase difference of $\pi$.\newline
The wall-normal velocity component drives the HFO mode and plays a profound role in the dynamics of the f\/low. The HFO mode exists at all of the investigated angles of attack and causes a global oscillation in the f\/low-f\/ield. The global f\/low oscillation around the aerofoil interacts with the laminar portion of the separated shear layer in the vicinity of the leading-edge and triggers an inviscid absolute instability that creates and sustains the TCV. When the UV of the TCV expands, it advects downstream and energies the HFO mode. The LFO mode 1, the LFO mode 2, and the HFO mode mutually strengthen each other until the LFO mode 1 and 2 overtake the HFO mode at the onset of stall. Consequently, the aerodynamic coeff\/icients start to oscillate at a low-frequency. At the angles of attack $9.25^{\circ} \leq \alpha \leq 9.6^{\circ}$, the LFO mode 2 is dominating the f\/low variables, and the low-frequency f\/low oscillation (LFO) exhibits a transition regime. At the angles of attack $9.7^{\circ} \leq \alpha \leq 10.0^{\circ}$, the f\/low variables are dominated by the LFO mode 1, and the LFO phenomenon is fully developed. At higher angles of attack, the HFO mode overtakes the LFO mode 2, again, and the aerofoil undergoes a full stall.\newline
F\/inally, the root causes that trigger the instability of the laminar separation bubble are determined and the mysterious underlying mechanism of its associated LFO phenomenon is revealed. Furthermore, the dynamics of the f\/low is discussed in detail. This opens the door for optimum design of aerofoils and other aerodynamic shapes, and smart control of the f\/low-f\/ield around them.

\section*{Acknowledgements}
All computations were performed on \verb"Aziz Supercomputer" at King Abdulaziz university's High Performance Computing Center (\url{http://hpc.kau.edu.sa/}). The authors would like to acknowledge the computer time and technical support provided by the center.

\bibliographystyle{jfm}
\bibliography{lfo3}

\newpage
\newpage
\newpage
\newpage
\newpage
\newpage
\newpage
\newpage
\newpage
\newpage
\newpage
\newpage
\newpage
\newpage
\newpage
\newpage
\newpage
\newpage

\end{document}